\documentclass[12pt]{ociamthesis}  

\usepackage{amssymb}
\usepackage{graphicx}
\usepackage{amsmath}
\usepackage[version=3]{mhchem}
\usepackage{slashed}
\usepackage{dcolumn}
\usepackage{caption}
\usepackage{subcaption}
\usepackage{verbatim}
\usepackage{amsfonts}
\usepackage{bm}
\usepackage{lscape}
\usepackage{multirow}
\usepackage{array}
\usepackage{bigints, relsize}
\usepackage{mathtools}
\usepackage{dsfont}
\usepackage{bbm}
\usepackage[utf8]{inputenc}
\usepackage[english]{babel}
\usepackage{relsize}

\usepackage[numbers,sort&compress]{natbib}
\usepackage{epigraph}
\usepackage[helvetica]{quotchap}  

\usepackage{hyperref}
\usepackage{color}
\hypersetup{
    colorlinks=true, 
    linktoc=all,     
    linktocpage = true, 
    linkcolor=blue,  
    citecolor=blue,
    urlcolor=blue
}

\usepackage{natbib}
\usepackage{amsthm}
\newtheorem{theorem}{Theorem}

\theoremstyle{definition}
\newtheorem{example}{Example}[section]
\newtheorem{definition}{Definition}[section]

\renewcommand{\arraystretch}{1.2}

\title{Holomorphic Yukawa Couplings \\[1ex]     
       in Heterotic String Theory}   

\author{Stefan Ionut Blesneag}             
\college{Wadham College}  

\degree{Doctor of Philosophy}     
\degreedate{Trinity 2021}         

\begin{document}

\baselineskip=18pt plus1pt

\setcounter{secnumdepth}{3}
\setcounter{tocdepth}{3}

\newcommand{\cp}[1]{{\mathbb C}{\mathbb P}^{#1}}
\newcommand{\mbf}[1]{\mathbf{#1}}
\newcommand{\cohclass}[1]{\left[{#1}\right]}
\newcommand{\rest}[1]{\left.{#1}\right|}

\newcommand{\nn}{\nonumber}
\newcommand{\ns}{\normalsize}

\newcommand{\tr}{\text{tr}}
\newcommand{\pt}{\partial}
\newcommand{\w}{\wedge}
\newcommand{\Ds}{\not\!\!D}

\newcommand{\HdR}{H_{\text{DR}}}

\newcommand{\CC}{{\mathbf{C}}}
\newcommand{\ZZ}{{\mathbf{Z}}}
\newcommand{\RR}{{\mathbf{R}}}
\newcommand{\PP}{{\mathbf{P}}}
\newcommand{\bN}{{\mathbf N}}
\newcommand{\bV}{{\mathbf V}}
\newcommand{\bX}{{\mathbf X}}
\newcommand{\bY}{{\matbbf Y}}
\newcommand{\bZ}{{\mathbf Z}}

\def\fnote#1#2{\begingroup\def\thefootnote{#1}\footnote{#2}
     \addtocounter{footnote}{-1}\endgroup}

\def\a{\alpha}
\def\b{\beta}
\def\g{\gamma}
\def\c{\chi}
\def\d{\delta}
\def\e{\epsilon}
\def\f{\phi}
\def\i{\iota}
\def\z{\psi}
\def\k{\kappa}
\def\l{\lambda}
\def\m{\mu}
\def\n{\nu}
\def\o{\omega}
\def\p{\pi}
\def\q{\theta}
\def\r{\rho}
\def\s{\sigma}
\def\t{\tau}
\def\u{\upsilon}
\def\x{\xi}
\def\z{\zeta}
\def\v{\varphi}

\def\D{\Delta}
\def\F{\Phi}
\def\G{\Gamma}
\def\J{\Psi}
\def\L{\Lambda}
\def\O{\Omega}
\def\P{\Pi}
\def\Q{\Theta}
\def\U{\Upsilon}
\def\X{\Xi}

\def\cA{{\mathcal A}}
\def\cB{{\mathcal B}}
\def\cC{{\mathcal C}}
\def\cE{{\mathcal E}}
\def\cF{{\mathcal F}}
\def\cH{{\mathcal H}}
\def\cL{{\mathcal L}}
\def\cK{{\mathcal K}}
\def\cM{{\mathcal M}}
\def\cN{{\mathcal N}}
\def\cO{{\mathcal O}}
\def\cP{{\mathcal P}}
\def\cR{{\mathcal R}}
\def\cV{{\mathcal V}}
\def\cS{{\mathcal S}}
\def \S {{\cal S}}

\def\ab{{\bar a}}
\def\bb{{\bar b}}

\def\gsim{ \lower .75ex \hbox{$\sim$} \llap{\raise .27ex \hbox{$>$}} }
\def\lsim{ \lower .75ex \hbox{$\sim$} \llap{\raise .27ex \hbox{$<$}} }
\def\be{\begin{equation}}
\def\ee{\end{equation}}
\def\bea{\begin{eqnarray}}
\def\eea{\end{eqnarray}}
\def\lf{\left (}
\def\rt{\right )}
\def \td {\tilde}
\def \rx {{\rm x}}

\def \V  {{\rm V}}
\def \ha {{1 \ov 2}}

\def \sql {{\sqrt{\l}}\ }
\def \del{\partial}
\def \a {\alpha}
\def \aa {{\a'}}
\def\ov{\over}
\def \ci {\cite}
\def \n {{\rm n}}

\def \foot {\footnote}
\def \bi{\bibitem}
\def\la{\label}\def\foot{\footnote}\newcommand{\rf}[1]{\eqref{#1}}

\def \ws {{\rm  A}}

\def \cD {{\cal D}}

\def \no {\nonumber}

\def \adss {$AdS_5 \times S^5\ $}

\def \a {\alpha }
\def \eps {\epsilon}

\def \GG  {{\cal G}}
\def \V  {{\rm V}}\def \UU {{v}}

\def\ket {\big\rangle}
\def\bra {\big\langle }
\def \VV {{\cal V}}

\def \oc {\Omega_c}
\def \tX {\tilde X}
\def \tvarphi {\tilde \varphi}
\def \tpsi {\tilde \psi}

\def \l {\lambda}
\def\foot{\footnote}
\def \tl  {{\tilde \l}}
\def \sql {{\sqrt \l}}
\def \adss {$AdS_5 \times S^5~$ }

\def \ov {\over}
\def \ci  {\cite}
\def \bi {\bibitem} 

\def \ed {\end{document}}

\newcommand{\varstr}[2]{\vrule height #1 depth #2 width0pt}

\maketitle                  
\include{dedication}        
\include{acknowledgements}  
\include{abstract}          

\begin{abstractseparate}
This thesis is concerned with heterotic $E_8 \times E_8$ string models that can produce quasi-realistic $N=1$ supersymmetric extensions of the Standard Model in the low-energy limit. We start rather generally by deriving the four-dimensional spectrum and Lagrangian terms from the ten-dimensional theory, through a process of compactification over six-dimensional Calabi-Yau manifolds, upon which holomorphic poly-stable vector bundles are defined. We then specialise to a class of heterotic string models for which the vector bundle is split into a sum of line bundles and the Calabi-Yau manifold is defined as a complete intersection in projective ambient spaces.

We develop a method for calculating holomorphic Yukawa couplings for such models, by relating bundle-valued forms on the Calabi-Yau manifold to their ambient space counterparts, so that the relevant integrals can be evaluated over projective spaces. The method is applicable for any of the $7890$ CICY manifolds known in the literature, and we show that it can be related to earlier algebraic techniques to compute holomorphic Yukawa couplings. We provide explicit calculations of the holomorphic Yukawa couplings for models compactified on the tetra-quadric and on a co-dimension two CICY. A vanishing theorem is formulated, showing that in some cases, topology rather than symmetry is responsible for the absence of certain trilinear couplings. In addition, some Yukawa matrices are found to be dependent on the complex structure moduli and their rank is reduced in certain regions of the moduli space.

In the final part, we focus on a method to evaluate the matter field K\"ahler potential without knowing the Ricci-flat Calabi-Yau metric. This is possible for large internal gauge fluxes, for which the normalisation integral localises around a point on the compactification manifold. We illustrate the procedure on CICYs embedded in the ambient space $\mathbb{P}^1 \times \mathbb{P}^3$, and express our result in terms of globally defined moduli.
\end{abstractseparate}

\begin{acknowledgements}
First of all, I would like to thank my supervisor, Andr\'e Lukas, for his patience in guiding me through some of the most intricate regions of string phenomenology. Without his encouragement and knowledge, I would not have been able to complete this thesis.

I would also like to thank our collaborators Evgeny Buchbinder, Andrei Constantin, Eran Palti, Philip Candelas, Fabian Ruehle, Callum Brodie, Rehan Deen and Andreas Braun, for their invaluable contributions and helpful conversations.

I am grateful to the Theoretical Physics department of Oxford for accepting me as a DPhil candidate and to the Science and Technology Facilities Council (STFC) for offering me a scholarship throughout my studies. Additionally, I am deeply grateful to the welfare and administration team from Wadham College, for their attentive interventions during a very rough period of my life.

Finally, I would like to thank my parents and my sister, for their unending support and love.
\end{acknowledgements}

\begin{romanpages}          
\tableofcontents            
\end{romanpages}            



\chapter{Introduction}

The search for a fundamental theory beyond the Standard Model of elementary particles has produced an impressive number of predictions and hypotheses. Perhaps the most important ones are supersymmetry, a symmetry which relates bosons and fermions, and grand unification, which implies the convergence of the three gauge couplings of the Standard Model into a single value at high energies, $M_{\textrm{GUT}}\sim 10^{16} \textrm{ GeV}$. The next step towards creating a Theory of Everything is to incorporate gravity in a quantum framework that is free of ultraviolet divergences and to propose a unified description of the four fundamental interactions. To date, string theory is the most successful attempt towards realising these goals.

In superstring theory, point-like particles are superseded by vibrating strings and the space-time is predicted to be ten-dimensional, with the six extra spatial dimensions remaining unobserved, given the scales currently probed. Furthermore, the cancellation of ten-dimensional gauge and gravitational anomalies is possible only for two select gauge groups -- $SO(32)$ and $E_8 \times E_8$ -- as it was originally demonstrated by Michael Green and John H. Schwarz in 1984~\cite{greenschwarzarticle}. Their discovery led to the development, in the following year, by David Gross, Jeffrey Harvey, Emil Martinec and Ryan Rohm \cite{stringquartet1,stringquartet2,stringquartet0} of the heterotic $E_8\times E_8$ string theory, which is phenomenologically one of the most promising superstring theories. In the low energy limit, this theory can give rise to $N=1$ supersymmetric models of chiral particles, for which the gauge group is embedded into one $E_8$ factor, while the other $E_8$ is interpreted as the ``hidden sector", where supersymmetry can be spontaneously broken. The goal of string phenomenology is to investigate under which circumstances the precise configuration of the Standard Model is obtained, so that string theory can be connected to measurable physics and the realm of falsifiable predictions. This could solve many problems that the Standard Model itself seems to present. For example, seemingly arbitrary features such as the number of particle families and the hierarchy of particle masses may arise from the underlying structure of the ten-dimensional theory.

In this thesis, we will pursue $E_8 \times E_8$ heterotic string model building by compactifying on Calabi-Yau manifolds~\cite{Candelas:1985en,Strominger:1985it,Witten:1985xc,greene1986}. The motivation for compactifying on such spaces is that we want to preserve $N=1$ supersymmetry at low energies, so that the Higgs mass can be stabilised\footnote{Even though supersymmetry has not been detected at LHC scales, so that the supersymmetric solution to the hierarchy problem is not completely natural anymore, it still helps bridging much of the hierarchy between the Planck and the electroweak scale.}. To be precise, we only want $N=1$ and not extended $N \geq 2$ supersymmetry in 4d, because those extended theories cannot produce chiral fermions. In the simplest case where the NS flux is set to zero, the Calabi-Yau manifolds turn out to be the only class of compact manifolds that satisfy the Killing spinor equations in the low energy limit, thus ensuring $N=1$ supersymmetry. Heterotic Calabi-Yau compactification models are well-known in the literature and have been shown to produce the spectrum of the Minimal Supersymmetric Standard Model (MSSM) ~\cite{Braun:2005ux, Braun:2005bw, Braun:2005nv, Bouchard:2005ag, Blumenhagen:2006ux, Blumenhagen:2006wj, Anderson:2007nc, Anderson:2008uw, Anderson:2009mh, Braun:2009qy, Braun:2011ni}. In recent years, a sizeable database of phenomenologically viable models has been produced through an algorithmic scan over large classes of compactifications \cite{Anderson:2011ns,Anderson:2012yf,Anderson:2013xka}. Given that compactifications with the correct spectrum can now be readily engineered, one of the most pressing problems is the calculation of Yukawa couplings for such models. 

As known from the Standard Model, Yukawa couplings describe interactions between fermions and the Higgs field, thus generating quark and lepton masses when the electroweak symmetry is spontaneously broken. Understanding these couplings as structurally connected to the geometry of the extra dimensions could prove that the masses we encounter in particle physics are not arbitrary, but rather reflect the inner geometry of the Universe. Unfortunately, the calculation of Yukawa couplings for geometric compactifications of the heterotic string
is not straightforward, even at the perturbative level, and relatively few
techniques and results are known \cite{Strominger:1985ks, Candelas:1987se, greene1, greene2, greene3, braunheovrut, bouchardcvetic, Anderson:2009ge, textures}. The task of computing the physical Yukawa couplings for such
models can be split up into two steps: the calculation of the holomorphic Yukawa couplings, that is, the
couplings in the superpotential, and the calculation of the matter field K\"ahler potential. The former relates
to a holomorphic quantity and can, therefore, to some extent be carried out algebraically, as explained in
Refs.~\cite{Candelas:1987se, Anderson:2009ge}. However, the matter field K\"ahler potential is non-holomorphic and its algebraic calculation
does not seem to be possible -- rather, it is likely that methods of differential geometry have to be used.\footnote{See Refs.~\cite{delaossahardy, candelasmetric, mcoristeffective} for recent progress in this direction.}
At present the matter field K\"ahler potential has not been worked out explicitly for any case other than
the standard embedding model (where it can be expressed in terms of the K\"ahler and complex structure moduli
space metrics).

The purpose of this thesis is to expand the knowledge that we have about holomorphic Yukawa couplings in heterotic string theory, by proposing a method to compute these couplings for a specific class of string models, namely for line bundle models on Complete Intersection Calabi-Yau (CICY) manifolds. The method is presented in Chapters~\ref{tetraquadricchapter} and \ref{chaptern>1codimension} in a generalised form and then applied to specific examples. As a secondary objective, we also develop in Chapter~\ref{kahlerchapter} a method for calculating the matter field K\"ahler potential in a very restricted scenario, where sufficiently large flux can lead to localisation of the matter field wave function. It still remains to be seen whether such an approach can meet all the phenomenological requirements. Nevertheless, we hope that our techniques will eventually lead to a framework where physical Yukawa couplings are reliably extracted from various geometrical models.

The structure of the thesis is as follows. In Chapter~\ref{odyssey}, we lay down the background material, starting with some relevant concepts from the Standard Model and the physics that is expected beyond it (supersymmetry, grand unification). Later on, we present the $E_8 \times E_8$ heterotic string theory and the mathematical apparatus that is used for compactification. The chapter then explains how to recover the four-dimensional spectrum and Lagrangian terms from the ten-dimensional theory, in order to enable a comparison with the observable physics.

In Chapter~\ref{tetraquadricchapter}, we develop techniques, based on differential geometry, to compute holomorphic Yukawa couplings for heterotic line bundle models on Calabi-Yau manifolds defined as hypersurfaces in products of projective
spaces. Our methods are based on constructing the required bundle-valued forms explicitly and evaluating the relevant integrals over the projective ambient space. It is shown that the rank of the Yukawa matrix can decrease at specific loci in complex structure moduli space. In particular, we compute the up Yukawa coupling and the singlet-Higgs-lepton trilinear coupling in the heterotic standard model described in Ref.~\cite{Buchbinder:2014qda}.

In Chapter~\ref{chaptern>1codimension}, we generalise the results of Chapter~\ref{tetraquadricchapter}, by applying similar techniques to higher co-dimension Complete Intersection Calabi-Yau manifolds. A vanishing theorem,
which we prove, implies that certain Yukawa couplings allowed by low-energy symmetries are zero due to topological reasons. To illustrate our methods, we calculate Yukawa couplings for $SU(5)$-based standard models on a co-dimension two complete intersection manifold.

In Chapter~\ref{kahlerchapter}, we propose an analytic method to calculate the matter field K\"ahler metric for line bundles models with large internal gauge flux. In this case, the integrals involved in the calculation localise around certain points on the compactification manifold and, hence, can be evaluated approximately without precise knowledge of the Ricci-flat Calabi-Yau metric. In a final step, we show how this local result can be expressed in terms of the global moduli of the Calabi-Yau manifold. The method is illustrated for the family of Calabi-Yau hypersurfaces embedded in $\mathbb{P}^1 \times \mathbb{P}^3$ and we obtain an explicit result for the matter field K\"ahler metric in this case.

The work presented in this thesis is drawn from three research papers. More precisely, Chapter~\ref{tetraquadricchapter} is based on 
\begin{itemize}
\item S.~Blesneag, E.~I.~Buchbinder, P.~Candelas and A.~Lukas, \textit{Holomorphic Yukawa Couplings in Heterotic String Theory}, JHEP {\bf 1601} (2016) 152.~\cite{paper1}
\end{itemize}
Chapter~\ref{chaptern>1codimension} is based on 
\begin{itemize}
\item S. Blesneag, E. I. Buchbinder and A. Lukas, \textit{Holomorphic Yukawa Couplings for Complete
Intersection Calabi-Yau Manifolds}, JHEP {\bf 1701} (2017) 119.~\cite{paper2}
\end{itemize}
Chapter~\ref{kahlerchapter} is based on 
\begin{itemize}
\item S. Blesneag, E. I. Buchbinder, A. Constantin, A. Lukas, E. Palti, \textit{Matter Field K\"ahler Metric in Heterotic String Theory from Localisation}, JHEP {\bf 1804} (2018) 139.~\cite{paper3}
\end{itemize} 
The Appendices~\ref{app:Kbundle},~\ref{coboundarymapappendix},~\ref{appendixPn},~\ref{appendixboundary} are also based on materials from the above-mentioned research papers and contain more technical background to support the calculations.  



\chapter{A String Theory Odyssey}
\label{odyssey}

\noindent The purpose of this chapter is to provide an overview of the background that is required for computing Yukawa couplings in heterotic string theory. We start with the Standard Model and what motivated physicists to search for a theory beyond it, then we continue with a review of supersymmetry and grand unification, culminating in the $E_8 \times E_8$ heterotic string theory and its ten-dimensional $N=1$ supergravity limit. At that stage, it will become evident that several tools from topology and complex geometry are needed. Therefore, we will provide a summary of the mathematics that is used throughout the thesis. Finally, the last part of the chapter is dedicated to the process of compactification, in an attempt to reconnect the high-energy ten-dimensional theory back to the four-dimensional physics from which we started. However, the search for the correct model is not free of phenomenological problems, the most obvious one being the presence of gauge-neutral moduli scalars in the spectrum of compactification. Such fields cause vacuum degeneracy, so they need to be stabilised (see, for example, Refs.~\cite{aglomoduli, aglomoduli2}). Moreover, correlating physical parameters to quantities from the 10d theory is particularly difficult, given that the latter depend on the unknown geometry of the extra dimensions. As it turns out, information about the spectrum and the \textit{holomorphic} Yukawa couplings can be extracted from less specific, quasi-topological properties of the internal manifold, while still leaving field normalisation unresolved. Proving this will become our main goal by the end of the chapter.

\section{Yukawa Couplings in the Standard Model}
\label{smsection}
\noindent Formulated in the early 1970s as a quantum field theory with symmetry group $G_{\textrm{SM}} = SU(3)\times SU(2)_L \times U(1)_Y$, the Standard Model has proven to be extremely successful for describing particle physics at energies up to the TeV scale. Its remarkable performance, demonstrated through decades of experimental testing, lies in its ability to accurately predict the interactions of all known elementary particles under three of the four fundamental forces  -- strong, weak and electromagnetic. More precisely, the $SU(3)$ part of the gauge group describes Quantum Chromodynamics, or the theory of the strong interaction, while the $SU(2)_L \times U(1)_Y$ part is responsible for the electroweak theory. In the Standard Model, these interactions are realised through the exchange of $12$ vector bosons ($8$ gluons and $4$ electroweak gauge bosons), which are in the adjoint representation of their corresponding gauge groups
\begin{equation}
\begin{array}{lllll}
\label{smgaugerep}
G^{1,...,8}=(\mathbf{8},\mathbf{1})_0, & \quad\quad & W^{1,2,3}=(\mathbf{1},\mathbf{3})_0, & \quad\quad & B=(\mathbf{1},\mathbf{1})_0 ,
\end{array}
\end{equation}
\noindent where the notation $(\mathbf{a},\mathbf{b})_c$ corresponds to $(SU(3),SU(2)_L)$ representations with subscript for the $U(1)$ charge. Matter is described by three generations of Weyl fermions (quarks and leptons), each carrying a representation of $G_{\textrm{SM}}$
\begin{align}
\label{representations}
Q^i &= \begin{pmatrix} u^i_L\\  d^i_L\end{pmatrix} = (\mathbf{3},\mathbf{2})_{1/6},  \quad \,\ \,\, u^i = (u^i_R)^c=(\overline{\mathbf{3}},\mathbf{1})_{-2/3},  \quad \,\, d^i = (d^i_R)^c=(\overline{\mathbf{3}},\mathbf{1})_{1/3}, \notag \\[1.5mm] L^i & = \begin{pmatrix} \nu_L^i \\  e_L \end{pmatrix} = (\mathbf{1},\mathbf{2})_{-1/2},  \quad \,\, e^i = (e^i_R)^c  =(\mathbf{1},\mathbf{1})_1,  \quad 
\end{align}
\noindent where generation number is labeled by $i = 1,2,3$ and, by convention, charge conjugation is applied on right-hand components in order to represent everything as left-handed. From these expressions one can see that the left- and right-hand Weyl fermions transform differently under the electroweak gauge group, which means that the Standard Model is a chiral theory. The Lagrangian used to encode all the information about the dynamics of the particles contains the following kinetic terms
\begin{align}
\mathcal{L}^{\textrm{SM}}_{\textrm{kin}} = & - \dfrac{1}{2} \textrm{Tr} \left[G_{\mu \nu} G^{\mu \nu}\right] - \dfrac{1}{2} \textrm{Tr} \left[W_{\mu \nu} W^{\mu \nu}\right]- \dfrac{1}{4} F_{\mu \nu} F^{\mu \nu}+ \notag \\ & + i \sum_{i=1}^3 \left[   \overline{Q}{}^i \slashed{D} Q^i +   \overline{u}{}^i \slashed{D} u^i +   \overline{d}{}^i \slashed{D} d^i+   \overline{L}{}^i \slashed{D} L^i+  \overline{e}{}^i \slashed{D} e^i\right],
\end{align}
\noindent where $G_{\mu\nu}$, $W_{\mu\nu}$ and $F_{\mu\nu}$ are the field strengths of the three gauge fields and $\slashed{D} = \overline{\sigma}{}^{\mu} D_{\mu}$ is the Dirac operator. As a consequence of gauge invariance, no explicit mass terms such as $m_{\psi} \overline{\psi}_L \psi_R$ or $ m^2_W W^{\mu} W_{\mu} $ are permitted, so the only way fermions and gauge bosons can acquire mass is through the spontaneous breaking of the electroweak symmetry. This is realised through the Higgs mechanism, with the introduction of a scalar Higgs field $H = (H^{0}, H^{-})^T$, which is a $(\mathbf{1},\mathbf{2})_{-1/2}$ representation of $G_{\textrm{SM}}$ and has the potential energy
\begin{eqnarray}
V(H) = - \mu^2 H^\dagger H + \lambda (H^\dagger H)^2.
\end{eqnarray} 
\noindent Because of its non-vanishing vacuum expectation value $\langle H \rangle = (v/\sqrt{2},0)^T$, the Higgs field breaks $SU(2)_L \times U(1)_Y \rightarrow U(1)_{\textrm{em}}$  at a scale of around $v = \mu/\sqrt{\lambda}= 246 $ GeV, such that the well-known electric charge is given by the combination $Q_{\textrm{em}} = T_3 + Y$, where $T_3 = \pm 1/2$ is the weak isospin of $SU(2)_L$ and $Y$ is the weak hypercharge. The kinetic term of the Higgs, $D^{\mu} H^{\dagger} D_{\mu} H$, is responsible for electroweak boson mass terms in the effective Lagrangian (three massive bosons $W^{\pm}$,  $Z$ and one massless photon). More explicitly, the three would-be Goldstone bosons of the spontaneously broken $SU(2)_L \times U(1)_Y$ symmetry are “eaten up” or absorbed to become the longitudinal components of the three massive gauge bosons. On the other hand, fermions acquire mass from interactions with the Higgs, described by the Yukawa terms
\begin{eqnarray}
\label{smyukawa}
\mathcal{L}^{\textrm{SM}}_{\textrm{Yuk}} = Y_u^{i j} Q^a_{i} u_{j} \bar{H}_a + Y_d^{i j} Q^a_{i} d_{j} \epsilon_{ab} H^b +  Y_e^{i j} L^a_{i} e_{j} \epsilon_{ab}H^b + \textrm{h.c.} 
\end{eqnarray}
\noindent where ``h.c." stands for ``Hermitian conjugate", $a,b=1,2$ label the $SU(2)_L$ components and $\epsilon_{a b}$ is the Levi-Civita tensor (see also Ref. \cite{srednicki}). As one can see from this formula, the Yukawa couplings $Y^{i j}_{u,d,e}$ dictate the structure of fermionic mass matrices $m^{ij}_f = Y^{ij}_f v / \sqrt{2}$ in the low energy limit. These matrices are expected to be non-diagonal, since weak interaction eigenstates act as a mixture of mass eigenstates both in the quark and lepton sector. For quarks, the mixing is realised through the unitary CKM (Cabibbo – Kobayashi – Maskawa) matrix, which is parametrised by three angles and one CP violation phase. In the case of the leptons, a similar PMNS (Pontecorvo–Maki–Nakagawa–Sakata) matrix is introduced, however it is important to note that in this case, the Standard Model has to be extended to include mechanisms that explain neutrino mass, usually by assuming the existence of heavy sterile right-handed neutrinos, $N^i$. In the Minimally Extended Standard Model for example, the $N^i$ generate effective 5-dimensional Weinberg operators $y^2 L \bar{H} \bar{H} L /M_N$, due to neutrino Yukawa interactions $y^{ij} L_i N_j \bar{H}$, thus inducing small Majorana neutrino masses $m_{\nu} \sim y^2 v^2 /M_N$, where $M_N \gg v$ is the mass scale of the $N^i$. This is just one of the many possible realisations of the so-called “seesaw mechanism”.

Finally, one remarkable feature of the Standard Model is the automatic cancellation of gauge anomalies. These anomalies, which are violations of gauge symmetries under quantum corrections, could in principle arise from triangle loops of chiral fermions contributing to triple gauge-boson vertices. For example, for the $[U(1)_Y]^3$ coupling, the anomaly is proportional to $\textrm{Tr}(Y^3)$, while for the non-abelian $[SU(n)]^{3}$ coupling it is proportional to $\textrm{Tr} (T_a \lbrace T_b, T_c \rbrace)$, where the $T_a$ are the generators of $SU(n)$. Further anomalies arising from $U(1)_Y [SU(n)]^2$ triangles are written as summations $\textrm{Tr}(Y)_L$ over the left-hand fermions ($n=2$) and $\textrm{Tr}(Y)_q$ over the quarks ($n=3$), while mixed anomalies, which involve both gauge and gravitational interactions, have a total factor of $\textrm{Tr}(Y)$. It can be shown that the structure of the Standard Model, as described by \eqref{representations}, ensures that all these anomalies are canceled.

\section{Beyond the Standard Model}
\noindent Despite its experimental success, the Standard Model cannot be the complete description of our Universe because of various compelling reasons. First of all, it neglects gravity, thus acting as an effective theory at energies much lower than the Planck scale. The established theory of gravity, General Relativity, is in fact incompatible with the quantum framework of the Standard Model and is proven to be non-renormalisable. Moreover, the Standard Model fails to provide an explanation for the existence of dark matter and dark energy, which constitute about 22\% and 74\% of the Universe. It fails to explain why the cosmological constant is so small ($\Lambda \sim 10^{-12} \textrm{eV}^4 $) compared to the quantum field theory estimation which is $120$ orders of magnitude higher.

Another factor that makes the Standard Model seem incomplete is the fact that its dynamics depends on $19$ free parameters (the three gauge couplings, the nine quark and lepton masses, not including neutrinos, the three CKM mixing angles, the CKM CP-violating phase $\delta$, the CP-violating strong-interaction parameter $\theta_{\textrm{QCD}}$, the Higgs vev $v$ and the Higgs self-coupling $\lambda$), all of which have to be derived from experiment, without any theoretical insight into their origin. If one were to accomodate neutrino oscillations, the resulting model would require nine additional parameters (masses, mixing angles and CP-violating phases), thus complicating the problem even further \cite{johnellis}. Overall, one does not know why the $U(1)_Y$ charges take the values that they do, or why there are precisely three generations of quarks and leptons.

Further problems come from what is called “naturalness” or the idea that physical parameters should naturally be of the same order, otherwise a reasonable explanation has to be given for their hierarchy. For example, the large discrepancy between the weak force and gravity constitutes the principal hierarchy problem of the SM. It is unknown why the electroweak scale $M_W \sim 10^2 \textrm{ GeV}$ is so many orders of magnitude smaller than the Planck scale $M_{\textrm{P}}\sim 10^{19}\textrm{ GeV}$. Given the fact that the Higgs mass can receive quantum loop corrections which are proportional to the cut-off scale, so that $\Delta m_H^2 = \mathcal{O}(\Lambda_{\textrm{cutoff}}^2)$, one would expect $m_H$ to be much closer to $M_{\textrm{P}}$, unless an underlying mechanism ensures that the sum of all these  corrections vanishes. It is also unclear why the strong CP-violating angle $\theta_{\textrm{QCD}}$ is so small compared to the CKM CP-violating phase. Experimental limits of the neutron electric dipole moment set $\vert\theta_{\textrm{QCD}}\vert < 10^{-10}$, which constitutes the strong CP problem. Last but not least, it is a mystery why the SM particles seem to obey a hierarchical pattern, with masses varying from around $1$ eV for the neutrinos and $0.5$ MeV for the electron to $173$ GeV for the top quark.

As a consequence of these unanswered questions, physicists are searching for extensions of the Standard Model at higher, yet unexplored energies.

\subsection{Supersymmetry and the MSSM}
\label{susymssm}

\noindent Supersymmetry (SUSY) is the only possible non-trivial\footnote{By non-trivial we mean an extension that is not simply a direct product between the Poincar\'e group and an internal group.} extension of the Poincar\'e algebra of special relativity, as it is inferred from the Haag-$\slashed{\textrm{L}}$opusza\' nski-Sohnius generalisation \cite{haaglopuszanski} of the Coleman-Mandula no-go theorem \cite{colemanmandula}. As such, supersymmetry looks promising as a possible symmetry of nature. Being a graded-Lie algebra, it is realised through the introduction of $N$ spinorial generators $Q^A_{\alpha}$, $A=1,...,N$, and their Hermitian conjugates $\overline{Q}{}^A_{\dot{\alpha}}$, which satisfy certain anti-commutation relations. However, in all future discussions, we will only consider the simple case, $N=1$, for which the algebra is given by
\begin{eqnarray}
\lbrace Q_{\alpha}, \overline{Q}_{\dot{\alpha}} \rbrace = 2  \sigma_{\alpha \dot{\alpha}}^{\mu} P_{\mu}, \,\,\,\,\,\,\,\,\,\,\,\,\,\,\,\,\,\,\,\,\,\,\,\,\, \lbrace Q_{\alpha}, Q_{\beta} \rbrace = 0 , \,\,\,\,\,\,\,\,\,\,\,\,\,\,\,\,\,\,\,\,\,\,\,\,\, \lbrace \overline{Q}_{\dot{\alpha}}, \overline{Q}_{\dot{\beta}} \rbrace=0,
\end{eqnarray}
\noindent where $P_{\mu}$ is the generator of spacetime translations and $\alpha$, $\dot{\alpha}=1,2$ are spinor indices.

In the framework of supersymmetry, every particle has a superpartner whose spin differs by half an integer, so that a fermion is transformed into a boson and vice versa, through the action of the supersymmetric generator. Schematically,
\begin{eqnarray}
\delta \phi \sim \overline{\epsilon} \psi, \,\,\,\,\,\,\,\,\,\,\,\,\,\,\,\,\,\,\,\,\,\,\,\,\,\,\,\,\,\,\,\,\,\,\,\,\,\,\,\,\,\,\,\,\,\,\,\,\,\, \delta \psi  \sim\epsilon \partial \phi ,
\end{eqnarray}
\noindent for a boson field $\phi$ and a fermion field $\psi$, where $\epsilon^{\alpha}$ is an infinitesimal, anticommuting, constant spinor, parametrising the transformation. Particles which are superpartners of each other are grouped together into supermultiplets (irreducible representations of the SUSY algebra) and have the same mass, although after supersymmetry is broken, some of them can become significantly heavier, thus explaining why they are not observed. There are different types of supermultiplets: those with spin $(1/2,0)$ are called chiral multiplets, because they contain chiral fermions and their superpartners, while those with spin $(1,1/2)$ are called vector multiplets, containing vector bosons and their superpartners. Any given supermultiplet can be written as a superfield, i.e. a function of the spacetime coordinates $x^{\mu}$ and some fermionic dimensions $\lbrace \theta^{\alpha}$, $\overline{\theta}{}^{\dot{\alpha}}\rbrace_{\alpha,\dot{\alpha}=1,2}$ called Grassmann numbers. Together, the coordinates $(x^{\mu}, \theta^{\alpha}$, $\overline{\theta}{}^{\dot{\alpha}})$ parametrise the eight-dimensional superspace. With these notations, the action governing the dynamics of $n$ chiral superfields $\Phi^I = (\phi^I,\psi^I)$, $I=1,...,n$, in a 4d $N=1$ SUSY theory, is generally expressed as
\begin{eqnarray}
S = \int d^4 x \int d^2 \theta d^2 \overline{\theta}  K(\Phi^I, \Phi^I{}^{\dagger}) + \left( \int d^4 x \int d^2 \theta W(\Phi^I) + \textrm{h.c.} \right),
\end{eqnarray}
\noindent where $W$ is the superpotential -- a holomorphic function providing Yukawa and mass terms
\begin{eqnarray}
\label{generalsuperpotential}
W = \lambda_{I J K} \Phi^I \Phi^J \Phi^J + M_{I J} \Phi^I \Phi^J,
\end{eqnarray}
\noindent while $K$ is the K\"ahler potential, a general real function, which gives rise to kinetic terms of the form $G_{IJ} \partial^{\mu} \phi^{I*} \partial_{\mu} \phi^J$ and $i G_{IJ} \overline{\psi}{}^I \slashed{D} \psi^J $, where $G_{IJ} = \tfrac {\partial^2 K} {\partial \Phi^{I \dagger} \partial \Phi^J}$. It is important to note here that the Yukawa couplings $\lambda_{I J K}$ of the superpotential and the physical Yukawa couplings $Y_{IJK}$ of Eq.~\eqref{smyukawa} can be identified only if the kinetic terms are brought to a canonical form, through an appropriate field redefinition, $\Phi^I \rightarrow \tilde{\Phi}^I = U^I{}_J \Phi^J$, where $U^I{}_J$ is a unitary matrix. Otherwise, if for example lack of knowledge about $G_{IJ}$ prevents such a redefinition to be applied, $\lambda_{I J K}$ are to be referred to as \textbf{holomorphic} Yukawa couplings, to distinguish them from $Y_{IJK}$. Returning to the supersymmetric action, the term corresponding to vector supermultiplets reads
\begin{eqnarray}
\label{gaugesusyaction}
S = \int d^4 x \int d^2 \theta f_{ab}(\Phi^I)\mathcal{W}^a  \mathcal{W}^b + \textrm{h.c.} \, ,
\end{eqnarray}
\noindent where $\mathcal{W}^a$ is the “field strength” chiral superfield and $f_{ab}$ is the holomorphic gauge kinetic function, with indices $a,b$ running over the adjoint representations of gauge groups in the theory. 

\vspace{3mm}

One of the main benefits of supersymmetry is that it solves the fine-tuning problem of the Higgs mass, by canceling UV divergences pair by pair, since bosonic and fermionic superpartners contribute with opposite signs. It is for this reason that the study of supersymmetry is so relevant for the Standard Model. Naturally preserving the electroweak scale at $10^2$ GeV  means that SUSY has to be encountered at energies of several TeV \cite{cohen}. The simplest extension, involving the smallest field content, is called the Minimal Supersymmetric Standard Model (MSSM). In this model, every SM particle is interpreted to be an element of an $N=1$ supermultiplet along with its yet undiscovered partner, so that $Q^i$, $u^i$, $d^i$ are redefined to represent quark-squark pairs, $L^i$ and $e^i$ denote lepton-slepton pairs, while $G$, $W$ and $B$ are the gauge boson-gaugino pairs. The fermionic partner of the Higgs, the higgsino, would upset the anomaly cancellation conditions $\textrm{Tr}(Y) \textrm{, } \textrm{Tr}(Y^3) \stackrel{!}{=} 0 $ and it is for this reason that supersymmetry has not one, but two Higgs supermultiplets $H_u = (H_u^{+}, H_u^0)$ and $H_d = (H_d^0, H_d^{-})$, with opposite hypercharges that cancel the anomalies. The MSSM superpotential containing the Yukawa couplings and the Higgs mass term reads  
\begin{eqnarray}
\label{mssmyukawa}
W = Y_u^{i j} Q_i u_{j} H_u + Y_d^{i j} Q_i d_{j} H_d + Y_e^{i j} L_i e_{j} H_d + \mu H_u H_d .
\end{eqnarray}
\noindent It can be shown that the holomorphicity prevents the superpotential from being renormalised at any order in perturbation theory \cite{grisaru1979}\cite{seiberg1993}. This is because all perturbative corrections are real (non-holomorphic), therefore they cannot modify a holomorphic quantity such as $W$. Another consequence of the holomorphicity of $W$ is that Yukawa terms involving the conjugate of the Higgs field are no longer permitted, as they were in Eq.~\eqref{smyukawa}, therefore the introduction of two Higgs superfields $H_u$ and $H_d$ is necessary to ensure all matter fields receive a mass \cite{wessandbagger}. As for the mass parameter $\mu$ responsible for electroweak symmetry breaking, it leads to a naturalness problem (the “$\mu$-problem”), when one tries to understand why $\mu \ll M_\textrm{P}$. In principle, the MSSM superpotential could also include other renormalisable terms of the form
\begin{eqnarray}
\label{protondecayterms}
W_{\textrm{RV}} = \beta^{i j k} L_i L_j e_{k} + \beta^{'}{}^{i j k} L_i Q_j d_{k} + \beta^{''}{}^{i j k} u_{i} d_{j} d_{k} + \alpha^i L_i H_u ,
\end{eqnarray}
\noindent where $\alpha$, $\beta$, $\beta'$ and $\beta''$ are the couplings of the interactions, however these terms would violate lepton and baryon number, thus allowing fast proton decay, so they should be ruled out. One way of doing this is by imposing a discrete global symmetry $R = (-1)^{3(B-L)+2s}$, which is known as R-parity, where $s$ is the spin of the particle and $B$ and $L$ are baryon and lepton numbers. For SM particles $R=1$, while for their superpartners $R=-1$. As R-parity is conserved, this would mean that the lightest superpartner (LSP) is a stable particle and in fact it is considered a candidate for dark matter. Typically the LSP is deemed to be the neutralino, a linear combination of the neutral electroweak gauginos and the neutral higgsinos.

\vspace{3mm}

On a final note, if supersymmetry is indeed a symmetry of nature it is clear that it must be broken at a certain energy scale $M_{\textrm{SUSY}}$, given that none of the predicted superpartners have been observed. In general, SUSY breaking translates into the requirement that certain auxiliary fields $F_i = - \partial W/\partial \phi^i$ and $D^a$, associated to chiral and vector supermultiplets respectively, acquire non-zero vacuum expectation values $\langle F_i\rangle$, $\langle D^a \rangle \neq 0$. This is equivalent to saying that the potential energy
\begin{eqnarray}
\label{scalarpotential}
V = \vert F_i \vert^2 + \dfrac{1}{2} (D^a)^2
\end{eqnarray}
\noindent is non-zero. In the MSSM however, supersymmetry cannot be broken spontaneously, otherwise the cancellation of quadratic divergences would not occur. Instead, one has to break supersymmetry “softly”, by introducing explicit SUSY-breaking terms in the Lagrangian. Such terms, like the scalar mass term $m_{\phi}^2 \phi^* \phi$, the gaugino mass term $M_{\lambda} \lambda \lambda$ and the bilinear and trilinear scalar couplings $b^{i j} \phi_i \phi_j$ and $a^{i j k} \phi_i \phi_j \phi_k$ are believed to be the effective result of an underlying SUSY breaking mechanism that occurs in a hidden sector and is communicated to the MSSM via some messenger fields. One can think of the hidden sector as a collection of singlets under $G_{\textrm{SM}}$ which interact with the SM particles very weakly, for example through gravity. In string theory, a common interpretation is that the hidden and visible sectors are geometrically separated -- they live on different branes separated by extra dimensions and the messenger fields propagate between them, in the bulk \cite{djhchung}. In any case, the great inconvenience of the soft SUSY breaking is that it introduces $105$ new free parameters (masses, phases, mixing angles) in addition to those already found in the Standard Model. This degree of arbitrariness in the Lagrangian makes the MSSM seem like an incomplete description of particle physics.

\subsection{Grand Unified Theories}
\label{gutsubsection}
\noindent Another natural way to extend the Standard Model is to presume that the non-semisimple gauge group $G_{\textrm{SM}}$ is embedded into a larger simple group $G_{\textrm{GUT}}$, so that the three gauge couplings $g_1$, $g_2$ and $g_3$, corresponding to $U(1)_Y$, $SU(2)_L$ and $SU(3)$ respectively, must equate a single coupling $g_{\textrm{GUT}}$ of a Grand Unified Theory (GUT). This unification is motivated by the observation that gauge couplings evolve with respect to energy scale, according to a set of equations called “the renormalisation group”. For example, $g_3$ is shown to decrease as the energy scale increases, while $g_1$ gets significantly larger. At around $10^{15} - 10^{16}\textrm{ GeV}$, the three gauge couplings reach very similar values, although they are not precisely equal. Equality is however acquired in the MSSM, which is one of the reasons why we will consider only supersymmetric versions of Grand Unified Theories, despite the fact that originally they were not built with supersymmetry in mind. Another reason why SUSY GUTs are favoured is because they can be embedded in superstring theories, thus paving the way for higher energy exploration. The phenomenological interpretation of a Grand Unified Theory is that $G_{\textrm{GUT}}$ is the underlying symmetry of nature, while the Standard Model is the effective theory resulting after $G_{\textrm{GUT}}$ has been broken at a certain high-energy scale. As such, all GUTs provide an explanation for the values of the $U(1)_Y$ charges, by embedding the hypercharge in a simple group. For example, the well-known condition $Q_{\textrm{proton}}+Q_{\textrm{electron}}=0$ follows from the fact that quarks and leptons are combined in the same GUT multiplets, and the GUT group generators are traceless \cite{ksbabu}. 

Since the maximum number of commuting generators (i.e. the rank) of $G_{\textrm{SM}}$ is 4, it is required that $\textrm{rank}(G_{\textrm{GUT}})\geq 4$. For $SU(n)$, the rank is $n-1$, therefore the smallest simple group that can contain $G_{\textrm{SM}}$ is $SU(5)$. In this GUT model, discovered by Georgi and Glashow \cite{georgiglashow}, each generation of SM particles fits into a $\overline{\mathbf{5}} \oplus \mathbf{10}$ representation of $SU(5)$, in the following way
\begin{eqnarray}
\overline{\mathbf{5}} \rightarrow (\overline{\mathbf{3}},\mathbf{1})_{1/3} \oplus (\mathbf{1},\mathbf{2})_{-1/2}, \,\,\,\,\,\,\,\,\, \mathbf{10} \rightarrow (\mathbf{3},\mathbf{2})_{1/6} \oplus(\overline{\mathbf{3}},\mathbf{1})_{-2/3} \oplus  (\mathbf{1},\mathbf{1})_1,
\end{eqnarray}
\noindent where $\overline{\mathbf{5}}$ is the anti-fundamental representation of $SU(5)$ and $\mathbf{10}$ is the antisymmetric compontent of the $\mathbf{5} \otimes \mathbf{5}$ matrix. Comparing this to Eq.~\eqref{representations} shows that $\overline{\mathbf{5}}^i$ contains $\lbrace d^i, L^i\rbrace$ and $\mathbf{10}^i$ contains $\lbrace Q^i, u^i, e^i \rbrace$, for $i=1,2,3$. The interaction of these matter fields is mediated by $n^2-1 = 24$ gauge fields, transforming in the adjoint representation of $SU(5)$, for which the SM decomposition reads
\begin{eqnarray}
\mathbf{24} \rightarrow (\mathbf{8},\mathbf{1})_0 \oplus (\mathbf{1},\mathbf{3})_0, \oplus (\mathbf{1},\mathbf{1})_0 \oplus (\mathbf{3},\mathbf{2})_{-5/6} \oplus (\overline{\mathbf{3}},\mathbf{2})_{5/6},
\end{eqnarray}
\noindent where the first three terms are recognisable as the $12$ SM gauge fields in Eq.~\eqref{smgaugerep}, and the last two terms are $12$ new bosons denoted by $\lbrace X^{\pm}, Y^{\pm}\rbrace_{1,2,3}$. These new bosons can mediate transitions  between quarks and leptons, thus violating lepton and baryon number as well as enabling proton decay modes such as $p\rightarrow \pi^0 e^+$ \cite{jhisano}. Consequently, when $SU(5)$ is broken, the $X$ and $Y$ bosons must acquire a mass of at least $10^{15} \textrm{ GeV}$ according to current experimental limits, a condition which is met by the SUSY version of the $SU(5)$ GUT, but not by the minimal non-supersymmetric model \cite{wdeboer}. The symmetry breaking mechanism is achieved by a GUT-Higgs scalar field $H_{\mathbf{24}}$ in the adjoint representation $\mathbf{24}$, having a non-vanishing vev $\langle H_{\mathbf{24}} \rangle = v_{24} \textrm{ diag} (2,2,2,-3,-3)$, which is proportional to the hypercharge generator and commutes with $G_{\textrm{SM}}$. The masses of the $X$ and $Y$ bosons are then given by $M_{X,Y}^2 \sim g_{\textrm{GUT}}^2 v^2_{24}$. On the other hand, the SM Higgs doublet belongs to a $\overline{\mathbf{5}}_H$ representation of $SU(5)$, along with three other states which form the colour-triplet Higgs $T$. Since the triplet can also mediate proton decay, this time through the mode $p\rightarrow \overline{\nu} K^+$, it must receive a heavy mass of at least $10^{15}$ GeV \cite{dboer}. The manner in which $H$ and $T$ acquire masses so different in orders of magnitude ($m_{H}/m_{T}\sim \mathcal{O}(10^{-13})$) is referred to as the double-triplet splitting problem. In the supersymmetric GUT, two Higgs doublets $H_d$ and $H_u$ are needed, which can sit in $\overline{\mathbf{5}}_H$  and $\mathbf{5}_H$ respectively, giving rise to the following types of Yukawa couplings
\begin{eqnarray}
W^{SU(5)}_{\textrm{Yuk}} = Y^{i,j}_u \Phi^i_{\mathbf{10}} \Phi^j_{\mathbf{10}} H^u_{\mathbf{5}} + Y^{i,j}_{d,l}  \Phi^i_{\mathbf{10}} \Phi^j_{\overline{\mathbf{5}}} H^d_{\overline{\mathbf{5}}} .
\end{eqnarray}
\noindent Here, each term is to be interpreted as a singlet, through an appropriate contraction of $SU(5)$ indices: $\epsilon_{abcde} \Phi^{ab}_{\mathbf{10}} \Phi^{cd}_{\mathbf{10}} H^{e}_{\mathbf{5}}$ and $\Phi^{a b}_{\mathbf{10}} \Phi_{\overline{\mathbf{5}} a} H_{\overline{\mathbf{5}} b}$. Also, note that in the $SU(5)$ theory, $Y_d$ and $Y_l$ are equal at the GUT scale, thus exemplifying Yukawa coupling unification.

Looking further, one can seek GUT groups with rank larger than $4$. For $\textrm{rank} = 5$, the only simple groups in which $G_{\textrm{SM}}$ can be embedded are $SU(6)$ and $SO(10)$, while for $\textrm{rank} = 6$, they are $SU(7)$ and $E_6$. We will only discuss the $SO(10)$ and $E_6$ GUT models, since the $SU(6)$ and $SU(7)$ cases are simply extensions of the $SU(5)$ theory and they add no interesting new features. The relevant chain of embeddings reads
\begin{eqnarray}
\label{gutgroups}
G_{\textrm{SM}} \subset SU(5)\subset SO(10) \subset E_6,
\end{eqnarray}
\noindent but in later chapters we will see that $E_6$ can be embedded further in larger groups such as $E_7$ and $E_8$, the latter being particularly significant because of the $E_8 \times E_8$ heterotic string theory. One must stress however that $E_7$ and $E_8$ do not qualify as 4d GUT groups, because they have no complex representations and therefore the theories would not be chiral.

In the $SO(10)$ GUT, one spinor representation $\mathbf{16}$ of $SO(10)$ contains all the fifteen particles of an SM family and an additional unknown particle, interpreted to be the right-handed neutrino. More explicitly, if we decompose $\mathbf{16} = \overline{\mathbf{5}} \oplus \mathbf{10} \oplus \mathbf{1}$  under the $SU(5)$ subgroup, we recognise the previously discussed $\overline{\mathbf{5}} \oplus \mathbf{10}$ representation in $SU(5)$ plus the extra singlet $\mathbf{1}$. The adjoint representation $\mathbf{45}$ contains the 12 SM gauge bosons and 33 extra gauge bosons with weak and colour charges. Unlike $SU(5)$, the $SO(10)$ group can be broken down to $G_{\textrm{SM}}$ either directly or through various intermediate steps, for example via the maximal subgroup decomposition: $SO(10) \rightarrow SU(5) \times U(1) \rightarrow G_{\textrm{SM}}$, or via the Pati-Salam model \cite{patisalam}: $SO(10) \rightarrow SU(4) \times SU(2)_L \times SU(2)_R\rightarrow G_{\textrm{SM}}$. Through their non-vanishing vevs, the scalar fields of the Higgs sector are the ones dictating which symmetry-breaking scenario occurs. For a direct decomposition, a $\mathbf{144}_H$ scalar multiplet is needed; for intermediate routes, combinations of scalars such as $\mathbf{45}_H$/$\mathbf{54}_H$/$\mathbf{210}_H$ (rank-preserving) and $\mathbf{16}_H$/$\mathbf{126}_H$ (rank-reducing) are introduced instead. The sheer complexity of the Higgs sector is often regarded as a problem of GUTs that have large unification groups. As for the MSSM Higgs doublets $H_u$ and $H_d$, they are contained in one fundamental representation $\mathbf{10}_H$ of $SO(10)$, which decomposes as $\mathbf{5}_H\oplus\overline{\mathbf{5}}_H$ under $SU(5)$. The only permitted Yukawa term reads
\begin{eqnarray}
W^{SO(10)}_{\textrm{Yuk}} = Y^{i,j} \Phi^i_{\mathbf{16}} \Phi^j_{\mathbf{16}} H_{\mathbf{10}},
\end{eqnarray}
\noindent such that all Yukawa coupling constants are unified at the $SO(10)$ GUT scale.

Finally, in the $E_6$ GUT, the $15+1$ matter fields of one SM generation are fitted inside the fundamental representation $\mathbf{27}$. The remaining $11$ states correspond to unknown heavy particles (non-chiral quarks and leptons), thus making the $E_6$ theory very cumbersome from a phenomenological perspective. Usually, the MSSM Higgs doublets also descend from one family $\mathbf{27}$ multiplet, while the other two $\mathbf{27}$s contain pairs of Higgs which are inert and do not get a vev \cite{pathron}. Well-known breaking patterns are $E_6 \rightarrow SO(10) \times U(1)$ with branching $\mathbf{27}\rightarrow\mathbf{16}_1\oplus\mathbf{10}_{-2}\oplus\mathbf{1}_4$ and $E_6 \rightarrow SU(3)_c \times SU(3)_L \times SU(3)_R$ (trinification model) with branching $\mathbf{27}\rightarrow (\mathbf{3},\mathbf{3}, \mathbf{1}) \oplus (\overline{\mathbf{3}}, \mathbf{1}, \overline{\mathbf{3}}) \oplus (\mathbf{1},\overline{\mathbf{3}},\mathbf{3})$. Again, the symmetry breaking is realised by scalar fields which must be added to the theory. The Yukawa coupling is of the form
\begin{eqnarray}
W^{E_6}_{\textrm{Yuk}} = Y^{i,j} \Phi^i_{\mathbf{27}} \Phi^j_{\mathbf{27}} H_{\mathbf{27}}.
\end{eqnarray}

Overall, Grand Unified Theories have an equal share of strengths and weaknesses. On the one hand, gauge coupling unification and the explanation for the values of hypercharges are extremely valuable features. On the other hand, no reason is given for the existence of three generations and many new Higgs particles have to be introduced by hand in order to explain symmetry breaking. Moreover, the double-triplet splitting problem has to be solved through mechanisms which avoid severe fine-tuning, but add new representations to the model. Examples of such mechanisms are the Dimopoulos–Wilczek (missing vev) mechanism in $SO(10)$ and the missing partner model in $SU(5)$ \cite{mohapatra, raby}. Last but not least, important predictions of GUTs such as proton decay and magnetic monopoles have not yet been confirmed by experiment.

\subsection{Supergravity}

\noindent Supergravity is an extension of supersymmetry, designed to include the principles of General Relativity. In order to make this possible, supersymmetry has to become local, with a spacetime-dependent spinor $\epsilon(x)$ parametrising the infinitesimal SUSY transformation. The key ingredient of supergravity is the graviton $h_{\mu \nu}$, a massless spin-2 elementary particle which couples to the stress-energy tensor, thus mediating gravitational interactions. Its fermionic, spin-$3/2$ partner, the gravitino $\psi^\alpha_\mu$, equipped both with a spinor index $\alpha$ and a spacetime index $\mu$, is the gauge field of local supersymmetry and becomes massive when SUSY is broken, by absorbing the emerging goldstino in the so-called super-Higgs mechanism. There are two ways in which the graviton can be related to the metric $g_{\mu \nu}$, either through an infinitesimal expansion $g_{\mu\nu} = \eta_{\mu\nu} + h_{\mu\nu}$ around the flat metric $\eta_{\mu \nu}$, or through the vielbein formalism. As is well-known from General Relativity, the metric (and implicitly the graviton) has to satisfy the Einstein's field equations, which, in a vacuum, correspond to minimising the Einstein-Hilbert action
\begin{eqnarray}
S_{\textrm{EH}}= \dfrac{1}{16\pi G_N} \int d^4 x \sqrt{-g} R \, ,
\end{eqnarray}
\noindent where $G_N$ is Newton’s constant.

As for the chiral and vector multiplets of the theory, they are taken into account when the superpotential $W$, the K\"ahler potential $K$ and the field strength superfield $\mathcal{W}$ are included in the total action, so that, for example, Eq.~\eqref{mssmyukawa} is reinterpreted in the context of supergravity. Of particular interest is the scalar potential
\begin{eqnarray}
\label{sugrapotential}
V = e^K \left(K^{I \overline{J}} D_I W  D_{\overline{J}} \overline{W} - 3 \vert W \vert^2\right) + \dfrac{1}{2} \mathcal{D}^2,  \\[1mm]
\quad \quad \quad \quad  \textrm{where} \,\, D_I W = \dfrac{\partial W}{\partial \Phi^I}  + \dfrac{\partial K}{\partial \Phi^I} W , \notag
\end{eqnarray}
\noindent which is expressed in terms of the auxiliary fields $F_I = e^{K/2} D_I W$ and $\mathcal{D}^a$. As in the global SUSY case, supersymmetry is spontaneously broken when at least one auxiliary field has a non-zero vev, however this time $V$ is no longer positive semidefinite, therefore solutions with $V=0$ that approximate our Universe can in principle be consistent with broken supersymmetry.

Another feature of supergravity is that it can be formulated in more than four dimensions, in a way that mimics the Kaluza-Klein theory, the primary motivation being the unification of gravity with the other three forces of nature. Being non-renormalisable however, supergravity has to be interpreted as the low-energy limit of a higher structure. In particular, superstring theories lead to effective supergravity theories in 10d, as we will see in later chapters. For this reason, it is useful to establish a specific string model before returning to supergravity and its implications to the phenomenology of particle physics.

\section{String Theory}  
\noindent String Theory is the leading candidate for a quantum theory that unifies all interactions of nature, including gravity. In the framework of string theory, elementary particles are interpreted to be one-dimensional objects called strings, which appear to be point-like only at energies much lower than the string scale $M_s$. These objects can be either open or closed and sweep a two-dimensional surface, called worldsheet, that is parametrised by a space coordinate $\sigma$ and a time coordinate $\tau$. As strings vibrate, each vibrational mode is identified with a fundamental particle, whose mass and quantum numbers are determined by the equations of the string. Overall, the infinite number of oscillation modes gives rise to an infinite tower of particles, but only the zero modes, i.e the massless particles are observable at energies way below $M_s$. In particular, the zero-mode spectrum of a closed string always includes the graviton, which means that gravity is automatically incorporated. Unlike supergravity however, string theory is UV-complete, since there is a UV/IR correspondence between strings at high and low energies, thus allowing all UV divergences to be reinterpreted as IR divergences. Moreover, the Lagrangian of string theory only contains one free parameter $\alpha'$ (historically called Regge slope), which defines the string length $l_s = \sqrt{\alpha '}$ and the string scale $M_s=1/\sqrt{\alpha '}$.

\begin{figure}[h]
\centering
\includegraphics[width=0.6\textwidth]{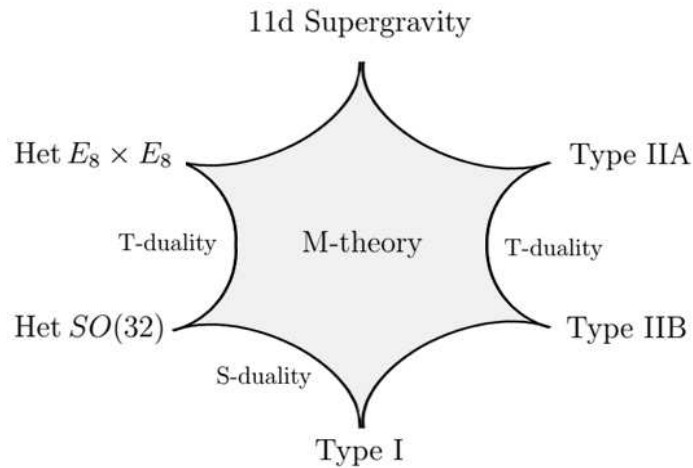}
\caption{The various types of superstring theories arising as lower limits of the M-theory. \cite{kiritsislecture}}
\label{figsuperstrings}
\end{figure}

The first attempt to build a string model was the bosonic string theory, only consistent in $D=26$ dimensions, the reason being conformal anomaly cancellation. Because its spectrum only contains bosons and because of the presence of tachyons (particles with negative mass squared), it was clear from the start that this theory could not represent a description of our Universe. Further development led to the appearance of superstring theory, where both bosons and fermions are present and where the dimension of the spacetime is restricted to $D=10$. No tachyons are present in superstring theory and the spectrum of excitations is governed by supersymmetry. Phenomenologically, this makes superstring theory a possible extension of physics at higher energies. After the first string revolution in 1984, five different superstring theories were developed, namely Type I, Type IIA, Type IIB, heterotic $SO(32)$ and heterotic $E_8 \times E_8$. Type I is based on both open and closed strings, while the other four only contain closed strings. During the second string revolution (1994), it was proved that these theories may be the lower limits of an underlying 11-dimensional theory, called the M-theory (see Figure~\ref{figsuperstrings}). Moreover, the five superstring theories are connected by dualities: for example, Type IIA and Type IIB are T-dual, and so are heterotic $SO(32)$ and heterotic $E_8 \times E_8$. There is also an S-duality between heterotic $SO(32)$ and Type I. We will further discuss in more detail what a heterotic string theory is, since they are the class of string models on which this thesis is based. The reason why they are so important in this context is because they connect with particle physics naturally. One $E_8$ factor of the gauge group $E_8 \times E_8$ contains all the GUT groups in \eqref{gutgroups}. By comparison, IIB has no gauge symmetry and IIA has only $U(1)$ gauge symmetry, thus making them more problematic from a phenomenological perspective, unless D-branes are introduced. 

\subsection{Heterotic string theory}
\noindent The heterotic string theory is one of the most promising candidates for a unified theory of physics. It was first introduced in 1985 as a theory of closed strings with decoupled left- and right-moving modes, where the left sector is defined in 26 dimensions as a bosonic string theory with spacetime coordinates $\lbrace X_L^i(\sigma + \tau)\rbrace_{i=0,...,25} $, while the right sector is defined in 10 dimensions as a superstring theory, with bosonic and fermionic coordinates denoted as $\lbrace X_R^i (\sigma-\tau)\rbrace_{i=0,...,9}$ and $\lbrace\psi^i_R(\sigma-\tau)\rbrace_{i=0,...,9}$ respectively, where $\tau$ and $\sigma$ are worldsheet coordinates of the string. The extra $16$ degrees of freedom in the left sector are regarded as dimensions of an internal compact space, namely a maximal $16\textrm{d}$ torus $T^{16}$ with critical radius $R = \sqrt{\alpha'}$. It is useful to re-label these parameters as $\lbrace X^I(\sigma + \tau)\rbrace_{I=10,...,25}$, and separate them from the rest of the bosonic left-movers $\lbrace X_L^i(\sigma + \tau) \rbrace_{i=0,...,9}$, which are to be combined with their right-moving counterparts in order to form the physical spacetime coordinates of the string in 10 dimensions 
\begin{eqnarray}
X^i(\sigma, \tau) = X_L^i(\sigma + \tau) + X_R^i(\sigma - \tau).
\end{eqnarray}
\noindent In conformity with Ref.~\cite{stringquartet1}, the worldsheet action which characterises the dynamics of the heterotic string can be written as 
\begin{eqnarray}
S = - \int d \tau \int_0^\pi d \sigma \dfrac{1}{4 \pi \alpha '} \left[ \partial_{\alpha} X^i \partial^{\alpha} X^i + \partial_{\alpha} X^I \partial^{\alpha} X^I + \overline{\psi}{}^i_R \Gamma^{\alpha} \partial_{\alpha} \psi_R^i \right],
\end{eqnarray}
\noindent where $\Gamma^\alpha$ ($\alpha=0,1$) are the two-dimensional Dirac matrices satisfying $\lbrace \Gamma^\alpha,\Gamma^\beta\rbrace = 2 \eta^{\alpha \beta}$.

Under the light-cone quantisation, which is introduced to remove all negative-norm states, the coordinates $X^{\pm} = (X^0 \pm X^{9})/\sqrt{2}$ and $\psi^{\pm}_R=(\psi_R^0 \pm \psi_R^{9})/\sqrt{2}$ are gauge-fixed. This has the apparent effect of breaking manifest $SO(1,9)$ Lorentz symmetry down to the rotational subgroup $SO(8)$ of the transverse coordinates $\lbrace X^i \rbrace_{i=1,...,8}$ and $\lbrace \psi_R^i \rbrace_{i=1,...,8}$. Furthermore, under an alternative formulation of heterotic string theory, known as the fermionic construction, the extra 16 degrees of freedom $\lbrace X^I\rbrace_{I=10,...,25}$ are redefined as 32 left-moving spin-$1/2$ Majorana fermions $\lbrace \lambda^A\rbrace_{A=1,...,32}$, whose action according to Ref.~\cite{bbschwarz} is of the form
\begin{eqnarray}
S \sim \int d^2 \sigma \overline{\lambda}{}^A \Gamma^{\alpha} \partial_{\alpha} \lambda^A.
\end{eqnarray}
\noindent Through this formulation, the entire action of the string acquires a global $SO(32)$ symmetry: fermions $\lambda^A$ transform in the fundamental representation, while coordinates $X^i$, $\psi_R^i$ are singlets. For this reason, left-movers can be specified as $SO(8) \times SO(32)$ multiplets, while right movers, which are only affected by the $SO(8)$ rotation, are labeled by their $SO(8)$ quantum numbers. It is intuitive to assume that the global $SO(32)$ symmetry descends locally to a gauge symmetry. In fact, it can be shown that heterotic string theories possess either gauge group $SO(32)$ or $E_8 \times E_8$, depending on the choice of GSO projection. The GSO projection is essential for realising space-time supersymmetry, because it truncates the spectrum on every mass level, so that the bosonic and fermionic degrees of freedom become equal. In particular, the tachyonic ground state of the bosonic sector is completely eliminated. Since we will solely focus on the $E_8\times E_8$ theory, the GSO projection involves splitting the fermions $\lambda^A$ into two groups and imposing a separate projection condition on each set, so that bosonic states $\lambda^A \lambda^B$ form the adjoint representation $(\mathbf{248}, \mathbf{1}) \oplus (\mathbf{1}, \mathbf{248})$ of $E_8 \times E_8$. As for the other heterotic string theory, it is often disregarded by phenomenology, because the adjoint of $SO(32)$ does not contain the spinor representation of $SO(10)$, therefore the connection to SUSY GUTs is harder to realise.

Now, in order to build the massless spectrum of the $E_8\times E_8$ theory, the left-moving and right-moving modes of zero mass have to be paired, so that their tensor product is interpreted as a physical state. The massless states of the right-moving sector are the vector $\mathbf{8}_V$ and the spinor $\mathbf{8}$ representations of $SO(8)$, while in the left-moving sector, we encounter a vector $(\mathbf{8}_V, \mathbf{1})$ and a tensor $(\mathbf{1},\textbf{Adj}_{E_8 \times E_8})$ representation of $SO(8) \times (E_8 \times E_8)$. The combination of the two sectors gives the following spectrum of particles 
\begin{align}
(\mathbf{8}_V,\mathbf{1}) \times (\mathbf{8}_V + \mathbf{8}) & = (\mathbf{1},\mathbf{1}) + (\mathbf{28},\mathbf{1}) + (\mathbf{35},\mathbf{1}) + (\mathbf{56},\mathbf{1}) + (\mathbf{8}',\mathbf{1}), \\
(\mathbf{1},\textbf{Adj}_{E_8 \times E_8})\times (\mathbf{8}_V + \mathbf{8}) & = (\mathbf{8}_V ,\textbf{Adj}_{E_8 \times E_8})+(\mathbf{8} ,\textbf{Adj}_{E_8 \times E_8}),
\end{align}
\noindent which are to be interpreted in the next section as the gravity multiplet $\lbrace \phi, B, g, \psi, \lambda \rbrace$ and the $E_8 \times E_8$ gauge multiplet $\lbrace A^a, \chi^a \rbrace$ of an effective $N=1$ supergravity theory.

\section{The 10d Heterotic $N = 1$ Supergravity}
\label{n=1susy}
\noindent In the low energy limit, where string excitations are much smaller than $M_s$, the heterotic string theory is described by a 10-dimensional $N = 1$ supergravity, which is coupled to a 10-dimensional $E_8 \ \times E_8$ super-Yang-Mills theory. The field content of this theory is given by a gravity multiplet, which contains the metric $g_{M N}$, the NS  two-form $B_{M N}$, and the dilaton $\phi$, as well as their fermionic superpartners, the gravitino $\psi_M$ and the dilatino $\lambda$, and an $E_8 \times E_8$ gauge multiplet, consisting of the vector potential $A_M^a$ and its superpartner, the gaugino $\chi^a$. The field strength associated to the gauge field is defined as $F = dA + A \wedge A$, while the spin connection $\omega$ gives rise to the curvature tensor $R= d\omega + \omega \wedge \omega$. With these notations, using Refs.~\cite{GSW},~\cite{bbschwarz} and~\cite{Polchinski}, the bosonic part of the 10d supergravity action is written up to first order in $\alpha '$ as
\begin{gather}
\label{10daction}
S = \dfrac{1}{2 \kappa^2} \int d^{10} x \sqrt{-g} e^{- 2 \phi} \left( R + 4 (\partial \phi)^2 - \dfrac{1}{2} H^2 - \dfrac{\alpha'}{4} \textrm{Tr} F^2 \right) , \\  \textrm{with }  H = d B - \dfrac{\alpha'}{4} (\omega_{\textrm{YM}} - \omega_{\textrm{L}})\, , \notag
\end{gather}
\noindent where $\kappa$ is the ten-dimensional gravitational coupling constant, while $\omega_{\textrm{YM}}$ and $\omega_{\textrm{L}}$ are the gauge and gravitational Chern-Simons forms, respectively. In a similar way, the fermionic terms of the action are given by Eq.~\eqref{fermionicaction}, however due to supersymmetry, they can be omitted. 

\begin{figure}[h]
\centering
\includegraphics[width=0.7\textwidth]{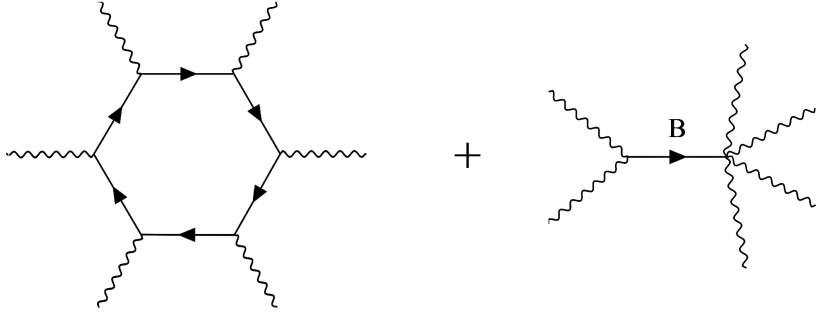}
\caption{Diagram depiction of the Green-Schwarz mechanism. Anomalies arising from chiral fermion loops are canceled by the tree-level exchange of a 2-form $B$.}
\label{figgreenschwarz}
\end{figure}

Before moving on with our discussion, a close examination of anomalies is in order. Since heterotic string theories are chiral, gauge and gravitational anomalies are expected to arise from hexagon loops of chiral fermions. These anomalies are analogous to the triangle diagrams in the Standard Model and their external fields are combinations of gauge bosons and gravitons. Mathematically, the 10d anomaly is represented by a 12-form polynomial, which factorises into a 4-form and an 8-form term for specific gauge groups such as $E_8\times E_8$ and $SO(32)$. In fact, those two groups are the only viable gauge groups for a consistent 10d super-Yang-Mills and supergravity theory primarily because of this factorisation, since anomalies of this form are reducible and can be canceled out completely \cite{fabianruehle}. This is realised through the famous Green-Schwarz mechanism, by introducing counter-terms corresponding to tree-level exchanges of a two-form $B$ between external fields (see Fig.~\ref{figgreenschwarz}). Here, $B$ is known to be the anti-symmetric component of the gravity multiplet, and $H$ is its field strength. It is important to note that the Chern-Simons forms $\omega_{\textrm{YM}} $ and $\omega_{\textrm{L}}$ are added to the expression of $H$ in order for the Green-Schwarz mechanism to take place. By definition, the Chern-Simons forms are expressed as
\begin{eqnarray}
\omega_{\textrm{YM}} =  \textrm{Tr} \left(A \wedge dA + \dfrac{2}{3} A \wedge A \wedge A\right), \,\,\,\,\,\,\,\,\,\,\,\, \omega_{\textrm{L}} =  \textrm{Tr} \left(\omega \wedge d \omega +  \dfrac{2}{3} \omega \wedge \omega \wedge \omega\right)
\end{eqnarray}
\noindent and satisfy $d\omega_{\textrm{YM}} = \textrm{Tr} (F\wedge F)$ and $d \omega_{\textrm{L}} =  \textrm{Tr}(R\wedge R) $. Because of this, the field strength $H$ must obey the modified Bianchi identity 
\begin{eqnarray}
\label{modifiedbianchi}
d H = \dfrac{\alpha'}{4}\left(\textrm{Tr} (R \wedge R) - \textrm{Tr} (F \wedge F)\right) .
\end{eqnarray}
\noindent With this setup, one can finally build an anomaly-free particle model.

In order for supersymmetry to be realised, it is required that the variation of all fields under the SUSY transformation is zero. More exactly, suppose $\epsilon(x^M)$, a spinor of $SO(1,9)$, is the parameter of local $N=1$ supersymmetry with corresponding supercharge $Q$.  Then the operator $Q$ must annihilate the vacuum state $\vert 0 \rangle$, so that $\langle \delta_{\epsilon} \Phi  \rangle \equiv \langle 0\vert \left[ \overline{\epsilon} Q, \Phi  \right]   \vert 0 \rangle   \stackrel{!}{=} 0$ for any field $\Phi $. If, $\Phi$ is bosonic, the equation is trivial, because the supersymmetric variation of a bosonic field is equal to a sum of fermionic fields, whose vevs always vanish, otherwise they would violate Lorentz invariance. Therefore, one only needs to ensure that the variation $\langle \delta_{\epsilon} \Phi \rangle$ vanishes in the case where $\Phi$ is fermionic. Following \cite{GSW}, the supersymmetry variations of the fermionic fields (the gravitino, the dilatino and the gaugino) are given by
\begin{align}
\label{killingspinoreq}
\delta \psi_M & = \dfrac{1}{\kappa} D_M \epsilon + \dfrac{1}{8 \sqrt{3} \kappa} e^{-\phi} \left( \Gamma_{M}{}^{NPQ} - 9 \delta_M^N \Gamma^{PQ}\right) \epsilon H_{NPQ}  + (\textrm{Fermi})^2 ,\notag \\
\delta \chi^a & = - \dfrac{\sqrt{\alpha'}}{4 \sqrt{2} \kappa } e^{-\phi/2} \Gamma^{M N} F_{MN}^a \epsilon + (\textrm{Fermi})^2 , \\
\delta \lambda & = - \dfrac{1}{\sqrt{2}} \left(\Gamma \cdot \partial \phi \right) \epsilon + \dfrac{1}{4 \sqrt{6} \kappa}e^{-\phi} \Gamma^{MNP} \epsilon H_{MNP}  + (\textrm{Fermi})^2. \notag
\end{align}
\noindent Therefore, the Killing spinor equations are obtained by imposing $\langle\delta \psi_M\rangle, \langle\delta \chi^a\rangle, \langle\delta \lambda\rangle \stackrel{!}{=} 0 $. 

\subsection{The compactification ansatz}
\label{compactificationansatz}
\noindent In the remainder of this section, we discuss, in a somewhat informal manner, the preliminary mathematical conditions for a realistic heterotic string model. More technical details regarding the mathematics used here are provided in Section~\ref{mathschapter}.

The goal of constructing 10d heterotic string theories is to connect them to observable 4d physics and, in particular, to $N=1$ SUSY extensions of the Standard Model. This is done by compactifying the six extra dimensions at a compactification scale $M_c=1/l_c$, large enough to escape detection (with $l_c$ being the typical length of the curled up dimensions). The simplest ansatz is to assume that the 10d background is a direct product $M_4 \times X$, where $M_4$ is a 4d maximally symmetric space (Minkowski, de Sitter or anti-de Sitter), as suggested by current cosmological models\footnote{Experimental evidence indicates that the Universe is de Sitter, however the cosmological constant is so small, compared to the string scale $M_s$, that string compactifications with Minkowski space are a good approximation. In addition, string models with AdS vacuum are also compatible with observation, provided that the vacuum is uplifted via the KKLT mechanism.}, and $X$ is a 6d compact manifold with tangent bundle $TX$. To avoid confusion, the 10d coordinates will be labeled as $\lbrace x^{M}\rbrace_{M=0,...,9}$, the 4d external coordinates as $\lbrace x^{\mu}\rbrace_{\mu=0,...,3}$ and the 6d internal coordinates as $\lbrace y^{m}\rbrace_{m=4,...,9}$. With these notations, every field can be decomposed explicitly into external and internal components. For example, the background metric (i.e. the vev of $g_{MN}$) becomes block diagonal
\begin{eqnarray}
d s^2 = g_{\mu \nu} dx^{\mu} dx^{\nu} + g^{(6)}_{m n} dy^m dy^n, 
\end{eqnarray}
\noindent as the Lorentz group $SO(1,9)$ breaks down to $SO(1,3)\times SO(6)$. Any non-diagonal perturbation $\delta g_{\mu n}$ is forbidden to acquire a vev, since it transforms as a vector under $SO(1,3)$ and would therefore violate Lorentz invariance. In more general cases, warp factors $A(y^m)$ are introduced by fluxes to modify the 4d metric to $e^{2 A(y)}g_{\mu \nu}$. Although useful in the context of moduli-stabilisation, such factors will not be considered here.

In addition to specifying a spacetime ansatz, one needs to break the gauge group $E_8 \times E_8$ down to the gauge group $H_{4d}$ of a 4d theory. This is achieved by turning on background values for the internal components of gauge fields $A^a_m$, which are interpreted as connections of a vector bundle $V \rightarrow X$, with structure group $G \subset E_8$. The effect is that one $E_8$ factor splits as $G \times H_{4d}$, provided that $H_{4d}$ is the commutant of $G$ in $E_8$. The other $E_8$ is considered to be in the ``hidden sector", which couples only gravitationally to the physical theory and therefore its effects are negligible. If $\tilde{V}\rightarrow X$ is the hidden sector vector bundle, with structure group in $E_8$, then the complete $E_8 \times E_8$ bundle over $X$ is given by $V \oplus \tilde{V}$. In most calculations however, we will assume $\tilde{V}$ to be trivial.

The choice of internal manifold $X$ and vector bundle $V$ is not arbitrary. As argued in Section~\ref{susymssm}, preserving $N=1$ SUSY in 4d is important for phenomenology and therefore certain restrictions have to be imposed. Supersymmetry is needed at low energies in order to stabilise the Higgs mass and we only want $N=1$, rather than $N \geq 2$ supersymmetry, because the theory has to contain chiral fermions. 

\subsection{Conditions for 4d $N = 1$ supersymmetry}
\label{conditionsforn=1susy}
\noindent Finding the conditions for 4d $N = 1$ SUSY amounts to applying the compactification ansatz $M_4 \times X$ to the Killing spinor equations in \eqref{killingspinoreq}. One looks for a solution which preserves precisely $1/4$ of the original $16$ supercharges. In order to simplify the discussion, we will assume that the vev of the dilaton $\phi$ is a constant and the vev of the field strength $H$ vanishes. Under these assumptions, the equation for the dilatino is automatically satisfied. However, the equations corresponding to the other two fermionic fields, the gravitino and the gaugino, are non-trivial and can be recast in the following form
\begin{align}
\label{covconst}
\langle \delta_{\epsilon} \psi_M \rangle & =  \nabla_M \epsilon \stackrel{!}{=} 0 \\
\label{gaugino}
 \langle \delta_{\epsilon} \chi^a\rangle & = \Gamma^{m n} F^a_{m n} \epsilon \stackrel{!}{=} 0 .
\end{align}
\noindent Here, the SUSY parameter $\epsilon$ is a $\mathbf{16}$ Majorana-Weyl spinor of $SO(1,9)$, which breaks into $(\mathbf{2}, \mathbf{4}) \oplus (\mathbf{2}', \overline{\mathbf{4}})$ under $SO(1,3)\times SO(6)$. It is convenient to express $\epsilon$ as $\eta \otimes \xi$, where $\eta$ is the external spinor $\mathbf{2}$ and $\xi$ is the internal spinor $\mathbf{4}$. The requirement of \eqref{covconst} that $\epsilon$ is covariantly constant means that both $\nabla_{\mu} \eta$ and $\nabla_m \xi$ must vanish. Using the relation
\begin{eqnarray}
\left[ \nabla_{\mu}, \nabla_{\nu}\right]\eta = \dfrac{1}{4} R_{\mu\nu\rho\sigma}\gamma^{\rho\sigma} \eta,
\end{eqnarray}
\noindent and the fact that, for a maximally symmetric space, the Riemann curvature tensor is
\begin{eqnarray}
R_{\mu\nu\rho\sigma} = \dfrac{R}{12} \left( g_{\mu\rho} g_{\nu\sigma} - g_{\mu\sigma} g_{\nu\rho} \right),
\end{eqnarray}
\noindent one can see that the only way for $\nabla_{\mu} \eta$ to be zero is when the 4d manifold $M_4$ is flat (Minkowski)\footnote{To our knowledge, no dS vacuum was ever achieved in heterotic models. On the other hand, up-lifting to dS is claimed to be done in KKLT IIB models using anti-D3-branes. \cite{kkltref}}.

In a similar way, the condition $\nabla_m \xi \stackrel{!}{=}0$ implies that $R_{m n p q} \Gamma^{p q} \xi$ must vanish, but without a maximally symmetric restriction, the internal manifold $X$ is not required to be flat, only Ricci-flat, i.e. $R_{m n} = 0$. In addition to that, the holonomy group of $X$ has to ensure that the spinor $\xi$ stays invariant under parallel transport. The most general holonomy of a 6d manifold is $SO(6) \simeq SU(4)$, however in our case it has to be a subgroup $\textrm{Hol}(X) \subset SU(4) $ such that every element $U \in \textrm{Hol}(X)$ satisfies $U \xi = \xi$. One can easily see that the largest subgroup with this property is $SU(3)$, as $\xi$ can be rotated to take the form $(0,0,0,\xi_0)^T$, so that $SU(3)$ transformations acting on the first three components leave $\xi$ as a singlet. In fact, $SU(3)$ and discrete subgroups of $SU(3)$, which will not be considered here\footnote{For simplicity of the model, we assume the holonomy group of $X$ is $SU(3)$, however in orbifold constructions, a discrete subgroup of $SU(3)$ can be considered.}, are the only viable picks, because smaller subgroups of $SU(4)$ would allow too many supercharges in the 4d theory. Therefore, in addition to being Ricci-flat, $X$ must have precisely $SU(3)$ holonomy. Finding manifolds with such properties is in principle not an easy task. Luckily, a conjecture by Eugenio Calabi \cite{calabiconj1, calabiconj2}, followed by a proof by Shing-Tung Yau \cite{yautheorem} led to the following theorem
\begin{theorem} [Yau's theorem]
\label{yautheorem}
A compact, 2n-dimensional K\"ahler manifold with vanishing first Chern class admits a unique Ricci-flat K\"ahler metric, for each given K\"ahler class.
\end{theorem}
\noindent Here, K\"ahler means complex manifold with holonomy $U(n)$ and the first Chern class $c_1(TX)$ is a topological invariant given by the cohomology class $[R]$ of the Ricci form
\begin{eqnarray}
\label{firstchernclass}
c_1 (TX) \equiv \dfrac{1}{2 \pi} [R] \stackrel{!}{=} 0.
\end{eqnarray}
\noindent Such a manifold is called Calabi-Yau (CY) and its holonomy group is proven to be contained in $SU(n)$. This is because K\"ahler manifolds already have reduced anomaly $U(n) \simeq SU(n) \times U(1)$. The $U(1)$ factor, generated by the Ricci tensor, then vanishes if the Ricci-flat condition is imposed. In Section~\ref{cysection}, we will study the properties of Calabi-Yau manifolds in more detail, in particular the Calabi-Yau threefolds, which have $SU(3)$ holonomy and are therefore suitable for compactification.

The other condition for unbroken supersymmetry, \eqref{gaugino}, can be recast in complex coordinates in the form of the Hermitian Yang-Mills equations
\begin{eqnarray}
\label{hermitianyangmills}
g^{a \overline{b}} F_{a \overline{b}} = 0, \,\,\,\,\,\,\,\,\,\,\,\,\, F_{a b} = F_{\overline{a} \overline{b}} = 0 .
\end{eqnarray}
\noindent The second equation, $ F_{a b} = F_{\overline{a} \overline{b}} \stackrel{!}{=} 0$, can be satisfied for a suitable connection $A$, if the vector bundle $V$ is holomorphic (i.e. the transition functions are holomorphic maps). According to the Donaldson-Uhlenbeck-Yau (DUY) theorem \cite{duycitation1, duycitation2}, there exists a connection for which $g^{a \overline{b}} F_{a \overline{b}} \stackrel{!}{=} 0$ also holds true, if $V$ is poly-stable and has vanishing slope. Mathematically, the notion of stability is introduced by defining the slope of a bundle $V$ as
\begin{eqnarray}
\label{slopedefinition}
\mu(V) \equiv \dfrac{1}{\textrm{rk}(V)} \int_X c_1(V) \wedge J \wedge J,
\end{eqnarray} 
\noindent where $J$ is the K\"ahler form on $X$ and $c_1(V)$ and $\textrm{rk}(V)$ are the first Chern class and the rank of the bundle. Then $V$ is called stable if the slope satisfies $\mu(F) < \mu(V)$, for all sub-sheafs $F \subset V$ with rank $0<\textrm{rk}(F)<\textrm{rk}(V)$, and poly-stable if it is decomposable as a direct sum of stable bundles $V= \bigoplus_{i=1}^n U_i$, with slopes $\mu(U_i) = \mu(V)$. 

As for the zero-slope condition, it is automatically satisfied if we assume that 
\begin{eqnarray}
\label{c1Viszero}
c_1 (V) \equiv \dfrac{i}{2 \pi} [\textrm{Tr} F] = 0,
\end{eqnarray}
\noindent which is equivalent to saying that the structure group of $V$ is special unitary. This is needed to ensure that the structure group of $V$ embeds into $E_8$, and in addition to that, special unitary groups such as $SU(5)$, $SU(4)$ and $SU(3)$ give rise to the GUT groups $SU(5)$, $SO(10)$ and $E_6$ respectively, in the 4d theory. In general however, $U(n)$ vector bundles with $c_1(V) \neq 0$ can also lead to phenomenologically interesting compactifications \cite{timoweigandunitary}, although they will not be the subject of this thesis.

In conclusion, in order for $N=1$ supersymmetry to be preserved at lower energies, the 4d space must be flat, the internal manifold must be Calabi-Yau and the vector bundle must be holomorphic and poly-stable. 

\subsection{Conditions for anomaly cancellation}
\noindent The background geometry is further constrained by the anomaly cancellation condition. Since the left-hand side of \eqref{modifiedbianchi} is exact, $\textrm{Tr} (R \wedge R)$ and $\textrm{Tr} (F \wedge F)$ must be in the same cohomology class, thus leading to a topological identity
\begin{eqnarray}
\label{anomaly2ndchern}
\textrm{ch}_2(TX) = \textrm{ch}_2 (V)+\textrm{ch}_2 (\tilde{V})
\end{eqnarray}
\noindent between the tangent bundle of $X$ and the $E_8 \times E_8$ gauge bundle (here $\textrm{ch}_2$ denotes the second Chern character). In certain theories, the vacuum is altered by the presence of 5-branes, so the anomaly condition becomes
\begin{eqnarray}
\label{ch2ch2ch2w}
\textrm{ch}_2(TX) - \textrm{ch}_2 (V) - \textrm{ch}_2 (\tilde{V}) = W,
\end{eqnarray}
\noindent with $W$ being the homology class of the two-cycles (curves) in $X$, around which the 5-branes wrap. Preservation of supersymmetry requires these cycles to be holomorphic, and consequently $W$ to be effective, i.e. an element of the Mori cone $\lbrace \sum_i a_i [C_i]$, where $C_i \subset X \textrm{ are holomorphic curve representatives}$, and $a_i \in \mathbb{R}^+ \rbrace$. Provided that the left-hand side of \eqref{ch2ch2ch2w} is effective, one can construct anomaly-free models by wrapping 5-branes on the relevant cycles. If we assume $\tilde{V}$ is trivial and use $c_1(TX) = c_1(V) = 0$, we obtain 
\begin{eqnarray}
\label{c2c2W}
c_2(TX) - c_2(V) = W,
\end{eqnarray}
\noindent for $W\in H_2(X,\mathbb{Z})$, an effective class.

\section{Mathematical Ingredients for Compactification}
\label{mathschapter}
\noindent In order to have a proper understanding of Calabi-Yau manifolds and holomorphic vector bundles (the main ingredients of compactification), one needs to be familiarised with concepts from complex geometry and Hodge theory. In this section we will briefly discuss these topics and then proceed to describe a particular class of CY manifolds that is used in this thesis, namely the Complete Intersection Calabi Yau manifolds (CICYs).

\subsection{Complex manifolds}
\label{complexmanifoldssection}
\noindent 
\begin{definition}
An $n$-dimensional complex manifold $M$ is a $2 n$-dimensional real manifold that resembles the complex flat space $\mathbb{C}^n$ locally. 
\end{definition}
\noindent In order to satisfy this, $M$ must be equipped with an atlas of charts $( U_{\alpha}, \phi_{\alpha})$, where $ \lbrace U_{\alpha} \rbrace$ are open subsets that cover $M$, and every map $\phi_{\alpha}: U_{\alpha} \rightarrow \mathbb{C}^n$ must be a homeomorphism. Moreover, for any two given subsets $U_{\alpha}$ and $U_{\beta}$ that satisfy $U_{\alpha}\cap U_{\beta} \neq \emptyset$, the associated transition map $\psi_{\beta \alpha } \equiv \phi_{\beta} \circ \phi^{-1}_{\alpha}$, \ $ \psi_{\beta \alpha }: \phi_{\alpha}(U_{\alpha}\cap U_{\beta})\rightarrow \phi_{\beta}(U_{\alpha}\cap U_{\beta})$ must be holomorphic. This is to ensure that the tangent space can be complexified with respect to some projection operators, such that $T_p M = T_p M^{+} \oplus T_p M^{-} $ and every vector field is decomposable into a holomorphic and an anti-holomorphic component. It has been shown that 
\begin{theorem}
An even-dimensional real manifold is complex if and only if it is endowed with a globally defined almost complex structure $I_{a}{}^{b}$, satisfying $I_{a}{}^{b} I_{b}{}^{c} = - \delta_{a}^{c}$, and the Niejenhuis tensor $N_{a b}{}^c \equiv I_{[a}{}^c{}_{;b]} - I_{[a}{}^d I_{b]}{}^e I^c{}_{d;e}$ vanishes. 
\end{theorem}
\noindent This makes it possible for local complex coordinates $z^a$ and $\overline{z}^{\overline{a}}$ to be introduced, so that on every patch, one has $I_{a}{}^{b} = i \delta_{a}{}^{b}$, $I_{\overline{a}}{}^{\overline{b}} = - i \delta_{\overline{a}}{}^{\overline{b}}$ and $I_{a}{}^{\overline{b}} = I_{\overline{a}}{}^{b}=0$. Any complex manifold admits a metric of the form $d s^2 = g_{a \overline{b}} d z^{a} d \overline{z}^{\overline{b}}$, which is called hermitian. It can be used to define the fundamental 2-form
\begin{eqnarray}
\label{kahlerform}
J = i g_{a \overline{b}} d z^{a} \wedge d \overline{z}^{\overline{b}},
\end{eqnarray}
\noindent by lowering one index of the complex structure, i.e. $J_{a \overline{b}} = I_a{}^c g_{c \overline{b}}$. 
\begin{definition}
A K\"ahler manifold is a complex manifold with hermitian metric, for which the form $J$ is closed.
\end{definition}
\noindent In this case, $J$ is called a K\"ahler form. The restriction $d J = 0$ ensures that the only non-zero coefficients of the Levi-Civita connection are $\Gamma_{a b}^c$ and $\Gamma_{\overline{a} \overline{b}}^{\overline{c}}$, thus preserving holomorphicity under parallel transport, so the holonomy group is $U(n)$. Another important feature of K\"ahler geometry is that on every patch $U_{\alpha}$, we can define a real-valued function $K_{\alpha}$, known as the K\"ahler potential, which specifies the metric
\begin{eqnarray}
\label{kahlermetricfrompotential}
g_{a \overline{b}} = \partial_a \partial_{\overline{b}} K_{\alpha}. 
\end{eqnarray}
\noindent The K\"ahler potential is not unique and, on the intersection of patches, two K\"ahler potentials are related by $K_{\alpha} (z, \overline{z}) = K_{\beta}(z, \overline{z}) + f_{\alpha \beta} (z) + \overline{f}_{\alpha \beta} (\overline{z})$, where $f$ and $\overline{f}$ are a holomorphic and an anti-holomorphic function. Further properties stem from the computation of curvature tensors and, in particular, the Ricci form $R \equiv i R_{a \overline{b}} d z^a \wedge d \overline{z}^b = i \partial \overline{\partial} \textrm{log} ({g}^{1/2}) $, which is closed, globally defined and determines the first Chern class. We conclude our discussion of K\"ahler manifolds with a short look on complex projective spaces.

\begin{example}
As a subclass of compact K\"ahler manifolds, complex projective spaces $\mathbb{CP}^n$ or, shortly, $\mathbb{P}^n$ will be the building blocks for our Calabi-Yau manifolds. These spaces are obtained by identifying points in $\mathbb{C}^{n+1}\setminus \lbrace 0 \rbrace$ according to the equivalence relation $(x_0,...,x_{n}) \sim \lambda (x_0,...,x_{n})$, for $\lambda \in \mathbb{C}^*$, so that every element $(x_0: ...: x_n) \in \mathbb{P}^n$ corresponds to a line through the origin. The numbers $x_\alpha$ are called homogeneous coordinates on $\mathbb{P}^n$. On each open patch $U_{\alpha} = \lbrace (x_0:...:x_{n}) \vert x_{\alpha} \neq 0 \rbrace$, we can define new parameters $\xi^{\alpha}_{\mu} \equiv x_{\mu}/x_{\alpha} $ ($\mu \neq \alpha$), which map $U_{\alpha}$ to $\mathbb{C}^n$ and are called inhomogeneous coordinates. The transition functions on $U_{\alpha}\cap U_{\beta}$ overlaps are simply multiplications by $x_{\alpha}/x_{\beta}$ so they are holomorphic. In line with \eqref{kahlermetricfrompotential}, the Fubini-Study K\"ahler potential is introduced on every patch $U_{\alpha}$
\begin{eqnarray}
K_{\alpha} = \dfrac{i}{2 \pi} \textrm{ln} \kappa_{\alpha} \,\textrm{,}\,\,\,\,\,\,\,\,\,\,\, \textrm{where } \kappa_{\alpha} \equiv \sum_{\substack{\mu = 0}}^{n} \vert \xi_{\mu}^{\alpha}\vert^2 ,
\end{eqnarray}
\noindent which allows one to calculate the metric and the K\"ahler form $J= i \partial \overline{\partial} K_{\alpha}$. In particular, the Ricci tensor $R=-(n+1)J$ shows that projective spaces are not Calabi-Yau, even though many well-known Calabi-Yau manifolds are submanifolds of projective spaces.
\end{example}

\subsection{Hodge theory}
\label{hodgetheorychapter}
\noindent Hodge theory is an area of algebraic geometry that studies cohomology groups. The reason why it is important to compactification is that there is a correlation between the topology of the internal manifold and the low-energy spectrum of particles. In this chapter we will analyse the most basic types of cohomologies: de Rham and Dolbeault. Section~\ref{vectorbundleschapter} will introduce vector bundle cohomologies, which are more elaborate. We start by defining $\Omega^p(M)$, as the set of all $p$-forms that live on a Riemannian $n$-dimensional manifold $M$, and we assume that $M$ is compact and without a boundary. The exterior derivative is an operator $d:\Omega^p(M)\rightarrow \Omega^{p+1}(M)$, so that 
\begin{eqnarray}
d \omega_p = \dfrac{1}{p!} \partial_{i_0} \omega_{i_1 ... i_p} dx^{i_0} \wedge dx^{i_1} \wedge ... \wedge dx^{i_p}, \,\,\,\,\,\,\,\,\,\, \omega_p \in \Omega^{p}(M).
\end{eqnarray}
\noindent A $p-$form $\omega_p$ is called closed if $d \omega_p = 0$ and exact if $\omega_{p} = d \nu_{p-1}$, for some $\nu_{p-1} \in \Omega^{p-1}(M)$. Since $d^2 = 0$, all exact forms are also closed, while closed forms are not necessarily exact (although on local patches, they can be expressed as $\omega_{p} = d \nu_{p-1}$).
\begin{definition}
Let $Z^{p}$ be the set of closed $p$-forms on $M$, and $B^{p}$ the set of exact $p$-forms on $M$. Then the quotient $H^p(M)=Z^p/B^p$ is called the $p$th de Rham cohomology group.
\end{definition}
\noindent The elements of $H^p(M)$ are cohomology classes $[\omega_p]$, obtained through setting the equivalence relation $\omega_p \sim \omega_p + d \nu_{p-1}$, and $\omega_p$ is called the representative of the class. The dimension of $H^p(M)$ is a topological invariant, referred to as the Betti number $b_p$. It is useful to define the Hodge star operator $\ast: \Omega^p(M) \rightarrow \Omega^{n-p}(M)$
\begin{eqnarray}
\ast (d x^{i_1} \wedge ... \wedge d x^{i_p}) = \dfrac{\sqrt{\vert g \vert}}{(n-p)!}\epsilon^{i_1...i_p}{}_{i_{p+1} ... i_{n}} dx^{i_{p+1}} \wedge ... \wedge dx^{i_n},
\end{eqnarray}
\noindent in order to introduce the inner product on $p$-forms
\begin{eqnarray}
(\alpha, \beta) = \int_M \alpha \wedge \ast \beta, \,\,\,\,\,\,\,\,\,\,\, \textrm{where} \,\,\,\, \alpha, \beta \in \Omega^p(M),
\end{eqnarray}
\noindent as well as the adjoint exterior derivative $d^{\dagger}:\Omega^p(M)\rightarrow \Omega^{p-1}(M)$, $d^{\dagger}= (-1)^{n p + n+1} \ast d \ast $, for which $(\alpha_p,d \beta_{p-1}) = (d^\dagger \alpha_p, \beta_{p-1})$, and the Laplace operator $\Delta = d d^{\dagger} + d^{\dagger} d$.
\begin{definition}
A form $\gamma_p$ is said to be harmonic if $\Delta \gamma_p = 0$. 
\end{definition}
\noindent Using the inner product, one can prove that on a compact manifold without a boundary, $\gamma_p$  is harmonic if and only if $d \gamma_p = d^{\dagger} \gamma_p = 0$. Moreover, the Hodge decomposition theorem states that for any $p$-form $\omega_p$, there is a unique decomposition $\omega_p = \gamma_p + d \alpha_{p-1} + d^{\dagger} \beta_{p+1}$, where $\gamma_p$ is the corresponding harmonic $p$-form. The consequence of this is that every cohomology class contains precisely one harmonic representative. The Betti number $b_p$ is therefore identical with the number of linearly independent harmonic $p$-forms on $M$. In particular, if $\gamma_p$ is harmonic, then $\ast \gamma_p$ is also harmonic, which means that $H^p(M)$ and $H^{n-p}(M)$ are isomorphic and $b_p=b_{n-p}$ (the Poincar\'e duality). 
\vspace{10mm}

  Moving on to complex manifolds, we define $M$ as an $n$-dimensional compact complex manifold and $\Omega^{p,q}(M)$ as the space of $(p,q)$-forms on $M$. The exterior derivative can be split into $d = \partial + \bar{\partial}$, where $\partial$ and $\bar{\partial}$ are called Dolbeault operators, acting separately as $\partial: \Omega^{p,q}(M)\rightarrow \Omega^{p+1,q}(M)$ and $\bar{\partial}: \Omega^{p,q}(M)\rightarrow \Omega^{p,q+1}(M)$. They satisfy $\partial^2 = \bar{\partial}^2 =0$. As before, a $(p,q)$-form $\omega_{p,q}$ is $\bar{\partial}$-closed if $\bar{\partial} \omega_{p,q} = 0$ and $\bar{\partial}$-exact if $ \omega_{p,q} = \bar{\partial} \nu_{p,q-1}$ for some $\nu_{p,q-1} \in \Omega^{p,q-1}(M)$.
\begin{definition}The $(p,q)$th Dolbeault cohomology is defined as the quotient $H_{\bar{\partial}}^{p,q}(M) = Z^{p,q}_{\bar{\partial}}(M)/B^{p,q}_{\bar{\partial}}(M)$, where $Z^{p,q}_{\bar{\partial}}(M)$ and $B^{p,q}_{\bar{\partial}}(M)$ are the sets of $\bar{\partial}$-closed and $\bar{\partial}$-exact $(p,q)$-forms, respectively.
\end{definition}
\noindent The dimension of $H_{\bar{\partial}}^{p,q}(M)$ is called the Hodge number $h^{p,q}$. On complex manifolds with hermitian metric, an inner product between $(p,q)$-forms is introduced
\begin{eqnarray}
(\alpha, \beta) = \int_M \alpha \wedge \overline{\ast} \beta, \,\,\,\,\,\,\,\,\,\,\, \textrm{where} \,\,\,\, \alpha, \beta \in \Omega^{p,q}(M) \,\,\,\,\,\textrm{and} \,\,\,\,\, \overline{\ast} \beta \equiv \ast \bar{\beta},
\end{eqnarray}
\noindent which allows one to define adjoint operators $\partial^{\dagger}$, $\bar{\partial}^{\dagger}$ and Laplace operators 
\begin{eqnarray}
\Delta_{\partial} = \partial\partial^{\dagger} + \partial^{\dagger} \partial, \,\,\,\,\,\,\,\,\,\,\,\,\,\,\,\,\,\,\Delta_{\bar{\partial}} = \bar{\partial} \bar{\partial}^{\dagger} + \bar{\partial}^{\dagger} \bar{\partial} .
\end{eqnarray}
\noindent A $(p,q)$-form $\gamma_{p,q}$ is said to be $\bar{\partial}$-harmonic, if it is annihilated by the Laplacian $\Delta_{\bar{\partial}}$. One can prove that this is equivalent to $\bar{\partial} \gamma_{p,q} = 0$ and $\bar{\partial}^{\dagger} \gamma_{p,q} = 0$. Moreover, $\bar{\partial}$-harmonic forms are in one-to-one correspondence with the cohomology classes of $H^{p,q}_{\overline{\partial}}(M)$, due to the unique Dolbeault decomposition $\omega_{p,q} = \gamma_{p,q} + \bar{\partial} \alpha_{p, q-1} + \bar{\partial}^{\dagger}\beta_{p, q+1}$ for any form $\omega_{p,q}$, where $\gamma_{p,q}$ is $\bar{\partial}$-harmonic. This is in analogy with the de Rham case, although in general the de Rham and Dolbeault cohomologies are not related: de Rham is purely topological, while Dolbeault depends on the specific choice of complex structure. Only on compact K\"ahler manifolds is there an explicit relation, namely through the identity $\Delta = 2\Delta_{\bar{\partial}}$, which ensures that
\begin{eqnarray}
H^r(M) = \bigoplus_{p+q=r} H^{p,q}_{\bar{\partial}}(M), \,\,\,\,\,\,\,\,\,\,\,\,\,\,\,\,\,\,\,\,\,\,\, b_{r} = \sum_{p+q=r} h^{p,q}. 
\end{eqnarray}
\noindent In this case, the Dolbeault cohomology is considered quasi-topological, because it only depends on complex structure and not on the choice of K\"ahler metric \cite{GSW} \cite{dominicjoyce}. The corresponding Hodge numbers are constrained to satisfy
\begin{eqnarray}
\label{hodgerule1}
  h^{p,q} = h^{q,p} \textrm{  (Hodge symmetry)},   \,\,\,\,\,\,\,\,\,\,\,\,\,\,\,\, h^{p,q} = h^{n-p,n-q} \textrm{  (Serre duality)}. 
\end{eqnarray}
\noindent Now, on a compact K\"ahler manifold, the K\"ahler form  \eqref{kahlerform} is not just closed, but can be chosen to be co-closed (and therefore harmonic), so it can be expanded in a basis of harmonic $(1,1)$-forms $\lbrace \omega_i \rbrace$ as
\begin{eqnarray}
\label{kahlercone}
J = \sum^{h^{1,1}}_{i=1} t^i \omega_i,
\end{eqnarray}
\noindent where the parameters $t^i$ give the K\"ahler cone, i.e. the set of possible K\"ahler forms on $M$. Such parameters are constrained by the requirement that $\int_M J^n > 0$, since $J^n$ is proportional to the volume element, and also by $\int_{C_1} J > 0$, $\dots$, $\int_{C_{n-1}} J^{n-1}>0$, where $C_i$ is an $i$-cycle. It is often possible and convenient to choose a basis $\lbrace \omega_i \rbrace$, such that the K\"ahler cone is $t^i > 0$. On a final note, the Ricci form is also closed, but not necessarily harmonic, and it defines the cohomology class $c_1(TM) \in H_{\bar{\partial}}^{1,1}(M)$. If $c_1(TM)$ is trivial (i.e. $R$ is exact), the compact K\"ahler manifold is called Calabi-Yau.

\subsection{Calabi-Yau manifolds}
\label{cysection}

As mentioned before, a Calabi-Yau $n$-fold is an $n$-dimensional compact K\"ahler manifold with vanishing first Chern class \eqref{firstchernclass}. On such a manifold, Theorem~\ref{yautheorem} (Yau's Theorem) guarantees the existence of a unique Ricci-flat metric. Moreover, the metric has special holonomy group $SU(n)$, which means that the manifold admits covariantly constant spinors $\xi$, $\overline{\xi}$ of opposite chirality. Simple examples of Calabi-Yau $n$-folds include the complex elliptic curve, i.e. the two-torus $T^2$ ($n=1$) and the $K3$ surfaces ($n=2$). However, in conformity with our compactification ansatz in Section~\ref{conditionsforn=1susy}, we require that $X$ is a Calabi-Yau threefold. Moreover, $X$ must have precisely $SU(3)$ holonomy and not a subgroup thereof, so trivial cases such as $T^6$ or $T^2 \times K3$ are excluded.
\begin{theorem}
A compact K\"ahler threefold is Calabi-Yau if and only if it admits a nowhere vanishing holomorphic $(3,0)$-form $\Omega$.
\end{theorem}
\noindent More explicitly, a holomorphic $(3,0)$-form can be written as $\Omega_{m n r} = f(z) \epsilon_{m n r}$ on every patch, where $\epsilon_{m n r}$ is the Levi-Civita symbol and $f(z)$ is a nowhere vanishing holomorphic function. This ensures that the Ricci form is exact, so the first Chern class vanishes. Conversely, for any Calabi-Yau manifold, one can construct a nowhere vanishing $(3,0)$-form $\Omega$, using the covariantly constant spinor $\xi$ and the $\gamma$-matrices
\begin{align}
\label{holomorphic30}
\Omega_{m n r} = \xi^{T}\gamma_{m n r} \xi. 
\end{align}
\noindent It can be proven that $\Omega$ is harmonic and (up to a constant) unique. The corresponding cohomology class $[\Omega]$ is therefore the only element in $H^{3,0}(X)$, so $h^{3,0} = 1$. In addition to that, the uniqueness of $\Omega$ sets a duality
\begin{eqnarray}
\label{hodgerule2}
h^{0,q} = h^{0,3-q}, \,\,\,\,\,\,\,\,\,\,\,\,\,\,\,\, h^{p,0} = h^{3-p,0},
\end{eqnarray}
 because for any given class $[\alpha] \in H^{0,q} (X)$, there exists a unique class $[\beta] \in H^{0,3-q}(X)$, such that $\int_X \Omega \wedge \alpha \wedge \beta  = 1$. The rules listed in \eqref{hodgerule1}, together with the fact that $h^{0,0} = 1$ for a connected manifold, and $h^{0,1} = 0$ for strictly $SU(3)$ holonomy,\footnote{The reason behind this is that spinors on a CY manifold correspond to $(0,k)$-forms, for $k=0,...,3$, due to Dirac matrices acting like creation and annihilation operators: $\lbrace \gamma^m, \gamma^n\rbrace = \lbrace \gamma^{\overline{m}}, \gamma^{\overline{n}}\rbrace=0$ and $\lbrace \gamma^{m}, \gamma^{\overline{n}}\rbrace = 2 g^{m\overline{n}}$. In particular, the states $\gamma^{\overline{m}}\xi$ and $\gamma^{\overline{m}}\gamma^{\overline{n}}\xi$ get multiplied to $(0,1)$ and $(0,2)$-forms respectively, and as they are not allowed to transform trivially under $SU(3)$, the restriction $h^{0,1}=h^{0,2}=0$ has to be imposed.} show that the only unconstrained Hodge numbers remain $h^{1,1}$ and $h^{2,1}$. This information is best encoded by the Hodge diamond diagram 
 \begin{eqnarray}
 \label{hodgediamond}
\begin{tabular}{p{1mm}p{1mm}p{1mm}p{1mm}p{1mm}p{1mm}p{1mm}p{1mm}p{1mm}}
& & & $h^{0,0}$ & & & \\
& &$h^{1,0}$ & & $h^{0,1}$ & & \\
&$h^{2,0}$ & & $h^{1,1}$ & &$h^{0,2}$ & \\
$h^{3,0}$ & &$h^{2,1}$ & & $h^{1,2}$ & & $h^{0,3}$ \\
&$h^{3,1}$ & & $h^{2,2}$ & &$h^{1,3}$ & \\
& &$h^{3,2}$ & & $h^{2,3}$ & & \\
& & & $h^{3,3}$ & & & \\
\end{tabular}
\,\,\,\,\,\,\,\,
=
\begin{tabular}{p{1mm}p{1mm}p{1mm}p{1mm}p{1mm}p{1mm}p{1mm}p{1mm}p{1mm}}
& & & $\,1$ & & & \\ 
& &$ \,0$ & & $\,0$ & & \\ 
&$\, 0$ & & $h^{1,1}$ & &$\,0$ & \\ 
$\, 1$ & &$h^{2,1}$ & & $h^{2,1}$ & & $\,1$ \\ 
&$\,0$ & & $h^{1,1}$ & &$\,0$ & \\ 
& &$\,0$ & & $\,0$ & & \\
& & & $\,1$ & & & \\
\end{tabular} \,\,.
\end{eqnarray}
\noindent Additionally, one can introduce the Euler number, which simplifies to 
\begin{eqnarray}
\chi \equiv \sum_{p=0}^6 (-1)^p b_p = 2 \left(h^{1,1}-h^{2,1}\right).
\end{eqnarray}
\noindent Being K\"ahler, Calabi-Yau manifolds admit a metric $d s^2 = g_{m \overline{n}} d z^{m} d \overline{z}^{\overline{n}}$ with a corresponding K\"ahler form $J = i g_{m \overline{n}}  d z^{m} \wedge d \overline{z}^{\overline{n}}$ that is closed and expanded in a basis of harmonic $(1,1)$-forms $\lbrace \omega_i \rbrace$ as in \eqref{kahlercone}. Integrating $J \wedge J \wedge J$ over $X$ gives the volume $\mathcal{V}$ of the manifold
\begin{eqnarray}
\label{cyvolume}
\int_X J \wedge J \wedge J   = d_{ijk} t^i t^j t^k = 6 \mathcal{V} ,
\end{eqnarray}
\noindent where $d_{i j k}$ are the intersection numbers
\begin{eqnarray}
\label{intersectionnumbers}
d_{ijk} = \int_X \omega_i \wedge \omega_j \wedge \omega_k.
\end{eqnarray}
\noindent It is obvious from \eqref{kahlercone} that the choice of K\"ahler form is determined by $h^{1,1}$ real parameters $t^i$. In a similar way, the complex structure is specified by $h^{2,1}$ complex parameters $z^a$, as seen by expanding the $(2,1)$-form $I_{m n\overline{r}} = \Omega_{mns} I^s{}_{\overline{r}}$ in a basis of harmonic forms. Consequently, the Hodge numbers $h^{1,1}$ and $h^{2,1}$ describe the topology of the manifold in two separate ways: $h^{1,1}$ counts the independent size deformations that keep the complex structure invariant, while  $h^{2,1}$  counts deformations of shape. An important observation, which led to the concept of mirror symmetry, was that Calabi-Yau manifolds come in pairs $(X,\tilde{X})$, satisfying $H^{1,1}(X) \simeq H^{2,1}(\tilde{X})$ and $H^{2,1}(X) \simeq H^{1,1}(\tilde{X})$. The type IIA string theory compactified on $X$ was proven to be dual to the type IIB string theory compactified on $\tilde{X}$. In the case of heterotic string theory however, mirror symmetry is not so well understood.

\subsection{Complete Intersection Calabi-Yau manifolds}
\label{cicysection}
\noindent One reliable method of constructing Calabi-Yau Manifolds is by embedding them in a product of projective spaces $\mathcal{A}= \mathbb{P}^{n_1} \times ... \times \mathbb{P}^{n_m}$, referred to as the ambient space. The reason why complex projective spaces are used instead of $\mathbb{C}^n$ is because none of the K\"ahler submanifolds of $\mathbb{C}^n$ are compact, while all analytic submanifolds of $\mathbb{P}^n$ are guaranteed to be K\"ahler and compact \cite{Candelas:1987kf}. A Complete Intersection Calabi-Yau Manifold (CICY) $X$ is defined as the intersection of $K$ hypersurfaces $\lbrace{M_j}\rbrace_{j=1,...,K}$,
\begin{eqnarray}
X = M_1 \cap ... \cap M_K,
\end{eqnarray}
\noindent where each $M_j$ is the zero locus of a polynomial $p_j$, with variables in the ambient space $\mathcal{A}$. In order for the intersection to be called complete, it is necessary that the $K$-form
\begin{eqnarray}
\Psi = d p_1 \wedge ... \wedge d p_K,
\end{eqnarray} 
\noindent is nowhere vanishing on $X$, to ensure that $X$ does not possess any singularities. Additionally, this imposes that the dimension of the CICY is equal to the dimension of the ambient space minus the number of polynomials. In the case where $X$ is a threefold, this means
\begin{eqnarray}
\sum^m_{i=1} n_i - K = 3 .
\end{eqnarray}
\noindent Now, suppose the homogenous coordinates of each projective space $\mathbb{P}^{n_i}$  are written as $\mathbf{x}^{(i)} = (x_0^{(i)}:x_1^{(i)}:...:x_{n_i}^{(i)} )$, so that $\mathcal{A}$ has projective coordinates $(\mathbf{x}^{(1)}, ..., \mathbf{x}^{(m)})$. Every polynomial $p_j$ defined on $\mathcal{A}$ is characterised by a multi-degree vector $\mathbf{q}_j = (q^1_j,...,q^m_j)$, where $q^i_j$ specifies the degree in the $\mathbf{x}^{(i)}$ coordinate.  It is useful to represent the corresponding CICY through the following configuration matrix
\begin{eqnarray}
\begin{bmatrix}
\mathbb{P}^{n_1}& \vline & q^1_1 & q^1_2 & \dots & q^1_K \\
\mathbb{P}^{n_2}&\vline &q^2_1 & q^2_2 & \dots & q^2_K \\
\vdots &\vline &\vdots & \vdots & \ddots & \vdots \\
\mathbb{P}^{n_m}& \vline &q^m_1 & q^m_2 & \dots & q^m_K
\end{bmatrix}^{h^{1,1},h^{2,1}}_{\chi} ,
\end{eqnarray}
\noindent where the condition $\sum_{j=1}^K q^i_j = n_i + 1$ needs to be imposed for every $i=1,...,m$, in order for $c_1(TX)$ to vanish. There is a finite number ($7890$, to be precise) of possible CICY configurations, as originally established in Refs. \cite{Candelas:1987kf} and \cite{7890}. Out of those, we are only interested in the favourable configurations, for which the K\"ahler form $J$ descends directly from the K\"ahler forms $J_i$ of the projective spaces $\mathbb{P}^{n_i}$, for $i=1,...,m$. These CICYs have $h^{1,1} = m$ and their K\"ahler cone \eqref{kahlercone} and intersection numbers \eqref{intersectionnumbers} are simply obtained by setting  the basis of $(1,1)$-forms $\lbrace\omega_i\rbrace$ to be $\omega_i = J_i\vert_X$. It is for this reason that favourable CICYs are preferred. Well-known examples include
\begin{eqnarray}
\label{examplesofcicys}
\begin{matrix}
\textrm{the quintic}  &&&  \textrm{the bicubic} &&& \textrm{the tetraquadric}
\\[0.15cm]
\,\,\,\ \left[ \mathbb{P}^{4} \, \vline \ 5 \right]^{1,101}_{-200},  &&&  \,\,\,\,\,\,\,\, \setlength\arraycolsep{2.5pt} \begin{bmatrix}
\mathbb{P}^{2}& \vline & 3  \\
\mathbb{P}^{2}&\vline & 3
\end{bmatrix}^{2,83}_{-162} , &&&  \,\,\,\,\,\,\,\, \setlength\arraycolsep{2.5pt} \begin{bmatrix}
\mathbb{P}^{1}& \vline &  2  \\
\mathbb{P}^{1}&\vline & 2 \\
\mathbb{P}^{1} &\vline &  2\\
\mathbb{P}^{1}& \vline &  2
\end{bmatrix}^{4,68}_{-128} . 
\end{matrix}
\end{eqnarray} 
\noindent In particular, the tetraquadric manifold will constitute the focus of Chapter~\ref{tetraquadricchapter}.

\subsection{Holomorphic vector bundles and their cohomologies}
\label{vectorbundleschapter}
We conclude our review of mathematical concepts with a short discussion of holomorphic vector bundles. More useful information can be found in Appendix~\ref{appendixvectorbundles}.
\begin{definition} A vector bundle $E$ over an $n$-dimensional complex manifold $M$ is called holomorphic if it is endowed with a holomorphic projection $\pi: E \rightarrow M$ and the local trivialisation maps $\phi_\alpha: \pi^{-1}(U_\alpha) \rightarrow U_\alpha \times \mathbb{C}^r$ are biholomorphic. 
\end{definition}
\noindent An equivalent statement is that on every overlap $U_{\alpha} \cap U_{\beta} \neq \emptyset$, the transition function $t_{\alpha \beta} \equiv \phi_{\alpha} \circ \phi^{-1}_{\beta}$, \ $ t_{\alpha \beta}: U_{\alpha} \cap U_{\beta} \rightarrow GL(r,\mathbb{C})$ is holomorphic. At every point $p \in M$, the fiber $E_p \equiv \pi^{-1}(p)$ is an $r$-dimensional complex vector space, thus giving the rank $r$ of the vector bundle. In particular, a vector bundle of rank $1$ is called a line bundle. Moreover, for each vector bundle $E$, one can define the dual bundle $E^*$ over $M$, whose fiber $E^*_p$ is the set of linear maps $f: E_p \rightarrow \mathbb{C}$.  

A local section is a map $\sigma: U_{\alpha} \rightarrow E$ that satisfies $\pi \circ \sigma = \textrm{id}_M$. It can be expanded as $\sigma = \sum_{i=1}^r \sigma^i s_i$ with respect to a local frame of  $r$ linearly independent sections $( s_1,...,s_r )$ that span the fiber $E_p$ at every point $p \in U_{\alpha}$. In a similar way, bundle-valued $(p,q)$-forms are written as $\alpha = \sum_{i=1}^r \alpha^i \otimes s_i$, where $\alpha^i \in \Omega^{p,q}(M)$. The space of these $(p,q)$-forms is denoted $\mathcal{A}^{p,q} (E)$.  
\begin{example}
Obvious examples of holomorphic vector bundles are the holomorphic tangent bundle $T^{1,0}M$ and its dual, the holomorphic cotangent bundle $T_{1,0}^*M$, whose sections are expanded by $\lbrace\partial_i \rbrace$ and $\lbrace d z_i \rbrace$ respectively, for $i=1,...,n$. Since their relation to the anti-holomorphic bundles $T^{0,1}M$ and $T^*_{0,1}M$ is isomorphic, through charge conjugation, it is customary to work only with $T^{1,0}M$ and $T_{1,0}^*M$ and simply refer to them as $TM$ and $T^*M$. In general, the set of $(p,q)$-forms $\Omega^{p,q}(M)$ represents the space of sections of the bundle $\wedge^p T_{1,0}^*M \otimes \wedge^q T_{0,1}^*M $.
\end{example}
In analogy to Section~\ref{hodgetheorychapter}, the operator $\overline{\partial}: \Omega^{p,q}(M) \rightarrow \Omega^{p,q+1}(M)$ can be generalised to act on $E$-valued forms as $\overline{\partial}_E: \mathcal{A}^{p,q}(E)\rightarrow\mathcal{A}^{p,q+1}(E)$, such that locally $\overline{\partial}_E \alpha = \sum_{i=1}^r \overline{\partial} \alpha^i \otimes s_i$. Essentially, the procedures of Hodge theory are repeated to reveal that $\overline{\partial}_E$-harmonic forms are in one-to-one correspondence with cohomology classes of 
\begin{eqnarray}
H^{p,q}(M,E)= \dfrac{\textrm{Ker}\left(\overline{\partial}_E: \mathcal{A}^{p,q}(E)\rightarrow\mathcal{A}^{p,q+1}(E) \right)}{\textrm{Im}\left(\overline{\partial}_E: \mathcal{A}^{p,q-1}(E)\rightarrow\mathcal{A}^{p,q}(E) \right)},
\end{eqnarray}
\noindent where $H^{p,q}(M,E) \simeq H^q(M,E \otimes \wedge^p T^*_{1,0}M)$.

On a Calabi-Yau threefold $X$ with poly-stable vector bundle $V$, several simplifications occur. Firstly, the fact that $TX \simeq \wedge^2 T^*X$ (due to the uniqueness of $\Omega$), implies that $H^1(X,TX) \simeq H^{2,1}(X)$ and $H^1(X,T^*X) \simeq H^{1,1}(X)$. Secondly, because the canonical bundle $K_X = \wedge^3 T^*_{1,0}X$  is trivial, the following version of the Serre duality holds
\begin{eqnarray}
H^q(X,V)\simeq H^{3-q}(X, V^*).
\end{eqnarray}
\noindent Finally, the index of $V$ is given by the Hirzebruch–Riemann–Roch theorem
\begin{eqnarray}
\label{atiyahsingerindex}
\textrm{ind}(V) \equiv \sum_{q=0}^{3} (-1)^q h^q(X,V) = \int_X \textrm{Td}(X)  \wedge \textrm{ch} (V) = \dfrac{1}{2} \int_X c_3 (V),
\end{eqnarray}
\noindent where $\textrm{Td}(X)$ and  $\textrm{ch}(V)$ are topological invariants of  $V$ and $X$, known as the Todd class and the Chern character respectively, while $c_3 (V)$ is the third Chern class of $V$. In particular, for a stable $SU(r)$ bundle, $h^0(X,V)=h^3(X,V)=0$, so $\textrm{ind}(V) = - h^1(X,V) + h^1(X,V^*)$.
\begin{definition} Let  us assume $X$ is a Complete Intersection Calabi-Yau manifold embedded in an ambient space $\mathcal{A}$. Then the normal bundle $\mathcal{N}$ on $\mathcal{A}$ is the quotient
\begin{eqnarray}
 \mathcal{N}_{\mathcal{A}\vert X} ={T \mathcal{A}\vert_X}\big/{TX},
\end{eqnarray}
\noindent where $T \mathcal{A}\vert_X$ is the restriction of the tangent bundle $T \mathcal{A}$ on $X$.
\end{definition}
\noindent The rank of $\mathcal{N}$ is precisely $K$, the co-dimension of $X$. If we denote by $i$ the natural injection of $TX$ into $T \mathcal{A}\vert_X$, one can construct a short exact sequence
\begin{eqnarray}
0 \longrightarrow TX \stackrel{i}{\longrightarrow} T \mathcal{A} \vert_X \stackrel{\nabla \mathbf{p}}{\longrightarrow} \mathcal{N}_{\mathcal{A}\vert X} \longrightarrow  0,
\end{eqnarray}
\noindent where $\mathbf{p} = (p_1,...,p_K)$ is the $K$-tuple of defining polynomials. In general, it is convenient to build short exact sequences of vector bundles, because they induce long exact sequences in cohomology, from which the cohomology groups can be calculated. More precisely, if $A$, $B$, $C$ are three vector bundles on an $n$-dimensional base space $M$, satisfying
\begin{eqnarray}
\label{veryshortformulacohomologiesveryshort}
0 \longrightarrow A \stackrel{f}{\longrightarrow} B \stackrel{g}{\longrightarrow} C \longrightarrow 0,
\end{eqnarray}
\noindent then the corresponding relation between cohomology groups is 
\begin{align}
\label{verylongformulaoncohomologiesverylong}
0 & \longrightarrow H^0(M,A)\stackrel{f}{\longrightarrow}  H^0(M,B)\stackrel{g}{\longrightarrow} H^0(M,C) \stackrel{\delta_0}{\longrightarrow} \notag \\
 & \longrightarrow H^1(M,A) \stackrel{f}{\longrightarrow}H^1(M,B)\stackrel{g}{\longrightarrow} H^1(M,C) \stackrel{\delta_1}{\longrightarrow } \notag \\
  &  \,\,\,\,\,\,\,\,\,\,\,\,\,\,\,\,\,\,\,\,\,\,\,\,\, \vdots \,\,\,\,\,\,\,\,\,\,\,\,\,\,\,\,\,\,\,\,\,\,\,\,\, \,\,\,\,\,\,\,\,\,\,\, \vdots  \,\,\,\,\,\,\,\,\,\,\,\,\,\,\,\,\,\,\,\,\,\,\,\,\,\,\,\,\,\,\,\,\,\,\,\, \vdots  \notag  \\ 
  & \longrightarrow H^ n (M,A) \stackrel{f}{\longrightarrow} H^n(M,B)\stackrel{g}{\longrightarrow} H^n(M,C) \longrightarrow 0 ,
\end{align}
\noindent where $\delta_i$ are the coboundary maps.\footnote{More information about the coboundary map can be found in Appendix~\ref{coboundarymapappendix}.} Of particular importance is the Koszul sequence, which relates a vector bundle $\mathcal{V}$ on the ambient space $\mathcal{A}$ to its restriction $V=\mathcal{V}\vert_X$ on the Calabi-Yau manifold, using the dual to the normal bundle $\mathcal{N}^*$,
\begin{eqnarray}
\label{Koszulseq1sttime}
0 \longrightarrow \wedge^K \mathcal{N}^* \otimes \mathcal{V} \longrightarrow ... \longrightarrow \wedge^2 \mathcal{N}^* \otimes \mathcal{V} \longrightarrow  \mathcal{N}^* \otimes \mathcal{V} \longrightarrow \mathcal{V} \longrightarrow V  \longrightarrow 0.
\end{eqnarray}
\noindent This sequence is short exact only if $K=1$, but even for higher co-dimensions we can split \eqref{Koszulseq1sttime} into short exact pieces in order to express cohomology groups $H^q(X,V)$ in terms of ambient space cohomologies.

In this thesis we will make a major simplification, by assuming models in which the vector bundle on $X$ is a Whitney sum of line bundles, i.e. $V=\bigoplus_{i=1}^r L_i$. Such vector bundles have structure group $S(U(1)^r)$, rather than $SU(r)$, and are motivated by the fact that cohomologies of line bundles are much easier to calculate. Moreover, line bundles are automatically stable,\footnote{By the definition of stability in Section~\ref{conditionsforn=1susy}, all line bundles are trivially stable because they have no proper subsheaf.} so $V$ is poly-stable, provided that the slope of each line bundle is $\mu(L_i) = \mu(V)  \stackrel{!}{=} 0$. For these reasons, it is proper to discuss here the line bundles that will serve as building blocks for our model.

\begin{example}
The tautological (or universal) line bundle on $\mathbb{P}^n$ is defined as a sub-bundle of $\mathbb{P}^n \times \mathbb{C}^{n+1}$ for which the fiber at every point $(x_0:...:x_n)\in \mathbb{P}^n$ is the line through the origin $\lbrace{ (\lambda x_0, ..., \lambda x_n), \lambda\in\mathbb{C}^*\rbrace}$ or, more formally written,
\begin{eqnarray}
\mathcal{O}_{\mathbb{P}^n}(-1)=\lbrace (l,v)\in \mathbb{P}^n \times \mathbb{C}^{n+1} \vert v \in l\rbrace .
\end{eqnarray}
\noindent Its dual, the hyperplane line bundle $\mathcal{O}_{\mathbb{P}^n}(1)$ is a sub-bundle of $\mathbb{P}^n \times (\mathbb{C}^{n+1})^*$, whose fiber at every point is the space of linear functionals $\sum_{i=0}^n \lambda_{i} x^{i} \in \mathbb{C}$ (hence it is a bundle of hyperplanes). More line bundles can be defined on $\mathbb{P}^n$ as tensor products $\mathcal{O}_{\mathbb{P}^n}(k) = \mathcal{O}_{\mathbb{P}^n}(1)^{\otimes k}$ and $\mathcal{O}_{\mathbb{P}^n}(-k) = \mathcal{O}_{\mathbb{P}^n}(-1)^{\otimes k}$, for $k\in \mathbb{Z}$. They are used to build line bundles on the ambient space $\mathcal{O}_{\mathcal{A}}(\mathbf{k}) = O_{\mathbb{P}^{n_1}}(k_1) \otimes ... \otimes O_{\mathbb{P}^{n_m}}(k_m)$, where $\mathbf{k} = (k_1,...,k_m)$, and through restriction, line bundles on the Calabi-Yau manifold $\mathcal{O}_X(\mathbf{k}) = \mathcal{O}_{\mathcal{A}}(\mathbf{k})\vert_X$. 
\end{example}
\noindent In particular, the cohomology groups of line bundles on $\mathcal{A}$ are related via the K\"unneth formula to cohomology groups of line bundles on individual projective spaces
\begin{eqnarray}
\label{kunneth}
H^{q}(\mathcal{A}, \mathcal{O}_{\mathcal{A}}(\mathbf{k})) = \bigoplus_{q_1+...+q_m=q} H^{q_1}(\mathbb{P}^{n_1},\mathcal{O}(k_1)) \times ... \times H^{q_m}(\mathbb{P}^{n_m},\mathcal{O}(k_m)).
\end{eqnarray}
\noindent In future chapters we will learn how exactly to calculate these cohomologies and how to use the Koszul sequence \eqref{Koszulseq1sttime} in order to determine $H^q(X,\mathcal{O}_X(\mathbf{k}))$.

\section{Dimensional Reduction of the 10d Theory}
\label{dimredchapter}
\noindent Now that we have the mathematical tools to proceed with the compactification, our goal is to dimensionally reduce the heterotic 10d theory down to 4d. There are several steps that need to be taken. The first is the dimensional reduction of the gravitational sector, namely the dilaton, $B$-field and Einstein-Hilbert terms of the bosonic action \eqref{10daction}. Next in line is the dimensional reduction of the matter sector, or the  $\alpha'$-dependent part, with the emergence of 4d matter multiplets in the resulting GUT group representations. In passing, we will also discuss what happens with the fermionic action and how to obtain holomorphic Yukawa couplings from the 10d theory. In the last part of this section, we will specify how to further break the GUT group down to the SM group via Wilson lines, as this is the final stage through which the heterotic string theory is connected to particle physics.

Most of the results in this chapter are well-known in the literature (\cite{GSW},\cite{Candelas:1990pi},\cite{bbschwarz},\cite{ibanezuranga}), however some of them, such as Eqs.~\eqref{crosstermaction}--\eqref{originalresults}, are based on original work. A more detailed description of these original formulae will be provided later, in Section~\ref{kahlersec2}.

\subsection{Dimensional reduction of the bosonic gravity sector}
\label{gravitysectorsection}
\noindent At the first stage of compactification, we can neglect all $\mathcal{O}(\alpha')$ contributions to the 10d bosonic action, in order to focus on the gravity sector. We start by expanding the bosonic fields of the gravity multiplet according to the compactification ansatz.

\paragraph{Expanding the dilaton}
\noindent Under the assumption that 10d fields are perturbations around the $M_4 \times X$ vacuum, any scalar field $\phi(x^M)$ can be expanded in terms of the external and the internal coordinates as $\phi(x,y) \simeq \phi(x) \phi(y)$, where $x$ and $y$ are short-hand notations for $x^{\mu}$ and $y^m$. The solution to the equation of motion $\Delta \phi = 0$ implies that $\phi(y)$ is a constant, which can be taken to be $1$, therefore the 10d dilaton is trivially expanded as $\phi(x,y)=\phi(x)$. Of particular significance is the background value of the dilaton, which determines the string coupling constant via $g_s = e^{\langle \phi \rangle}$. 

\paragraph{Expanding the $B$ field} For the case of the $B$ field, it is again necessary to analyse the equations of motion. In order for $B$ to be physical, it must be invariant under a gauge transformation $\delta B = d \Lambda$ which decouples the time-like modes of $B$ (these modes are responsible for negative norm states). The usual gauge choice, $d^\dagger B = 0$, together with the equation $d^\dagger d B = 0$, derived from the minimisation of action,\footnote{Writing the action of $B$ as an inner product $( H, H ) = \int H \wedge \ast H$, where $H=dB + \mathcal{O}(\alpha')$ is the field strength, and then imposing $\delta (H,H) \stackrel{!}{=}$ for small variations of $B$ leads to the result $d^\dagger d B = 0$.} show that $\Delta B= 0$, i.e. $B$ is harmonic. Now, on a vacuum space $M_4 \times X$, the Laplacian splits into $\Delta = \Delta_4 + \Delta_X$, which means that massless fields\footnote{In the equation $\Delta_4 B + \Delta_X B = 0$, $\Delta_X B$ has the role of a mass-squared term for the effective 4d field, however non-zero eigenvalues of $\Delta_X$ are too large to be measured, as they are proportional to $1/l_c^2 \gg 1$, where $l_c$ is the typical length scale of the internal manifold. It is for this reason that only massless fields are relevant at low energy.} of the 4d theory correspond to zero modes of the Laplace operator $\Delta_X$, and therefore to harmonic forms representing classes in $H^{p,q}(X)$. Applying what we learned about cohomology groups of Calabi-Yau manifolds, the expansion of $B$ reads
\begin{eqnarray}
\label{decompbfield}
B(x,y) = B(x) + \sum_{i=1}^{h^{1,1}} \tau^i (x) \omega_i(y) ,
\end{eqnarray}
\noindent where $B(x)$ is the rank-2 tensor field in the 4d theory, $\lbrace \tau^i\rbrace$ is a set of  $h^{1,1}$ scalar fields, known as moduli, and $\lbrace \omega_i \rbrace$ is a basis of harmonic $(1,1)$-forms. In particular, $B(x)$ has a single degree of freedom and is a pseudoscalar, so it can be dualised to a 4d real scalar field $\gamma (x)$
\begin{eqnarray}
d B = \ast_4 d \gamma .
\end{eqnarray}

\paragraph{Expanding the metric} Following \cite{Candelas:1990pi}, the internal manifold remains Ricci-flat upon $\delta g$ deformations of the metric, provided that the Lichnerowicz equation is satisfied
\begin{equation}
\nabla^F \nabla_F \delta g_{A B} + 2 R_A{}^C{}_B{}^D \delta g_{CD}=0.
\end{equation}
\noindent Hence the solution is of the form $\delta g = \delta g_{\overline{m} \overline{n}} dy^{\overline{m}} dy^{\overline{n}} + \delta g_{m \overline{n}} dy^{m} dy^{\overline{n}} + \textrm{c.c.}$, where
\begin{align}
\label{ansatzg1}
\delta g_{m \overline{n}} & = - i \sum_{i=1}^{h^{1,1}} t^{i}(x) \omega_{i \, m \overline{n}}(y), \\
\label{ansatzg2}
\delta g_{\overline{m} \overline{n}} & =  - \dfrac{1}{\Vert\Omega\Vert^2}  \overline{\Omega}_{\overline{m}}{}^{p q}  \sum_{a=1}^{h^{2,1}} z^{a}(x) \rho_{a \, p q \overline{n}}(y) 
\end{align}
\noindent are expansions in bases of harmonic $(1,1)$-forms $\lbrace \omega_i \rbrace$ and harmonic $(2,1)$-forms $\lbrace \rho_a \rbrace$, respectively. The 4d real scalar fields $t^i$ are called K\"ahler moduli and parametrise size deformations, while the 4d complex scalar fields $z^a$ are the complex structure moduli parametrising deformations of shape.

It is now possible to apply the field expansions \eqref{decompbfield},~\eqref{ansatzg1} and \eqref{ansatzg2} to the 10d action \eqref{10daction}, but while doing so, we want to remove the exponential prefactor $e^{- 2 \phi}$ from the Ricci scalar term. This involves moving from the string frame of tree-level string interactions to the Einstein frame through a Weyl rescaling of the external metric
\begin{eqnarray}
g_{\mu\nu} \rightarrow e^{2 \phi}  g_{\mu\nu} \, .
\end{eqnarray}
\noindent Under this transformation, the 4d bosonic action at zeroth order ($\alpha' \approx 0$) reads
\begin{eqnarray}
\label{0thorderbosonicaction}
\!\!\!\! S_0 = \dfrac{1}{2\kappa_4^2} \int \! d^4 x \sqrt{- g} \left( R - \! \dfrac{\!\!2}{(S\! + \! \overline{S})^2}\partial^{\mu} S \partial_{\mu} \overline{S} - 2 G_{i j} \partial^{\mu} T^i \partial_{\mu} \overline{T}^j \! - \!  2 G_{a \overline{b}} \partial^\mu z^a \partial_{\mu} \overline{z}^{\overline{b}}\right)\!,
\end{eqnarray}
\noindent where 
\begin{eqnarray}
\label{modulispacemetrics}
G_{ij} = \dfrac{1}{4 \mathcal{V}}\int_X \omega_i \wedge \ast_X \omega_j \, , \,\,\,\,\,\,\,\,\,\,\,\,\,\,\,\,\,\,\,\,\,\,\,\,\, G_{a\overline{b}} = \dfrac{\int_X \rho^a \wedge \overline{\rho}^{\overline{b}}}{\int_X  \Omega \wedge \overline{\Omega} }
\end{eqnarray}
\noindent are the K\"ahler and complex structure moduli space metrics respectively, and
\begin{equation}
\label{modulifields}
S= e^{-2 \phi} + i \gamma , \,\,\,\,\,\,\,\,\,\,\,\,\,\,\,\,\,\,\,\,\,\,\,\,\,\,\,\,\,\,\, T^i = t^i + i \tau^i 
\end{equation}
\noindent are defined as 4d complex moduli fields, together with $z^a$. In total, the gravity sector contains $h^{1,1}+h^{2,1}+1$ such fields, sitting in $N = 1$ chiral supermultiplets. It is also useful to introduce K\"ahler potentials
\begin{eqnarray}
K^{(J)} = - \textrm{ln}\left(\dfrac{4}{3}\int_X J \wedge J \wedge J\right),\,\,\,\,\,\,\,\,\,\,\,\,\,\,
K^{(\textrm{CS})}= - \textrm{ln}\left( i \int_X \Omega \wedge \overline{\Omega} \right),
\end{eqnarray}
\noindent from which the moduli space metrics can be derived, using $G_{i j} = \partial_{i}\partial_{j} K^{(J)}$ and $G_{\overline{a} b} = \partial_{\overline{a}} \partial_b K^{(\textrm{CS})}$. Finally, note that in the expression \eqref{0thorderbosonicaction}, $\kappa_4$ is assumed to be the 4d gravitational coupling, so that $\kappa^2_4=8 \pi G_N$, where $G_N$ is Newton's constant. The relation between $\kappa_4^2$ and its 10d counterpart $\kappa^2 \sim g^2_s l_s^8$ is given by $\kappa^2 = \kappa_4^2 \mathcal{V}$, where $\mathcal{V}$ is the volume of the compact manifold. Interestingly, this also relates the string length and the Planck length via $l_{\textrm{P}} \sim g_s l_s^4/\sqrt{\mathcal{V}}$, showing that $l_{\textrm{P}} < l_s$ for adequate values of $g_s$ and $\mathcal{V}$. With these observations, we conclude our discussion of the gravity sector.

\subsection{Dimensional reduction of the bosonic matter sector}
\label{mattersectorsection}

\noindent Compactifying the $\alpha'$-dependent part of the bosonic action~\eqref{10daction} involves expanding the gauge field as $A = A^{(0)} + A^{(1)}$, where $A^{(0)}$ is the non-zero vacuum expectation value on $X$ and $A^{(1)}$ is an infinitesimal fluctuation. Being in the adjoint representation $\mathbf{248}$ of $E_8$, $A^{(1)}$ must split as 
\begin{equation}
\label{248repdecomposition} 
 {\bf 248}\rightarrow \left[({\rm Adj}_G,{\bf 1})\oplus ({\bf 1},{\rm Adj}_H)\oplus\bigoplus ({\cal R}_G, {\cal R}_H)\right]_{G\times H}
\end{equation}
\noindent under the symmetry breaking $E_8 \rightarrow G\times H$, where $G$ is the group structure of the holomorphic vector bundle $V$ and $H$ is the effective 4d GUT group. In particular, the piece transforming in the adjoint of $H$ is interpreted to be the 4d gauge boson $A(x)=A_{\mu}(x) dx^{\mu}$, while for the other components the ansatz is 
\begin{eqnarray}
\label{expansiona}
A^{(1)}_{(\mathcal{R}_G,\mathcal{R}_H)} = C^I(x) \nu_{I\overline{m}}(y) d y^{\overline{m}} + \overline{D}{}^{P}(x) \overline{\sigma}_{P m}(y) dy^m,
\end{eqnarray}
\noindent where $C^I$ and $D^P$ are matter fields in the representations $\mathcal{R}_H$ and $\overline{\mathcal{R}}_H$ respectively, while $\nu_{I} \in H^1(X, V_{\mathcal{R}_G})$ and $\sigma_P \in H^1(X, V^*_{\mathcal{R}_G})$ are harmonic $(0,1)$-forms on the bundles $V_{\mathcal{R}_G}$ and $V^*_{\mathcal{R}_G}$, associated to representation $\mathcal{R}_G$ of $G$. This is in line with the requirement that $A^{(1)}$ is harmonic, which comes from the Yang-Mills condition $d_A \ast F = 0$ and the gauge choice $d_A \ast A^{(1)} = 0$, with $d_A$ being the covariant derivative on the bundle. With the ansatz \eqref{expansiona} and redefining $F$ as the 4d field strength of $A(x)$, the Yang-Mills action becomes\footnote{One notices that representations $(\mathcal{R}_G,\mathcal{R}_H)$ and $(\overline{\mathcal{R}}_G,\overline{\mathcal{R}}_H)$ are both present in the decomposition of $\mathbf{248}$ and the relation $A^{(1)}_{(\overline{\mathcal{R}}_G,\overline{\mathcal{R}}_H)} = A^{(1)*}_{(\mathcal{R}_G,\mathcal{R}_H)}$ is implied by complex conjugation.}
\begin{eqnarray}
\label{reducedyangmillsaction}
S_{\textrm{YM}} = - \dfrac{\alpha'}{2\kappa_4^2} \int d^4 x  \sqrt{-g}\left(\dfrac{1}{4} \textrm{Re} (f) \, \textrm{Tr}F^2 + 2 G_{I J} D_{\mu} \overline{C}{}^I D^{\mu} C^J +...\right),
\end{eqnarray}
\noindent where $f=S$ is the gauge kinetic function, $G_{IJ}$ is the matter field K\"ahler metric
\begin{eqnarray}
\label{matterfieldmetric}
G_{I J} = \dfrac{1}{2 \mathcal{V}} \int_X d^6 y \sqrt{g^{(6)}} g^{ (6)\overline{m} n} \nu_{I \overline{m}} \overline{\nu}_{J n}  = \dfrac{1}{2 \mathcal{V}} \int_X \nu_I \wedge \overline{\ast}_V \nu_J \, ,
\end{eqnarray}
\noindent and by the ellipsis we indicate that further kinetic terms must be added, one for each type of $\mathcal{R}_H$-multiplets in the decomposition \eqref{248repdecomposition}. The number of families in a representation $\mathcal{R}_H$ is given by $n_{\mathcal{R}_H} = h^1(X, V_{\mathcal{R}_G})$, while the number of anti-families is $n_{\overline{\mathcal{R}}_H} = h^1(X, V^*_{\mathcal{R}_G})$. This means that the net number $n_{\mathcal{R}_H} - n_{\overline{\mathcal{R}}_H}$ of $\mathcal{R}_H$-multiplets is a topological invariant of the bundle, namely the index $\textrm{ind}(V_{\mathcal{R}_G})$, as defined in \eqref{atiyahsingerindex}. In the particular case of the representation $({\rm Adj}_G, \mathbf{1})_{G\times H}$, the resulting 4d scalars are uncharged under the GUT group $H$, so they are called vector bundle moduli. As the corresponding bundle for ${\rm Adj}_G$ is $V \otimes V^*$, the number of bundle moduli is $n_{\mathbf{1}}=h^1({X,V \otimes V^*})$. For the other representations of $G$, $V_{\mathcal{R}_G}$ descends from the principal bundle $V$, either as $V$ itself, if $\mathcal{R}_G$ is the fundamental representation, or as the dual $V^*$ or the wedge products $\wedge^2 V$, $\wedge^2 V^*$. Quoting \cite{Anderson:2009ge}, the main results for vector bundles with structure groups $G = SU(3)$, $SU(4)$ and $SU(5)$ are presented in Table~\ref{mytablenow}. All three cases lead to a corresponding GUT that was discussed in Section~\ref{gutsubsection}. Finally, the coefficient in front of the Yang-Mills action \eqref{reducedyangmillsaction} can be identified with the GUT coupling $1/g_{\textrm{GUT}}^2$, thus giving a proportionality of the form $g_{\textrm{GUT}} \sim g_s l_s^3 /\sqrt{\mathcal{V}}$ between $g_{\textrm{GUT}}$ and the parameters of the 10d theory.

\begin{table}[h]
\begin{center}
\begin{footnotesize}
\begin{tabular}{|l|l|l|}\hline
$E_8 \rightarrow G \times H$ & Decomposition of $\mathbf{248}$& Spectrum \\\hline\hline
& & $n_{\textbf{27}} = h^1(V)$ \\ $SU(3)\times E_6$& $({\bf 8},{\bf 1})\oplus ({\bf 1},{\bf 78})\oplus ({\bf 3},{\bf 27})\oplus (\overline{\bf 3},\overline{\bf 27})$ & $n_{\overline{\textbf{27}}} = h^1(V^*)$ \\ & & $n_{\textbf{1}} = h^1(V\otimes V^*)$\\\hline
& & $n_{\textbf{16}} = h^1(V)$ \\ & & $n_{\overline{\textbf{16}}} = h^1(V^*)$ \\ $SU(4) \times SO(10)$& $({\bf 15},{\bf 1})\oplus ({\bf 1},{\bf 45})\oplus ({\bf 4},{\bf 16})\oplus (\overline{\bf 4},\overline{\bf 16})\oplus ({\bf 6},{\bf 10})$ & $n_{\textbf{10}} = h^1(\wedge^2 V)$ \\ & &  $n_{\textbf{1}} = h^1(V\otimes V^*)$\\\hline
& & $n_{\textbf{10}} = h^1(V^*)$ \\& & $n_{\overline{\textbf{10}}} = h^1(V)$ \\$SU(5)\times SU(5)$&  $({\bf 24},{\bf 1})\oplus ({\bf 1},{\bf 24})\oplus ({\bf 5},\overline{\bf 10})\oplus (\overline{\bf 5},{\bf 10})\oplus ({\bf 10},{\bf 5})\oplus (\overline{\bf 10},\overline{\bf 5})$ & $n_{\overline{\textbf{5}}}= h^1(\wedge^2 V^*)$  \\& & $n_{\textbf{5}} = h^1(\wedge^2 V)$ \\& & $n_{\textbf{1}} = h^1(V\otimes V^*)$\\\hline
\end{tabular}
\end{footnotesize}
\end{center}
\caption{\it Different symmetry breaking scenarios for the $E_8$ gauge group, as dictated by the choice of a vector bundle $V$ with structure group $G = SU(3)$, $SU(4)$ or $SU(5)$. The resulting spectrum of particles (matter fields and bundle moduli) is determined by the cohomology groups of the bundle.}\label{mytablenow}
\end{table}

As a side note, one of the original methods of heterotic compactification was the standard embedding, through which the vector bundle $V$ is chosen to be identical with the holomorphic tangent bundle $TX$, so that the anomaly cancellation condition \eqref{c2c2W} is automatically satisfied. Since the holonomy of $X$ is $SU(3)$, the spin connection is embedded in the gauge group $E_8$ as the gauge connection of an $SU(3)$ subgroup. Thus, the standard embedding is a case of $SU(3)\times E_6$ compactification and has a net number of families $\vert n_{\textbf{27}}-n_{\overline{\textbf{27}}}\vert =  \vert h^{2,1}(X) -  h^{1,1}(X) \vert = \vert \chi \vert/2$. Despite its computational triumphs (4d physical Yukawa couplings, in particular), the model is too restrictive, therefore we look for models with a more general embedding.

To conclude this section, the compactification of the $\alpha'$-dependent part of the bosonic action does not only give rise to matter field kinetic terms. There are also several cross-terms arising from the $\vert H \vert^2$ action. This is seen by expanding $B$ in terms of the moduli $\tau^i$ as in \eqref{decompbfield} and the Chern-Simons form $\omega_{\textrm{YM}}$ in terms of the matter fields $C^I$ to obtain
\begin{eqnarray}
\label{crosstermaction}
S_{\textrm{cross terms}} = \dfrac{\alpha'}{2 \kappa_4^2} \int d^4 x \sqrt{-g} \left(\dfrac{1}{2} \Lambda_{i I J} \partial^{\mu} \tau^i \overline{C}{}^I D_{\mu} C^J + \textrm{c.c.}\right),
\end{eqnarray}
\noindent where $\Lambda_{i I J}$ is a coupling
\begin{eqnarray}
\Lambda_{i I J} = \dfrac{1}{2\mathcal{V}} \int_X (\ast_X \omega_i) \wedge \nu_I \wedge (\mathcal{H} \overline{\nu}_J),
\end{eqnarray}
\noindent and $\mathcal{H}$ is the Hermitian structure on the bundle. It is important to note that in the presence of matter fields, the moduli fields in Eq.~\eqref{modulifields} get modified to an expression of the form
\begin{eqnarray}
T^i = t^i + i \tau^i + \alpha' \, \Gamma^{i}_{I J} C^I \overline{C}{}^J,
\end{eqnarray}
\noindent for certain coefficients $\Gamma^i_{I J}$. Using this and the K\"ahler potential to compute the terms $K_{T^i \overline{C}{}^J} \partial^{\mu} T^i D_{\mu} \overline{C}{}^J$, one identifies
\begin{eqnarray}
\label{originalresults}
\Lambda_{i I J} = - i \dfrac{\partial G_{IJ}}{\partial t^i}, \,\,\,\,\,\,\,\,\,\,\,\,\,\,\,\,\,\,\, \Gamma^{i}_{I J} = - \dfrac{1}{4} G^{i j} \dfrac{\partial G_{I J}}{\partial t^j}.
\end{eqnarray}
\noindent where $G^{i j}$ is the inverse K\"ahler moduli metric and $G_{I J}$ is the matter field metric.\footnote{It is easy to see that $G_{IJ}$ is indeed $t$-moduli dependent, by expressing it as
\begin{eqnarray}
G_{IJ} = - \dfrac{i}{4 \mathcal{V}}\int_X J \wedge J \wedge \nu_I \wedge (\mathcal{H} \overline{\nu}_J) \notag, \,\,\,\,\,\, \textrm{where} \,\,\,\,\,\, J=t^i \omega_i , \,\,\,\,\,\, \mathcal{V} = \dfrac{1}{6} d_{i j k} t^i t^j t^k .
\end{eqnarray}}

\subsection{Dimensional reduction of the fermionic sector}
\label{fermionicsection}
\noindent Due to supersymmetry, all the fermionic fields in the 4d theory are expected to be the superpartners of the bosonic fields derived in Sections~\ref{gravitysectorsection} and~\ref{mattersectorsection}. For this reason, compactifying the fermionic sector may seem redundant and is mainly useful as a consistency check. As a reminder, the 10d fermions are the spin-$3/2$ gravitino $\psi_M$, the spin-$1/2$ dilatino $\lambda$ and the spin-$1/2$ gaugino $\chi$, which transforms in the adjoint representation of $E_8$. They have the following kinetic terms 
\begin{eqnarray}
\label{fermionicaction}
\resizebox{0.89\hsize}{!}{$S_{\textrm{f}} =  - \dfrac{1}{2 \kappa^2} \mathop{\mathlarger{\int}} d^{10}x \sqrt{-g} e^{-2 \phi}\bigg [\overline{\psi}_M\Gamma^{MNP}D_N \psi_P + \dfrac{1}{2} \overline{\lambda} \Gamma^M D_M \lambda + \dfrac{\alpha'}{2} \textrm{Tr}(\overline{\chi}\Gamma^M D_M\chi) \bigg ],$}
\end{eqnarray}
\noindent where all $\Gamma^N$'s are 10d $\Gamma$-matrices and their antisymmetrised product is given by
\begin{eqnarray}
\Gamma^{N_1 N_2 ... N_n} = \dfrac{1}{n!} \Gamma^{[N_1}\Gamma^{N_2}...\Gamma^{N_n]} \, .
\end{eqnarray}
\noindent Constructing the Clifford algebra in various dimensions is thoroughly discussed in sources like Ref.~\cite{Polchinski} (Appendix B). For our particular case,
\begin{eqnarray}
\Gamma^{\mu} = \gamma^{\mu} \otimes \mathbbmss{1}, \,\,\,\,\,\,\,\,\,\,\,\,\,\,\,\,\,\,\,\,\,\,\,\,\,\,\,\,\, \Gamma^{m} = \gamma^5 \otimes \gamma^{m}, \,\,\,\,\,\,\,\,\,\,\,\,\,\,\,\,\,\,\,\,\,\,\,\,\,\,\,\,\, \Gamma^{\overline{m}} = \gamma^5 \otimes \gamma^{\overline{m}}, 
\end{eqnarray}
\noindent where $\gamma^{\mu}$ are the standard Dirac matrices in 4d, $\gamma^5$ is their corresponding chirality operator, and $\gamma^{m}$, $\gamma^{\overline{m}}$ are internal manifold gamma-matrices satisfying
\begin{eqnarray}
\lbrace \gamma^{m},\gamma^{\overline{n}}\rbrace = 2 g^{m\overline{n}}, \,\,\,\,\,\,\,\,\,\,\,\,\,\,\,\,\,\,\,\,\,\,\,\,\,\,\,  \lbrace \gamma^{m},\gamma^{n}\rbrace = \lbrace\gamma^{\overline{m}}, \gamma^{\overline{n}} \rbrace=0.
\end{eqnarray}

\noindent Spinors are defined in $D$ dimensions as $2^{[D/2]}$-dimensional representations of the Lorentz group $SO(1,D-1)$, by taking generators $\sigma^{MN} = \frac{1}{4}[\Gamma^M, \Gamma^N]$. If $D$ is even, a Dirac spinor can be split into two irreducible Weyl representations and if a constraint $\Psi^\dagger = \Psi^T C$ can be applied, where $C$ is the charge conjugation matrix, then the spinor $\Psi$ is said to be Majorana. In 10d, the Majorana-Weyl spinor $\mathbf{16}$ of $SO(1,9)$ decomposes under $SO(1,9) \rightarrow SO(1,3) \times SO(6)$ as 
\begin{eqnarray}
\label{decomp2424}
\mathbf{16} \rightarrow (\mathbf{2},\mathbf{4}) \oplus (\mathbf{2}',\overline{\mathbf{4}}),
\end{eqnarray}
\noindent from which one can see that the compactification ansatz $M_4 \times X$ renders $\Psi(x,y) = \Psi^{4d}(x) \otimes \Psi^X(y) + \overline{\Psi}{}^{4d}(x) \otimes \overline{\Psi}{}^X(y)$ for any spinor field $\Psi$. The representation $\mathbf{4}$ of $SO(6)$ is further split into $\mathbf{1} \oplus \mathbf{3}$ under the reduced $SU(3)$ holonomy, so in general $\Psi^X$ can be expanded as 
\begin{eqnarray}
\Psi^X(y) = a(y) \overline{\xi} + b_{\overline{m}}(y) \gamma^{\overline{m}} \xi,
\end{eqnarray}
\noindent where $a$ is a smooth function and $b$ is a $(0,1)$-form, both being defined on $X$ or a bundle thereof, depending on the gauge representation of $\Psi^X$, while $\gamma^{\overline{m}}$ and $\gamma^{m}$ act as raising and lowering operators, and $\xi$, $\overline{\xi}$ are the covariantly constant spinors of opposite chirality, satisfying $\gamma^m \xi =0$ and $\gamma^{\overline{m}} \overline{\xi} =0$ (see Ref. \cite{Candelas:1987is}).

In terms of the equations of motion, the gravitino, being a vector-spinor field,\footnote{Note that unlike spinors, vector-spinor fields have $ 2^{[D/2]-1}(D-3)$ components in $D$ dimensions, if they are massless.} respects the Rarita-Schwinger equation $\Gamma^{MNP} D_N \psi_P =0$, while the dilatino and the gaugino satisfy Dirac equations $\slashed{D} \lambda =0$ and $\slashed{D} \chi=0$ respectively, where the 10d Dirac operator $\slashed{D}$ is $\Gamma^M D_M$ and the covariant derivative $D_M$ is defined with respect to the spin connection and (in the case of the gaugino) the background gauge fields. Now, the Dirac operator splits as $\slashed{D} = \slashed{D}_4 + \slashed{D}_X$, when acting on the  $(\mathbf{2},\mathbf{4})$ component of $\mathbf{16}$, and as $\slashed{D} = \slashed{D}^\dagger_4 + \slashed{D}^\dagger_X$, when acting on $(\overline{\mathbf{2}},\overline{\mathbf{4}})$. Therefore, 4d massless left-handed fermions correspond to zero modes of $\slashed{D}_X$, while 4d massless right-handed fermions correspond to zero modes of $\slashed{D}_X^{\dagger}$. The net chirality is given by the index $\textrm{ind}(\slashed{D}_X) = \textrm{dim} \, \textrm{ker} \slashed{D}_X - \textrm{dim} \, \textrm{ker} \slashed{D}^\dagger_X$, which was shown to be a topological invariant of the manifold via the Atiyah-Singer index theorem,
\begin{eqnarray}
\textrm{ind}(\slashed{D}_X, V)= \int_X \hat{A} (X) \wedge \textrm{ch}(V),
\end{eqnarray}
\noindent where $\hat{A} (X)$ is the $A$-roof genus, a topological quantity of $X$ isomorphically related to the Todd class $\textrm{Td}(X)$, so that on an almost complex manifold $\textrm{Td}(X) = e^{c_1(X)/2} \hat{A} (X)$ \cite{dsfreed}. This identifies $\textrm{ind}(\slashed{D}_X, V)$ with the index of the Hirzebruch–Riemann–Roch theorem \eqref{atiyahsingerindex} that was used to describe the bosonic spectrum, thus equating the number of bosonic and fermionic degrees of freedom. For example, 4d matter fermions $\tilde{C}^I$ (the superpartners of bosonic matter fields $C^I$) arise from the dimensional reduction of 10d gauginos, having a number of $\mathcal{R}_H$-multiplets that is equal to $\textrm{ind}(\slashed{D}_X, V_{\mathcal{R}_G})$, where representations $\mathcal{R}_G$ and $\mathcal{R}_H$ are defined as in \eqref{248repdecomposition}. The $\text{Adj}_H$ component of $\chi(x)$ corresponds to the 4d gaugino and together with the gauge field $A_{\mu}(x)$, it forms the vector supermultiplet. On the other hand, the 10d dilatino only gives rise to a 4d Weyl spinor (the 4d dilatino), which is the superpartner of the modulus field $S$. Altogether, the explicit decompositions for the gaugino and the dilatino read\footnote{Note that the Majorana condition ${\chi}^{\dagger}_{(\mathcal{R}_G, \mathcal{R}_H)} = \chi^T_{(\overline{\mathcal{R}}_G, \overline{\mathcal{R}}_H)} C$ is satisfied if $\chi_{(\overline{\mathcal{R}}_G,\overline{\mathcal{R}}_H)}$ is similarly expanded in terms of $\tilde{D}^P$ and $\overline{\tilde{C}}{}^I$.}
\begin{align}
\label{decompgaugino}
\chi_{(\mathcal{R}_G,\mathcal{R}_H)} & = \tilde{C}^I(x) \otimes \nu_{I\overline{m}}(y) \gamma^{\overline{m}} \xi + \overline{\tilde{D}}{}^P(x) \otimes \overline{\sigma}_{P m}(y) \gamma^m \overline{\xi}, \\
\label{decompdilatino}
\lambda (x,y) & = \lambda(x) \otimes \overline{\xi} + \overline{\lambda}(x) \otimes \xi,
\end{align}
\noindent where we used the same notations as in \eqref{expansiona}.
  
The compactification of the gravitino is roughly similar, although more elaborate. The components $\psi_{\mu}$, with an external index, transform as a vector-spinor of $SO(1,3)$ and a spinor of $SO(6)$, but only $\psi^{4d}_{\mu} \otimes \overline{\xi}$ is massless, where $\psi^{4d}_{\mu}$ is interpreted as the massless 4d gravitino. The internal component $\psi_m = \psi^{4d}_m \otimes \psi^{X}_m$ however transforms as a spinor of $SO(1,3)$ and a vector-spinor of $SO(6)$, and its 4d massless fields correspond to solutions of the Dirac equation $\slashed{D}_X \psi^{X}_m = 0$, thus giving (see for example Ref. \cite{benmachiche})
\begin{eqnarray}
\label{decompgravitino}
\psi_m = \tilde{T}^i \otimes (\omega_i){}_{m\overline{n}}\gamma^{\overline{n}}\xi+\dfrac{1}{\Vert\Omega\Vert^2}\overline{\tilde{z}}{}^a\otimes(\overline{\rho}_{a}){}_{m \overline{n}\overline{p}}\Omega_q{}^{\overline{n}\overline{p}}\gamma^q \overline{\xi},
\end{eqnarray}
\noindent where $\tilde{T}^i(x)$ and $\tilde{z}^a(x)$ are Weyl fermions that carry no gauge charges, while $\omega_i \in H^{1,1}(X)$ and $\rho_a \in H^{2,1}(X)$ are the same bases of harmonic forms that we used in Eqs. \eqref{ansatzg1}--\eqref{ansatzg2}. Thus, $\tilde{T}^i$ and $\tilde{z}^a$ are indeed the superpartners of moduli fields $T^i$ and $z^a$, with which they form chiral multiplets. All in all, the $N=1$ spectrum in the low-energy theory is summarised in Table~\ref{tablesummary}
\begin{table}[h]
\begin{center}
\begin{tabular}{|l|l|l|l|}\hline
Multiplet & $H_{\textrm{GUT}}$ rep.& Field content & Number \\\hline\hline
gravitational & $\mathbf{1}$ &  $(g_{\mu \nu}, \psi_{\mu})$ & $1$ \\ \hline
vector & $\textrm{Adj}_H$ & $(A_{\mu}, \chi)$ & $1$\\ \hline
linear & $\mathbf{1}$ & $(S, \lambda)$ & $1$ \\ \hline
K\"ahler moduli& $\mathbf{1}$ & $(T^i,\tilde{T}^i)$ & $h^{1,1}(X)$\\ \hline
complex structure moduli & $\mathbf{1}$ & $(z^a, \tilde{z}^a)$ & $h^{2,1}(X)$\\ \hline
matter chiral & $\mathcal{R}_H$ & $(C^I,\tilde{C}^I)$ & $ \textrm{ind} (V_{\mathcal{R}_G})$ \\ \hline
\end{tabular}
\end{center}
\caption{\it The 4d $N=1$ supermultiplets obtained in heterotic compactification.}\label{tablesummary}
\end{table}

Using the ans\"atze \eqref{decompgaugino}--\eqref{decompgravitino} and after performing a Weyl rescaling of the 4d metric, as well as other rescalings such as $\lambda \rightarrow e^{2 \phi}\lambda$ and $\chi \rightarrow e^{- \phi} \chi$, the 4d fermionic action is brought to the form
\begin{multline}
\label{reducedfermionaction}
S_{\textrm{f}} = - \dfrac{i}{\kappa_4^2}\int d^4 x \sqrt{-g} \bigg [ i\epsilon^{\mu\rho\nu\lambda} \overline{\psi}_{\mu} \overline{\sigma}_{\lambda}D_{\rho} \psi_{\nu} +  \dfrac{\alpha'}{2} \textrm{Re}(f) \textrm{Tr}(\overline{\chi} \overline{\sigma}^{\mu} D_{\mu} \chi) + \dfrac{\!\!1}{(S\!+\!\overline{S})^2}\overline{\lambda} \overline{\sigma}^{\mu} D_{\mu} \lambda + \\ + G_{ij} \overline{\tilde{T}}{}^i \overline{\sigma}^{\mu}D_{\mu}\tilde{T}^j+ G_{\overline{a} b} \overline{\tilde{z}}{}^a \overline{\sigma}^{\mu} D_{\mu} \tilde{z}^b + \alpha' G_{IJ} \overline{\tilde{C}}{}^I \overline{\sigma}^{\mu}D_{\mu}\tilde{C}^J \bigg ],
\end{multline}
\noindent with the metrics $G_{ij}$, $G_{a \overline{b}}$ and $G_{I J}$ defined as in \eqref{modulispacemetrics} and \eqref{matterfieldmetric}.\footnote{In the last stage of compactification, we assumed $\xi$ is normalised and we used the identities $\xi^{\dagger} \gamma^m \gamma^{\overline{n}} \xi = 2 g^{m \overline{n}}$ and $\xi^{\dagger}\gamma^m \gamma^{\overline{n}} \gamma^p \gamma^{\overline{q}}\xi = 4 g^{m \overline{n}} g^{p \overline{q}}$.}

\subsection{Holomorphic Yukawa couplings and other interaction terms}
\noindent As we saw in Section~\ref{smsection}, Yukawa interactions occur when two fermions are coupled to a scalar field, or in our notation, when $\tilde{C}{}^I \tilde{C}^J C^K$ terms are present in the 4d action. Such terms are obtained from the 10d gaugino kinetic term $\textrm{Tr}(\overline{\chi}\Gamma^M D_M\chi)$, namely from its $\overline{\chi} \gamma^m A_m \chi$ part, by expressing the covariant derivative as $D_M \chi^a = \partial_{M} \chi^a +  f^a{}_{b c}A^b_{M}\chi^c$ (where $f^a{}_{b c}$ are the structure constants of  $E_8$), and extracting the fields $\tilde{C}{}^I$ and $C^I$ as low-energy massless modes of $\chi$ and $A_m$, respectively. Now, the manner in which $\chi$ and $A_m$ are decomposed is described by representations $(\mathcal{R}^i_{G}, \mathcal{R}^i_{H})$ of $G\times H$, where the index $i=1,2,3$ refers to each of the three fields involved in the coupling and $\mathcal{R}^1_G \otimes \mathcal{R}^2_G \otimes \mathcal{R}^3_G$ forms a singlet under $G$. With the expansion of $\chi$ and $A_{\overline{m}}$ into bases of harmonic $(0,1)$-forms $\nu_{i,I} \in H^1(X, V_{\mathcal{R}^i_G})$,  as in \eqref{expansiona} and \eqref{decompgaugino}, and using the definition \eqref{holomorphic30} of $\Omega$, one obtains the compactified formula
\begin{eqnarray}
\label{yukcouplingdimred}
S_{\textrm{Yuk}} \sim \int_{M_4} \lambda_{I J K}\tilde{C}{}_1^I \tilde{C}_2^J C_3^K, \,\ \textrm{where} \,\ \lambda_{I J K} =\int_X \Omega \wedge \nu^a_{1,I} \wedge \nu^b_{2,J} \wedge \nu^c_{3,K} f_{a b c},
\end{eqnarray}
\noindent with bundle indices $a,b,c$ running over the dimension of each representation $\mathcal{R}_G^i$. Overall, the structure constants $f_{abc}$ ensure that $\lambda_{IJK}$ is invariant under $G$. The implications of Eq.~\eqref{yukcouplingdimred} are profound. Since $\lambda_{I J K}$ is a quasi-topological quantity, it can be evaluated without knowledge of the internal metric $g_{m \overline{n}}$ or the connections on the bundle. However, one has to know the specific values of harmonic forms $\nu_{i,I}$, which is in general not easily achievable. By contrast, the matter field K\"ahler metric $G_{IJ}$ depends on $g_{m \overline{n}}$, as seen from Eq.~\eqref{matterfieldmetric}, so without a precise geometrical description of $X$, the fields cannot be canonically normalised. In the next chapter, we will continue the discussion on holomorphic Yukawa couplings in more depth, as they are the main focus of this thesis. 

Having established the holomorphic Yukawa couplings, the superpotential and the matter field K\"ahler potential are given by $W = \lambda^{(ijk)}_{IJK} C^I_i C^J_j C^K_k$ and $K^{(m)} = G^{(i)}_{IJ} C_i^I \overline{C}{}^J_i $, respectively (where by $C^I_i$ we now refer to chiral superfields), and one can see that $W$ is in agreement with the Gukov-Vafa-Witten expression
\begin{eqnarray}
W = \int_X \Omega \wedge H \, .
\end{eqnarray}
\noindent Finally, the heterotic compactification is not complete without dimensional reduction of  the 10d interaction terms
\begin{multline}
S_{\textrm{int}} = - \dfrac{1}{2 \kappa^2}\int d^{10}x \sqrt{-g} e^{-2\phi} \bigg[\dfrac{1}{12}\big(\overline{\psi}_M\Gamma^{MNPQR}\psi_R+6 \overline{\psi}{}^N\Gamma^P \psi^Q - \\ - \sqrt{2} \overline{\psi}_M \Gamma^{NPQ}\Gamma^M \lambda\big) H_{NPQ} + \dfrac{1}{\sqrt{2}} \overline{\psi}_M \Gamma^N \Gamma^M \lambda \partial_N \phi - \\ - \dfrac{\alpha'}{8} \textrm{Tr}\big(\overline{\chi} \Gamma^{MNP} \chi\big) H_{MNP} - \dfrac{\alpha'}{2} \textrm{Tr} \big(\overline{\chi}\Gamma^M \Gamma^{NP} \big(\psi_M+\dfrac{\sqrt{2}}{12}\Gamma_M \lambda\big) F_{NP}\big)\bigg],
\end{multline}
\noindent wherein the first component is responsible for the 4d gravitino mass term 
\begin{eqnarray}
S \sim \int_{M_4} \overline{\psi}_{\mu} \overline{\sigma}^{\mu \nu} \psi_{\nu} e^{K/2} W \, ,
\end{eqnarray}
\noindent while other components of the action give gravitino-fermion interactions and D-terms. 

\subsection{Moduli stabilisation}
As stated at the beginning of this chapter, the presence of moduli fields in the low-energy spectrum is one of the most pressing problems of string compactification. From a phenomenological standpoint, the moduli affect the predictivity of the theory, because they have no potential, so their vevs can vary continuously and arbitrarily, as time-dependent parameters. Moreover, moduli fields can mediate certain long-range forces, for which there is no experimental evidence. For these reasons, moduli must be lifted from the low-energy spectrum.

In the context of heterotic compactification, solutions to the moduli problem have been given in Refs.~\cite{aglomoduli, aglomoduli2}, where all geometrical moduli are stabilised, with the exception of one linear combination. The guiding idea is that the Hermitian Yang-Mills equations \eqref{hermitianyangmills} constrain the moduli space by requiring certain F- and D-terms in the effective 4d potential to vanish. Such terms explicitly descend from the 10d action, namely from a component of it that reads\footnote{Here, $S_{\textrm{part.}}$ can be obtained by using the integrability condition on the modified Bianchi identity~\eqref{modifiedbianchi} and then substituting the result into the $\alpha'$-dependent part of the action.}
\begin{eqnarray}
S_{\textrm{part.}} = - \dfrac{1}{2 \kappa^2}\dfrac{\alpha'}{4} \int d^{10} x \sqrt{-g} \left( -\dfrac{1}{2}\textrm{Tr}(g^{m \overline{n}}F_{m \overline{n}})^2 + \textrm{Tr}(g^{m \overline{m}}g^{n \overline{n}}F_{m n}F_{\overline{m}\overline{n}})\right).
\end{eqnarray} 
\noindent Whenever a deformation of complex structure fails to preserve $F_{mn} = F_{\overline{m}\overline{n}}=0$, at least one F-term becomes non-zero, thus signaling that the modulus of the deformation should not belong to the low-energy theory. Similarly, the failure of metric deformations to satisfy $g^{m \overline{n}} F_{m\overline{n}}=0$ is correlated to non-vanishing D-terms, which stabilise K\"ahler moduli. In general, supersymmetric and non-supersymmetric regions of the K\"ahler cone are separated by  “walls of stability”, on which the bundle $V$ must split into direct sums of smaller components, in order to preserve supersymmetry. These regions in particular provide important applications for model building and moduli stabilisation. Other stabilisation techniques involve non-perturbative effects such as gaugino condensation and membrane instantons. In the end, a viable theory would have to ensure stabilisation of all moduli, including the remaining linear combination and the $h^1(V \otimes V^*)$ bundle moduli.

As a work on string phenomenology, this thesis is concerned to some extent with the problem of moduli stabilisation, however instead of addressing the problem explicitly, we focus on the dependence of Yukawa couplings to moduli. In certain cases, the Yukawa couplings vanish and this behavior is independent of moduli. When they do not vanish, the Yukawa couplings are expressed as functions of moduli, which can be combined with results from moduli stabilisation. It is also possible to reverse the logic. One might figure out for which moduli values reasonable Yukawa couplings are obtained, and these are the values at which the moduli need to be stabilised. For example,  light Higgs usually only occurs for special complex structure choices, which can be inferred from our results.

\subsection{Wilson lines}

\noindent The problem of descending from a GUT group $H_{\textrm{GUT}}$ down to the Standard Model is resolved in heterotic string theory by turning on Wilson lines. By definition, Wilson lines are configurations of internal gauge fields with vanishing field strength, but non-vanishing parallel transport around non-contractible loops. 	For example, if $\gamma$ is a homotopically non-trivial loop in the fundamental group $\pi_1(X)$, then the Wilson line induces a homomorphism $\varphi: \pi_1(X) \rightarrow H_{\textrm{GUT}}$ through the path-ordered exponential
\begin{eqnarray}
U_{\gamma} = P \,\textrm{exp}\left( \oint_{\gamma} A_m d y^m \right),
\end{eqnarray}
\noindent thus embedding gauge-invariant observables into the GUT group. At low energies, $H_{\textrm{GUT}}$ is broken by the vevs $\langle A_m \rangle$ down to the subgroup commuting with the image of $\varphi$. The advantage of using Wilson lines compared to conventional symmetry breaking is that no additional Higgs bosons are introduced, instead the necessary ingredients are already found in the topology of the internal manifold.

The issue however is that the CICYs we defined so far in Section~\ref{cicysection} are simply connected, i.e. they have a vanishing fundamental group. Nevertheless, given a simply-connected manifold $X$, it is possible to construct a non-simply connected space by dividing $X$ by a freely acting\footnote{The action of $\Gamma$ is called free if it has no fixed points in $X$.} discrete symmetry $\Gamma$. In this case, the fundamental group of $X/\Gamma$ is $\Gamma$ and the application of the Lefschetz fixed point theorem ensures that  $X/\Gamma$ is indeed a Calabi-Yau manifold.\footnote{More precisely, for any element $g \in \Gamma$, the induced map $g^*$ on cohomology preserves the holomorphic $(3,0)$-form $\Omega$, due to the Lefschetz fixed point theorem $\sum_{q=0}^3 (-1)^q \textrm{Tr} \left(g^* \vert_{H^{q,0}(X)}\right) = 0$.} If $\vert \Gamma \vert$ is the order of the discrete group, then the Euler number of $X/\Gamma$ is $\chi(X)/\vert \Gamma \vert$, and similarly, the index of the bundle $\tilde{V} \rightarrow X/\Gamma$ descending from $V$ is given by $\textrm{ind}(V)/\vert \Gamma \vert$. This means that for a realistic GUT model with three generations of particles, the requirement is that
\begin{eqnarray}
\vert \textrm{ind}(V) \vert = 3 \vert \Gamma \vert,
\end{eqnarray}
\noindent where we employ here the data from Table~\ref{mytablenow} and use the fact that for an $SU(5)$ bundle, $\textrm{ind}(\wedge^2 V)=\textrm{ind}(V)$. Altogether, the topological constraints for the vector bundle $V$ are summarised in Table~\ref{tablerestrictions}.
\begin{table}[h]
\begin{center}
\begin{tabular}{|l|l|l|l|}\hline
Physical requirement & Topological constraint \\\hline\hline
GUT group & $c_1(V)=0$ \\ \hline
Anomaly cancellation &$c_2(TX) - c_2(V) \geq 0 $  \\ \hline
Supersymmetry & $V$ is poly-stable  \\ \hline
Three generations & $\vert \textrm{ind}(V) \vert = 3 \vert \Gamma \vert$ \\ \hline
\end{tabular}
\end{center}
\caption{\it Phenomenological constraints for a holomorphic vector bundle $V$ on a Calabi-Yau threefold $X$, in the context of heterotic compactification.}\label{tablerestrictions}
\end{table}
In the next chapter we will show how these conditions are implemented if $V$ is a direct sum of line bundles. By concluding this review, we can now proceed to the main topic of the thesis.

\chapter{Yukawa Couplings in the Heterotic Tetra-quadric Model}
\label{tetraquadricchapter}

In the last sections of Chapter~\ref{odyssey}, we checked that heterotic Calabi-Yau compactifications can indeed produce terms which are identifiable with the standard $N = 1$ supergravity action in four dimensions \cite{wessandbagger}. This was mainly provided by Eqs.~\eqref{0thorderbosonicaction}, \eqref{reducedyangmillsaction} and \eqref{reducedfermionaction}, with the proviso that moduli fields describing the size and shape of extra dimensions have to be stabilised. The natural step forward would be to search for realistic models with the correct spectrum of particles and physical Yukawa couplings matching experimental observation.

Over the past two decades, string model building based on heterotic Calabi-Yau compactifications has seen considerable progress~\cite{Braun:2005ux, Braun:2005bw, Braun:2005nv, Bouchard:2005ag, Blumenhagen:2006ux, Blumenhagen:2006wj, Anderson:2007nc, Anderson:2008uw, Anderson:2009mh, Braun:2009qy, Braun:2011ni} and large classes of models with the MSSM spectrum can now be constructed using algorithmic approaches~\cite{Anderson:2011ns,Anderson:2012yf,Anderson:2013xka}. The other problem however, involving the calculation of Yukawa couplings for such models, has remained largely unaddressed, both in terms of general techniques and actual specific results. In this chapter, we will attempt to make some progress in this direction and develop new methods, mainly based on 
differential geometry, to calculate holomorphic Yukawa couplings for heterotic line bundle models. 

Calculating the physical Yukawa couplings of a supersymmetric string compactification comes in two parts: the calculation 
of the holomorphic Yukawa couplings, that is, the couplings in the superpotential, and the calculation of the matter field K\"ahler metric, in order to 
work out the field normalisation. The holomorphic Yukawa couplings are quasi-topological -- they do not depend on the Calabi-Yau metric or the Hermitian Yang-Mills 
connection on the bundle -- and they can, therefore, in principle, be calculated with algebraic methods. 
The situation is very different for the K\"ahler metric which does depend on the metric and the bundle connection. 
It is unlikely that an algebraic method for its calculation can be found and, hence,  methods of differential geometry will be required. 

At present, a full calculation of the physical (perturbative) Yukawa couplings is only understood for heterotic Calabi-Yau models with standard embedding. In this case, the holomorphic Yukawa couplings for the $(1,1)$ matter fields are given by the Calabi-Yau triple intersection numbers~\cite{Strominger:1985ks}, while the holomorphic $(2,1)$ Yukawa couplings have been worked out in Ref.~\cite{Candelas:1987se}. The matter field K\"ahler metrics are identified with the corresponding moduli space metrics \eqref{modulispacemetrics}, as explained in Ref.~\cite{Candelas:1990pi}. Further, in Ref.~\cite{Candelas:1987se}, the relation between the analytic calculation of $(2,1)$ holomorphic Yukawa couplings and the algebraic approach has been worked out in detail.

Much less is known for heterotic Calabi-Yau models with general vector bundles. An algebraic approach for the calculation of holomorphic Yukawa couplings for such ``non-standard embedding" models has been outlined and applied to examples in Ref.~\cite{Anderson:2009ge}. However, the matter field K\"ahler metric has not been computed for any non-standard embedding model on a Calabi-Yau manifold and no clear method for its computation has been formulated. 

The purpose of this chapter is two-fold. First, we would like to develop explicit methods based on differential geometry to compute the holomorphic Yukawa couplings for heterotic models with non-standard embedding. Secondly, we would like to understand how these methods relate to the algebraic ones pioneered in Ref.~\cite{Candelas:1987se} and further developed in Ref.~\cite{Anderson:2009ge}. Apart from occasional remarks, we will not be concerned with the matter field K\"ahler metric until Chapter~\ref{kahlerchapter}.  For ease of terminology,  the term ``Yukawa couplings" will refer to the holomorphic Yukawa couplings in the remainder of the thesis.

The present work will be carried out within the context of heterotic line bundle models~\cite{Anderson:2011ns,Anderson:2012yf,Anderson:2013xka}, perhaps the simplest class of heterotic Calabi-Yau models with non-standard embedding. For those models, the gauge bundle has an Abelian structure group $G= S(U(1)^r)$ and is realised by a sum of line bundles, a feature which makes explicit calculations of bundle properties significantly more accessible. Furthermore, we will work within perhaps the simplest class of Calabi-Yau manifolds, namely complete intersections in products of projective spaces~\cite{Green:1986ck,Candelas:1987kf,Hubsch:1992nu} (CICYs). More specifically, we focus on hyper-surfaces in products of projective spaces and the tetra-quadric in the ambient space ${\cal A}=\mathbb{P}^1\times\mathbb{P}^1\times\mathbb{P}^1\times\mathbb{P}^1$ in particular. On the one hand, 
the simplicity of the set-up facilitates developing new and explicit methods to calculate Yukawa couplings. On the other hand, it is known~\cite{Anderson:2011ns,Anderson:2012yf} that this class contains interesting models with a low-energy MSSM spectrum, so that we will be able to apply our methods to quasi-realistic examples. 

Picking a vector bundle with Abelian structure group means that the low energy theory is governed by the gauge group $H = H_{\textrm{GUT}}\times S(U(1)^r)$, commuting with $G$. One could naively ask whether such a model is anomaly-free, given that the extra $U(1)$ bosons could in principle contribute to anomalous triangle loops. The answer however is that such anomalies are canceled by the 4d version of the Green-Schwarz mechanism, which ensures that the extra $U(1)$ bosons acquire St\"uckelberg masses close in magnitude to the compactification scale. Thus, in the 4d theory, the additional $U(1)$ symmetries are to be interpreted as global and, in fact, their presence may be rather beneficial for phenomenology. In conjunction with topology, these global symmetries could potentially explain the structure of Yukawa matrices and why certain terms, such as \eqref{protondecayterms}, that trigger fast proton decay, are forbidden.

The plan of this chapter is as follows. In the next section, we will lay the ground by reviewing some of the basics, including the general structure of heterotic Yukawa couplings, heterotic line bundle models and complete intersection Calabi-Yau manifolds. Since our main focus will be on the tetra-quadric Calabi-Yau manifold, we need to understand in some detail the differential geometry of $\mathbb{P}^1$ and line bundles thereon. This will be developed in Section~\ref{forms}. General results for Yukawa couplings on the tetra-quadric and some toy examples are given in Section~\ref{toyex}. Section~\ref{realex} presents a complete calculation of the Yukawa couplings for a quasi-realistic model~\cite{Anderson:2011ns,Anderson:2012yf, Buchbinder:2013dna, Buchbinder:2014qda,Buchbinder:2014sya,Buchbinder:2014qca} with MSSM spectrum on the tetra-quadric. We conclude in Section~\ref{c1conclusions}.

Some related matters and technical issues have been deferred to the Appendices.

\section{Yukawa couplings in line bundle models}


\subsection{General properties of Yukawa couplings in heterotic Calabi--Yau compactifications}


We start with an overview of holomorphic Yukawa couplings in the context of the $E_8 \times E_8$ heterotic string theory on a 
Calabi--Yau manifold (see, for example, Ref.~\cite{GSW}). As seen in Section~\ref{dimredchapter}, the matter fields originate from the $E_8\times E_8$ gauge fields $A$ and 
the associated gauginos. Here we focus on one $E_8$ factor (``the visible sector") and assume that the Calabi-Yau manifold $X$ carries a 
principal bundle with structure group $G$ embedded into $E_8$. The (visible) low-energy gauge group $H$ is then the 
commutant of $G$ within $E_8$ and the types of matter multiplets can be read off from the branching
\begin{equation} 
 {\bf 248}\rightarrow \left[({\rm Adj}_G,{\bf 1})\oplus ({\bf 1},{\rm Adj}_H)\oplus\bigoplus ({\cal R}_G, {\cal R}_H)\right]_{G\times H} \label{genbranch}
\end{equation}
of the ${\bf 248}$ adjoint representation of $E_8$ under $G\times H$. Specifically, for the above branching, the low-energy theory can contain  matter multiplets transforming as representations ${\cal R}_H$ under $H$. These multiplets descend from harmonic bundle valued (0,1)-forms $\nu\in H^1(X,V)$, where $V$ is a vector bundle associated to the principal bundle via the $G$ representations ${\cal R}_G$. Consider three representations $({\cal R}_G^i,{\cal R}_H^i)$, where $i=1,2,3$, which appear in the decomposition~\eqref{genbranch}, such that ${\cal R}_G^1\otimes {\cal R}_G^2\otimes {\cal R}_G^3$ contains a singlet. The three associated vector bundles are denoted as $V_i$ with harmonic bundle-valued (0,1)-forms $\nu_i\in H^1(X,V_i)$. Then, the associated holomorphic Yukawa couplings can be computed from
\begin{equation}
\lambda(\nu_1,\nu_2,\nu_3)=\int_X\Omega\wedge\nu_1\wedge\nu_2\wedge\nu_3\; , \label{Yukgen}
\end{equation}
where $\Omega$ is the holomorphic $(3,0)$ form on $X$ and an appropriate contraction over the bundle indices in $\nu_i$ onto the singlet direction is implied. Let us introduce sets of basis forms, $\nu_{i,I}$, where $I=1,\ldots ,h^1(X,V_i)$, for the cohomologies $H^1(X,V_i)$ and define $\lambda_{IJK}=\lambda(\nu_{1,I},\nu_{2,J},\nu_{3,K})$. The four-dimensional $N=1$ chiral superfields associated to $\nu_{i,I}$ are denoted $C_i^I$ and these fields transform as ${\cal R}_H^i$ under the gauge group $H$. The superpotential for these fields can be written as
\begin{equation}
 W=\lambda_{I J K}C_1^I C_2^J C_3^K\; .
\end{equation} 
Here, we are mainly interested in the phenomenologically promising structure groups $G=SU(3)$, $SU(4)$, $SU(5)$ (and their maximal rank sub-groups), which lead to the low-energy gauge groups $H=E_6$, $SO(10)$, $SU(5)$ (times possible $U(1)$ factors), respectively. For these three groups, the decomposition~\eqref{genbranch} takes the form
\begin{align}
&{\bf 248} \,\rightarrow \, \left[({\bf 8},{\bf 1})\oplus ({\bf 1},{\bf 78})\oplus ({\bf 3},{\bf 27})\oplus (\overline{\bf 3},\overline{\bf 27})\right]_{SU(3)\times E_6}
\label{E6dec}\\
& {\bf 248}\,\rightarrow \,\left[({\bf 15},{\bf 1})\oplus ({\bf 1},{\bf 45})\oplus ({\bf 4},{\bf 16})\oplus (\overline{\bf 4},\overline{\bf 16})\oplus ({\bf 6},{\bf 10})\right]_{SU(4)\times SO(10)}
\label{SO10dec}\\
& \resizebox{0.91\hsize}{!}{$
{\bf 248} \,\rightarrow\, \left[({\bf 24},{\bf 1})\oplus ({\bf 1},{\bf 24})\oplus ({\bf 5},{\bf 10})\oplus (\overline{\bf 5},\overline{\bf 10})\oplus ({\bf 10},\overline{\bf 5})\oplus (\overline{\bf 10},{\bf 5})\right]_{SU(5)\times SU(5)}$
} 
\label{SU5dec}
\end{align}

For $G=SU(3)$ we have matter multiplets in representations ${\bf 27}$, $\overline{\bf 27}$ and ${\bf 1}$ of the low-energy gauge group $H=E_6$ and possible Yukawa couplings of type ${\bf 27}^3$, $\overline{\bf 27}^3$, ${\bf 1}\,{\bf 27}^2$ and ${\bf 1}\,\overline{\bf 27}^2$.

For $G=SU(4)$, the families come in ${\bf 16}$ representations and the anti-families  in $\overline{\bf 16}$ representations of $H=SO(10)$. Higgs multiplets reside in ${\bf 10}$ representations and bundle moduli in singlets, ${\bf 1}$. Possible Yukawa couplings are of type ${\bf 10}\,{\bf 16}^2$, ${\bf 10}\,\overline{\bf 16}^2$, ${\bf 1}\,{\bf 16}\,\overline{\bf 16}$ and ${\bf 1}\,{\bf 10}^2$.

Finally, for $G=SU(5)$ and low-energy gauge group $H=SU(5)$ we have families in $\overline{\bf 5}\oplus {\bf 10}$, anti-families in ${\bf 5}\oplus\overline{\bf 10}$ and bundle moduli singlets, ${\bf 1}$. Allowed Yukawa couplings include the up-type Yukawa couplings ${\bf 5}\,{\bf 10}^2$, the down-type Yukawa couplings $\overline{\bf 5}\,\overline{\bf 5}\,{\bf 10}$ as well as the singlet couplings ${\bf 1}\,{\bf 5}\,\overline{\bf 5}$, ${\bf 1}\,{\bf 10}\,\overline{\bf 10}$.

While Eq.~\eqref{Yukgen} has been, initially, written down in terms of the harmonic representatives $\nu_i$ of the cohomologies $H^1(X,V_i)$, it is important to note that the expression is, in fact, independent of the choice of representatives. 
To see this, perform the transformation\footnote{Here and in the following, we will often denote the derivative $\bar{\partial}_E$ on differential forms taking values in the vector bundle $E$ simply by $\bar{\partial}$ to avoid cluttering the notation.} $\nu_i\rightarrow \nu_i+\bar{\partial}\xi_i$ on Eq.~\eqref{Yukgen}, where $\xi_i$ are sections of $V_i$. Then, integrating by parts and using $\bar{\partial}\nu_i=0$, $\bar{\partial}\Omega=0$ and $\bar{\partial}^2=0$, it follows immediately that
\begin{equation}
 \lambda(\nu_1+\bar{\partial}\xi_1,\nu_2+\bar{\partial}\xi_2,\nu_3+\bar{\partial}\xi_3)=\lambda(\nu_1,\nu_2,\nu_3)\; .
\end{equation}
This quasi-topological property of the holomorphic Yukawa couplings means that they can, in principle, be computed purely algebraically, as has been noted in Refs.~\cite{Candelas:1987se,Anderson:2009ge}. To recall how this works we focus on the case $G=SU(3)$ and low-energy gauge group $H=E_6$. The families in ${\bf 27}$ descend from bundle-valued (0,1)-forms $\nu,\mu,\rho \in H^1(X,V)$, where $V$ is the associated vector bundle in the fundamental representation, ${\bf 3}$, of $SU(3)$. Since $c_1(V)=0$ it follows that $\wedge^3 V\cong{\cal O}_X$ and we have a map
\be 
H^1 (X, V) \times H^1 (X, V) \times H^1 (X, V) \to H^3 (X, \wedge^3 V)\simeq H^3 (X, {\cal O}_X) \simeq {\mathbb C}\,. 
\label{1.4}
\ee
More explicitly, this can be expressed by the cup product
\begin{equation}
 \nu\wedge\mu\wedge\rho=\tau(\nu,\mu,\rho)\,\overline{\Omega}\; ,\label{cup}
\end{equation}
where $\tau(\nu,\mu,\rho)$ is a complex number and $\overline{\Omega}$ is the unique harmonic representative of the cohomology group $H^3 (X, {\cal O}_X)$. Inserting into Eq.~\eqref{Yukgen}, it follows that the complex number $\tau(\nu,\mu,\rho)$ is proportional to the Yukawa couplings via
\begin{equation}
 \lambda(\mu,\nu,\rho)=\tau(\nu,\mu,\rho)\int_X\Omega\wedge\overline{\Omega}\; .
\end{equation} 
This means that the ${\bf 27}^3$ Yukawa couplings, up to an overall constant, can be computed algebraically, by performing a (cup) product between three cohomology representatives. Similar arguments can be made for the other Yukawa couplings in the $SU(3)$ case and indeed for other bundle structure groups $G$.

Such an algebraic calculation has been carried out for certain examples in Refs.~\cite{Candelas:1987se,Anderson:2009ge}. While it is elegant and avoids the evaluation of integrals, it also has a number of drawbacks. As a practical matter, the relevant cohomologies are not always directly known, but are merely represented by certain isomorphic cohomologies. In this case, it is not always obvious how the cup product should be carried out. Perhaps more significantly, computing the physical (rather than just the holomorphic) Yukawa couplings also requires knowledge of the matter field K\"ahler metric \eqref{matterfieldmetric} which is proportional to the inner product
\begin{equation}
 (\nu,\omega)=\int_X\nu\wedge \bar{\star}_E\,\omega
\end{equation}
between two harmonic $(0,1)$-forms $\nu$, $\omega$ representing cohomologies in $H^1(X,V)$. Unlike the holomorphic Yukawa couplings, this expression is not independent of the choice of representatives due to the presence of the complex conjugation, as can be seen by performing a transformation $\nu\rightarrow\nu+\bar{\partial}\alpha$, $\omega\rightarrow\omega+\bar{\partial}\beta$. It needs to be computed with the harmonic $(0,1)$-forms and requires knowledge of the Ricci-flat Calabi-Yau metric.  Consequently, a full calculation of the physical Yukawa couplings  will have to rely on differential geometry. One purpose of this thesis is to develop such differential geometry methods, for the immediate purpose of calculating the holomorphic Yukawa couplings, but in view of a full calculation of the physical couplings in the future.


\subsection{A review  of line bundle models}


Perhaps the simplest heterotic compactifications for which to calculate Yukawa couplings, 
apart from models with standard embedding, are line bundle models. In the remainder of the thesis, we will focus on calculating holomorphic Yukawa 
couplings for such line bundle models and, in the present sub-section, we begin by reviewing their general structure, following 
Refs.~\cite{Anderson:2011ns, Anderson:2012yf}. 

Heterotic line bundle models rely on a gauge bundle with (visible) Abelian structure group $G=S(U(1)^n)$, which can be described by a line bundle sum
\be 
V = \bigoplus_{a=1}^n L_a\quad\mbox{with}\quad c_1(V)=0\; ,
\label{Vdef}
\ee
where $L_a\rightarrow X$ are line bundles over the Calabi-Yau manifold $X$. Here, the condition $c_1(V)=0$ ensures that the structure group of $V$ is indeed special unitary, rather than merely unitary. 

As every heterotic model, line bundle models need to satisfy two basic consistency conditions. Firstly, the bundle $V$ needs to be supersymmetric, which is equivalent to requiring vanishing slopes
\begin{equation}
 \mu(L_a)\equiv\int_X c_1(L_a)\wedge J\wedge J\stackrel{!}{=}0
\end{equation}
for all line bundles $L_a$, where $J$ is the K\"ahler form of the Calabi-Yau manifold $X$. The slope-zero conditions 
are constraints in K\"ahler moduli space which have to be solved simultaneously for all line bundles in order for the bundle $V$ to preserve supersymmetry. 
Secondly, we need to be able to satisfy the heterotic anomaly condition which is guaranteed if we require that
\begin{equation}
 c_2(TX)-c_2(V)\geq 0 \; ,
\end{equation} 
\noindent or, equivalently, that $c_2(TX)-c_2(V)$ is in the Mori cone of $X$. In this case, the anomaly condition can always be satisfied by adding five-branes to the 
model (although other completions involving a non-trivial hidden bundle or a combination of hidden bundle and five-branes are usually possible).\\[2mm]
Of particular interest are line bundle sums with rank $n=3,4,5$ for which the associated (visible) 
low-energy gauge groups are $H=E_6\times S(U(1)^3)$, $H=SO(10)\times S(U(1)^4)$ and $SU(5)\times S(U(1)^5)$, respectively. 
For the non-Abelian part of these gauge groups, the multiplet structure of the low-energy theory can be read off 
from Eqs.~\eqref{E6dec}--\eqref{SU5dec}. In addition, multiplets carry charges under the Abelian part, $S(U(1)^n)$, of the gauge group. 
It is convenient to describe these charges by an integer vector  ${\bf q} =(q_1, q_2, \dots, q_n)$. Since we would like to label 
representations of $S (U(1)^n)$, rather than of $U(1)^n$, two such vectors ${\bf q}$ and ${\bf \tilde{q}}$ have to be identified 
if ${\bf q} - {\bf \tilde{q}} = {\mathbb Z} (1, 1, \dots, 1)$. This charge vector will be attached as a subscript to the representation 
of the non-Abelian part. The number of each type of multiplet equals the dimension of the cohomology $H^1(X, K)$ for a certain 
line bundle $K$, which is either one of the line bundles $L_a$ or a tensor product thereof. 
The precise list of multiplets for the three cases $n=3,4,5$, together with the associated line bundles $K$ is provided in 
Tables~\ref{tab:e6}, \ref{tab:so10} and \ref{tab:su5}.
\renewcommand{\arraystretch}{1.2}
\begin{table}[h]
\begin{center}
\begin{tabular}{|l|l|l|l|}\hline
multiplet&indices&line bundle $K$&interpretation\\\hline\hline
${\bf 27}_{{\bf e}_a}$&$a=1,2,3$&$L_a$&families/Higgs\\\hline
$\overline{\bf 27}_{-{\bf e}_a}$&$a=1,2,3$&$L_a^*$&mirror-families/Higgs\\\hline
${\bf 1}_{{\bf e}_a-{\bf e}_b}$&$a,b=1,2,3\,,\;a\neq b$&$L_a\otimes L_b^*$&bundle moduli\\\hline
\end{tabular}
\end{center}
\caption{\it Multiplets and associated line bundles for bundle structure group $G=S(U(1)^3)$ and low-energy gauge group $H=E_6\times S(U(1)^3)$.}\label{tab:e6}
\end{table}
\begin{table}[h]
\begin{center}
\begin{tabular}{|l|l|l|l|}\hline
multiplet&indices&line bundle $K$&interpretation\\\hline\hline
${\bf 16}_{{\bf e}_a}$&$a=1,2,3,4$&$L_a$&families\\\hline
$\overline{\bf 16}_{-{\bf e}_a}$&$a=1,2,3,4$&$L_a^*$&mirror-families\\\hline
${\bf 10}_{{\bf e}_a+{\bf e}_b}$&$a=1,2,3,4\,,\;a<b$&$L_a\otimes L_b$&Higgs\\\hline
${\bf 1}_{{\bf e}_a-{\bf e}_b}$&$a,b=1,2,3,4\,,\;a\neq b$&$L_a\otimes L_b^*$&bundle moduli\\\hline
\end{tabular}
\end{center}
\caption{\it Multiplets and associated line bundles for bundle structure group $G=S(U(1)^4)$ and low-energy gauge group $H=SO(10)\times S(U(1)^4)$.}\label{tab:so10}
\end{table}
\begin{table}[h]
\begin{center}
\begin{tabular}{|l|l|l|l|}\hline
multiplet&indices&line bundle $K$&interpretation\\\hline\hline
${\bf 10}_{{\bf e}_a}$&$a=1,2,3,4,5$&$L_a$&$(Q,u,e)$ families\\\hline
$\overline{\bf 10}_{-{\bf e}_a}$&$a=1,2,3,4,5$&$L_a^*$&$(\tilde{Q},\tilde{u},\tilde{e})$ mirror-families\\\hline
$\overline{\bf 5}_{{\bf e}_a+{\bf e}_b}$&$a,b=1,2,3,4,5\,,\;a<b$&$L_a\otimes L_b$&$(L,d)$ families\\\hline
${\bf 5}_{-{\bf e}_a-{\bf e}_b}$&$a,b=1,2,3,4,5\,,\;a<b$&$L_a^*\otimes L_b^*$&$(\tilde{L},\tilde{d})$ mirror-families\\\hline
${\bf 1}_{{\bf e}_a-{\bf e}_b}$&$a,b=1,2,3,4,5\,,\;a\neq b$&$L_a\otimes L_b^*$&bundle moduli\\\hline
\end{tabular}
\end{center}
\caption{\it Multiplets and associated line bundles for bundle structure group $G=S(U(1)^5)$ and low-energy gauge group 
$H=SU(5)\times S(U(1)^5)$. 
}\label{tab:su5}
\end{table}
As is clear from the tables, all relevant $S(U(1)^n)$ charges can be expressed easily in terms of the $n$-dimensional standard unit vectors ${\bf e}_a$. Frequently, in order to simplify the notation for multiplets, we will replace the subscripts ${\bf e}_a$ simply by $a$.  For example, in the $SO(10)\times S(U(1)^4)$ case, the multiplet ${\bf 16}_{{\bf e}_a}$ becomes ${\bf 16}_a$ or the multiplet ${\bf 10}_{{\bf e}_a+{\bf e}_b}$ becomes ${\bf 10}_{a,b}$.

For all three cases, the low-energy spectrum contains fields ${\bf 1}_{a,b}$ which are singlets under the non-Abelian part of the 
gauge group, but are charged under $S(U(1)^n)$. These fields should be interpreted as bundle moduli which parametrise 
deformations away from a line bundle sum to bundles with non-Abelian structure group. For many models of interest these bundle moduli 
are present in the low-energy spectrum and, in such cases, the Abelian bundle is embedded in a moduli space of generically non-Abelian bundles. 
Much can be learned about non-Abelian bundles by such deformations away from the Abelian locus. This is one of the reasons why studying Yukawa 
couplings for line bundle models can yield insights into the structure of Yukawa couplings for non-Abelian bundles. Another reason is more technical. 
In practice, non-Abelian bundles are often constructed from line bundles, for example via extension or monad sequences, and, hence, some of the 
methods developed for line bundles will be useful to tackle the non-Abelian case.
\vskip 2mm
So far, we have considered the ``upstairs" theory with a GUT-type gauge group. In order to break this theory to the 
standard-model group we require a freely-acting symmetry $\Gamma$ on the Calabi-Yau manifold $X$. The line bundle sum $V$ should descend to the 
quotient Calabi-Yau $X/\Gamma$, that is, it should have a $\Gamma$-equivariant structure. 
Downstairs, on the manifold $X/\Gamma$, we should include a Wilson line, defined by a representation $W$ of $\Gamma$ into 
the (hypercharge direction of the) GUT group. As a result, each downstairs multiplet, $\psi$, acquires an induced $\Gamma$-representation 
denoted $\chi_\psi$. Luckily, the resulting downstairs spectrum can be computed in a simple group-theoretical fashion from the upstairs spectrum. 
Consider a certain type of upstairs multiplet with associated line bundle $K$. 
By virtue of the $\Gamma$-equivariant structure of $V$, the cohomology $H^1(X, K)$, associated to the upstairs multiplet, 
becomes a $\Gamma$-representation.\footnote{In more complicated cases, line bundles might not be equivariant individually, 
but several line bundles may form an equivariant block. However, the computation of downstairs cohomology for such 
cases proceeds in a similar group-theoretical fashion.} To compute the spectrum of a certain type, $\psi$, of downstairs multiplet 
contained in $H^1(X, K)$, we should determine the $\Gamma$-singlet part of 
\begin{equation}
 H^1(X, K)\otimes \chi_\psi\; . \label{equivcoh}
\end{equation} 
Fortunately, the computation of Yukawa couplings relates to this Wilson line breaking mechanism in a straightforward way. 
We can obtain the downstairs (holomorphic) Yukawa couplings by basically extracting the relevant $\Gamma$-singlet directions of the upstairs 
Yukawa couplings.

In our later examples, we will consider Wilson line breaking for the gauge group $SU(5)$. In this case, the Wilson line can be conveniently described in terms of two one-dimensional $\Gamma$-representations $\chi_2$, $\chi_3$, satisfying $\chi_2^2\otimes\chi_3^3=1$ and with at least one of them non-trivial. Such a Wilson line, embedded into the hypercharge direction, breaks $SU(5)$ to the standard model group. The $\Gamma$-representations $\chi_\psi$ of the various standard model multiplets, which enter Eq.~\eqref{equivcoh}, are then explicitly given by
\begin{equation}
 \chi_Q=\chi_2\otimes\chi_3\,,\quad \chi_u= \chi^2_3\,,\quad \chi_e=\chi_2^2\,,\quad \chi_d=\chi_3^*\,,\quad \chi_L=\chi_2^*\,,\quad
 \chi_H=\chi_2^*\,,\quad \chi_{\bar{H}}=\chi_2\;. \label{WLcharges}
\end{equation} 


\subsection{Holomorphic Yukawa couplings for line bundle models}
\label{examples}

For heterotic line bundle models, the $(0,1)$-forms $\nu_1$, $\nu_2$ and $\nu_3$, contained in the general expression~\eqref{Yukgen} for the Yukawa couplings, represent the first cohomologies of certain line bundles, denoted by $K_1$, $K_2$ and $K_3$, so that $\nu_i\in H^1(X,K_i)$. The structure of the integral~\eqref{Yukgen} (or, equivalently, four-dimensional gauge symmetry) means that such a line bundle Yukawa coupling can be non-zero only if
\begin{equation}
 K_1 \otimes K_2 \otimes K_3 = {\cal O}_X\; .
\end{equation}
Provided this is the case, the Yukawa coupling is given by 
\begin{equation}
\lambda(\nu_1,\nu_2,\nu_3)=\int_X\Omega\wedge\nu_1\wedge\nu_2\wedge\nu_3\; , \label{Yuklb}
\end{equation}
\begin{table}[h]
\begin{footnotesize}
\begin{center}
\begin{tabular}{|l||c|c|c|c|c|}\hline
Gauge group&Yukawa coupling&$K_1$&$K_2$&$K_3$&index constraint\\\hline\hline
\multirow{3}{*}{$E_6\times S(U(1)^3)$}
&${\bf 27}_a\,{\bf 27}_b\,{\bf 27}_c$&$L_a$&$L_b$&$L_c$&$a,b,c$ all different\\\cline{2-6}
&$\overline{\bf 27}_a\,\overline{\bf 27}_b\,\overline{\bf 27}_c$&$L_a^*$&$L_b^*$&$L_c^*$&$a,b,c$ all different\\\cline{2-6}
&${\bf 1}_{a,b}\,{\bf 27}_b\,\overline{\bf 27}_a$&$L_a\otimes L_b^*$&$L_b$&$L_a^*$&$a\neq b$\\\hline\hline
\multirow{3}{*}{$SO(10)\times S(U(1)^4)$}
&${\bf 10}_{a,b}\,{\bf 16}_c\,{\bf 16}_d$&$L_a\otimes L_b$&$L_c$&$L_d$&$a,b,c,d$ all different\\\cline{2-6}
&${\bf 10}_{a,b}\,\overline{\bf 16}_a\, \overline{\bf 16}_b$&$L_a\otimes L_b$&$L_a^*$&$L_b^*$&$a\neq b$\\\cline{2-6}
&${\bf 1}_{a,b}\,{\bf 16}_b\, \overline{\bf 16}_a$&$L_a\otimes L_b^*$&$L_b$&$L_a^*$&$a\neq b$\\\hline\hline
\multirow{6}{*}{$SU(5)\times S(U(1)^5)$}
&$\overline{\bf 5}_{a,b}\,\overline{\bf 5}_{c,d}\,{\bf 10}_e$&$L_a\otimes L_b$&$L_c\otimes L_d$&$L_e$&$a,b,c,d,e$ all different\\\cline{2-6}
&${\bf 5}_{a,b}\,{\bf 10}_a\,{\bf 10}_b$&$L_a^*\otimes L_b^*$&$L_a$&$L_b$&$a\neq b$\\\cline{2-6}
&${\bf 5}_{a,b}\,{\bf 5}_{c,d}\,\overline{\bf 10}_e$&$L_a^*\otimes L_b^*$&$L_c^*\otimes L_d^*$&$L_e^*$&$a,b,c,d,e$ all different\\\cline{2-6}
&$\overline{\bf 5}_{a,b}\,\overline{\bf 10}_a\,\overline{\bf 10}_b$&$L_a\otimes L_b$&$L_a^*$&$L_b^*$&$a\neq b$\\\cline{2-6}
&${\bf 1}_{a,b}\,{\bf 5}_{a,c}\,\overline{\bf 5}_{b,c}$&$L_a\otimes L_b^*$&$L_a^*\otimes L_c^*$&$L_b\otimes L_c$&$a\neq b\,,a\neq c\,,b\neq c$\\\cline{2-6}
&${\bf 1}_{a,b}\,{\bf 10}_b\,\overline{\bf 10}_a$&$L_a\otimes L_b^*$&$L_b$&$L_a^*$&$a\neq b$\\\hline
\end{tabular}
\caption{Relation between the line bundles $K_i$ which enter the expression~\eqref{Yuklb} for the Yukawa couplings and the line bundles $L_a$ which define the vector bundle $V$ in Eq.~\eqref{Vdef}. Note that $K_1\otimes K_2\otimes K_3={\cal O}_X$ always follows, 
in some cases due to $c_1(V)=0$, which imples $L_1\otimes\cdots\otimes L_n={\cal O}_X$.}\label{tab:KLrel}
\end{center}
\end{footnotesize}
\end{table}
an expression similar to Eq.~\eqref{Yukgen}, but with the $(0,1)$-forms $\nu_i$ now taking values in the line bundles $K_i$. The precise relation between the line bundles $K_i$ and the line bundles $L_a$ in Eq.~\eqref{Vdef} which define the vector bundle $V$ depends on the low-energy gauge group and the type of Yukawa coupling under consideration. For the three gauge groups of interest and the relevant types of Yukawa couplings these relations are summarised in Table~\ref{tab:KLrel}.
From Eq.~\eqref{Yuklb} it is clear that the Yukawa couplings can depend on the complex structure moduli of the Calabi-Yau manifold $X$. Later, we will see examples with and without explicit complex structure dependence. Given that individual line bundles have no moduli, line bundle Yukawa couplings do not depend on bundle moduli. However, as discussed earlier, line bundle models often reside in a larger moduli space of non-Abelian bundles and Yukawa couplings on this larger moduli space will, in general, display bundle moduli dependence. In this context, our results for line bundle models can be interpreted as a leading-order expressions which are exact at the line bundle locus and provide a good approximation for small deformations away from the line bundle locus.
 

\subsection{Projective ambient spaces}
\label{coboundary}


So far, our discussion applies to line bundle models on any Calabi-Yau manifold. In this sub-section and from now on we will specialise to what is perhaps the simplest class of Calabi-Yau manifolds, namely Calabi-Yau hyper-surfaces in products of projective spaces. Restricting to this class allows us to take the first steps towards evaluating the Yukawa integral~\eqref{Yuklb} and, later on, to explicitly construct the relevant cohomology representatives and compute the integral.

Concretely, we will consider ambient spaces of the form
\be 
{\cal A}= {\mathbb P}^{n_1} \times {\mathbb P}^{n_2} \times \dots \times {\mathbb P}^{n_m} \,, 
\label{Adef}
\ee
where $n_1+n_2+...+ n_m=4$. The Calabi-Yau hyper-surface $X$ in ${\cal A}$ is defined as the zero-locus of a homogeneous polynomial $p$ with multi-degree $(n_1+1,n_2+1,\ldots ,n_m+1)$, which can be thought of as a section of the normal bundle ${\cal N}={\cal O}_{\cal A}(n_1+1,n_2+1,\ldots ,n_m+1)$. Examples in this class include  the quintic in ${\mathbb P}^4$, the bicubic in 
${\mathbb P}^2 \times {\mathbb P}^2 $ and the tetra-quadric in ${\mathbb P}^1 \times {\mathbb P}^1 \times {\mathbb P}^1 \times {\mathbb P}^1$. 

To evaluate the Yukawa couplings for such Calabi-Yau hyper-surfaces, we first assume that the relevant $(0,1)$-forms $\nu_i$ and the $(3,0)$-form $\Omega$ on $X$ can be obtained as restrictions of ambient space counterparts $\hat{\nu}_i$ and $\hat{\Omega}$. Under this assumption and by inserting an appropriate delta-function current~\cite{Candelas:1987se}, we can re-write Eq.~\eqref{Yuklb} as the ambient space integral
\be 
\l(\nu_1,\nu_2,\nu_3) =- \frac{1}{2 i}\int_{{\cal A}} \hat{\Omega} \wedge  \hat{\nu}_{1}  \wedge  \hat{\nu}_{2}   \wedge  \hat{\nu}_{3} \wedge  \d^2 (p) dp \wedge d {\bar p}\,. 
\label{Yukamb}
\ee
Note that this expression contains the imaginary prefactor $i$, which is completely acceptable for holomorphic Yukawa couplings. One expects the Yukawa couplings to become real (except for the small CP violating phase in the CKM matrix) only after field normalisation is performed. The construction of $\Omega$ and $\hat{\Omega}$ for Calabi--Yau hyper-surfaces in products of projective spaces is 
well known~\cite{Witten:1985xc, Strominger:1985it, Candelas:1987se, Candelas:1987kf} and we will simply present the result. To this end, we introduce the forms 
\be 
\mu_i = \frac{1}{n_i!} \eps_{\a_0 \a_1 \dots \a_{n_i}} x^{\a_0}_i d x^{\a_1}_i \wedge \dots \wedge dx^{\a_{n_i}}_i\;,\quad
\mu =\mu_1 \wedge \mu_2 \wedge \dots \wedge \mu_m\,. 
\label{10.2}
\ee
where $x^\a_i$ are the homogeneous coordinates on $\mathbb{P}^{n_i}$. With these definitions, the form $\hat{\Omega}$ satisfies 
\be 
\hat{\Omega} \wedge d p = \mu\,. 
\label{10.4}
\ee
Combining this relation with the current identity
\be 
\delta^2 (p) d \bar p = \frac{1}{\pi} {\bar \pt} \Big( \frac{1}{p}\Big)
\label{10.5}
\ee
leads to the following expression
\be 
\l(\nu_1,\nu_2,\nu_3) =- \frac{1}{2 \pi i}\int_{{\cal A}} \frac{\mu}{p} \wedge \Big[ {\bar \pt} \hat{\nu}_1 \wedge \hat{\nu}_2 \wedge \hat{\nu}_3-
 \hat{\nu}_1 \wedge {\bar \pt} \hat{\nu}_2 \wedge \hat{\nu}_3 +\hat{\nu}_1 \wedge  \hat{\nu}_2 \wedge {\bar \pt} \hat{\nu}_3
 \Big]\,. 
 \label{Yukamb1}
 \ee
for the Yukawa couplings. In deriving this expression, we have performed an integration by parts and ignored the boundary term. This boundary term will be more closely examined in Appendix~\ref{appendixboundary} and we will show that it vanishes in all cases of interest. 

To understand the implications of this result we need to analyse the relation between the ambient space forms $\hat{\nu}_i$ and their restrictions, $\nu_i$, to the Calabi-Yau manifold $X$. Let $K$ be any of the line bundles $K_1$, $K_2$, $K_3$ and ${\cal K}$ its ambient space counterpart, so that $K={\cal K}|_X$. For a given cohomology representative $\nu\in H^1(X,K)$, we would like to construct an ambient space form $\hat{\nu}$ with $\nu=\hat{\nu}|_X$. The line bundles $K$ and ${\cal K}$ are related by the Koszul sequence
\be 
0 \longrightarrow {\cal N}^* \otimes {\cal K} \stackrel{p}{\longrightarrow} {\cal K}  \stackrel{r}{\longrightarrow} K  \longrightarrow 0\; , 
\label{10.8}
\ee
a short exact sequence with $p$, the defining polynomial of the Calabi-Yau manifold, and $r$, the restriction map. This short exact sequence leads to an associated long exact sequence in cohomology, whose relevant part is given by
\bea
 \cdots&\longrightarrow &  H^1 ({\cal A}, {\cal N}^* \otimes {\cal K}) \stackrel{p}{\longrightarrow} H^1 ({\cal A}, {\cal K}) \stackrel{r}{\longrightarrow} H^1 (X, K)\nonumber \\
&\stackrel{\delta}{\longrightarrow} &H^2 ({\cal A}, {\cal N}^* \otimes {\cal K}) \stackrel{p}{\longrightarrow} H^2 ({\cal A}, {\cal K}) \stackrel{r}{\longrightarrow} H^2 (X, K) 
\longrightarrow \dots\; ,
\label{longex}
\eea
where $\delta$ is the co-boundary map. This sequence allows us to relate the cohomology $H^1(X,K)$ to ambient space cohomologies, namely
\bea
H^1 (X, K) & = & r \Big( {\rm Coker} \Big( H^1 ({\cal A}, {\cal N}^* \otimes {\cal K}) \stackrel{p}{\rightarrow}
H^1 ({\cal A}, {\cal K})\Big) \Big)  \notag \\  & \oplus & 
\d^{-1} \Big( {\rm Ker} \Big( H^2 ({\cal A}, {\cal N}^* \otimes {\cal K}) \stackrel{p}{\rightarrow}
H^2 ({\cal A}, {\cal K})\Big) \Big) \; .
\label{H1eq}
\eea
Evidently, $H^1(X,K)$ can receive two contributions, one from $H^1({\cal A},{\cal K})$ (modulo identifications) and the other from (the kernel in) $H^2({\cal A},{\cal N}^*\otimes{\cal K})$. Let us discuss these two contributions separately, keeping in mind that the general case is a sum of these.\\[2mm]
{\bf Type 1}: If $\nu$ descends from $H^1({\cal A},{\cal K})$, we refer to it as ``type 1". In this case, we have a $(0,1)$-form $\hat{\nu}\in H^1({\cal A},{\cal K})$ which, under the map $r$, restricts to $\nu\in H^1(X,K)$. What is more, since $\hat{\nu}$ represents an ambient space cohomology it is closed, so
\begin{equation}
 \bar{\partial}\hat{\nu}=0\; .
\end{equation} 
{\bf Type 2:} The situation is somewhat more involved if $\nu$ descends from $H^2({\cal A},{\cal N}^*\otimes{\cal K})$, a situation we refer to as ``type 2". In this case, we can start with an ambient space $(0,2)$-form $\hat{\omega}=\delta(\nu)\in H^2({\cal A},{\cal N}^*\otimes{\cal K})$, which is the image of $\nu$ under the co-boundary map. The definition of the co-boundary map from Appendix~\ref{coboundarymapappendix} tells us that, in this case, $\nu$ can be obtained as the restriction to $X$ of an ambient space $(0,1)$-form $\hat{\nu}$ which is related to $\hat{\omega}$ by
\be
{\bar \partial} \hat{\nu}=p \hat{\omega}  \,. 
\label{coboundarymap}
\ee
Unlike in the previous case, the form $\hat{\nu}$ is no longer closed.\\[2mm]
The Yukawa coupling~\eqref{Yukamb1} involves three $(0,1)$-forms, $\hat{\nu}_1$, $\hat{\nu}_2$ and $\hat{\nu}_3$, each of which can be either of type 1 or type 2 (or a combination of both types), so that a variety of possibilities ensues. Perhaps the simplest possibility arises when all three forms are of type 1, so that $\bar{\partial}\hat{\nu}_i=0$ for $i=1,2,3$. Then, Eq.~\eqref{Yukamb1} shows that the Yukawa coupling vanishes,
\begin{equation}
 \l(\nu_1,\nu_2,\nu_3)=0\;.
\end{equation} 
 This vanishing is quasi-topological and related to the cohomology structure for $K_1$, $K_2$ and $K_3$ in the sequence~\eqref{longex} -- there is no expectation that it can be explained in terms of a symmetry in the four-dimensional theory. An explicit example of this case will be presented later.
 
The next simplest possibility is for two of the forms, say $\hat{\nu}_1$ and $\hat{\nu}_2$, to be of type 1, so that $\bar{\partial}\hat{\nu}_1= \bar{\partial}\hat{\nu}_2=0$, while $\hat{\nu}_3$ is of type 2, so that $\bar{\partial}\hat{\nu}_3=p\hat{\omega}_3$ for some $(0,2)$-form $\hat{\omega}_3$. Inserting into Eq.~\eqref{Yukamb1}, the Yukawa coupling now reduces to the simple expression
\be
\l(\nu_1,\nu_2,\nu_3) = - \frac{1}{2 \pi i}\int_{{\cal A}} \mu  \wedge \hat{\nu}_1 \wedge  \hat{\nu}_2 \wedge  \hat{\omega}_3\,. 
\label{Yuk112}
\ee
As we will see, this formula is very useful since it is expressed in terms of ambient space forms, which can often be readily written down. 
When more than one of the forms is of type 2, the general formula~\eqref{Yukamb1} needs to 
be used and working out all the required forms becomes more complicated. We will study  examples for all these cases later on.
 

\section{Line bundle valued harmonic forms } 
\label{forms}
From hereon we will focus on tetra-quadric Calabi-Yau manifolds in the ambient space ${\cal A}={\mathbb P}^1\times{\mathbb P}^1\times{\mathbb P}^1\times{\mathbb P}^1$. Besides the general usefulness of working with a concrete example, the tetra-quadric offers a number of additional advantages. Firstly, the ambient space consists of ${\mathbb P}^1$ factors only and is, therefore, particularly simple to handle. Moreover, it is known~\cite{Anderson:2011ns,Anderson:2012yf} that quasi-realistic line bundle standard models exist on the tetra-quadric, so we will be able to apply our methods for calculating Yukawa couplings to physically relevant models. However, the methods we develop in the context of the tetra-quadric can be generalised to other Calabi-Yau hypersurfaces in products of projective spaces and even to higher co-dimension CICYs, as will be seen in Chapter~\ref{chaptern>1codimension}.

The main purpose of this section is to set out the relevant differential geometry for $\mathbb{P}^1$, find the harmonic 
bundle-valued forms for all line bundles on $\mathbb{P}^1$ and apply the results to the full ambient space ${\cal A}$. In particular, 
we will work out a multiplication rule for bundle-valued harmonic forms which will be crucial in order to establish the relation between the 
algebraic and analytic methods for calculating holomorphic Yukawa couplings. Since Yukawa couplings depend only on the cohomology classes of the corresponding
forms, we are free to use any non-trivial representatives. For our calculation, we will rely on forms which
are harmonic relative to the Fubini-Study metric on ${\cal A}$. As we will see, these can be explicitly constructed. For easier accessibility, this chapter is kept somewhat informal. A review of some relevant mathematical background, mostly following Ref.~\cite{H}, can be 
found in Appendix~\ref{app:Kbundle}. The proof of the multiplication rule for harmonic forms on projective space is contained in Appendix~\ref{appendixPn}.


\subsection{Construction of line bundle valued harmonic forms on ${\mathbb P}^1$}
\label{p1}
We begin by collecting some well-known properties of ${\mathbb P}^1$. Homogeneous coordinates on ${\mathbb P}^1$ are denoted by $x^\a$, where $\a=0,1$, and we introduce the standard open patches $U_{(\a)}=\{[x^0:x^1]\,|\, x^\a\neq 0\}$ with affine coordinates $z=x^1/x^0$ on $U_{(0)}$ and $w=x^0/x^1$ on $U_{(1)}$.  The transition function on the overlap is given by $w=1/z$. For convenience, subsequent formulae will usually be written on the patch $U_{(0)}$ and in terms of the coordinate $z$.

The K\"ahler potential for the Fubini--Study metric on ${\mathbb P}^1$ reads
\be 
\mathfrak{K}= \frac{i}{2 \pi} \log \kappa\,, \qquad \kappa= 1+ |z|^2\,,
\label{2.1}
\ee
with associated K\"ahler form and K\"ahler metric given by
\be
J= \pt {\bar \pt}\mathfrak{K}= \frac{i}{2 \pi \kappa^2} dz \wedge d {\bar z}\,, \qquad  g_{z \bar z}= -i J_{z \bar z} =\frac{1}{2 \pi \kappa^2}\,. 
\label{2.2}
\ee
Note that the normalisation of $\mathfrak{K}$ has been chosen such that $\int_{{\mathbb P}^1} J=1$.

Line bundles on ${\mathbb P}^1$ are classified by an integer $k$ and are denoted by ${\cal O}_{\mathbb{P}^1}(k)$. They can be explicitly constructed by dualising and taking tensor powers of the universal bundle ${\cal O}_{\mathbb{P}^1}(-1)$. With  the above covering of ${\mathbb P}^1$ and and the fiber coordinate $v$, the transition function of $ {\cal O}_{{\mathbb P}^1} (k)$ can be written as
\be 
\phi_{01} (z, v)=  (1/z, z^{k} v)\,. 
\label{transfct}
\ee
This means that a section of $ {\cal O}_{{\mathbb P}^1} (k)$  given by $s_{(0)}$ on  $U_{(0)}$ and  $s_{(1)}$ on $U_{(1)}$  transforms as $s_{(0)}(z)= z^k s_{(1)}(1/z)$. 

A hermitian structure $H$ on ${\cal L}={\cal O}_{\mathbb{P}^1}(k)$ can be introduced by
\be 
H= \kappa^{-k}\; , 
\label{2.5}
\ee
and the associated Chern connection, $\nabla^{0,1}= \bar \pt$ and $\nabla^{1,0}= \pt+ A$, with gauge potential $A= {\bar H}^{-1} \pt {\bar H} = \pt \log {\bar H}$ and curvature $F= d A= {\bar \pt} {\pt} \log {\bar H}$ is explicitly specified by
\be 
A= -\frac{k \bar z}{\kappa} dz\,, \quad  F=- 2 \pi i k J\,. 
\label{2.6}
\ee
The last result for the field strength allows the calculation of the first Chern class of ${\cal L}$, which is given by
\be
c_1 ({\cal L})= \frac{i}{2 \pi}  F =k J\,, \quad \int_{{\mathbb P}^1} c_1 ({\cal L}) =k\,. 
\label{2.8}
\ee
Having introduced a hermitian structure and a connection on the line bundles ${\cal L}$, we can now turn to a discussion of their cohomology and their associated harmonic bundle-valued forms. As explained in Appendix~\ref{app:Kbundle}, a  ${\cal L}$-valued harmonic form $\a$ is characterised by the equations
\be 
{\bar \pt} \a =0\,, \quad \pt ({\bar H} \star \a)=0\,, 
\label{harmeqs}
\ee
where $\star$ is the Hodge star on $\mathbb{P}^1$ with respect to the Fubini-Study metric. The first of these equations simply asserts the $\bar{\partial}$-closure of $\a$, which is already sufficient to obtain representatives for cohomology. However, $\bar{\partial}$-closed forms which differ by a $\bar{\partial}$-exact form describe the same cohomology class and such a redundant description of cohomology is not convenient for our purposes. For this reason, we will solve both equations~\eqref{harmeqs} and work with the resulting harmonic representatives, which are in one-to-one correspondence with the relevant cohomology. 

The cohomology of ${\cal L}={\cal O}_{\mathbb{P}^1}(k)$ is obtained from the Bott formula and we should distinguish three qualitatively different cases. 
For $k\geq 0$ only the zeroth cohomology is non-vanishing, while for $k\leq -2$ only the first cohomology is non-vanishing. For $k=-1$ the cohomology is entirely trivial. We will now discuss these three cases in turn and explicitly compute the bundle-valued harmonic forms by solving Eqs.~\eqref{harmeqs}.\\[2mm]
{\bf Case 1)} $k\geq 0$: In this case, the Bott formula implies that $h^0(\mathbb{P}^1,{\cal L})=k+1$ and $h^1(\mathbb{P}^1,{\cal L})=0$. Hence, we are looking for sections or bundle-valued $(0,0)$-forms of ${\cal L}$. In this case, the second equation~\eqref{harmeqs} is automatically satisfied, while the first one implies that the section is holomorphic, so $\alpha=\alpha(z)$. For a monomial $\alpha=z^l $, a transformation to the other patch gives $z^l=w^{-l}=z^kw^{k-l}$, with the $z^k$ factor being the desired transition function. This means that the section is holomorphic in both patches, only if $l=0,\ldots ,k$. This leads to the well-known result that the sections are given by degree $k$ polynomials, that is,
\begin{equation}
 \alpha=P_{(k)}(z)\; .
 \end{equation}
Note that the space of these polynomials is indeed $k+1$-dimensional, as required.\\[2mm]
{\bf Case 2)} $k=-1$: In this case, all cohomologies of ${\cal L}$ vanish and there are no forms to be determined.\\[2mm]
{\bf Case 3)} $k\leq -2$: Now, $h^1(\mathbb{P}^1,{\cal L})=-k-1$ and $h^0(\mathbb{P}^1,{\cal L})=0$. Hence, we are looking for harmonic $(0,1)$-forms $\alpha=f(z,\bar{z})d\bar{z}$. Clearly, the first equation~\eqref{harmeqs} is automatically satisfied for such $\alpha$. Using $\star d\bar{z}=-id\bar{z}$ and $\star\alpha =-i\alpha$, the second equation can be written as $\partial(\bar{H}\alpha)=0$, which leads to the general solution $\alpha = \kappa^kg(\bar{z})d\bar{z}$, with a general anti-holomorphic function $g(\bar{z})$. For a monomial $g(\bar{z})=\bar{z}^l$, this transforms to the other patch as
\begin{equation}
 \alpha=(1+|z|^2)^k\bar{z}^ld\bar{z}=-z^k(1+|w|^2)^k\bar{w}^{-k-l-2}d\bar{w}\; . \label{harm01}
\end{equation} 
For holomorphy in both patches, we should therefore have $l=0,\ldots ,-k-2$, so $g(\bar{z})$ is a general polynomial of degree $-k-2$ in $\bar{z}$. It will be convenient to denote such a polynomial of degree $-k-2$ by $P_{(k)}$, with the understanding that the negative degree subscript implies a dependence on $\bar{z}$, rather than $z$. With this notation, the full solution takes the form
\begin{equation}
 \alpha=\kappa^kP_{(k)}(\bar{z})d\bar{z}\; . \label{1forms}
\end{equation} 
Note that the space of degree $-k-2$ polynomials has indeed dimension $-k-1$, as required.\\[2mm]
 

\subsection{Maps between line bundle cohomology on ${\mathbb P}^1$}
\label{maps}
Calculating Yukawa couplings requires performing a wedge product of bundle-valued forms. It is therefore natural to study how the harmonic forms on $\mathbb{P}^1$ found in the previous sub-section multiply. Recall that we have harmonic $(0,0)$-forms taking values in ${\cal O}_{\mathbb{P}^1}(k)$ for $k\geq 0$ and harmonic $(0,1)$-forms taking values in ${\cal O}_{\mathbb{P}^1}(k)$ for $k\leq -2$. 

Multiplying two harmonic $(0,0)$-forms, representing classes in $H^0({\mathbb P}^1,{\cal O}_{{\mathbb P}^1}(k))$ and $H^0({\mathbb P}^0,{\cal O}_{{\mathbb P}^1}(l))$ respectively, is straightforward and it leads to another harmonic $(0,0)$-form which represents a class in $H^0({\mathbb P}^1,{\cal O}_{{\mathbb P}^1}(k+l))$. 

The only other non-trivial case -- the multiplication of a harmonic $(0,0)$-form with a harmonic $(0,1)$-form -- is less straightforward. To be concrete, for $k\leq -2$ and $\delta>0$, we consider a harmonic $(0,1)$-form $\a_{(k-\d)} \in H^1 ({\mathbb P}^1,  {\cal O}_{{\mathbb P}^1} (k-\d))$ and a degree $\delta$ polynomial $p_{(\d)}$, representing a class in $H^0 ({\mathbb P}^1,  {\cal O}_{{\mathbb P}^1} (\d))$. The product $p_{(\d)}\a_{(k-\d)}$ is a $(0,1)$-form which represent a class in $H^1 ({\mathbb P}^1,  {\cal O}_{{\mathbb P}^1} (k))$, but is not of the form~\eqref{harm01} and, hence, is not harmonic. We would, therefore, like to work out the harmonic representative, denoted $\a_{(k)}\in H^1 ({\mathbb P}^1,  {\cal O}_{{\mathbb P}^1} (k))$, which is equivalent in cohomology to this product $p_{(\d)}\a_{(k-\d)}$. This means we should solve the equation
\be 
p_{(\d)} \a_{(k-\d)} + {\bar \pt} s= \a_{(k)}\; ,
\label{prodeqgen}
\ee
where $s$ is a suitable section of ${\cal O}_{\mathbb{P}^1}(k)$. In general, the section $s$ can be cast into the form
\be 
s= \sum_{m \geq -k} S_{(k+m, -m-2)} (z, \bar z) \kappa^{-m}\,, 
\label{2.14}
\ee
where $S_{(k+m, -m-2)} (z, \bar z)$ is a polynomial of degree $k+m$ in $z$ and of degree $m$ in $\bar z$.  This can be seen by demanding the correct transformation under the transition function~\eqref{transfct}. It turns out that in order to solve Eq.~\eqref{prodeqgen}, we only require the single term with $m=-k+\delta-1$ in this sum for $s$. Using this observation and the general formula~\eqref{1forms} for harmonic $(0,1)$-forms, we insert the following expressions
\begin{eqnarray}
\resizebox{0.89\hsize}{!}{$\a_{(k-\d)} =\kappa^{k-\d} P_{(k- \d)} (\bar z) d \bar z\;, \quad   \a_{(k)} =\kappa^{k} Q_{(k)} (\bar z) d \bar z
\;,\quad s= \kappa^{k-\d+1} S_{(\d-1, k -\d+1)} (z, \bar z)\,.$}
\end{eqnarray}
into Eq.~\eqref{prodeqgen} to cast it into the more explicit form
\be
p P +\k \pt_{\bar z} S - (-k+\d-1) z S = \k^{\d} Q\,.
\label{prodeq}
\ee
Here, for simplicity of notation, we have dropped the subscripts indicating degrees. Eq.~\eqref{prodeq} determines the polynomials $Q$ and $S$ for given $p$ and $P$ and can be solved by comparing monomial coefficients. This is relatively easy to do for low degrees and we will discuss a few explicit examples below. For arbitrary degrees, Eq.~\eqref{prodeq} seems surprisingly complicated and it is, therefore, remarkable that a closed solution for $Q$ can be written down. To formulate this solution, we introduce the homogeneous counterparts of the polynomials $p$, $P$, $Q$ and $S$, which we denote as $\tilde{p}, \tilde{P}$, $\tilde{Q}$ and $\tilde{S}$. They depend on the homogeneous coordinates $x^0$, $x^1$ and are obtained from the original polynomials by replacing $z=x^1/x^0$ and multiplying with the appropriate powers of $x^0$ and $\bar{x}^0$. Then, the polynomial $\tilde{Q}$  which solves Eq.~\eqref{prodeq} can be written as
\be 
\tilde{Q} ({\bar x}^0, {\bar x}^1) = c_{k-\d, \d} \ \tilde{p} (\pt_{ {\bar x}^0}, \pt_{ {\bar x}^1})  \tilde{P}  ({\bar x}^0, {\bar x}^1)\,, \quad
c_{k-\d, \d} =\frac{(-k-1)!}{(\d-k-1)!}\,. 
\label{prodsol}
\ee
Here  $\tilde{p} (\pt_{ {\bar x}^0}, \pt_{ {\bar x}^1})$ denotes the polynomial $\tilde{p}$ with the coordinates replaced by the 
corresponding partial derivatives. These derivatives act on the polynomial $\tilde{P}$ in the usual way and thereby lower the degree to the one expected for $\tilde{Q}$. The proof of Eq.~\eqref{prodsol} is given in Appendix~\ref{appendixPn}. Unfortunately, we are not aware at present of a similar closed solution for the polynomial $S$.\\[2mm]
While this discussion may have been somewhat technical, the final result is relatively simple and can be summarised as follows. For $k\geq 0$, the harmonic $(0,0)$-forms representing the cohomology $H^0(\mathbb{P}^1,{\cal O}_{\mathbb{P}^1}(k))$ are given by degree $k$ polynomials $P_{(k)}(z)$, which depend on the coordinate $z$. For $k\leq -2$, the harmonic $(0,1)$-forms representing the cohomology $H^1(\mathbb{P}^1,{\cal O}_{\mathbb{P}^1}(k))$ can be identified with degree $-k-2$ polynomials, denoted as $P_{(k)}(\bar{z})$, which depend on $\bar{z}$. The product of two $(0,0)$-forms is simply given by polynomial multiplication, while the product of a $(0,0)$-form and a $(0,1)$-form is performed by using the homogeneous versions of these polynomials and converting the coordinates in the former to partial derivatives which act on the latter.  Let us finish this subsection by illustrating the above discussion with two explicit examples.\\[2mm]
{\bf Example 1:} Consider the case $k=-3$ and $\d=1$, so that the relevant forms and associated polynomials are explicitly given by
\begin{eqnarray}
 \alpha_{(-4)}&=&\kappa^{-4}P_{(-4)}(\bar{z})d\bar{z}\;,\quad P_{(-4)}=a_0+a_1\bar{z}+a_2\bar{z}^2 \;, \notag\\
 \alpha_{(-3)}&=&\kappa^{-3}Q_{(-3)}d\bar{z}\;,\qquad Q_{(-3)}=b_0+b_1\bar{z} \;, \\
 s&=&\k^{-3}S_{(0,-5)}\,,\qquad\; S_{(0,-5)}=c_{0,0}+c_{0,1}\bar{z}+c_{0,2}\bar{z}^2+c_{0,3}\bar{z}^3\; , \notag \\
  p_{(1)}&=&f_0+f_1 z \;, \notag
\end{eqnarray} 
where $a_i$, $b_i$, $f_i$ and $c_{i,j}$ are constants.  Inserting these polynomials into Eq.~\eqref{prodeq}, comparing coefficients for same monomials and solving for the $b_i$ and $c_{i,j}$ in terms of the $a_i$ and $f_i$ results in
\begin{eqnarray}
Q_{(-3)}&=&\frac{1}{3} \left( 2 a_0 f_0+a_1  f_1+ \left(a_1 f_0+2 a_2 f_1\right)\bar{z}\right) \;,\label{Qres2}\\
S_{(0,-5)}&=&\frac{1}{3}\left( -a_2 f_0 \bar{z}^3+  \left(a_2 f_1-a_1   f_0\right)\bar{z}^2+  \left(a_1 f_1-a_0 f_0\right)\bar{z}+a_0 f_1\right) \;.
\end{eqnarray}
For the algebraic calculation based on Eq.~\eqref{prodsol}, we start with the homogeneous polynomials
\begin{equation}
\tilde{p}=f_0 x_0+f_1 x_1\;,\quad \tilde{P}=a_0\bar{x}_0^2+a_1 \bar{x}_0\bar{x}_1+a_2\bar{x}_1^2\;,\quad \tilde{S}=c_{0,0}\bar{x}_0^3+c_{0,1}\bar{x}_0^2\bar{x}_1+c_{0,2}\bar{x}_0\bar{x}_1^2+c_{0,3}\bar{x}_1^3\; .
\end{equation}
Inserting these into Eq.~\eqref{prodsol} gives
\begin{equation}
\tilde{Q}=\frac{1}{3} \left( (2 a_0 f_0+a_1  f_1)\bar{x}_0+ \left(a_1 f_0+2 a_2 f_1\right)\bar{x}_1\right)\; ,
\end{equation}
which is indeed the homogeneous version of the polynomial $Q_{(-3)}$ in Eq.~\eqref{Qres2}.\\[2mm]
{\bf Example 2:} Let us choose $k=-1$ and $\d=2$. Since there are no harmonic forms for $k=-1$, we have $Q=0$, while the other forms and polynomials are given by
\bea
\a_{(-3)} &=& \k^{-3} P_{(-3)} (\bar z) d \bar z\,, \quad P_{(-3)}= a_0 + a_1 \bar z \,, \notag \\
s &=&\k^{-2}S_{(2,-4)}\,,\qquad\; S_{(2,-4)}=c_{0,0} +c_{0,1} \bar z + c_{0,2} {\bar z}^2 + c_{1,0} z + c_{1,1} |z|^2 + c_{1,2} {\bar z} |z|^2 \,, \notag \\
p_{(2)}&=&p_0 +p_1 z +p_2 z^2 \,.
\eea
We note that, from~\eqref{prodeqgen}, we now need to solve the equation $p_{(2)} \a_{(-3)} =-{\bar \pt} s$, which is similar in structure to Eq.~\eqref{coboundarymap} that determines the co-boundary map. Indeed, we will later find the present example useful to explicitly work out a co-boundary map. Inserting the above polynomials into Eq.~\eqref{prodeq} and comparing coefficients as before leads to
\be 
S_{(2,-4)} =\frac{1}{2} (p_1 a_0 + p_2 a_1) - p_0 a_0 \bar z -\frac{1}{2} p_0 a_1 {\bar z}^2 +\frac{1}{2} p_2 a_0 z
+p_2 a_1 |z|^2 -\frac{1}{2} (p_0 a_0 + p_1 a_1){ \bar z} |z|^2\,.  \label{coboundres}
\ee


\subsection{Line bundle valued harmonic forms on ${\mathbb P}^1 \times {\mathbb P}^1\times {\mathbb P}^1  \times {\mathbb P}^1 $}
\label{maps1}


In this sub-section, we generalise the above results for $\mathbb{P}^1$ to the ambient space ${\cal A}= {\mathbb P}^1 \times {\mathbb P}^1\times {\mathbb P}^1  \times {\mathbb P}^1 $. On each ${\mathbb P}^1$, we introduce homogeneous coordinates $(x^0_i, x^1_i)$, where $i=1, \dots, 4$ and cover each ${\mathbb P}^1$ with two standard open sets $U_{(i,\a)}=\{[x_i^0:x_i^1]\,|\,x_i^\a\neq 0\}$. Further, we introduce affine coordinates $z_i= x^1_i/x^0_i$ on $U_{(i,0)}$ and $w_i= x^0_i/x^1_i$ on $U_{(i,1)}$. On the intersection of $U_{(i,0)}$ and $U_{(i,1)}$
we have $z_i= 1/w_i$. An open cover for the entire space ${\cal A}$ is given by the $16$ sets $U_{(1,\a_1)}\times \cdots \times U_{(4,\a_4)}$. For practical purposes, we will usually work on the set $U_{(1,0)}\times \cdots \times U_{(4,0)}$ with coordinates $z_1,\ldots,z_4$.

For each $\mathbb{P}^1$, we have a Fubini--Study K\"ahler potential and K\"ahler form given by
\be 
\mathfrak{K}_i=\frac{i}{2 \pi} \log \kappa_i \,, \quad \kappa_i= 1+ |z_i|^2\,,\quad J_i = \frac{i}{2 \pi \kappa_i^2} dz_i \wedge d {\bar z}_i\
\ee
and the K\"ahler cone of ${\cal A}$ is parametrised by $J=\sum_{i=1}^4 t^iJ_i$, with all $t^i>0$. 

The line bundles on ${\cal A}$ are obtained as the tensor products
\begin{equation}
 {\cal O}_{\cal A}({\bf k})={\cal O}_{\mathbb{P}^1}(k^1)\otimes\cdots\otimes {\cal O}_{\mathbb{P}^1}(k^4)
\end{equation} 
and are, hence, labeled by a four-dimensional integer vector ${\bf k} =  (k^1, k^2, k^3, k^4)$. Straightforwardly generalising Eq.~\eqref{2.5}, we can introduce a Hermitian structure
\be 
H= \prod_{i=1}^4 \kappa_i^{-k^i}\,. 
\label{20.4}
\ee
on these line bundles. The gauge field and gauge field strength for the associated Chern connection
\be 
A= {\bar H}^{-1} \pt {\bar H}=- \sum_{i=1}^4 k^i \pt \log \kappa_i\;,\quad F ={\bar \pt} A =- 2 \pi i  \sum_{i=1}^4 k^i J_i\,
\label{20.5}
\ee
lead to the first Chern class
\be
c_1 \left( {\cal O}_{\cal A}({\bf k})\right)= \frac{i}{2 \pi} F = \sum_{i=1}^4 k^i J_i\,.
\label{20.5.2}
\ee
The cohomology for ${\cal K}={\cal O}_{\cal A}({\bf k})$ can be obtained by combining the Bott formula for cohomology on $\mathbb{P}^1$ with the K\"unneth formula \eqref{kunneth}. If any of the integers $k^i$ equals $-1$ all cohomologies of ${\cal K}$ vanish. In all other cases, precisely one cohomology, $H^q({\cal A},{\cal K})$, is non-zero, and $q$ equals the number of negative integers $k^i$. The dimension of this non-vanishing cohomology is given by
\begin{equation}
 h^q({\cal A},{\cal K})=\prod_{i:k^i\geq 0}(k^i+1)\prod_{i:k^i\leq -2}(-k^i-1)\; .
\end{equation} 
Generalising our results for $\mathbb{P}^1$, the harmonic $(0,q)$-forms representing this cohomology can be written as
\be 
\a_{({\bf k})} = P_{({\bf k})} \prod_{i: k^i \leq -2} \k_i^{k^i} d {\bar z}_i\,,
\label{akres}
\ee
where $P_{({\bf k})} $ is a polynomial of degree $k^i$ in $z_i$, provided $k^i \geq 0$, and of degree $-k^i-2$ in ${\bar z}_i$, if $k^i \leq -2$. It is also useful to write down a homogeneous version of these forms, which is given by
\be 
\a_{({\bf k})} = \tilde{P}_{({\bf k})} \prod_{i: k^i \leq -2} \s_i^{k^i}  {\bar \m}_i\,,
\label{akreshom}
\ee
where
\be
\s_i=|x_i^0|^2+|x_i^1|^2\,,\qquad \mu_i=\e_{\a\b}x_i^\a x_i^\b\;,
\ee
and $\tilde{P}_{(\bf k)}$ denotes the homogeneous counterpart of $P_{(\bf k)}$. 

We would now like to generalise our rule for the multiplication of forms obtained on $\mathbb{P}^1$. In general, we have a map
\be
H^q ({\cal A}, {\cal O}_{{\cal A}} ({\bf k}))\times 
H^p ({\cal A}, {\cal O}_{{\cal A}} ({\bf l}))  \to  H^{q+p} ({\cal A}, {\cal O}_{{\cal A}} ({\bf k}+{\bf l}))
\label{20.10.1}
\ee
between cohomologies induced by the wedge product and we would like to work out this map for the above harmonic representatives. For a harmonic $(0,q)$-form $\a_{({\bf k})}\in H^q({\cal A},{\cal O}_{\cal A}({\bf k}))$ with associated polynomial $P_{({\bf k})}$ and a harmonic $(0,p)$-form $\b_{({\bf l})}$ with associated polynomial $R_{({\bf l})}$, the wedge product $\a_{({\bf k})}\wedge\b_{({\bf l})}$ is equivalent in cohomology to a harmonic $(0,q+p)$-form, which we denote by $\g_{({\bf k}+{\bf l})}\in H^{q+p} ({\cal A}, {\cal O}_{{\cal A}} ({\bf k}+{\bf l}))$ with associated polynomial $Q_{({\bf k}+{\bf l})}$. In general, the relation between those forms can be written as
\be 
\a_{({\bf k})}  \wedge \b_{({\bf l})}  +{\bar \pt} s = \g_{({\bf k} +{\bf l})} \, ,
\label{20.12}
\ee
for a suitable $(0,p+q-1)$-form $s$ taking values in ${\cal O}_{\cal A}({\bf k}+{\bf l})$. Our earlier results for $\mathbb{P}^1$ show that the polynomial
$Q_{({\bf k}+{\bf l})}$ which determines $\g_{({\bf k} +{\bf l})}$ can be directly obtained from $P_{({\bf k})}$ and $R_{({\bf l})}$ by the formula
\be 
\tilde{Q}= c_{{\bf k}, {\bf l}} \tilde{ P} \tilde {R}\,,
\ee
where, as before, $\tilde{ P}, \tilde{R}, \tilde{Q}$ are the  homogeneous counterparts of $P, R, Q$ and $c_{{\bf k}, {\bf l}}$ is the appropriate product of numerical factors in Eq.~\eqref{prodsol}. The understanding is that positive degrees in a particular $\mathbb{P}^1$, represented by powers of $x_i^\a$ should be converted into derivatives $\partial_{\bar{x}^i_\a}$ whenever they act on negative degrees in the same $\mathbb{P}^1$, represented by $\bar{x}_i^\a$.  When both degrees in $\tilde{P}$ and $\tilde{R}$ are positive for a given $\mathbb{P}^1$, a simple polynomial multiplication should be carried out. Finally, for two negative degrees in the same $\mathbb{P}^1$, the resulting $\tilde{Q}$ vanishes (since there will be a term $d\bar{z}^i \wedge d\bar{z}^i$ in the corresponding wedge product of the forms).


\subsection{Line bundles and cohomology on the tetra-quadric}
\label{relations}


As the final step in our discussion of line bundles and harmonic forms, we need to consider line bundles on the tetra-quadric $X$. Recall that a tetra-quadric resides in the ambient space ${\cal A}=\mathbb{P}^1\times\mathbb{P}^1\times \mathbb{P}^1\times \mathbb{P}^1$  and is defined as the zero locus of a polynomial $p$ of multi-degree $(2,2,2,2)$, which can be seen as a section of the normal bundle 
\be 
{\cal N}= {\cal O}_{{\cal A}} ({\bf q})\,, \quad  {\bf q} = (2, 2, 2, 2)\,. 
\label{20.14}
\ee
The tetra-quadric has Hodge numbers $h^{1,1}(X)=4$ and $h^{2,1}(X)=68$. Later, we will use the freely-acting $\Gamma=\mathbb{Z}_2\times\mathbb{Z}_2$ symmetry of the tetra-quadric, whose generators are given by
\begin{equation}
 g_1=\left(\begin{array}{cc}1&0\\0&-1\end{array}\right)\;,\quad g_2=\left(\begin{array}{cc}0&1\\1&0\end{array}\right)\; . \label{g1g2}
 \end{equation}
These matrices act simultaneously on all four pairs of homogeneous coordinates. The quotient $\tilde{X}=X/\Gamma$ is a Calabi-Yau manifold with Hodge numbers $h^{1,1}(\tilde{X})=4$ (since all four K\"ahler forms $J_i$ are $\Gamma$-invariant) and $h^{2,1}(\tilde{X})=20$ (using divisibility of the Euler number).

All line bundles on the tetra-quadric can be obtained as restrictions of line bundles on ${\cal A}$, that is
\begin{equation}
 {\cal O}_X({\bf k})={\cal O}_{{\cal A}}({\bf k})|_X\; .
\end{equation} 
As discussed in Section~\ref{coboundary}, the Koszul sequence and its associated long exact sequence provide a close relationship between line bundle cohomology on ${\cal A}$ and $X$, which is summarised by Eq.~\eqref{H1eq}. This equation shows that the cohomology of a line bundle $K={\cal O}_X({\bf k})$  depends on the first and second cohomologies of the ambient space line bundles ${\cal K}={\cal O}_{\cal A}({\bf k})$ and ${\cal N}^*\otimes{\cal K}={\cal O}_{\cal A}({\bf k}-{\bf q})$. As discussed earlier, line bundles on ${\cal A}$ have at most one non-vanishing cohomology and, hence, ${\cal K}$ and ${\cal N}^*\otimes{\cal K}$ have at most one non-zero cohomology each. This leads to the following four cases:
\begin{enumerate}
\item[1)]\underline{$H^2 ({\cal A}, {\cal N}^* \otimes {\cal K})=0$  and $H^2 ({\cal A}, {\cal K})=0$}\\
In this case, $H^1 (X, K)$ is given by $(0,1)$-forms $\a_{({\bf k})}$, as in Eq.~\eqref{akres}, with associated polynomials
$P_{({\bf k})}$ and, in the terminology of Section~\ref{coboundary}, the cohomology representatives are of type 1. If $H^1({\cal A},{\cal N}^*\otimes {\cal K})$ is non-trivial we have to compute the co-kernel in Eq.~\eqref{H1eq}, which amounts to imposing the identification $\tilde{P}_{({\bf k})} \sim \tilde{P}_{({\bf k})} +\tilde{p} \tilde{Q}_{({\bf k}- {\bf q})}$ for arbitrary polynomials $\tilde{Q}_{({\bf k}- {\bf q})}$ of multi-degree ${\bf k}-{\bf q}$. Recall that the tilde denotes the homogeneous version of the polynomials and that coordinates appearing with positive degree have to be converted into derivatives whenever they act on negative degree coordinates, as discussed at the end of the last sub-section. Since the coefficients of $p$ depend on the complex structure, this identification leads to complex structure dependence of the representatives.
\item[2)] \underline{$H^1 ({\cal A}, {\cal N}^* \otimes {\cal K})=0$  and $H^1 ({\cal A}, {\cal K})=0$}\\
In this case, $H^1 (X, K)$ is represented by $(0,2)$-forms $\a_{({\bf k}- {\bf q})}$, with  associated polynomials $P_{({\bf k}-{\bf q})}$, satisfying 
$p  \a_{({\bf k}- {\bf q})}= {\bar \pt} \b_{({\bf k})}$ for a suitable $(0,1)$-form $\b_{({\bf k})}$. Using the terminology of Section~\ref{coboundary}, this corresponds to type 2 representatives. If $H^2 ({\cal A}, {\cal K})\neq 0$, we have to work out the kernel in Eq.~\eqref{H1eq}, which amounts to imposing the condition $\tilde{p}\tilde{P}_{({\bf k}-{\bf q})}=0$. This leads to explicit complex structure dependence of the representatives.
\item[3)] \underline{$H^1 ({\cal A}, {\cal N}^* \otimes {\cal K})=0$  and $H^2 ({\cal A}, {\cal K})=0$}\\
This is a combination of the previous two cases, where $H^1 (X, K)$ is a direct sum of type 1 and type 2 contributions.
\item[4)] \underline{$H^2 ({\cal A}, {\cal N}^* \otimes {\cal K})=0$   and $H^1 ({\cal A}, {\cal K})=0$}\\
In this case, $H^1(X, K)=0$. 
\end{enumerate}


\section{Yukawa couplings on the tetra-quadric and some toy examples}\label{toyex}


We have now collected all relevant technical details on line bundles and harmonic bundle-valued forms on the tetra-quadric and are ready to apply these to concrete calculations of Yukawa couplings. To begin, we collect some general statements on Yukawa couplings on the tetra-quadric -- including the precise relation between an explicit analytic calculation of the integral and a corresponding algebraic calculation -- and then move on to work out Yukawa couplings for a number of toy examples. In the next section, we compute the Yukawa couplings for a quasi-realistic standard model on the tetra-quadric.


\subsection{General properties of Yukawa couplings}
\label{comments}
As we have discussed earlier, we can distinguish two types of harmonic bundle-valued $(0,1)$-forms on the tetra-quadric: forms of type 1, which descend from harmonic $(0,1)$-forms on the ambient space, and forms of type 2, which descend from harmonic $(0,2)$-forms on the ambient space. 
The Yukawa couplings involve three harmonic $(0,1)$-forms and, as shown in Section~\ref{coboundary}, their structure depends on the types of these $(0,1)$-forms.

Let us consider a line bundle model on the tetra-quadric, specified by line bundles $L_a$, where $a=1,\ldots ,n$, and a 
Yukawa coupling with three associated line bundles $K_1={\cal O}_X({\bf k}_1)$, $K_2={\cal O}_X({\bf k}_2)$ and $K_3={\cal O}_X({\bf k}_3)$, 
which are related to $L_a$ as in Table~\ref{tab:KLrel}. Consider three harmonic $(0,1)$-forms $\nu_i\in H^1(X,K_i)$. 
We have seen that the Yukawa coupling vanishes if these three forms are of type 1. 
The next simplest case, when two of the forms, say $\nu_1$ and $\nu_2$, are of type 1 and 
descend from ambient space harmonic $(0,1)$-forms $\hat{\nu}_1\in H^1({\cal A},{\cal O}_{\cal A}({\bf k}_1))$ 
and $\hat{\nu}_2\in H^1({\cal A},{\cal O}_{\cal A}({\bf k}_2))$, while $\nu_3$ is of type 2 and descends from a harmonic 
ambient space $(0,2)$-form $\hat{\omega}_3\in H^2 ({\cal A}, {\cal O}_{\cal A} ({\bf k}_3 -{\bf q}))$, leads to the particularly simple formula
\be 
\l(\nu_1,\nu_2,\nu_3) = - \frac{1}{2 \pi i}\int_{{\mathbb C}^4} d^4 z  \wedge \hat{\nu}_1 \wedge  \hat{\nu}_2 \wedge  \hat{\omega}_3\,, 
\label{Yuk112copy}
\ee
for the Yukawa coupling. This follows from Eq.~\eqref{Yuk112}, together with Eqs.~\eqref{10.2}, which show that the form $\mu$ is given by
\be
\mu = d z_1 \wedge dz_2 \wedge dz_3 \wedge dz_4 = d^4 z\,. 
\label{muform}
\ee
The integral over ${\cal A}$ can then be thought of as the integral over ${\mathbb C}^4$, provided the forms
$ \hat{\nu}_1,   \hat{\nu}_2,  \hat{\omega}_3$ transform to the other patches as sections of the appropriate line bundles. 
Since $\hat{\nu}_1$ and $\hat{\nu}_2$ are $(0,1)$-forms, the vectors ${\bf k}_1$ and ${\bf k}_2$ should contain 
precisely one entry $\leq -2$ each, while the vector ${\bf k}_3$ contains precisely two entries $\leq 0$, in line with $\hat{\o}_3$ 
being a $(0,2)$-form. Further, recall from Table~\ref{tab:KLrel} that ${\cal K}_1\otimes{\cal K}_2\otimes{\cal K}_3={\cal O}_{\cal A}$ and, 
hence, ${\bf k}_1+{\bf k}_2+{\bf k}_3=0$. This means that the four non-positive entries in these vectors must all arise in different $\mathbb{P}^1$ directions. 
Hence, we can assume, possibly after re-ordering, that $k_1^1 \leq -2$, $k_2^2 \leq -2$ and $k_3^3, k_3^4 \leq 0$, while all other entries are positive. 
With these conventions, we can apply Eq.~\eqref{akres} to write down the relevant forms as
\be 
\hat{\nu}_1 = \k_1^{k_1^1} P_{({\bf k}_1)} d {\bar z}_1\,, \qquad 
\hat{\nu}_2 = \k_2^{k_2^2} R_{({\bf k}_2)} d {\bar z}_2\,, \qquad 
\hat{\o}_3 = \k_3^{k_3^3 -2 } \k_4^{k_3^4 -2 } T_{({\bf k}_3  -{\bf q} )} d {\bar z}_3 \wedge  d {\bar z}_4\,. 
\ee
Inserting these forms into Eq.~\eqref{Yuk112copy} leads to the integral
\be 
\l(\nu_1,\nu_2,\nu_3)  = - \frac{1}{2 \pi i}\int_{{\mathbb C}^4} d^4 z \ d^4 {\bar z}  \ \k_1^{k_1^1} \k_2^{k_2^2}\k_3^{k_3^3 -2 } \k_4^{k_3^4 -2 } 
P_{({\bf k}_1)} R_{({\bf k}_2)} T_{({\bf k}_3  -{\bf q} )}  \,. 
\label{Yuk112spec}
\ee
There are two ways of evaluating this integral. Firstly, we can explicitly insert the factors $\kappa_i=1+|z_i|^2$ and the polynomials and simply integrate, using polar coordinates in each $\mathbb{C}$ plane. All terms with non-matching powers of $z_i$ and $\bar{z}_i$ vanish due to the angular integration. The remaining terms all reduce to the standard integrals
\begin{eqnarray}
 \int_{\mathbb{C}}\frac{|z|^{2q}}{\kappa^p}dz\,d\bar{z}=2\pi i I_{p,q}\;, \,\,\,\, I_{p,q}=2\int_0^\infty dr\frac{ r^{2q+1}}{(1+r^2)^p}=\frac{q!}{(p-1)\cdots (p-q-1)} \;,\label{stdint}
\end{eqnarray}
where $p \geq q+2$ is a requirement satisfied for all the cases in use. Alternatively, we can work out the integral~\eqref{Yuk112spec} ``algebraically". To do this, we first note that the integrand $\hat{\nu}_1 \wedge  \hat{\nu}_2 \wedge  \hat{\omega}_3$ represents an element of the one-dimensional cohomology $H^4({\cal A},{\cal N}^*)$. It can, therefore, be written as  $\mu (P, R, T)\k_1^{-2} \k_2^{-2} \k_3^{-2} \k_4^{-2} d^4 {\bar z}$, where
\be
 \mu (P, R, T)= \tilde{P} \tilde{R} \tilde{T}  \label{Yukalg}
\ee
is the product of the three associated polynomials (carried out as discussed in Section~\ref{maps1}) and simply a complex number. Inserting this into Eq.~\eqref{Yuk112copy} shows that 
\be 
\l(\nu_1,\nu_2,\nu_3) =  8 i \pi^3  c \mu (P, R, T)\; , \label{lmurel}
\ee
where the numerical factor $c$ follows from Eq.~\eqref{prodsol} and is explicitly given by
\be 
c=c_{k_1^1, -k_1^1-2} \ c_{k_2^2, -k_2^2-2}\  c_{k_3^3-2, -k_3^3} \ c_{k_3^4-2, -k_3^4}\, . \label{cgen}
\ee
In conclusion, up to an overall numerical (and explicitly computed) factor, the Yukawa couplings are simply given by Eq.~\eqref{Yukalg} and can, hence, be obtained by a multiplication of the associated polynomials.

In the general case,  the Yukawa couplings are given by the integral~\eqref{Yukamb1} which can be written as
\be 
\l(\nu_1,\nu_2,\nu_3) =- \frac{1}{2 \pi i}\int_{{\mathbb C}^4} d^4 z \wedge [\hat{\o}_1 \wedge \hat{\nu}_2 \wedge \hat{\nu}_3-
 \hat{\nu}_1 \wedge \hat{\o}_2 \wedge \hat{\nu}_3 +\hat{\nu}_1 \wedge  \hat{\nu}_2 \wedge  \hat{\o}_3 ]\, , 
 \label{Yukgen4}
 \ee
with the $(0,1)$-forms $\hat{\nu}_i$ and the $(0,2)$-forms $\hat{\o}_i$ in this expression related by
\be
{\bar \pt} \hat{\nu}_i=p \hat{\o}_i \,. 
\ee
If the Yukawa coupling depends on more than one form of type 2, we have to solve this last equation for some of the $\hat{\nu}_i$ in terms of $\hat{\omega}_i$. This can be done explicitly for specific examples, as we will demonstrate later, but as discussed in Section~\ref{maps}, we are currently not aware of a general solution.


\subsection{An example with vanishing Yukawa couplings}
\label{vanishing}


We would like to consider a rank four line bundle sum on the tetra-quadric specified by the line bundles 
\begin{equation}
\begin{array}{ll}
 L_1 = {\cal O}_X  (-1, 0, 0, 1)\,, \quad & L_2= {\cal O}_X  (0, -2, 1, 3)\,, \\ L_3 = {\cal O}_X  (0, 0, 1, -3)\,, \quad & L_4= {\cal O}_X  (1, 2, -2, -1)\,.
\end{array}
\end{equation}
This bundle leads to a four-dimensional theory with gauge group $SO(10)\times S(U(1)^4)$. Table~\ref{tab:so10} contains the basic information required to determine the multiplet content of such a theory, and together with the cohomology results
\begin{equation}
\begin{array}{llllll}
 h^{^{\!\bullet}}(X,L_2)&=&(0,8,0,0)\, ,&h^{^{\!\bullet}}(X,L_3)&=&(0,4,0,0) \, ,\\[3pt]
 h^{^{\!\bullet}}(X,L_1\otimes L_4)&=&(0,3,3,0)\, ,&h^{^{\!\bullet}}(X,L_2\otimes L_3)&=&(0,3,3,0) \, ,\\[3pt]
 h^{^{\!\bullet}}(X,L_1\otimes L_2^*)&=&(0,0,12,0) \, ,&h^{^{\!\bullet}}(X,L_1\otimes L_3^*)&=&(0,0,12,0) \, ,\\[3pt]
 h^{^{\!\bullet}}(X,L_2\otimes L_3^*)&=&(0,7,15,0)\, ,&h^{^{\!\bullet}}(X,L_2\otimes L_4^*)&=&(0,60,0,0)\, , \\[3pt]
 h^{^{\!\bullet}}(X,L_3\otimes L_4^*)&=&(0,0,36,0)\, , &
\end{array} 
\end{equation}
we find the upstairs spectrum
\be 
\resizebox{1\hsize}{!}{$
8 \ {\bf 16}_2\,, \  4 \ {\bf 16}_3\,, \ 3 \ {\bf 10}_{1,4}\,, \ 3 \ {\bf 10}_{2,3}\,,  \ 12 \ {\bf 1}_{2,-1}\,, \ 12 \ {\bf 1}_{3,-1}\,, \ 
7 \ {\bf 1}_{2,-3}\,, \ 15 \ {\bf 1}_{3,-2}\,, \ 60 \ {\bf 1}_{2,-4}\,, \ 36 \ {\bf 1}_{4,-3}\,.$}
\ee
This spectrum is designed to produce a standard-model with three families upon dividing by a freely-acting symmetry of order four. Such symmetries are indeed available for the tetra-quadric however, unfortunately, for group-theoretical reasons these symmetries cannot break the $SO(10)$ gauge group to the standard model group. For this reason, the above model should be considered a toy example.

Nevertheless, it is useful to calculate the Yukawa couplings for this model, in order to gain some experience with our formalism. Specifically, we are interested in couplings of the type
\be 
\l_{IJK} {\bf 10}^{(I)}_{1, 4} {\bf 16}^{(J)}_{2} {\bf 16}^{(K)}_{3} \,. 
\ee
which are allowed by the $SO(10)\times S(U(1)^4)$ gauge symmetry. Following Table~\ref{tab:KLrel}, the required harmonic forms are contained in the first cohomologies of the line bundles
\be
\resizebox{1.01\hsize}{!}{$ 
K_1= L_1 \otimes L_4 = {\cal O}_X (0, 2, -2, 0)\,,\quad K_2= L_2 ={\cal O}_X  (0, -2, 1, 3)\,, \quad
K_3= L_3 ={\cal O}_X  (0, 0, 1, -3)\,. $}
\ee
These line bundles satisfy $H^1 (X, K_i)\cong H^1 ({\cal A}, {\cal K}_i)$ and $H^2 ({\cal A}, {\cal N}^* \otimes {\cal K}_i)=0$, where ${\cal  K}_i$ are the corresponding ambient space line bundles with $K_i = {\cal K}_i|_X$. This shows (see Section~\ref{coboundary}) that all three harmonic forms which enter the Yukawa integral are of type 1. From our general arguments, this means that the Yukawa couplings vanish, so
\be
 \l_{IJK} =0\; .
\ee 
Note that this vanishing is, apparently, not caused by a symmetry in the low-energy theory, but happens due to quasi-topological reasons related to the cohomology of the line bundles involved. (However, we do not rule out that a symmetry which explains this vanishing result may be found.)

\subsection{An $E_6$ example} \label{E6example}
For a simple example with gauge group $E_6\times S(U(1)^3)$, consider the following choice of line bundles
\be
\resizebox{1.01\hsize}{!}{$ 
 L_1=K_1={\cal O}_X(-2,0,1,0)\,,\quad L_2=K_2={\cal O}_X(0,-2,0,1)\,,\quad L_3=K_3={\cal O}_X(2,2,-1,-1)\; .$}
\ee
The above line bundles $K_i$ may also arise as appropriate tensor products for other gauge groups, see Table~\ref{tab:KLrel}, and the subsequent calculation also applies to these cases. However, for definiteness, we will focus on $E_6\times S(U(1)^3)$ and the corresponding multiplets, as summarised in Table~\ref{tab:e6}. The cohomology results
\be
h^\bullet(K_1)=(0,2,0,0)\,,\quad h^\bullet(K_2)= (0,2,0,0)\,,\quad h^\bullet(K_3)=(0,4,0,0)
\ee
show that we have a spectrum
\be
2\; {\bf 27}_1\,,\;2\;{\bf 27}_2\,,\;4\;{\bf 27}_3
\ee
plus $E_6$ singlets which are irrelevant to the present discussion. We are interested in the Yukawa couplings
\be
 \l_{IJK}{\bf 27}_1^{(I)}\,{\bf 27}_2^{(J)}\,{\bf 27}_3^{(K)}\; .
\ee 
Clearly, the first two line bundles are of type 1 with the corresponding harmonic $(0,1)$-forms contained in $H^1({\cal A},{\cal K}_1)$ and $H^1({\cal A},{\cal K}_2)$. However, $K_3$ is of type 2 and the associated harmonic $(0,2)$-forms represent the cohomology $H^2({\cal A},{\cal N}^*\otimes{\cal K}_2)$. Altogether, using Eq.~\eqref{akres}, this means the relevant harmonic forms and polynomials are
\begin{equation}
\begin{array}{lll}
 \hat{\nu}_1=\kappa_1^{-2}P_{(-2,0,1,0)}d\bar{z}_1 \, ,&\quad&P_{(-2,0,1,0)}=p_0+p_1z_3 \, , \\
 \hat{\nu}_2=\kappa_2^{-2}Q_{(0,-2,0,1)}d\bar{z}_2 \, ,&\quad&Q_{(0,-2,0,1)}=q_0+q_1z_4 \, ,\\
 \hat{\omega}_3=\kappa_3^{-3}\kappa_4^{-3}R_{(0,0,-3,-3)}d\bar{z}_3\wedge d\bar{z}_4 \, , &\quad&
 R_{(0,0,-3,-3)}=r_0+r_1\bar{z}_3+r_2\bar{z}_4+r_3\bar{z}_3\bar{z}_4 \, ,
 \end{array}
 \end{equation}
where $p_I$, $q_I$ and $r_I$ are complex coefficients parametrising the various ${\bf 27}$ multiplets. Multiplying the three polynomials and discarding terms with different powers of $z_i$ and $\bar{z}_i$ gives
\begin{equation}
 PQR=p_0q_0r_0+p_0q_1r_2|z_4|^2+p_1q_0r_1|z_3|^2+p_1q_1r_3|z_3|^2|z_4|^2+\mbox{ non-matching terms} \, .
\end{equation}
This can be directly inserted into the integral~\eqref{Yuk112spec} and, together with the standard integrals~\eqref{stdint} (specifically, $I_{2,0}=1$, $I_{3,0}=1/2$, $I_{3,1}=1/2$), we find
\begin{equation}
 \lambda(P,Q,R)=2 i \pi^3\left(p_0q_0r_0+p_0q_1r_2+p_1q_0r_1+p_1q_1r_3\right)\; . \label{exresint1}
\end{equation} 
Alternatively, we can use the algebraic calculation method based on Eq.~\eqref{Yukalg}. For simplicity of notation, we denote the four sets of homogeneous ambient space coordinates by $(x_i^\a)=((x_0,x_1),(y_0,y_1),(u_0,u_1),(v_0,v_1))$ from hereon. Then, the homogeneous versions of the three polynomials read explicitly
\begin{equation}
 \tilde{P}=p_0u_{0}+p_1u_{1}\;,\quad \tilde{Q}=q_0v_{0}+q_1v_{1}\;,\quad \tilde{R}=r_0\bar{u}_{0}\bar{v}_{0}+r_1\bar{v}_{0}\bar{u}_{1}+r_2\bar{u}_{0}\bar{v}_{1}+r_3\bar{u}_{1}\bar{v}_{1}\; .
\end{equation} 
Their product is given by
\begin{eqnarray}
 \mu(P,Q,R)&=&\left(p_0{\partial}_{\bar{u}_0}+p_1{\partial}_{\bar{u}_1}\right)\left(q_0{\partial}_{\bar{v}_0}+q_1{\partial}_{\bar{v}_1}\right)
 \left(r_0\bar{u}_{0}\bar{v}_{0}+r_1\bar{v}_{0}\bar{u}_{1}+r_2\bar{u}_{0}\bar{v}_{1}+r_3\bar{u}_{1}\bar{v}_{1}\right) \notag \\
 &=&p_0q_0r_0+p_0q_1r_2+p_1q_0r_1+p_1q_1r_3\; ,
\end{eqnarray} 
where we have converted the coordinates in $\tilde{P}$ and $\tilde{Q}$ into derivatives, as required by our general rules. Inserting the correct numerical coefficient from Eqs.~\eqref{lmurel} and \eqref{cgen}, this indeed coincides with the result~\eqref{exresint1} from direct evaluation of the integral. If we choose a standard basis where each of the coefficients $p_I$, $q_I$ and $r_I$ equals one while all others vanish, we can write down the explicit Yukawa matrices
\begin{equation}
( \l_{1JK})=2 i \pi^3 \left(\begin{array}{llll}1&0&0&0\\0&0&1&0\end{array}\right)\,,\quad
( \l_{2JK})=2 i \pi^3 \left(\begin{array}{llll}0&1&0&0\\0&0&0&1\end{array}\right)\; .
\end{equation}
Both matrices have maximal rank and are independent of complex structure.

\subsection{An example with complex structure dependence}\label{csexample}
We would like to discuss the Yukawa couplings related to the three line bundles
\be
 K_1={\cal O}_X(0,-2,1,1)\,,\quad K_2={\cal O}_X(-4,0,1,1)\,,\quad K_3={\cal O}_X(4,2,-2,-2)\; ,
\ee 
with cohomologies
\be
 h^\bullet(K_1)=(0,4,0,0)\,,\quad h^\bullet(K_2)=(0,12,0,0)\,,\quad h^\bullet(K_3)=(0,12,0,0)\; .
\ee 
It will be convenient to think about this situation as arising from an $SU(5)\times S(U(1)^5)$ model, defined by five line bundles $L_a$, with $K_1=L_1\otimes L_2$ and $K_2=L_3\otimes L_4$ and $K_3=L_5$. Then, using the correspondence from Table~\ref{tab:KLrel}, the $SU(5)\times S(U(1)^5)$ spectrum related to $K_1$, $K_2$ and $K_3$ is
\be
 4\;\overline{\bf 5}_{1,2}\;,\quad12\;\overline{\bf 5}_{3,4}\;,\quad 12\;{\bf 10}_5\; .
\ee 

\vspace{-2mm}

\noindent We will later introduce a $\mathbb{Z}_2\times\mathbb{Z}_2$ Wilson line to break to the standard model group, in which case, as we will see, the above spectrum reduces to
\be
 H_{1,2}\;,\quad 3\;d_{3,4}\;,\quad 3\;Q_5\; . 
\ee

\vspace{-2mm}

\noindent We are interested in computing the d-quark Yukawa couplings
\begin{equation}
 \l^{(d)}_{JK}H_{1,2}d_{3,4}^JQ_5^K\; .
\end{equation} 

\vspace{-2mm}
  
\noindent However, for now, we construct the relevant bundle-valued forms in the upstairs theory, and restrict to the $\mathbb{Z}_2\times\mathbb{Z}_2$-quotient later. The line bundles $K_1$ and $K_2$ are both of type 1, with $H^1(X,K_1)\cong H^1(X,{\cal K}_1)$ and $H^1(X,K_2)\cong H^1(X,{\cal K}_2)$, while $K_3$ is of type 2 and
\be 
H^1(X,K_3)\cong {\rm Ker}(H^2({\cal A},{\cal N}^*\otimes{\cal K}_3)\stackrel{p}{\rightarrow }H^2({\cal A},{\cal K}_3) )\; . \label{ex3ker}
\ee
Hence, following Eq.~\eqref{akreshom}, the relevant ambient space forms and polynomials can be written in terms of homogeneous coordinates as
\begin{equation}
\!\!
\resizebox{1.05\hsize}{!}{$
\begin{array}{llll}
4 \ \overline{\mathbf{5}}_{1,2} &\!\!\rightarrow & \!\!\hat{\nu}_1=\sigma_2^{-2}\tilde{Q}_{(0,-2,1,1)}\bar{\mu}_2 \, , & \tilde{Q}\in{\rm Span}(u_0v_0,u_0v_1,u_1v_0,u_1v_1) \, ,\\[1mm]
12 \ \overline{\mathbf{5}}_{3,4} &\!\!\rightarrow &\!\! \hat{\nu}_2=\sigma_1^{-4}\tilde{R}_{(-4,0,1,1)}\bar{\mu}_1 \, , & \tilde{R}\in{\rm Span}(\bar{x}_0^2,\bar{x}_0\bar{x}_1,\bar{x}_1^2)\,{\rm Span}(u_0,u_1)\,{\rm Span}(v_0,v_1) \, , \\[1mm]
12 \ \mathbf{10}_5 &\!\!\rightarrow& \!\!\hat{\omega}_3=\sigma_3^{-2}\sigma_4^{-2}\tilde{S}_{(2,0,-4,-4)}\bar{\mu}_3\wedge \bar{\mu}_4 \, , &\tilde{S}\in{\rm Span}(x_0^2,x_0x_1,x_1^2)\,{\rm Span}(\bar{u}_0^2,\bar{u}_0\bar{u}_1,\bar{u}_1^2)\\[1mm]
 &&& \;\;\;\;\;\;\;{\rm Span}(\bar{v}_0^2,\bar{v}_0\bar{v}_1,\bar{v}_1^2)\; ,
 \end{array}$}
\end{equation}
where by Span() we mean the ideal generated by polynomials, with complex number coefficients. The polynomial $\tilde{S}$ lies in a $27$-dimensional space which, in line with Eq.~\eqref{ex3ker}, is mapped into the $15$-dimensional space
\begin{equation}
 {\rm Span}(x_0^4,x_0^3x_1,x_0^2x_1^2,x_0x_1^3,x_1^4)\,{\rm Span}(y_0^2,y_0y_1,y_1^2)\; .
\end{equation}
We have to ensure that $\tilde{S}$ resides in the kernel of this map, which amounts to imposing the condition
\be
 \tilde{p}\tilde{S}=0\; . \label{pS0}
\ee
This leads to a $12$-dimensional space, as expected.

These results are quite complicated due to the large number of multiplets. To simplify matters, it is useful to quotient by the freely-acting $\Gamma=\mathbb{Z}_2\times\mathbb{Z}_2$ symmetry with generators~\eqref{g1g2}. Representations of this symmetry are denoted by a pair of charges, $(q_1,q_2)$, where $q_i\in\{0,1\}$.  We choose a trivial equivariant structure for all line bundles and, following the discussion around Eq.~\eqref{WLcharges}, a Wilson line specified by $\chi_2=(1,1)$, $\chi_3=(0,0)$ with associated multiplet charges
\begin{equation}
 \chi_H=\chi_2=(1,1)\;,\quad \chi_d=\chi_3^*=(0,0)\;,\quad \chi_Q=\chi_2\otimes \chi_3=(1,1)\; .
\end{equation}
Taking into account that the differentials $\mu_i$ carry charge $(1,1)$ under the $\mathbb{Z}_2\times\mathbb{Z}_2$ symmetry, this choice means we should project onto the $(0,0)$ states for $\tilde{Q}$, and the $(1,1)$ states for $\tilde{R}$ and $\tilde{S}$. This leads to to the explicit $\mathbb{Z}_2\times\mathbb{Z}_2$-equivariant polynomials
\begin{equation}
\resizebox{1.02\hsize}{!}{$ 
\begin{array}{lll}
\tilde{Q}&=&u_0v_0+u_1v_1 \, ,\\[1.5mm]
 \tilde{R}&=&a_3 \left(u_0 v_0 \bar{x}_0 \bar{x}_1-u_1 v_1 \bar{x}_0 \bar{x}_1\right)+a_1 \left(u_0
   v_1 \bar{x}_0^2-u_1 v_0 \bar{x}_1^2\right)+a_2 \left(u_1 v_0 \bar{x}_0^2-u_0 v_1
   \bar{x}_1^2\right) \, , \\[1.5mm]
   \tilde{S}&=&b_4 \left(x_0^2 \bar{u}_1^2 \bar{v}_0 \bar{v}_1-x_1^2 \bar{u}_0^2 \bar{v}_0
   \bar{v}_1\right)+b_1 \left(x_0^2 \bar{u}_0^2 \bar{v}_0 \bar{v}_1-x_1^2 \bar{u}_1^2
   \bar{v}_0 \bar{v}_1\right)+b_6 \left(x_0 x_1 \bar{u}_0^2 \bar{v}_1^2-x_0 x_1
   \bar{u}_1^2 \bar{v}_0^2\right)+ \\[1.5mm]
   && b_3 \left(x_0^2 \bar{u}_0 \bar{u}_1 \bar{v}_1^2-x_1^2
   \bar{u}_0 \bar{u}_1 \bar{v}_0^2\right)+b_2 \left(x_0^2 \bar{u}_0 \bar{u}_1
   \bar{v}_0^2-x_1^2 \bar{u}_0 \bar{u}_1 \bar{v}_1^2\right)+b_5 \left(x_0 x_1 \bar{u}_0^2
   \bar{v}_0^2-x_0 x_1 \bar{u}_1^2 \bar{v}_1^2\right)\, .
\end{array}$}
\end{equation}
Hence, we are left with a single Higgs multiplet, $H_{1,2}$, three d-quarks, $d_{3,4}^I$, with parameters ${\bf a}=(a_I)$ and six left-handed quarks $Q_5^J$ with parameters ${\bf b}=(b_J)$. In terms of these parameters, the Yukawa couplings are given by
\begin{equation}
 \mu(Q,R,S)=\tilde{Q}\tilde{R}\tilde{S}=8 \left(a_1 \left(b_1+b_3\right)+a_2 \left(b_2+b_4\right)+a_3 b_5\right)\; . \label{yukex3}
\end{equation} 
However, for the ``physical" result we still have to find the kernel~\eqref{ex3ker}, that is, compute the vectors ${\bf b}$ which satisfy Eq.~\eqref{pS0}. To this end, we write down the most general tetra-quadric polynomial consistent with the $\Gamma=\mathbb{Z}_2\times \mathbb{Z}_2$ symmetry.
\begin{eqnarray}
\tilde{p}&=&C_1 u_0 u_1 v_0 v_1 x_0 x_1 y_0 y_1+C_2 (u_1^2 x_0 x_1 y_0 y_1 v_0^2+u_0^2 v_1^2 x_0x_1 y_0 y_1)+\nonumber\\
   &&C_3 (u_0^2 x_0 x_1 y_0 y_1 v_0^2+u_1^2 v_1^2 x_0 x_1 y_0 y_1)+C_{14}(u_0 u_1 v_1^2 y_0 y_1 x_0^2+u_0 u_1 v_0^2 x_1^2 y_0 y_1)+\nonumber\\
   &&C_{13} (u_1^2 v_0 v_1 y_0 y_1 x_0^2+u_0^2 v_0 v_1 x_1^2 y_0y_1)+C_{16} (u_0^2 v_0 v_1 y_0 y_1 x_0^2+u_1^2 v_0 v_1 x_1^2 y_0 y_1)+\nonumber\\
   &&C_{15} (u_0 u_1 v_0^2 y_0 y_1 x_0^2+u_0 u_1 v_1^2 x_1^2 y_0 y_1)+C_{12} (u_1^2 v_1^2 x_1^2 y_0^2+u_0^2 v_0^2 x_0^2 y_1^2)+\nonumber\\
   &&C_9(u_0^2 v_1^2 x_1^2 y_0^2+u_1^2 v_0^2 x_0^2 y_1^2)+C_{10} (u_0 u_1 v_0 v_1 x_1^2 y_0^2+u_0 u_1 v_0 v_1 x_0^2 y_1^2)+\nonumber\\
   &&C_{11}(u_1^2 v_0^2 x_1^2 y_0^2+u_0^2 v_1^2 x_0^2 y_1^2)+ C_8 (u_0^2 v_0^2 x_1^2 y_0^2+u_1^2 v_1^2  x_0^2 y_1^2)+\nonumber\\
   &&C_5(u_0 u_1 v_1^2 x_0 x_1 y_0^2+u_0 u_1 v_0^2 x_0 x_1 y_1^2)+C_4 (u_1^2 v_0 v_1 x_0 x_1 y_0^2+u_0^2 v_0 v_1 x_0 x_1y_1^2)+\nonumber\\
   &&C_7(u_0^2 v_0 v_1 x_0 x_1 y_0^2+u_1^2 v_0 v_1 x_0 x_1 y_1^2)+C_6 (u_0 u_1 v_0^2 x_0 x_1 y_0^2+u_0 u_1 v_1^2 x_0 x_1y_1^2)+\nonumber\\
   &&C_{17} (u_1^2 v_1^2 x_0^2 y_0^2+u_0^2 v_0^2 x_1^2 y_1^2)+C_{20} (u_0^2 v_1^2 x_0^2 y_0^2+u_1^2 v_0^2 x_1^2y_1^2)+\nonumber\\
   &&C_{19}(u_0 u_1 v_0 v_1 x_0^2 y_0^2+u_0 u_1 v_0 v_1 x_1^2 y_1^2)+C_{18} (u_1^2 v_0^2 x_0^2 y_0^2+u_0^2 v_1^2 x_1^2 y_1^2)+\nonumber\\
   &&C_{21} (u_0^2 v_0^2 x_0^2 y_0^2+u_1^2 v_1^2 x_1^2 y_1^2)\, .\label{pgen}
\end{eqnarray}
The dimension of the complex structure moduli space for $\tilde{X}=X/(\mathbb{Z}_2\times \mathbb{Z}_2)$ is given by $h^{2,1}(\tilde{X})=20$. The $21$ coefficients $C_i$ in the above polynomial provide projective (local) coordinates on this moduli space. Using this polynomial, Eq.~\eqref{pS0} is solved by vectors ${\bf b}$ satisfying
\be
 M{\bf b}=0\,,\qquad 
 M=\left(
\begin{array}{cccccc}
 \frac{C_{16}}{2} & \frac{C_{15}}{2} & \frac{C_{14}}{2} & \frac{C_{13}}{2} & 0 & 0 \\
 \frac{C_7}{2} & \frac{C_6}{2} & \frac{C_5}{2} & \frac{C_4}{2} & C_{21}-C_{17} &  C_{20}-C_{18} \\
 \frac{C_4}{2} & \frac{C_5}{2} & \frac{C_6}{2} & \frac{C_7}{2} & C_{12}-C_8 & C_{11}-C_9  \\
\end{array}
\right)\; .
\ee
The matrix $M$ has indeed a (generically) three-dimensional kernel, but its basis vectors ${\bf v}_I$, where $I=1,2,3$, are very complicated functions of the complex structure moduli. In principle, this basis can be computed, and ${\bf b}$ can then be written as
\begin{equation}
 {\bf b}=\sum_{I}\beta_I{\bf v}_I \, , \label{bspec}
\end{equation}
where the three $\beta_I$ now parametrise  the three left-handed quark families. Inserting this result into Eq.~\eqref{yukex3} gives the desired result for the Yukawa couplings, and it can be shown that the rank of the Yukawa matrix $\l_{IJ}^{(d)}$ is three at generic loci in the complex structure moduli space.

In order to obtain a more explicit result, we restrict to a five-dimensional sub-locus of our $20$-dimensional complex structure moduli space, described by polynomials of the form
\begin{eqnarray}
\tilde{p}_s&=&c_1 u_0 u_1 v_0 v_1 x_0 x_1 y_0 y_1+c_2 (u_0^2 v_0 v_1 x_0^2 y_0 y_1+u_1^2 v_0 v_1x_0^2 y_0 y_1+u_0 u_1 v_0^2 x_1 x_0 y_0^2+\nonumber\\
   &&u_0 u_1 v_1^2 x_1 x_0 y_0^2+u_0 u_1 v_0^2   x_1 x_0 y_1^2+u_0 u_1 v_1^2 x_1 x_0 y_1^2+u_0^2 v_0 v_1 x_1^2 y_0 y_1 + \nonumber\\
   && u_1^2 v_0 v_1
   x_1^2 y_0 y_1)+c_5 (u_0^2 v_1^2 x_0^2 y_0^2+u_0^2 v_0^2 x_1^2 y_0^2+u_1^2
   v_1^2 x_0^2 y_1^2+u_1^2 v_0^2 x_1^2 y_1^2)+\nonumber\\ && c_4 (u_0^2 v_0^2 x_0 x_1 y_0 y_1 -  u_1^2 v_0^2 x_0 x_1 y_0 y_1+
   u_0 u_1 v_1 v_0 x_0^2 y_0^2-u_0 u_1 v_1 v_0 x_1^2y_0^2- \nonumber\\ && u_0 u_1 v_1 v_0 x_0^2 y_1^2+u_0 u_1 v_1 v_0 x_1^2 y_1^2-u_0^2 v_1^2 x_0 x_1 y_0
   y_1+u_1^2 v_1^2 x_0 x_1 y_0 y_1)+\nonumber\\
   &&c_3(u_1^2 v_0^2 x_0^2 y_0^2+u_1^2 v_1^2 x_1^2 y_0^2+u_0^2 v_0^2 x_0^2 y_1^2+u_0^2 v_1^2 x_1^2 y_1^2)+c_6 (u_0^2
   v_0^2 x_0^2 y_0^2+\nonumber\\
   &&u_1^2 v_1^2 x_0^2 y_0^2+u_1^2 v_0^2 x_1^2 y_0^2+u_0^2 v_1^2 x_1^2 y_0^2+u_1^2 v_0^2 x_0^2 y_1^2+u_0^2 v_1^2 x_0^2 y_1^2+\nonumber\\ && u_0^2 v_0^2 x_1^2 y_1^2+u_1^2v_1^2 x_1^2 y_1^2)\label{pspec}\; .
 \end{eqnarray}  
In fact, this polynomial is the most general consistent with the freely-acting $\mathbb{Z}_4\times\mathbb{Z}_4$ symmetry of the tetra-quadric, which contains the $\mathbb{Z}_2\times \mathbb{Z}_2$ symmetry used previously as a sub-group. The equation $\tilde{p}_s\tilde{S}=0$ for the kernel now reads
\begin{equation}\label{Mex3}
 M{\bf b}=0\,,\qquad 
 M=\left(
\begin{array}{cccccc}
 c_2 & 0 & 0 & c_2 & 0 & 0 \\
 0 & c_2 & c_2 & 0 & 0 & 2 c_5-2 c_3 \\
 0 & c_2 & c_2 & 0 & 2 c_3-2 c_5 & 0 \\
\end{array}
\right)\; .
\end{equation}
Generically, the dimension of this kernel is three and a basis can be readily found as
\begin{equation}
\begin{array}{c}
{\bf v}_1=\frac{1}{8}\left(0,2 \left(c_3-c_5\right),0,0,-c_2,c_2\right)^T, \quad
{\bf v}_2=\frac{1}{8}\left(-c_2,0,0,c_2,0,0\right)^T,  \\[1.5mm]
{\bf v}_3=\frac{1}{8}\left(0,-c_2,c_2,0,0,0\right)^T.
\end{array}
\end{equation}
Inserting these vectors into Eq.~\eqref{bspec} and \eqref{yukex3} and choosing a standard basis for the coefficients ${\bf a}$ and ${\boldsymbol\beta}$ then gives the Yukawa couplings
\begin{equation}
\lambda^{(d)}=i \pi^3c
\left(
\begin{array}{ccc}
 0 & -c_2 & c_2 \\
 2 c_3-2 c_5 & c_2 & -c_2 \\
 -c_2 & 0 & 0 \\
\end{array}
\right)\, ,\label{yukex4}
\end{equation}
where $c$ is the numerical factor from Eq.~\eqref{lmurel}. Evidently, the generic rank of this matrix is two. This shows that the rank of the Yukawa matrix can vary in complex structure moduli space and can reduce at specific loci. In the present case, it is generically of rank three in the $20$-dimensional complex structure moduli space described by the polynomials~\eqref{pgen}. On the five-dimensional sub-locus, described by the polynomials~\eqref{pspec}, the rank reduces to two.

If we specialise further to the four-dimensional locus where $c_2=0$, the rank of \eqref{yukex4} reduces to one. It turns out that the tetra-quadric~\eqref{pspec} remains generically smooth on this sub-locus. However, we have to be careful since the rank of the matrix M in Eq.~\eqref{Mex3} also depends on the complex structure. In fact, for $c_2=0$ the rank of $M$ reduces to two so that the dimension of the kernel increases from three to four. Hence, on this sub-locus the spectrum in the low-energy theory enhances from three left-handed quark multiplets to four (plus one mirror left-handed quark multiplet, since the index remains unchanged). A basis of the kernel is then given by ${\bf v}_I={\bf e}_I/8$, where $I=1,\ldots ,4$, and ${\bf e}_I$ are the six-dimensional standard unit vectors. From Eq.~\eqref{bspec} and \eqref{yukex3}, this leads to the Yukawa couplings
\be
\l^{(d)}=i \pi^3 c
 \left(
\begin{array}{cccc}
 1 & 0 & 1 & 0 \\
 0 & 1 & 0 & 1 \\
 0 & 0 & 0 & 0 \\
\end{array}
\right)\, .
\ee
Hence, after properly including the additional multiplet, the rank of the Yukawa matrix remains two.



\section{Yukawa couplings in a quasi-realistic model on the tetra-quadric}\label{realex}
In the previous section, we have applied our methods to a number of toy examples and we have seen cases with vanishing and non-vanishing Yukawa couplings, both with and without complex-structure dependence. We would now like to calculate Yukawa couplings in a quasi-realistic model on the tetra-quadric, that is, a model with gauge group $SU(3)\times SU(2)\times U(1)$ (plus additional $U(1)$ symmetries which are Green-Schwarz anomalous or can be spontaneously broken) and the exact MSSM spectrum (plus moduli fields uncharged under the standard model group, including bundle moduli singlets). 
This model appears in the standard model data base~\cite{Anderson:2011ns,Anderson:2012yf} and has been further analysed in 
Refs.~\cite{Buchbinder:2013dna, Buchbinder:2014qda,Buchbinder:2014sya,Buchbinder:2014qca}. 
We begin by reviewing the basic structure of this model and then calculate the two types of non-vanishing Yukawa couplings which arise, that is, the standard up-quark Yukawa couplings and the singlet Yukawa couplings of the form $SL\overline{H}$, with bundle moduli singlets $S$. 


\subsection{The model}
The upstairs model is based on a rank five line bundle sum, $V=\bigoplus_{a=1}^5L_a$, on the tetra-quadric, with the five line bundles explicitly given by
\begin{equation}\label{lbs5}
 \begin{array}{lll}
 L_1={\cal O}_X(-1,0,0,1) \, ,& L_2= {\cal O}_X(-1,-3,2,2) \, , & L_3 ={\cal O}_X(0,1,-1,0) \, ,\\
 L_4 ={\cal O}_X(1,1,-1,-1) \, ,& L_5={\cal O}_X(1,1,0,-2)\; . &
 \end{array}
\end{equation}
Hence, the low-energy GUT group is $SU(5)\times S(U(1)^5)$. The non-zero cohomologies of line bundles appearing in $V$, $\wedge^2V$ and $V\otimes V^*$ are
\begin{equation}
\begin{array}{llllll}
 h^{^{\!\bullet}}(X,L_2)&=&(0,8,0,0)\, ,&h^{^{\!\bullet}}(X,L_5)&=&(0,4,0,0) \, ,\\[3pt]
 h^{^{\!\bullet}}(X,L_2\otimes L_4)&=&(0,4,0,0)\, ,&h^{^{\!\bullet}}(X,L_2\otimes L_5)&=&(0,3,3,0) \, ,\\[3pt]
 h^{^{\!\bullet}}(X,L_4\otimes L_5)&=&(0,8,0,0) \, ,&h^{^{\!\bullet}}(X,L_1\otimes L_2^*)&=&(0,0,12,0) \, ,\\[3pt]
 h^{^{\!\bullet}}(X,L_1\otimes L_5^*)&=&(0,0,12,0)\, ,&h^{^{\!\bullet}}(X,L_2\otimes L_3^*)&=&(0,20,0,0)\, , \\[3pt]
 h^{^{\!\bullet}}(X,L_2\otimes L_4^*)&=&(0,12,0,0)\, , &h^{^{\!\bullet}}(X,L_3\otimes L_5^*)&=&(0,0,4,0)\; .
\end{array} 
\end{equation}
Following Table~\ref{tab:su5}, these cohomologies give rise to the GUT spectrum
\begin{equation}
 8\, {\bf 10}_2\,,\; 4\,{\bf 10}_5\,,\;4\,\overline{\bf 5}_{2,4}\,,\;3\,\overline{\bf 5}_{2,5}^H\,,\;8\,\overline{\bf 5}_{4,5}\,,\; 3{\bf 5}_{2,5}^{\overline{H}}\,,\;
 12\,{\bf 1}_{2,1}\,,\;12\,{\bf 1}_{5,1}\,,\;20\,{\bf 1}_{2,3}\,,\;12\,{\bf 1}_{2,4}\,,\;4\,{\bf 1}_{5,3}\; .\label{gutspec}
\end{equation}
At the GUT level, the only superpotential terms allowed by the gauge symmetry are
\begin{equation}
 W=\lambda_{IJK} {\bf 5}_{2,5}^{(I)}{\bf 10}_2^{(J)}{\bf 10}_5^{(K)}+\rho_{IJK} {\bf 1}_{2,4}^{(I)}\overline{\bf 5}_{4,5}^{(J)}{\bf 5}_{2,5}^{(K)}\; ,
 \label{Wgut}
\end{equation} 
where the indices $I,J,K\ldots $ run over various ranges, as indicated by the multiplicities in the spectrum~\eqref{gutspec}, and $\lambda_{IJK}$ and $\rho_{IJK}$ are the couplings we would like to calculate.

Evidently, the above GUT model has $12$ families of quarks and leptons, three vector-like ${\bf 5}^{\overline{H}}$--$\overline{\bf 5}^H$ pairs, which can account for the Higgs multiplets, and a spectrum of bundle moduli singlets. This is a promising upstairs spectrum which may lead to a downstairs standard model upon dividing by a freely-acting symmetry of order four. Indeed, this can be accomplished using the $\mathbb{Z}_2\times\mathbb{Z}_2$ symmetry with generators~\eqref{g1g2}, a choice of Wilson line specified by $\chi_2=(0,1)$ and $\chi_3=(0,0)$ and a trivial equivariant structure for all line bundles. The relevant GUT multiplets branch as ${\bf 10}\rightarrow (Q,u,e)$, $\overline{\bf 5}\rightarrow (d,L)$, $\overline{\bf 5}^H\rightarrow (T,H)$ and  ${\bf 5}^{\overline{H}}\rightarrow (\bar{T},\bar{H})$ (where $T$ and $\bar{T}$ are the Higgs triplets, to be projected out). From Eq.~\eqref{WLcharges}, these standard model multiplets carry the Wilson line charges
 \be
 \begin{array}{lllll}
 \chi_Q=\chi_2\otimes\chi_3=(0,1) \, , &\quad&\chi_u=\chi_3^2=(0,0) \, , &\quad&\chi_e=\chi_2^2=(0,0) \, ,\\
 \chi_d=\chi_3^*=(0,0) \, , &\quad&\chi_L=\chi_2^*=(0,1) \, ,&\quad&\chi_H=\chi_2^*=(0,1)\, ,\\
 \chi_{\overline{H}}=\chi_2=(0,1) \, ,&\quad&\chi_T=\chi_3^*=(0,0)\, ,&\quad&\chi_{\overline{T}}=\chi_3=(0,0)\; .
 \end{array}
\ee 
Applying the rule~\eqref{equivcoh} for this choice of charges then leads to the downstairs spectrum
\begin{equation}
\resizebox{1\hsize}{!}{$
 2\, (Q,u,e)_2,\; (Q,u,e)_5,\;(d,L)_{2,4},\;2\,(d,L)_{4,5},\; H_{2,5},\;\overline{H}_{2,5},\;
 3\,{\bf 1}_{2,1},\;3\,{\bf 1}_{5,1},\;5\,{\bf 1}_{2,3},\;3\,{\bf 1}_{2,4},\;{\bf 1}_{5,3},$} \label{smspec}
\end{equation} 
a perfect MSSM spectrum plus additional bundle moduli singlets. Ordering the quarks as $(Q^{(I)})=(Q_2^1,Q_2^2,Q_5)$ and $(u^{(I)})=(u_2^1,u_2^2,u_5)$, the downstairs analogue of the superpotential~\eqref{Wgut} can be written as
\be
 W=\l_{IJ}^{(u)}\overline{H}_{2,5}u^{(I)}Q^{(J)}+\rho_{IJ} {\bf 1}_{2,4}^{(I)}L_{4,5}^{(J)}\overline{H}_{2,5}\; . \label{Wsm}
\ee
The up-Yukawa matrix $\l^{(u)}$ is further constrained by the $S(U(1)^5)$ symmetry and must be of the form
\begin{equation}
 \l^{(u)}=\left(\begin{array}{lll}0&0&a\\0&0&b\\a'&b'&0\end{array}\right)\; . \label{uppattern}
\end{equation} 
However, it is not yet clear that the entries $a$, $b$, $a'$, $b'$ of this matrix are non-zero and that the rank of  the up-Yukawa matrix is indeed two, as the pattern of \eqref{uppattern} suggests. This is the question we will answer in the next sub-section. The $3\times 2$ singlet coupling matrix $\rho$ is unconstrained by gauge symmetry and evidently plays an important role for the existence of a massless Higgs doublet pair, away from the line bundle locus. More precisely, if
\be
 \langle \rho_{IJ}{\bf 1}_{2,4}^{(I)}\rangle
\ee
is non-zero, then the Higgs pair (where a combination of the lepton multiplets plays the role of the down Higgs) receives a large mass and disappears 
from the spectrum. At the line bundle locus, we have $\langle  {\bf 1}_{2,4}^{(I)}\rangle=0$, and the Higgs pair is massless, consistent with 
the result of our cohomology calculation. However, once we move away from the line bundle locus such that $\langle  {\bf 1}_{2,4}^{(I)}\rangle\neq 0$,\footnote{Note that we 
can turn on all the available singlets except ${\bf 1}_{2, 4}^{(I)}$ and keep the Higgs pair massless. As was shown in~Ref.~\cite{Buchbinder:2014qda}, 
this deformation leads to a standard model with global $B-L$ symmetry.}
 the Higgs pair may become massive, depending on the structure of the couplings $\rho_{IJ}$. In fact, in Ref.~\cite{Buchbinder:2014qda} we have 
verified -- by performing a cohomology calculation for the associated non-Abelian bundles -- that the Higgs pair does indeed become massive for generic complex 
structure, once $\langle  {\bf 1}_{2,4}^{(I)}\rangle\neq 0$. This suggests that at least some of the singlet couplings $\rho_{IJ}$ are non-zero, generically.  Below, we will confirm this expectation by explicitly calculating
the couplings $\rho_{IJ}$.


\subsection{Up Yukawa coupling}
To calculate the up Yukawa couplings, we begin with the upstairs GUT model and focus on the first term in the superpotential~\eqref{Wgut}. The line bundles and ambient space harmonic forms (see Eq.~\eqref{akreshom}) for these multiplets are
\begin{equation}
 \begin{array}{lllll}
 3\;{\bf 5}^H_{2,5}&\longrightarrow&K_1=L_2^*\otimes L_5^* \, ,&\quad&\hat{\nu}_1=\s_3^{-2}\tilde{Q}_{(0,2,-2,0)}\bar{\m}_3 \, ,\\
 4\;{\bf 10}_2&\longrightarrow&K_2=L_5 \, , &\quad&\hat{\nu}_2=\s_4^{-2}\tilde{R}_{(1,1,0,-2)}\bar{\m}_4 \, , \\
 8\;{\bf 10}_5&\longrightarrow&K_3=L_2 \, , &\quad&\hat{\omega}=\s_1^{-3}\s_2^{-5}\tilde{S}_{(-3,-5,0,0)}\bar{\m}_1\wedge \bar{\m}_2\; ,
 \end{array}
 \end{equation}
with associated polynomials
\begin{eqnarray}
 \tilde{Q}&=&q_0y_0^2+q_1y_0y_1+q_2y_1^2 \, , \label{Qt}\\
 \tilde{R}&=&r_0x_0y_0+r_1x_1y_0+r_2x_0y_1+r_3x_1y_1 \, ,\\
 \tilde{S}&=&s_0\bar{x}_0\bar{y}_0^3+s_1\bar{x}_0\bar{y}_0^2\bar{y}_1+s_2\bar{x}_0\bar{y}_0\bar{y}_1^2+s_3\bar{x}_0\bar{y}_1^3+s_4\bar{x}_1\bar{y}_0^3+  s_5\bar{x}_1\bar{y}_0^2\bar{y}_1+ \notag \\ && s_6\bar{x}_1\bar{y}_0\bar{y}_1^2+s_7\bar{x}_1\bar{y}_1^3\; ,
 \end{eqnarray} 
 and coefficients $q_I$, $r_I$ and $s_I$ parametrising the multiplets. Evidently, $K_1$ and $K_2$ are of type 1, while $K_3$ is of type 2, so we can proceed with the algebraic calculation explained in Section~\ref{comments}. Converting everything to holomorphic coordinates for simplicity of notation, we have
\begin{eqnarray}
 \mu(Q,R,S)&=&\left(q_0\partial_{y_0}^2+q_1\partial_{y_0}\partial_{y_1}+q_2\partial_{y_1}^2\right)
                          \left(r_0\partial_{x_0}\partial_{y_0}+r_1\partial_{x_1}\partial_{y_0}+r_2\partial_{x_0}\partial_{y_1}+r_3\partial_{x_1}\partial_{y_1}\right)\nonumber\\
&& \big(s_0x_0y_0^3+s_1x_0y_0^2y_1+s_2x_0y_0y_1^2+s_3x_0y_1^3+s_4x_1y_0^3+
s_5x_1y_0^2y_1+ \nonumber\\ && \,\,\, s_6x_1y_0y_1^2+s_7x_1y_1^3\big) \nonumber\\
&=&2\left[3q_0r_0s_0+3q_0r_1s_4+q_0r_2s_1+q_0r_3s_5+q_1r_0s_1+q_1r_1s_5+\right.\nonumber\\
&&\left.\quad q_1r_2s_2+q_1r_3s_6+q_2r_0s_2+q_2r_1s_6+3q_2r_2s_3+3q_2r_3s_7\right]\; .
 \end{eqnarray} 
Inserting standard choices for the coefficients then leads to the couplings $\l_{IJK}$ in the superpotential~\eqref{Wgut}. In particular, we see that these couplings are just numbers, that is, they are independent of complex structure.\\[2mm]

For a simpler and physically more meaningful result, we should consider the downstairs theory. This means we have to extract, from the above polynomials $\tilde{Q}$, $\tilde{R}$ and $\tilde{S}$, the $\mathbb{Z}_2\times\mathbb{Z}_2$ equivariant parts. Remembering that the differentials $\mu_i$ carry charge $(1,1)$ under $\mathbb{Z}_2\times\mathbb{Z}_2$, while the $\s_i$ are invariant, this leads to
\begin{eqnarray}
\bar{H}&:&\tilde{Q}_{\bar{H}}=y_0y_1 \, ,\label{b1}\\
Q_5&:&\tilde{R}_{Q_5}=y_0x_1+y_1x_0 \, ,\\
u_5&:&\tilde{R}_{u_5}=y_0x_1-y_1x_0 \, ,\\
Q_2^\alpha&:&\tilde{S}_{Q_2}=-x_0y_0^2+x_1y_1^3\,,\;-x_0y_0y_1^2+x_1y_1y_0^2 \, ,\\
u_2^\alpha&:&\tilde{S}_{u_2}=x_0y_0^3+x_1y_1^3\,,\;x_0y_0y_1^2+x_1y_1y_0^2 \, .\label{b5}
\end{eqnarray}
To carry out the algebraic calculation, we first note that
\begin{equation}
 \lambda(Q,R,S)=\frac{i \pi^3}{24}\mu(Q,R,S)\; , 
\end{equation}  
where the additional factor of $1/4$ relative to Eq.~\eqref{lmurel} accounts for the fact that we are integrating over the upstairs manifold $X$, while the actual calculation should be carried out on the quotient $X/\Gamma$. We find
\begin{equation}
 \mu(\bar{H},u_5,Q_2^\alpha)=\left(\partial_{y_0}\partial_{y_1}\right)\left(\partial_{y_0}\partial_{x_1}-\partial_{y_1}\partial_{x_0}\right)
\left( \begin{array}{l}-x_0y_0^3+x_1y_1^3\\-x_0y_0y_1^2+x_1y_1y_0^2\end{array}\right)=\left(\begin{array}{l}0\\4\end{array}\right)
\end{equation}
and
\begin{equation}
 \mu(\bar{H},u_2^\alpha,Q_5)=\left(\partial_{y_0}\partial_{y_1}\right)\left(\partial_{y_0}\partial_{x_1}+\partial_{y_1}\partial_{x_0}\right)
\left( \begin{array}{l}x_0y_0^3+x_1y_1^3\\x_0y_0y_1^2+x_1y_1y_0^2\end{array}\right)=\left(\begin{array}{l}0\\4\end{array}\right) \, .
\end{equation}
Combining these results leads to the up Yukawa matrix
\begin{equation}
\label{upyukawamatrix}
 \lambda^{(u)}=\frac{i \pi^3}{6}\left(\begin{array}{lll}0&0&0\\0&0&1\\0&1&0\end{array}\right)\; .
\end{equation} 
We have, therefore, shown that the up Yukawa matrix has indeed rank 2, as suggested by the general structure~\eqref{uppattern}. In addition, we see that these Yukawa couplings are independent of complex structure. This happens because the cohomologies of the above line bundles $K_i$ have a simple representation in terms of ambient space cohomologies, without any kernel or co-kernel operations required.

It is interesting to point out that the Yukawa matrix obtained in Eq.~\eqref{upyukawamatrix} does not match its counterpart from particle physics very accurately (for example, in Eq.~\eqref{upyukawamatrix}, the charm- and top-quarks seem to have the same mass.) This is not necessarily an indication that the model has failed as, first of all, the fields are not normalised, and also non-perturbative effects were not yet taken into account. Such effects are generated by instantonic strings wrapped around curves and they are proportional to $\textrm{exp}(-n_i T^i)$, where $T^i$ are K\"ahler moduli and $n_i$ are positive integers parametrising the curve. The very small but non-zero mass of the up-quark could be obtained from such non-perturbative effects. Similarly, these effects could explain the great disparity between the masses of the charm-quark and top-quark.


\subsection{Singlet-Higgs-lepton coupling}
To calculate the singlet Yukawa coupling, we start with the upstairs theory as before and focus on the second term in the superpotential~\eqref{Wgut}. The relevant line bundles and forms are
\be\label{ex4forms}
\begin{array}{lllll}
 12\;{\bf 1}_{2,4}&\longrightarrow&K_1=L_2\otimes L_4^* \, , &\quad& \hat{\omega}_1=\kappa_1^{-4}\kappa_2^{-6}Q_{(-4,-6,1,1)}d\bar{z}_1\wedge d\bar{z}_2 \, , \\[1mm]
 8\;\overline{\bf 5}_{4,5}&\longrightarrow&K_2=L_4\otimes L_5 \, , &\quad&\hat{\omega}_2=\kappa_3^{-3}\kappa_4^{-5}R_{(0,0,-3,-5)}d\bar{z}_3\wedge d\bar{z}_4 \, ,\\[1mm]
 4\;{\bf 5}_{2,5}^{\overline{H}}&\longrightarrow&K_3=L_2^*\otimes L_5^* \, , &\quad& \hat{\nu}_3=\kappa_3^{-2}S_{(0,2,-2,0)}d\bar{z}_3\; .
\end{array}
\ee 
There are two additional complications, compared to the previous calculation, evident from this list of forms. First of all, the singlet space is defined as the kernel 
\be
 {\rm Ker}\left(H^2({\cal A},{\cal N}^*\otimes {\cal K}_1)\stackrel{p}{\rightarrow} H^2({\cal A},{\cal K}_1)\right)
\ee 
of a map between a $60$ and a $48$-dimensional space. These dimensions are quite large, but we will improve on this shortly by taking the $\mathbb{Z}_2\times\mathbb{Z}_2$ quotient. At any rate, we should impose the constraint $\tilde{p}\tilde{Q}=0$ on the polynomials $Q$ in order to work out this kernel, and this will lead to complex structure dependence.

Secondly, two line bundles, $K_1$ and $K_2$, are of type 2, which means that we will have to work with the more general Eq.~\eqref{Yukgen4} for the Yukawa couplings. Given the differentials $d\bar{z}_i$ which appear in  \eqref{ex4forms}, only the term proportional to $\hat{\omega}_1\wedge\hat{\nu}_2\wedge\hat{\nu}_3$ can contribute to the integral~\eqref{Yukgen4}. This means we need to determine the $(0,1)$-forms $\hat{\nu}_2$ satisfying
\be
\bar{\partial}\hat{\nu}_2=p\hat{\omega}_2\; .
\ee
To do this, we write down the two relevant polynomials
\begin{equation}
 R_{(0,0,-3,-5)}=r_0+r_1\bar{z}_3\;,\quad p=p_0+p_1z_3+p_2z_3^2 \; , \label{Rsplit}
\end{equation}
with the $z_3$-dependence made explicit and apply the result~\eqref{coboundres}, which reads
\begin{equation}
 {\cal R}=-\frac{1}{2}(p_1r_0+p_2r_1)+p_0r_0\bar{z}_3+\frac{1}{2}p_0r_1\bar{z}_3^2-\frac{1}{2}p_2r_0z_3-p_2r_1|z_3|^2+\frac{1}{2}(p_0r_0+p_1r_1)\bar{z}_3|z_3|^2 \; .\label{Rdef}
\end{equation} 
Then, the desired $(0,1)$-form $\hat{\nu}_2$ can be written as
\begin{equation}
 \hat{\nu}_2=\kappa_3^{-2}\kappa_4^{-5}{\cal R}d\bar{z}_4\; .
\end{equation}
Using these results for the forms in the basic formula~\eqref{Yukgen4} for the Yukawa couplings, we find
\begin{equation}
 \lambda(\nu_1,\nu_2,\nu_3)=-\frac{1}{2 \pi i}\int_{\mathbb{C}^4}\frac{Q{\cal R}S}{\kappa_1^4\kappa_2^6\kappa_3^4\kappa_4^5}d^4z\,d^4\bar{z}\; . \label{yukexample3}
\end{equation}  
To simplify the calculation, we descend to the downstairs theory and divide by the $\mathbb{Z}_2\times\mathbb{Z}_2$ with generators~\eqref{g1g2}. The polynomials $Q$, $R$ and $S$ then simplify to
\begin{eqnarray}
 Q&=&a_{14} \left(z_3 \bar{z}_1 \bar{z}_2^2+z_4 \bar{z}_1 \bar{z}_2^2\right)+a_5
   \left(\bar{z}_1^2 \bar{z}_2^2+z_3 z_4 \bar{z}_2^2\right)+a_4 \left(z_3 z_4 \bar{z}_1^2
   \bar{z}_2^2+\bar{z}_2^2\right)+ \nonumber \\ && a_7 \left(z_3 \bar{z}_2^3+z_4 \bar{z}_1^2
   \bar{z}_2\right)+ a_6 \left(z_4 \bar{z}_2^3+z_3 \bar{z}_1^2 \bar{z}_2\right)+a_{13}
   \left(\bar{z}_1 \bar{z}_2^3+z_3 z_4 \bar{z}_1 \bar{z}_2\right)+ \nonumber\\
   && a_{12} \left(z_3 z_4
   \bar{z}_1 \bar{z}_2^3+\bar{z}_1 \bar{z}_2\right)+a_2 \left(z_3 \bar{z}_1^2
   \bar{z}_2^3+z_4 \bar{z}_2\right)+ a_3 \left(z_4 \bar{z}_1^2 \bar{z}_2^3+z_3
   \bar{z}_2\right)+ \nonumber\\
   && a_8 \left(\bar{z}_2^4+z_3 z_4 \bar{z}_1^2\right)+a_9 \left(z_3 z_4
   \bar{z}_2^4+\bar{z}_1^2\right)+a_{10} \left(z_3 \bar{z}_1 \bar{z}_2^4+z_4
   \bar{z}_1\right)+\nonumber\\
   &&a_{11} \left(z_4 \bar{z}_1 \bar{z}_2^4+z_3 \bar{z}_1\right)+a_1\left(\bar{z}_1^2 \bar{z}_2^4+z_3 z_4\right)+a_0 \left(z_3 z_4 \bar{z}_1^2\bar{z}_2^4+1\right) \, ,\\
   R&=&b_1 \left(\bar{z}_4^2-\bar{z}_3 \bar{z}_4\right)+b_0 \left(1-\bar{z}_3 \bar{z}_4^3\right) \, ,\\
   S&=&z_2\; .
\end{eqnarray}
We still have to impose the condition $\tilde{p}\tilde{Q}=0$, which reduces the $15$ parameters ${\bf a}=(a_I)$ down to a generic number of three, corresponding to the three singlets ${\bf 1}_{2,4}$. The two coefficients ${\bf b}=(b_0,b_1)$ parametrize the leptons $L_{4,5}$, while $S=z_2$ represents the Higgs $\overline{H}_{2,5}$. From Eq.~\eqref{Rdef} and using the five-parameters $\mathbb{Z}_4\times \mathbb{Z}_4$-invariant family of tetra-quadrics~\eqref{pspec}, in order to make the calculation manageable, we can explicitly work out the polynomial ${\cal R}$. Then, inserting into Eq.~\eqref{yukexample3}, gives
\begin{eqnarray}
\label{Yukawa_5.3}
\lambda({\bf a},{\bf b})&=&\frac{i \pi^3}{6480}\big(2 a_{14} b_1 c_1+9 a_{12} b_0 c_2+9 a_{13} b_0 c_2-8 a_4 b_1 c_2-8 a_5 b_1 c_2+3 a_{12}
   b_1 c_2+ \nonumber \\ && \qquad \,\,\,\, 3 a_{13} b_1 c_2-36 a_7 b_0 c_3-
   12 a_2 b_1 c_3-12 a_{14} b_0 c_4+6 a_2 b_1c_4+6 a_3 b_1 c_4- \nonumber \\ && \qquad \,\,\,\, 6 a_6 b_1 c_4-6 a_7 b_1 c_4+4 a_{14} b_1 c_4-36 a_6 b_0 c_5- 12 a_3 b_1 c_5-36 a_2 b_0 c_6- \nonumber \\ && \qquad \,\,\,\, 36 a_3 b_0 c_6-12 a_6 b_1 c_6-12 a_7 b_1 c_6\big) \, . \label{yukex3gen}
\end{eqnarray} 
We still have to impose the kernel condition on the vector ${\bf a}$, and as before, we use the five-parameter family of tetra-quadrics~\eqref{pspec}. This condition can then be written as $M{\bf a}=0$, where
\begin{align}
&M=  \\
&\!\!{\scriptsize
\left(\arraycolsep=1.4pt\def\arraystretch{1.5}
\begin{array}{ccccccccccccccc}
 24 c_6 & 0 & 0 & 0 & 4 c_3 & 4 c_6 & 0 & 0 & 0 & 24 c_5 & 0 & 0 & 3 c_4 & 0 & 0 \\
 24 c_5 & 0 & 6 c_2 & 0 & 4 c_6 & 4 c_3 & 0 & 6 c_2 & 0 & 24 c_6 & 0 & 0 & -3 c_4 & 0 & 0
   \\
 24 c_4 & 24 c_6 & 0 & 6 c_2 & 4 c_6 \!\!- \!\!4 c_4 & 4 c_3\!\!+\!\!4 c_4 & 6 c_2 & 0 & 24 c_5 & -24 c_4 &
   12 c_2 & 0 & 3 c_1 & 3 c_4 & 2 c_2 \\
 0 & 24 c_5 & 0 & 0 & 4 c_3 & 4 c_6 & 0 & 0 & 24 c_6 & 0 & 12 c_2 & 0 & 0 & -3 c_4 & 2
   c_2 \\
 24 c_3 & 0 & 0 & 0 & 4 c_6 & 4 c_5 & 0 & 0 & 0 & 24 c_6 & 0 & 12 c_2 & -3 c_4 & 0 & 2
   c_2 \\
 24 c_6 & 24 c_4 & 6 c_2 & 0 & 4 c_4\!\!+\!\!4 c_5 & 4 c_6\!\!-\!\!4 c_4 & 0 & 6 c_2 & -24 c_4 & 24 c_3 &
   0 & 12 c_2 & 3 c_4 & 3 c_1 & 2 c_2 \\
 0 & 24 c_3 & 0 & 6 c_2 & 4 c_5 & 4 c_6 & 6 c_2 & 0 & 24 c_6 & 0 & 0 & 0 & 0 & -3 c_4 & 0
   \\
 0 & 24 c_6 & 0 & 0 & 4 c_6 & 4 c_5 & 0 & 0 & 24 c_3 & 0 & 0 & 0 & 0 & 3 c_4 & 0 \\
 0 & 0 & 12 c_6 & 12 c_6 & 8 c_2 & 8 c_2 & 12 c_3 & 12 c_5 & 0 & 0 & 0 & 0 & 0 & 0 & 4
   c_4 \\
 0 & 0 & 12 c_5 & 12 c_3 & 0 & 0 & 12 c_6 & 12 c_6 & 0 & 0 & 0 & 0 & 0 & 0 & -4 c_4 \\
 0 & 0 & 12 c_6 & 12 c_6 & 0 & 0 & 12 c_5 & 12 c_3 & 0 & 0 & 0 & 0 & 6 c_2 & 6 c_2 & 4
   c_4 \\
 0 & 0 & 12 c_3\!\!+\!\!12 c_4 & 12 c_4\!\!+\!\!12 c_5 & 8 c_2 & 8 c_2 & 12 c_6\!\!-\!\!12 c_4 & 12 c_6\!\!-\!\!12 c_4 &
   0 & 0 & 0 & 0 & 6 c_2 & 6 c_2 & 4 c_1\!\!-\!\!4 c_4 \\
\end{array}
\right)}\nonumber
\end{align}
This matrix has generic rank $12$ and, hence, a three-dimensional kernel, spanned by vector ${\bf v}_I$. We can write
\begin{equation}
 {\bf a}=\sum_{I}\alpha_I{\bf v}_I\; , \label{aexp}
\end{equation}
with the three coefficients $\alpha_I$ describing the three singlets $S^I$. Unfortunately, even for our 5-parameter family~\eqref{pspec} of tetra-quadrics, the ${\bf v}_I$ contain very complicated functions of the complex structure moduli, which make an analytic calculation impractical. Instead, we choose random numerical values for the complex structure moduli $c_1,\ldots ,c_6$, calculate a basis of ${\rm Ker}(M)$ for this choice and then work out the Yukawa matrix by inserting into Eqs.~\eqref{aexp} and \eqref{yukex3gen}. In this way, we obtain an explicit numerical $3\times 2$ Yukawa matrix $\rho$, valid at this specific point in complex structure moduli space. This calculation leads to a Yukawa matrix $\rho$ with rank two, and this should be considered the generic result in complex structure moduli space. 

An analytic calculation can be carried out by restricting to the 4-parameter sub-family with $c_2=0$. In this case, the kernel basis vectors are
\be
\resizebox{1.03\hsize}{!}{$ 
\begin{array}{ll}
{\bf v}_1=&\left(0,0,-c_1 \left(c_3^2+c_5 c_3-2 c_6^2\right)-c_4 \left(-c_3^2+\left(3 c_4-2
   c_6\right) c_3+c_5 \left(c_5+2 c_6\right)+c_4 \left(c_5+4 c_6\right)\right),\right.\\[1mm]
   &c_4c_3^2+\left(c_4^2+2 c_6 c_4+c_1 c_5\right) c_3-c_4 c_5 \left(c_5+2 c_6\right)+c_4^2
   \left(3 c_5+4 c_6\right)+c_1 \left(c_5^2-2 c_6^2\right),0,0, \\[1mm]
   & -\left(c_3-c_5\right)
   \left(c_4^2+c_3 c_4+\left(c_5+2 c_6\right) c_4-c_1 c_6\right),-\left(c_3-c_5\right)
   \left(c_4^2+c_3 c_4+\left(c_5+2 c_6\right) c_4-c_1 c_6\right),\\[1mm]
   &\left. 0,0,0,0,0,0,3\left(c_3-c_5\right) \left(c_3+c_4+c_5-2 c_6\right) \left(c_3+c_5+2 c_6\right)\right)^T \, ,\\[1mm]
{\bf v}_2=&\left(0,0,0,0,0,0,0,0,0,0,0,3 \left(c_3-c_5\right) \left(c_3+c_4+c_5-2 c_6\right)
   \left(c_3+c_5+2 c_6\right),0,0,0\right)^T \, , \\[1mm]
{\bf v}_3=&\left(0,0,0,0,0,0,0,0,0,0,3 \left(c_3-c_5\right) \left(c_3+c_4+c_5-2 c_6\right)
   \left(c_3+c_5+2 c_6\right),0,0,0,0\right)^T\, .
\end{array}$}
\ee
Inserting these vectors into Eq.~\eqref{aexp} and then into the general form \eqref{yukex3gen} of the Yukawa couplings leads to
\begin{equation}
\lambda({\boldsymbol\alpha},{\bf b})=\frac{i \pi^3}{360} \alpha _1 b_1 \left(c_3-c_5\right) \left(4 c_4^2+c_1 \left(c_3+c_5-2
   c_6\right)\right) \left(c_3+c_5+2 c_6\right)\; .
\end{equation}
For the Yukawa matrix $\rho$ in the superpotential~\eqref{Wsm}, this means
\be
\rho=\frac{i \pi^3}{360}\left(
\begin{array}{cc}0&\left(c_3-c_5\right) \left(4 c_4^2+c_1 \left(c_3+c_5-2
   c_6\right)\right) \left(c_3+c_5+2 c_6\right)\\0&0\\0&0\end{array}\right) \, . \label{Yuksing}
\ee   
The matrix has rank one, which is reduced from the generic value two that we have found for the five-dimensional family~\eqref{pspec}. Hence, we have found another example of a Yukawa coupling with rank varying as a function of complex structure. In addition, our results show that, for generic complex structure, the Higgs pair receives a mass whenever $\langle {\bf 1}_{2,4}\rangle\neq 0$, in agreement with the results in Ref.~\cite{Buchbinder:2014qda}. 

For special sub-loci of our four-parameter family of tetra-quadrics, characterised by the vanishing of one of the factors in Eq.~\eqref{Yuksing}, the Yukawa matrix vanishes entirely. However, as before, we have to be careful, since the kernel of the matrix $M$ might also change in these case. Let us begin by imposing $c_3=c_5$, in addition to $c_2=0$, on the family of polynomials~\eqref{pspec}. In this case, the dimension of ${\rm Ker}(M)$ turns out to be six and a basis is given by
\be
\resizebox{1.01\hsize}{!}{$
\begin{array}{ll}
\mathbf{v}_1=(0, 0, 0, 0, 0, 0, 0, 0, 0, 0, 0, 1, 0, 0, 0)^T, &\mathbf{v}_2=(0, 0, 0, 0, 0, 0, 0, 0, 0, 0, 1, 0, 0, 0, 0)^T, \\[1mm]
\mathbf{v}_3=(0, 1, 0, 0, -6, 0, 0, 0, 0, 1, 0, 0, 0, 0, 0)^T, &\mathbf{v}_4= (1, 0, 0, 0, 0, -6, 0, 0, 1, 0, 0, 0, 0, 0, 0)^T, \\[1mm]
\mathbf{v}_5=(0, 0, 0, 0, 0, 0, -1, 1, 0, 0, 0, 0, 0, 0, 0)^T, &\mathbf{v}_6= (0, 0, -1, 1, 0, 0, 0, 0, 0, 0, 0, 0, 0, 0, 0)^T.
\end{array}$}
\ee
Using these six vectors in Eqs.~\eqref{aexp} and \eqref{Yukawa_5.3} leads to a $6\times 2$ Yukawa matrix which vanishes entirely. Similar results are obtained for other sub-loci of interest. If $4 c_4^2+c_1 (c_3+c_5-2c_6)=0$, in addition to $c_2=0$, the dimension of the kernel becomes four and the $4\times 2$ Yukawa matrix vanishes entirely. The same statements hold for $c_3+c_5+2c_6=0$. This shows that there are specific loci in complex structure moduli space where the Higgs pair remains massless, even in the presence of generic bundle moduli VEVs. 


\section{Final remarks}
\label{c1conclusions}
In this chapter, we have developed methods to calculate holomorphic Yukawa couplings for heterotic line bundle models, focusing on Calabi-Yau manifolds defined as hypersurfaces in products of projective spaces and the tetra-quadric in $\mathbb{P}^1\times\mathbb{P}^1\times\mathbb{P}^1\times\mathbb{P}^1$ in particular.  While our approach is based on  differential geometry, we also make contact with the algebraic methods in Refs.~\cite{Candelas:1987se,Anderson:2009ge}.

We provide explicit rules for writing down the relevant bundle-valued harmonic forms which enter the Yukawa couplings. These forms can be identified with polynomials of certain multi-degrees which are the key players in the algebraic calculation. It turns out that these forms can be of different topological types, which we have referred to as type 1 and type 2 (as well as mixed type). If all three forms involved in a Yukawa coupling are of type 1, it turns out that the Yukawa coupling vanishes. This vanishing is topological in nature and is not, apparently, due to a symmetry in the low-energy theory. Our most explicit results, see for example Eq.~\eqref{Yuk112copy}, are for Yukawa couplings which involve two forms of type 1 and one form of type 2. We also show how to compute Yukawa couplings which involve more than one form of type 2, by explicitly working out co-boundary maps. 

The various cases are illustrated with explicit toy examples on the tetra-quadric. In Section~\ref{vanishing}, we have provided an example, based on the gauge group $SO(10)$, of a ${\bf 10}\,{\bf 16}\,{\bf 16}$ Yukawa coupling with topological vanishing, due to all three relevant forms being of type 1. An example of a complex structure independent ${\bf 27}^3$ Yukawa coupling for gauge group $E_6$ and a standard pattern of two forms of type 1 and one form of type 2 has been provided in Section~\ref{E6example}. Finally, Section~\ref{csexample} contains an example with gauge group $SU(5)$ which leads to a complex structure dependent d-quark Yukawa coupling.

In Section~\ref{realex} we have computed all Yukawa couplings allowed by the gauge symmetry for a line bundle standard model on the tetra-quadric. The up-quark Yukawa matrix turns out to be complex structure independent and of rank two, while the singlet coupling to $L\overline{H}$ is complex structure dependent. The latter involves two forms of type 2 and requires an explicit calculation of a co-boundary map as well as a kernel of a map in cohomology. 

For two of our examples, we have explicitly calculated the complex structure dependence of the Yukawa matrix, only for a sub-locus in complex structure moduli space. 
To our knowledge, this is the first such calculation for heterotic Calabi-Yau models with non-standard embedding. 
The detailed complex structure dependence of these Yukawa matrices is not necessarily physical since the matter 
field K\"ahler metric will also typically depend on complex structure. However, the rank of the Yukawa matrices is not affected by 
the field normalisation and has to be considered a physical quantity. We have shown that this rank can vary in complex structure moduli space. 

Our calculations were applied only to one quasi-realistic tetraquadric model in the database \cite{Anderson:2011ns}. An interesting direction for future research would be to calculate the holomorphic Yukawa couplings for all the other MSSM models, in order to see how the results vary from model to model. Extending our calculation methods to all the known models on the tetraquadric would be useful to see a distribution of the Yukawa couplings among the models.

The results of this chapter are limited to a relatively narrow class of Calabi-Yau manifolds and bundles with Abelian structure group. However, the methods we have developed point to and facilitate a number of generalisations. We expect that suitable generalisations of our approach can be used to calculate Yukawa couplings for more general classes of Calabi-Yau manifolds, notably higher co-dimension CICYs (as will be seen in Chapter~\ref{chaptern>1codimension}) and hypersurfaces in toric varieties. Non-Abelian bundles are frequently constructed from line bundles, for example via monad or extension sequences. The results for line bundles obtained in this chapter could be useful to calculate Yukawa couplings for such non-Abelian bundles.

The most pressing problem remains the calculation of the matter field K\"ahler metric which is essential in order to determine the physical Yukawa couplings. While we have not addressed this problem yet, it is clear that it requires an approach based on differential geometry. Our hope is that the methods developed in this 
chapter will eventually lead to a framework for such a calculation.

\chapter{Holomorphic Yukawa Couplings for Co-dimension $k \geq 2$ Complete Intersection Calabi-Yau Manifolds}
\label{chaptern>1codimension}

In Chapter~\ref{tetraquadricchapter}, we have presented a new approach to calculating the holomorphic Yukawa couplings,
based entirely on methods of differential geometry. This approach was developed in the context of the simplest class of Calabi-Yau manifolds -- hypersurfaces in products of projective spaces and the tetraquadric manifold in a product of four $\mathbb{P}^1$’s in particular -- and for bundles with Abelian structure groups. In its original form, as presented in Chapter~\ref{tetraquadricchapter}, this method is only applicable to a handful of Calabi-Yau manifolds. The purpose of this chapter is to present a significant generalisation to all complete
intersection Calabi-Yau manifolds. Hence, we will show that our approach is not restricted to specific manifolds but can in fact be applied to large classes, in this case to the almost 8000 CICY manifolds classified in Refs. \cite{Candelas:1987kf, candelascicy2}, as well as to their quotients \cite{braunquotients}. We would also like
to relate our method to the earlier algebraic ones \cite{Candelas:1987se, Anderson:2009ge} and demonstrate that the two approaches are equivalent. Although, in the present chapter, we will only discuss the holomorphic Yukawa couplings, we hope that the insight gained in this context will ultimately also be of use for the calculation of the matter
field K\"ahler potential and the physical Yukawa couplings.

As in the co-dimension one case, we start from the general expression of holomorphic Yukawa couplings for a line bundle model on a Calabi-Yau manifold $X$
\begin{eqnarray}
\label{Yukgen2}
\lambda(\nu_1, \nu_2, \nu_3) = \int_X \Omega \wedge \nu_1 \wedge \nu_2 \wedge \nu_3 \, .
\end{eqnarray}
Here, $\Omega$ is the holomorphic $(3, 0)$–form on $X$ and $\nu_i \in H^1(X, K_i)$ are closed $(0, 1)$-forms, taking values in certain line bundles $K_i$ on $X$, which represent the three types of matter multiplets involved in the
corresponding superpotential term. Consistency of Eq.~\eqref{Yukgen2} requires that $K_1 \otimes K_2 \otimes K_3 = \mathcal{O}_X$ , where $\mathcal{O}_X$ is the trivial bundle on $X$. The ambient space on which the CICY is defined is generally expressed as $\mathcal{A}=\mathbb{P}^{n_1} \times ... \times \mathbb{P}^{n_m}$, as explained in Section~\ref{cicysection}. Provided that the line bundles $K_i$ are obtained as restrictions of ambient space line bundles $\mathcal{K}_i \rightarrow \mathcal{A}$
to $X$, we will show that the $(0, 1)$-forms $\nu_i$ can be obtained from certain forms on the ambient space $\mathcal{A}$ and that the integral \eqref{Yukgen2} can be evaluated explicitly by converting it to an integral over the ambient
space.

More precisely, we find that a closed $(0, 1)$-form $\nu_i$
is, in general, related to an entire chain of ambient
space $(0, a)$-forms, $\hat{\nu}_{i,a}$, where $a = 1, . . . , k + 1$ and $k$ is the co-dimension of $X$ in $\mathcal{A}$. The integral \eqref{Yukgen2} can then be re-written as an integral over $\mathcal{A}$ which, in general, involves all forms $\hat{\nu}_{i,a}$. For a given $\nu_i$, the
associated chain may terminate early, in the sense that, for a certain $\tau_i$, we have $\hat{\nu}_{i,\tau_i} \neq 0$ and $\hat{\nu}_{i,a} = 0$ for all $a > \tau_i$. In this case we say that $\nu_i$ is of type $\tau_i$. One of our most important results is the vanishing theorem
\begin{eqnarray}
\tau_1+\tau_2+\tau_3 < \textrm{dim}(\mathcal{A}) \quad\Rightarrow \quad \lambda (\nu_1,\nu_2,\nu_3) = 0 \, .
\end{eqnarray}
\noindent  Particularly for high co-dimension and corresponding large ambient space dimension $\textrm{dim}(\mathcal{A})$, this statement implies the vanishing of many Yukawa couplings, since cases with large types $\tau_i$ are relatively rare. The vanishing due to this theorem can not be explained by an obvious symmetry of the effective four-dimensional theory and is topological in nature.

The outline of the chapter is as follows. In Section~\ref{codim2}, we generalise
to co-dimension two CICYs and in Section~\ref{higher}, we deal with the general case of arbitrary co-dimension. In
Section~\ref{chapter3examples}, our method is illustrated with several explicit examples and we conclude in Section~\ref{chapter3conc}. A number
of technical issues have been moved to the appendices. Of particular importance is Appendix~\ref{appendixPn}, which
explains the multiplication of harmonic forms on $\mathbb{P}^n$, the key ingredient required to relate our approach
to the earlier algebraic methods \cite{Candelas:1987se, Anderson:2009ge} for calculating holomorphic Yukawa couplings.

\section{Yukawa couplings for co-dimension two CICYs}
\label{codim2}


We start by discussing the case of co-dimension two manifolds. This consists of two main steps: first, showing how $(0,1)$-forms on the CICY can be obtained from ambient space forms and secondly, expressing the Yukawa couplings as integrals over the ambient space.


\subsection{Lifting forms to the ambient space}


As before, the ambient space ${\cal A}$ is given by a product of projective spaces
\be  
{\cal A}= {\mathbb P}^{n_1} \times {\mathbb P}^{n_2} \times ... \times  {\mathbb P}^{n_m}\,, 
\label{3.1}
\ee
but now we require that $n_1+...+n_m = 5$. The Calabi-Yau manifold $X$ is given by the common zero locus of two polynomials $p=(p_1,p_2)$ with multi-degrees $\mathbf{q}_1 = (q_1^1, ..., q_1^m)$ and $\mathbf{q}_2 = (q_2^1, ..., q_2^m)$, respectively. The Calabi-Yau condition, $c_1(TX) = 0 $, translates into
\begin{eqnarray}
q_1^r + q_2^r = n_r +1, 
\end{eqnarray}
\noindent for all $r=1,...,m$. We can also view $p$ as a global, holomorphic section of the normal bundle of $X$,
\be 
{\cal N}= {\cal O}_{{\cal A}} ({\bf q}_{1}) \oplus  {\cal O}_{{\cal A}} ({\bf q}_{2})\,, 
\label{3.3}
\ee
\noindent which is now a rank-$2$ vector bundle in $\mathcal{A}$. Naturally, the wedge product $\Lambda^2 \mathcal{N} = {\cal O}_{{\cal A}} ({\bf q}_{1} + {\bf q}_{2})$ is a line bundle.

As in the previous chapter, we would like to understand the relation between closed line-bundle valued $(0, 1)$-forms on $X$ and certain forms on the ambient space $\mathcal{A}$. We start with a line bundle $K \rightarrow X$, its ambient space
counterpart $\mathcal{K} \rightarrow  \mathcal{A}$ such that $K = \mathcal{K}\vert_X$ and a closed $K$-valued $(0, 1)$-form $\nu \in H^1 (X, K)$, which represents
any of the three forms $\nu_i$ entering the integral \eqref{Yukgen2} for the holomorphic Yukawa couplings. The relation
between $K$ and $\mathcal{K}$ is still described by the Koszul sequence which, due to $X$ being defined at co-dimension
two, is no longer short-exact but given by the four-term sequence
\be 
0 \longrightarrow \Lambda^2 {\cal N}^* \otimes {\cal K}
\stackrel{q}{\longrightarrow} {\cal N}^* \otimes {\cal K} \stackrel{p}{\longrightarrow} {\cal K}  \stackrel{r}{\longrightarrow} K  \longrightarrow 0\;.
\label{3.10.1}
\ee
As before, the map $p=(p_1,p_2)$ acts by multiplication and $r$ is the restriction map. The map $q$ is fixed by exactness of the sequence, that is $p \circ q =0$, and by matching polynomial degrees. As a result, it is given, up to an overall, irrelevant constant, by
\be 
q =
\left( \begin{array}{cc}
-p_2 \\ \,\,\ p_1 
\end{array} \right)\,.
\label{3.12}
\ee
\noindent In practice, the four-term sequence~\eqref{3.10.1} is best dealt with by splitting it up into the two short exact sequences
\begin{equation}
\begin{array}{l}
0 \longrightarrow \Lambda^2 {\cal N}^* \otimes {\cal K}
\stackrel{q}{\longrightarrow} {\cal N}^* \otimes {\cal K} \stackrel{g_1}{\longrightarrow} {\cal C}\longrightarrow 0\,, 
\\[1mm] 
0 \longrightarrow  {\cal C} \stackrel{g_2}{\longrightarrow} {\cal K}  \stackrel{r}{\longrightarrow} K  \longrightarrow 0
\end{array}
\label{3.13}
\end{equation}
\noindent where $\mathcal{C}$ is a suitable co-kernel and $g_1$, $g_2$ are maps satisfying $g_2 \circ g_1 = p$. These quantities are determined by exactness of the above two sequences but will, fortunately, not be required explicitly. The relevant
parts of the two long exact sequences associated to the short exact sequences~\eqref{3.13} read
\bea
 \cdots&\longrightarrow &  H^1 ({\cal A},  {\cal C}) \stackrel{g_2}{\longrightarrow} H^1 ({\cal A}, {\cal K}) \stackrel{r}{\longrightarrow} H^1 (X, K)\nonumber \\
&\stackrel{\delta_1}{\longrightarrow} &H^2 ({\cal A},  {\cal C}) \stackrel{g_2}{\longrightarrow} H^2 ({\cal A}, {\cal K}) 
\longrightarrow \dots\; ,
\label{3.15}
\eea
and
\bea
 \cdots&\longrightarrow &  H^1 ({\cal A}, \Lambda^2{\cal N}^* \otimes {\cal K}) \stackrel{q}{\longrightarrow} H^1 ({\cal A}, {\cal N}^* \otimes{\cal K}) 
 \stackrel{g_1}{\longrightarrow} H^1 ({\cal A},  {\cal C})\nonumber \\
&\stackrel{\delta_2}{\longrightarrow} &H^2 ({\cal A}, \Lambda^2{\cal N}^* \otimes {\cal K}) \stackrel{q}{\longrightarrow} 
H^2 ({\cal A}, {\cal N}^* \otimes {\cal K}) \stackrel{g_1}{\longrightarrow} H^2 ({\cal A},  {\cal C}) \nonumber \\
&\stackrel{\delta_3}{\longrightarrow} &H^3 ({\cal A}, \Lambda^2{\cal N}^* \otimes {\cal K}) \stackrel{q}{\longrightarrow} 
H^3 ({\cal A}, {\cal N}^* \otimes {\cal K}) 
\longrightarrow \dots\;.
\label{3.16}
\eea
\noindent Our goal is to obtain an expression for $H^1(X, K)$ in terms of ambient space cohomologies and from~\eqref{3.15} we find that
\bea 
H^1 (X, K) &=&  r \Big( {\rm Coker} \Big( H^1 ({\cal A},  {\cal C}) \stackrel{g_2}{\rightarrow}
H^1 ({\cal A}, {\cal K})\Big) \Big) \notag \\ &\oplus& 
\d_1^{-1} \Big( {\rm Ker} \Big( H^2 ({\cal A}, {\cal C}) \stackrel{g_2}{\rightarrow}
H^2 ({\cal A}, {\cal K})\Big) \Big) \; .
\label{3.17}
\eea
This expression is analogous to Eq.~\eqref{H1eq} obtained in the co-dimension one case, but here we still have to work out $H^1 ({\cal A},  {\cal C})$ and $H^2 ({\cal A},  {\cal C})$, which are obtained from the sequence~\eqref{3.16}
\bea 
H^1 ({\cal A},  {\cal C})& = & g_1 \Big( {\rm Coker} \Big( H^1 ({\cal A}, \Lambda^2{\cal N}^* \otimes {\cal K}) \stackrel{q}{\rightarrow}
H^1 ({\cal A}, {\cal N}^* \otimes {\cal K})\Big) \Big) \nonumber \\
 & \oplus & 
\d_2^{-1} \Big( {\rm Ker} \Big( H^2 ({\cal A}, \Lambda^2 {\cal N}^* \otimes {\cal K}) \stackrel{q}{\rightarrow}
H^2 ({\cal A},  {\cal N}^* \otimes {\cal K})\Big) \Big) \; , 
\label{3.18.1}
\eea
\bea 
H^2 ({\cal A},  {\cal C})& = & g_1 \Big( {\rm Coker} \Big( H^2 ({\cal A}, \Lambda^2{\cal N}^* \otimes {\cal K}) \stackrel{q}{\rightarrow}
H^2 ({\cal A}, {\cal N}^* \otimes {\cal K})\Big) \Big) \nonumber \\
 & \oplus & 
\d_3^{-1} \Big( {\rm Ker} \Big( H^3 ({\cal A}, \Lambda^2 {\cal N}^* \otimes {\cal K}) \stackrel{q}{\rightarrow}
H^3 ({\cal A},  {\cal N}^* \otimes {\cal K})\Big) \Big) \;. 
\label{3.18.2}
\eea
Substituting Eqs. ~\eqref{3.18.1} and \eqref{3.18.2} into Eq.~\eqref{3.17} gives the desired formula for $H^1 (X, K)$ in terms of ambient space cohomology. Despite its apparent complexity, we will see that it is possible to get to a
simple generalisation of the structure derived in the co-dimension one case.

We begin by observing that $H^1(X, K)$ receives contributions from three ambient space cohomologies, namely from $H^1 ({\cal A}, {\cal K})$, 
$H^2 ({\cal A}, {\cal N}^* \otimes {\cal K})$ and $H^3 ({\cal A}, \Lambda^2{\cal N}^* \otimes {\cal K})$ (or, more accurately, from kernels or
quotients within these cohomologies). This means that a given closed $(0,1)$-form $\nu \in H^1(X,K)$ descends, in general, from three ambient space forms, a $(0,1)$-form $\hat{\nu}$, a $(0,2)$-form $\hat{\omega}$ and a $(0,3)$-form $\hat{\rho}$. However,
a specific $\nu \in H^1(X,K)$ might not receive all three contributions. We call a $\nu \in H^1(X,K)$ “type 1” if the associated $\hat{\omega}$ and $\hat{\rho}$ vanish and, hence, if it is determined by the $(0,1)$-form $\hat{\nu}$ only. Likewise, $\nu \in H^1(X,K)$ is called “type 2” if the associated $\hat{\rho}$ vanishes and it is determined by the $(0,2)$-form $\hat{\omega}$. If $\nu \in H^1(X,K)$ is determined by $\hat{\rho}$, it is called “type 3”. In general, a  $\nu \in H^1(X,K)$ is a linear combination of these three types, but the discussion is much simplified if we focus on each type separately. In fact, it is always possible to choose a basis of $ H^1(X,K)$, such that every basis element has a definite type. Let us now be
more precise and discuss each of these three types in turn.

\vspace{2mm}
\noindent {\bf Type 1}: We will refer to $\nu \in H^1(X,K)$ as “type 1” if it descends from $H^1 (\mathcal{A}, \mathcal{K})$, that is, if there is a $(0, 1)$-form $\hat{\nu} \in H^1(\mathcal{A},\mathcal{K})$ on the ambient space with
\begin{equation}
\begin{array}{ll}
 \nu =\hat{\nu}|_X\, , \qquad   \qquad \qquad  & \nu \in H^1(X,K) \, ,
\\[1mm]
\bar \pt \hat{\nu} =0\, , \qquad   \qquad \qquad  & \hat{\nu} \in H^1(\mathcal{A},\mathcal{K}) . 
\end{array}
\end{equation}

\vspace{2mm} 
\noindent {\bf Type 2}:  We will refer to $\nu \in H^1 (X, K)$ as “type 2” if it descends from a closed $(0, 2)$-form $\hat{\omega} \in H^2 (\mathcal{A}, \mathcal{N} ^* \otimes \mathcal{K})$. To understand the relation between $\nu$ and $\hat{\omega}$, we need to chase through Eqs.~\eqref{3.17} and \eqref{3.18.2}. Starting with Eq.~\eqref{3.17} and setting $\hat{\gamma} = \delta_1(\nu) \in H^2
(\mathcal{A}, \mathcal{C})$, we know from the definition of the
co-boundary map $\delta_1$ (see Appendix~\ref{coboundarymapappendix} for a review) that there is a $(0, 1)$-form $\hat{\nu} \in \Omega^1(\mathcal{A},\mathcal{K})$, such that
\begin{equation}
\overline{\partial}\hat{\nu} = g_2 \hat{\gamma}\, , \qquad   \qquad \qquad  \nu = \hat{\nu}\vert_X \, .
\end{equation}
\noindent Further, from Eq.~\eqref{3.18.2}, there is a $\hat{\omega} \in H^2 (\mathcal{A},\mathcal{N}^* \otimes \mathcal{K})$ with
\begin{equation}
\hat{\gamma} = g_1 \hat{\omega} \, .
\end{equation}
\noindent Combining these last two equations, together with $g_2 \circ g_1 = p$ then leads to
\begin{eqnarray}
\overline{\partial}\hat{\nu} = (g_2 \circ g_1) \hat{\omega} = p \hat{\omega} \, .
\end{eqnarray}
\noindent To summarise this discussion, we can write down the following chain of equations
\begin{equation}
\begin{array}{ll}
 \nu =\hat{\nu}|_X\, , \qquad   \qquad \qquad  & \nu \in H^1(X,K) \, ,
\\[1mm]
\bar \pt \hat{\nu} = p \hat{\omega}\, , \qquad   \qquad \qquad  & \hat{\nu} \in \Omega^1(\mathcal{A},\mathcal{K}) , \\[1mm]
\bar \pt \hat{\omega} = 0 \, , \qquad   \qquad \qquad  &  \hat{\omega} \in H^2 (\mathcal{A},\mathcal{N}^* \otimes \mathcal{K}) \, ,
\end{array}
\end{equation}
\noindent which describes the relation between $\nu$ and the $(0,2)$-form $\hat{\omega}$ from which it descends.

\vspace{2mm} 
\noindent {\bf Type 3}: We will refer to $\nu$ as “type 3” if it descends from a closed $(0, 3)$-form $\hat{\rho} \in H^3 (\mathcal{A}, \Lambda^2 \mathcal{N}^* \otimes \mathcal{K})$ and we need to understand the relation between $\nu$ and $\hat{\rho}$. As in the case of type 2, we start with Eq.~\eqref{3.17} and define $\hat{\gamma} = \delta_1(\nu)\in H^2(\mathcal{A},\mathcal{C})$ and a $(0,1)$-form $\hat{\nu}\in \Omega^1(\mathcal{A},\mathcal{K})$, such that 
\begin{equation}
\overline{\partial}\hat{\nu} = g_2 \hat{\gamma}\, , \qquad   \qquad \qquad  \nu = \hat{\nu}\vert_X \, .
\end{equation}
\noindent From surjectivity of $g_1$ in the first sequence \eqref{3.13}, we can write $\hat{\gamma} = g_1 \hat{\omega}$ for an $\hat{\omega} \in \Omega^2(\mathcal{A}, \mathcal{N}^* \otimes \mathcal{K})$ and
combining this with the previous equation leads to
\begin{eqnarray}
\overline{\partial} \hat{\nu} = p \hat{\omega} \, ,
\end{eqnarray}
\noindent as in the type 2 case. However, unlike for the type 2 case, $\hat{\omega}$ is no longer closed and we need to carry out one more step. To this end, we consider Eq.~\eqref{3.18.2} and define the closed $(0, 3)$-form $\hat{\rho} = \delta_3(\hat{\gamma}) \in H^3 (\mathcal{A}, \Lambda^2 \mathcal{N}^* \otimes \mathcal{K})$. Writing out the co-boundary map $\delta_3$ (see Appendix \ref{coboundarymapappendix}) now leads to
\be 
\bar \pt \hat{\o} = q \hat{\rho}\,, \qquad   \qquad \qquad  \bar \pt \hat{\rho} =0\, .
\label{3.37}
\ee
\noindent Altogether, this gives the following chain of equations
\begin{equation}
\begin{array}{ll}
\nu =\hat{\nu}|_X \,,     \qquad   \qquad \qquad  & \nu \in H^1(X, K)\,, \\[1mm]
{\bar \pt} \hat{\nu} = p  \hat{\omega} \,,    \qquad   \qquad \qquad   &  \hat{\nu} \in \Omega^1 ({\cal A}, {\cal K})\,, \\[1mm]
\bar \pt \hat{\o}   = q \hat{\rho}\,,    \qquad   \qquad \qquad &  \hat{\o} \in \Omega^2 ({\cal A}, {\cal N}^* \otimes {\cal K})\,, \\[1mm]
\bar \pt \hat{\rho} =0 \,,   \qquad   \qquad \qquad & \hat{\rho} \in H^3 ({\cal A}, \Lambda^{2} {\cal N}^* \otimes {\cal K})\, ,
\end{array}
\label{3.37.2}
\end{equation}
\noindent which describes the relation between $\nu$ and the $(0, 3)$-form $\hat{\rho}$ from which it descends.

In fact, the system of equations \eqref{3.37.2} describes the general relationship between $\nu$ and the three ambient space forms $\hat{\nu}$, $\hat{\omega}$ and $\hat{\rho}$. For a given $\nu$, solving the equations \eqref{3.37.2} gives the associated ambient space forms which, in general, are all non-zero. The three types discussed above arise from Eq.~\eqref{3.37.2} as special cases. If $\hat{\omega} = \hat{\rho} = 0$ for a given $\nu$, then $\hat{\nu}$ is closed and $\nu$ is of type 1. If $\hat{\omega} \neq 0$  but $\hat{\rho} = 0$ (and $\hat{\nu}$ does
not have a closed part which would correspond to a type 1 component) then $\hat{\omega}$ is closed and $\nu$ is of type
2. Finally, if $\hat{\rho} \neq 0$ (and $\hat{\nu}$, $\hat{\omega}$ do not have closed parts which would correspond to type 1 and type 2
components, respectively) then $\nu$ is of type 3.

Let us point out that, in general, the set of all forms $\hat{\nu}$, $\hat{\omega}$,   $\hat{\rho}$ is not always identified with the entire spaces in the second column of \eqref{3.37.2} but, rather, with kernels and co-kernels of the maps $p$ and $q$ within
those spaces. In each particular case, these kernels and co-kernels can be found from Eqs.~\eqref{3.17}, \eqref{3.18.1} and \eqref{3.18.2}. 

Our goal now is to express the Yukawa couplings \eqref{Yukgen2} in terms of the ambient space forms $\hat{\nu}$, $\hat{\omega}$, $\hat{\rho}$. If $\nu$ is of a specific type, the highest non-vanishing form which appears in the Eqs.~\eqref{3.37.2} represents an
ambient space cohomology and can be written down explicitly, following the rules explained in Appendix~\ref{appendixPn}.
The lower-degree forms then have to be obtained by solving the Eqs.~\eqref{3.37.2}. In this way, all relevant
ambient space forms can be calculated explicitly. 


\subsection{A derivation of   Yukawa couplings}
\label{derivation}


We will now derive the formula for the Yukawa couplings \eqref{Yukgen2} in terms of ambient space forms. For each $(0, 1)$-form $\nu_i \in H^1 (X, K_i)$ involved, we have an associated chain of ambient space forms $\hat{\nu}_i$, $\hat{\omega}_i$ and $\hat{\rho}_i$, in line with the Eqs.~\eqref{3.37.2}. The forms $\hat{\omega}_i$ take values in the rank-2 line bundle sum $\mathcal{N}^* \otimes \mathcal{K}_i =
\mathcal{O}_{\mathcal{A}}(-\mathbf{q}_1) \otimes \mathcal{K}_i \oplus \mathcal{O}_{\mathcal{A}}(-\mathbf{q}_2) \otimes \mathcal{K}_i$  and we denote the two corresponding components by $\hat{\omega}_i^a$, where $a=1,2$. Starting with Eq. \eqref{Yukgen2}, we insert two delta-function currents
\be 
\l (\nu_1, \nu_2, \nu_3)= \frac{1}{(2 \pi i)^2} \int_{{\cal A}} \hat{\Omega} \wedge 
\hat{\nu}_1 \wedge \hat{\nu}_2 \wedge \hat{\nu}_3 \wedge dp_1 \wedge {\bar \pt}
\Big( \frac{1}{p_1}\Big) \wedge dp_2 \wedge {\bar \pt}
\Big( \frac{1}{p_2}\Big) \, ,
\label{3.6}
\ee
which converts the integral to one over the ambient space. Using the standard formula (see~\cite{Candelas:1987se, Strominger:1985it, Witten:1985xc, Candelas:1987kf})
\be 
\hat{\Omega} \wedge  dp_1 \wedge dp_2 =\mu \, ,
\label{3.7}
\ee
\noindent where $\mu$ has been defined in Eq.~\eqref{10.2}, we obtain
\be 
\l (\nu_1, \nu_2, \nu_3)= \frac{1}{(2 \pi )^2} \int_{{\cal A}} \mu \wedge 
\hat{\nu}_1 \wedge \hat{\nu}_2 \wedge \hat{\nu}_3   \wedge {\bar \pt}
\Big( \frac{1}{p_1}\Big)  \wedge {\bar \pt}
\Big( \frac{1}{p_2}\Big) \,. 
\label{3.8}
\ee
Now we have to integrate by parts twice, ignoring the boundary integrals, which do not contribute (see Appendix~\ref{appendixboundary}). 
After the first integration, we obtain
\be 
\l(\nu_1,\nu_2,\nu_3) = \frac{1}{(2 \pi )^2}\int_{{\cal A}} \frac{\mu}{p_1} \wedge \Big[\bar \pt \hat{\nu}_1 \wedge \hat{\nu}_2 \wedge \hat{\nu}_3-
 \hat{\nu}_1 \wedge \bar \pt \hat{\nu}_2 \wedge \hat{\nu}_3 +\hat{\nu}_1 \wedge  \hat{\nu}_2 \wedge  \bar \pt \hat{\nu}_3
 \Big] \wedge {\bar \pt}
\Big( \frac{1}{p_2}\Big) \,. 
 \label{3.9}
 \ee
 The derivatives of $\hat{\nu}_i$ can be evaluated using~\eqref{3.37.2}. This leads to
 \be 
\bar \pt \hat{\nu}_1 \wedge \hat{\nu}_2 \wedge \hat{\nu}_3-
 \hat{\nu}_1 \wedge \bar \pt \hat{\nu}_2 \wedge \hat{\nu}_3 +\hat{\nu}_1 \wedge  \hat{\nu}_2 \wedge  \bar \pt \hat{\nu}_3 := p\hat{\b} =
 p_1 \hat{\b}^{1}  +p_2 \hat{\b}^{2} \,, 
 \label{3.100.1}
 \ee
 where $\hat{\b}$ is a vector with components given by
\begin{equation}
\begin{array}{l}
\hat{\b}^{1} =\hat{\o}_1^{1}  \wedge \hat{\nu}_2 \wedge \hat{\nu}_3-
\hat{\nu}_1 \wedge \hat{\o}_2^{1} \wedge \hat{\nu}_3 +\hat{\nu}_1 \wedge  \hat{\nu}_2 \wedge  \hat{\o}_3^{1}\,, 
\\[1mm]
\hat{\b}^{2} =\hat{\o}_1^{2}  \wedge \hat{\nu}_2 \wedge \hat{\nu}_3-
\hat{\nu}_1 \wedge \hat{\o}_2^{2} \wedge \hat{\nu}_3 +\hat{\nu}_1 \wedge  \hat{\nu}_2 \wedge  \hat{\o}_3^{2}\,.
\end{array}
\label{3.10.2}
\end{equation}
Substituting these expressions back into the integral \eqref{3.9}, we note that the term $ p_2 \hat{\b}^{2}$ does not contribute, since $p_2 \bar \pt  \Big( \frac{1}{p_2}\Big) \sim p_2 \d^2 (p_2) d {\bar p}_2=0$ and that we are, hence, left with
\be 
\l(\nu_1,\nu_2,\nu_3) = \frac{1}{(2 \pi )^2}\int_{{\cal A}} \mu \wedge \hat{\b}^{1} \wedge {\bar \pt}
\Big( \frac{1}{p_2}\Big) =-
\frac{1}{(2 \pi )^2}\int_{{\cal A}} \frac{\mu }{p_2} \wedge \bar \pt \hat{\b}^{1}\,. 
 \label{3.40}
 \ee
Using Eqs.~\eqref{3.37.2}, we obtain that $\bar \pt \hat{\b}^{1} =- p_2 \hat{\eta}$, where $\hat{\eta}$ is given by
\begin{equation}
\begin{array}{lll}
\hat{\eta} &=& \hat{\rho}_1 \wedge \hat{\nu}_2 \wedge \hat{\nu}_3 + \hat{\nu}_1 \wedge \hat{\rho}_2 \wedge \hat{\nu}_3+
\hat{\nu}_1 \wedge \hat{\nu}_2 \wedge \hat{\rho}_3
 \\[1mm]
&+& \hat{\nu}_1 \wedge \hat{\o}_2^{2} \wedge \hat{\o}_3^{1}-  \hat{\nu}_1 \wedge \hat{\o}_2^{1} \wedge \hat{\o}_3^{2}
\\[1mm]
&+&   \hat{\o}_1^{1} \wedge \hat{\nu}_2 \wedge  \hat{\o}_3^{2}  -   \hat{\o}_1^{2} \wedge \hat{\nu}_2 \wedge  \hat{\o}_3^{1}
\\[1mm]
&+&   \hat{\o}_1^{2} \wedge  \hat{\o}_2^{1}  \wedge \hat{\nu}_3-    \hat{\o}_1^{1} \wedge  \hat{\o}_2^{2}  \wedge \hat{\nu}_3\,.
\end{array}
\label{3.42}
\end{equation}
Hence, the final expression for the Yukawa coupling is 
\be 
\l(\nu_1,\nu_2,\nu_3) =
\frac{1}{(2 \pi )^2}\int_{{\cal A}} \mu  \wedge \hat{\eta} \, ,
 \label{3.43}
 \ee
with $\hat{\eta}$ given in~\eqref{3.42}. Eq.~\eqref{3.43} together with Eq.~\eqref{3.42} is our main general result for the co-dimension two case. As we will see in Section~\ref{chapter3examples}, this result, together with the expressions for ambient space harmonic forms in Appendix~\ref{appendixPn} and Eq.~\eqref{3.37.2}, allows for an explicit calculation of the holomorphic Yukawa couplings.

It is worth discussing a number of special cases. If all three forms $\nu_i$ are of type 1, then $\hat{\o}_i = \hat{\rho}_i = 0$, for $i=1,2,3$ and as a result $\hat{\eta}$ in Eq.~\eqref{3.42} is zero and, hence, the Yukawa coupling vanishes. Now suppose two of the forms $\nu_i$, say $\nu_1$ and $\nu_2$, are of type 1, while $\nu_3$ is of type 2. In this case we have $\hat{\o}_i = \hat{\rho}_i =0$ for $i=1,2$ and $\hat{\rho}_3=0$, so that $\hat{\eta}=0$ in Eq.~\eqref{3.42} and the Yukawa coupling still vanishes. These observations can be summarised by the following

\vspace{2mm}

\noindent {\bf Theorem}: Assume that the forms $\nu_i$ which enter the integral \eqref{Yukgen2} for the Yukawa couplings are of type $\tau_i$, where $i=1,2,3$. Then
\begin{equation}
\label{theoremdim5}
\tau_1+\tau_2+\tau_3<{\rm dim} ({\cal A}) =5 \qquad \Longrightarrow \qquad \lambda(\nu_1,\nu_2,\nu_3) = 0 \, .
\end{equation}
\noindent For co-dimension one we have observed that the Yukawa coupling vanishes if all three forms $\nu_i$ are of type 1. The above vanishing theorem generalises this statement to the case of co-dimension two.

There are two special cases for which the expression \eqref{3.43} simplifies considerably. Firstly, assume that the types of the $(0, 1)$-forms $\nu_i$ are given by $(\tau_1,\tau_2,\tau_3) = (1, 1, 3)$. Then we have from Eqs.~\eqref{3.43} and \eqref{3.42}
\be 
\l(\nu_1,\nu_2,\nu_3) =
\frac{1}{(2 \pi )^2}\int_{{\cal A}} \mu  \wedge \hat{\nu}_1 \wedge \hat{\nu}_2  \wedge \hat{\rho}_3 \, ,
 \label{yuk113}
 \ee
\noindent and all three bundle-valued forms in the integrand represent ambient space cohomologies. The other simple case arises for types $(\tau_1,\tau_2,\tau_3) = (1, 2, 2)$, where Eq.~\eqref{3.43} becomes
\be 
\l(\nu_1,\nu_2,\nu_3) =
\frac{1}{(2 \pi )^2}\int_{{\cal A}} \mu  \wedge \hat{\nu}_1 \wedge \hat{\omega}_2 \wedge \hat{\omega}_3 \, ,
 \label{yuk122}
 \ee
\noindent with an anti-symmetric contraction of the bundle indices for $ \hat{\omega}_i$ understood. Again, all three forms in the
integrand represent ambient space cohomologies.

We will now proceed to arbitrary co-dimension and show that analogous statements can be obtained in the general case.

\section{Generalisations to higher co-dimensions}
\label{higher}


\subsection{Lifting forms to the ambient space}


We will now tackle the case of arbitrary co-dimension starting, as before, with the problem of writing
closed line bundle-valued $(0, 1)$-forms on the Calabi-Yau manifold in terms of ambient space forms. Our
ambient space remains the product of projective spaces
\be 
{\cal A}= {\mathbb P}^{n_1} \times {\mathbb P}^{n_2} \times ...  \times {\mathbb P}^{n_m}\, ,
\label{4.1}
\ee
where now $n_1+...+n_m = 3+k$, and $k$ is the co-dimension. The CICY manifold $X \subset \mathcal{A}$ is defined as the common zero locus of $k$ homogeneous polynomials $p_a$ with multi-degrees $\mathbf{q}_a = (q_a^1, ..., q_a^m)$, where $a=1,...,k$. The Calabi-Yau condition, $c_1(TX)=0$ now reads
\begin{eqnarray}
\sum_{a=1}^k q_a^r = n_r+1 \, ,
\end{eqnarray}
\noindent for all $r=1,...,m$. As before, we combine these polynomials into the row vector $p=(p_1,...,p_k)$, which can be viewed as a section of the line bundle sum
\begin{equation}
{\cal N} = {\cal O}_{{\cal A}} ({\bf q}_{1}) \oplus ... \oplus {\cal O}_{{\cal A}} ({\bf q}_{k}) \, .
\end{equation}
\noindent The relation between a line bundle $K \rightarrow X$ and its ambient space counterpart ${\cal K} \rightarrow {\cal A}$ (such that $K = {\cal K}\vert_X$ ) is again governed by the Koszul sequence
\be 
0 \longrightarrow \Lambda^k {\cal N}^* \otimes {\cal K}  \stackrel{q_k}{\longrightarrow} 
 \Lambda^{k-1} {\cal N}^* \otimes {\cal K}  \stackrel{q_{k-1}}{\longrightarrow}   \dots 
 \stackrel{q_{2}}{\longrightarrow}
{\cal N}^* \otimes {\cal K} \stackrel{q_1=p}{\longrightarrow} {\cal K}  \stackrel{q_0 = r}{\longrightarrow} K  \longrightarrow 0\;, 
\label{4.7}
\ee
\noindent which now consists of $k + 2$ terms and contains maps $q_a$ satisfying $q_a \circ q_{a+1} = 0$ for all $a = 0, ... , k-1$. As previously, $q_0 = r$ is the restriction map, $q_1 = p$ is the map acting by multiplication with the polynomial
vector $p$ and the higher maps $q_a$ for $a > 1$ are the obvious tensor maps induced by $p$.  An $(a + 1)$–form $\hat{\nu}$ taking values in $\Lambda^{a} {\cal N}^* \otimes {\cal K}$ has components $\hat{\nu}^{b_1,...,b_a}$ with completely anti-symmetrised upper indices, and the action of $q_a$ on this form can be explicitly written as
\begin{eqnarray}
(q_a \hat{\nu})^{b_1,...,b_{a-1}} = p_b \hat{\nu}^{b_1...b_{a-1}b} \, .
\label{eq4.5}
\end{eqnarray}
\noindent Splitting \eqref{4.7} up into $k$ short exact sequences and chasing through the associated long exact sequences shows that $H^1(X, K)$ can now receive contributions from the $k+1$ ambient space cohomologies $H^1(\mathcal{A}, {\cal K})$, $H^2(\mathcal{A}, {\cal N}^* \otimes {\cal K} )$, ..., $H^{k}(\mathcal{A}, \Lambda^{k-1} {\cal N}^* \otimes {\cal K})$, $H^{k+1}(\mathcal{A}, \Lambda^{k} {\cal N}^* \otimes {\cal K})$.  A closed $K$-valued $(0, 1)$-form $\nu \in H^1(X,K)$ is, therefore, related to a chain of $k+1$ ambient space $(0,a)$-forms $\hat{\nu}_a$, where $a=1,...,k+1$. The precise relationship between $\nu$ and $\hat{\nu}_a$ can be derived by a straightforward generalisation of the co-dimension two case discussed in the previous section. The result is
\begin{equation}
\begin{array}{llllll}
\nu & \!\!\! = & \! \hat{\nu}_1|_X \,,      \qquad \qquad \ \ & \nu & \!\!\! \in & \!\! H^1(X, K)\,, \\[1mm]
{\bar \pt} \hat{\nu}_1 &\!\!\! = & \! q_1  \hat{\nu}_2 \,,    \qquad   \qquad   \ \ \  & \hat{\nu}_1 & \!\!\! \in & \! \Omega^1 ({\cal A}, {\cal K})\,,  \\[1mm]
\bar \pt \hat{\nu}_2   & \!\!\! = & \! q_2 \hat{\nu}_3\,,    \qquad \qquad  \ \ \ &\hat{\nu}_2 & \!\!\! \in & \! \Omega^2 ({\cal A}, {\cal N}^* \otimes {\cal K})\,, \\[1mm]
& \!\!\! \, \vdots & \! \qquad \qquad & & \!\!\! \, \vdots & \! \\[1mm]
\bar \pt \hat{\nu}_{k} &  \!\!\! = & \! q_{k} \hat{\nu}_{k+1}\,,    \qquad   \qquad   & \hat{\nu}_{k} &
\!\!\! \in & \! \Omega^k ({\cal A}, \Lambda^{k-1}{\cal N}^* \otimes {\cal K})\,, \\[1mm]
\bar \pt \hat{\nu}_{k+1} & \!\!\! = & \! 0 \,,   \qquad   \qquad \ \ \  & \hat{\nu}_{k+1} &\!\!\! \in & \!\! H^{k+1} ({\cal A}, \Lambda^k{\cal N}^* \otimes {\cal K})\,.
\end{array}
\label{4.10}
\end{equation}
\noindent Note that, just like in the co-dimension two case, the forms $\hat{\nu}_a$ should be thought of as elements of certain kernels and co-kernels of the maps $q_a$ within the spaces on the right-hand side of Eq.~\eqref{4.10}. For a given $\nu \in H^1(X,K)$, the associated chain of ambient space forms is obtained by solving the above equations and, in general, this leads to $k + 1$ non-trivial forms $\hat{\nu}_a$. However, as before, it is useful to introduce the type $\tau$ of $\nu$, which can now take the values $\tau \in \lbrace 1, ... , k+1\rbrace$. We say that $\nu$ is of type $\tau$ if $\hat{\nu}_{\tau} \neq 0$, $\hat{\nu}_a = 0$ for all $a > \tau$ and all $\hat{\nu}_a$ for $a < \tau$ do not contain any $\overline{\partial}$-closed parts. In this case, $\nu$ descends, via the Eqs.~\eqref{4.10}, from the $\overline{\partial}$-closed $(0, \tau)$-form $\hat{\nu}_{\tau}$, which defines an element of $H^{\tau}(\mathcal{A}, \Lambda^{\tau-1} \mathcal{N}^* \otimes \mathcal{K})$.


\subsection{The structure of Yukawa couplings and a vanishing theorem}


Each of the three forms $\nu_i \in H^1(X,K_i)$ involved in the Yukawa coupling has, from Eq.~\eqref{4.10} an associated chain of ambient space forms which we denote by $\hat{\nu}_{i,a}$, where $a=1,...,k+1$. To derive the general expression for the Yukawa couplings we start with \eqref{Yukgen2}, insert $k$ delta-function currents and use the standard formula (see~\cite{Candelas:1987se, Strominger:1985it, Witten:1985xc, Candelas:1987kf})
\begin{eqnarray}
\hat{\Omega} \wedge d p_1 \wedge ... \wedge d p_k = \mu \, ,
\end{eqnarray}
\noindent where $\mu$ has been defined in Eq.~\eqref{10.2}. This leads to
\begin{align}
\lambda (\nu_1, \nu_2, \nu_3) &= 
\Big(- \frac{1}{2 \pi i}\Big)^k \int_{{\cal A}} \hat{\Omega} \wedge 
\hat{\nu}_{1,1} \wedge \hat{\nu}_{2,1} \wedge \hat{\nu}_{3,1} \wedge dp_1 \wedge {\bar \pt}
\Big( \frac{1}{p_1}\Big) \wedge  ... \wedge d p_k \wedge {\bar \pt} \Big( \frac{1}{p_k} \Big) 
\nonumber \\
& =
\frac{\tilde{C}_k}{(2 \pi)^k k!} \epsilon_{b_1 ... b_k} \int_{{\cal A}} \mu \wedge 
\hat{\nu}_{1,1} \wedge \hat{\nu}_{2,1} \wedge \hat{\nu}_{3,1}  \wedge {\bar \pt}
\Big( \frac{1}{p_{b_1}}\Big)  \wedge ... \wedge {\bar \pt} \Big( \frac{1}{p_{b_k}} \Big) \, ,
\label{exhibit1}
\end{align}
\noindent where $\tilde{C}_k=(-1)^{k(k+1)/2} \, i^k$. 
is a phase factor. Integrating the first $\overline{\partial}$ operator by parts (ignoring the boundary terms, whose vanishing can be shown in the same way as in Appendix~\ref{appendixboundary}) and using Eqs.~\eqref{eq4.5}, \eqref{4.10}, this turns into
\begin{align}
\lambda (\nu_1, \nu_2, \nu_3) = \frac{\tilde{C}_k}{(2 \pi)^k k!}  \epsilon_{b_1 ... b_k} & \int_{{\cal A}} \mu \wedge  \big(\hat{\nu}^{b_1}_{1,2} \wedge \hat{\nu}_{2,1} \wedge \hat{\nu}_{3,1} - \hat{\nu}_{1,1} \wedge \hat{\nu}^{b_1}_{2,2} \wedge \hat{\nu}_{3,1} + \notag \\ & \;\, + \hat{\nu}_{1,1} \wedge \hat{\nu}_{2,1} \wedge \hat{\nu}^{b_1}_{3,2}   \big) \wedge {\bar \pt}
\Big( \frac{1}{p_{b_2}}\Big)  \wedge ... \wedge {\bar \pt} \Big( \frac{1}{p_{b_k}} \Big) \, .
\label{exhibit2}
\end{align}
Here, the relation
\begin{eqnarray}
p_b \overline{\partial}\Big( \frac{1}{p_{b}} \Big) = 0
\end{eqnarray}
\noindent has led to the insertion of $\delta_b^{b_1}$ from Eq.~\eqref{eq4.5}, so that we remain with a sum over $b_1$, as indicated above (while the resulting factor $p_{b_1}$ from Eq.~\eqref{eq4.5} cancels against $1/p_{b_1}$). We can now continue integrating by parts until all factors of the form $\overline{\partial}(1/p_{b})$ are used up.  Each of these factors leads to a partial differentiation of all forms $\hat{\nu}_{i,a}^{b_1 ... b_{a-1}}$ which appear in the integral, effectively replacing them by the forms $\hat{\nu}_{i,a+1}^{b_1...b_{a-1}b}$, which
are found one step lower down in the chain \eqref{4.10}. Since there are $k$ such partial integrations to be performed, starting with three $(0, 1)$-forms, the end result is a sum which contains all products of three forms whose degree sums up to $\textrm{dim}(\mathcal{A}) = 3 + k$. This leads to
\begin{eqnarray}
\lambda (\nu_1, \nu_2, \nu_3) = \dfrac{C_k}{(2\pi)^k} \sum^{k+1}_{\substack{a_1,a_2,a_3=1 \\ a_1+a_2+a_3 = \textrm{dim}(\mathcal{A})}} (-1)^{s(a_1,a_2,a_3)} \int_{\mathcal{A}} \mu \wedge \hat{\nu}_{1,a_1} \wedge \hat{\nu}_{2,a_2} \wedge \hat{\nu}_{3,a_3} \, ,
\label{eq4.11}
\end{eqnarray}
\noindent where $s\,(a_1, a_2, a_3) = (a_1 + 1)a_2 + a_1 a_3 + a_2 a_3$ determines the relative signs of the terms and $C_k =(-1)^{k(k+1)/2}(-1)^{[(k+1)/2]} i^k$ is another phase. In this formula, the bundle indices have been suppressed so
the wedge product should be understood as including an appropriate tensoring of the bundle directions
to form a singlet, via anti-symmetrisation by $\epsilon_{b_1 ... b_k}$. The anti-symmetrisation is achieved by summing
in every case as many terms with permuted indices as required for complete anti-symmetry, each with a
factor $1$ or $-1$ and no additional overall normalisation. This means that, for example, $\hat{\nu}_{1,2} \wedge \hat{\nu}_{2,2} \wedge \hat{\nu}_{3,1}= \epsilon_{b_1 b_2} \hat{\nu}^{b_1}_{1,2} \wedge \hat{\nu}^{b_2}_{2,2} \wedge \hat{\nu}_{3,1}$, while $\hat{\nu}_{1,3} \wedge \hat{\nu}_{2,1} \wedge \hat{\nu}_{3,1} = \tfrac{1}{2} \epsilon_{b_1 b_2} \hat{\nu}^{b_1,b_2}_{1,3} \wedge \hat{\nu}_{2,1} \wedge \hat{\nu}_{3,1}$.

Eq.~\eqref{eq4.11} is our main general result for the holomorphic Yukawa couplings. All the ambient space
forms $\hat{\nu}_{i,a}$ can be constructed explicitly, starting with Appendix \ref{appendixPn} in order to write down (harmonic) representatives for ambient space cohomology for the highest degree non-trivial forms in the chain~\eqref{4.10} and then solving these equations to find all associated lower-degree forms. With these forms inserted, the integral \eqref{eq4.11} can be carried out explicitly, as we will demonstrate for the examples in Section \ref{chapter3examples}.

As before, it is useful to discuss some special cases. First assume, that the $(0, 1)$-forms $\nu_i$ are of type $\tau_i$, so that $\hat{\nu}_{i,a} = 0$ for all $a>\tau_i$. If the $\tau_i$ sum up to less than the ambient space dimension $\textrm{dim}(\mathcal{A})$, then all terms in Eq.~\eqref{eq4.11} vanish due to the summation constraint. As a result, the Yukawa coupling vanishes.
Let us formulate this concisely:

\vspace{2mm}

\noindent {\bf Theorem}:  Assume that the forms $\nu_i$ which enter the integral \eqref{Yukgen2} for the Yukawa couplings are of type $\tau_i$, where $i = 1, 2, 3$. Then
\be 
\tau_1+ \tau_2 +\tau_3 < {\rm dim} ({\cal A}) \qquad \Longrightarrow \qquad \l (\nu_1,\nu_2,\nu_3) =0 \, .
\label{4.12}
\ee

\vspace{2mm}

\noindent This is the general version of the vanishing theorem we have already seen for co-dimensions one and two
in previous sections. As we have discussed, the type $\tau$ of a form $\nu \in H^1(X,K)$ is determined by the cohomology $H^{\tau}(\mathcal{A}, \Lambda^{\tau-1} \mathcal{N}^* \otimes \mathcal{K})$, from which it descends via successive co-boundary maps. As a rule of thumb, large $\tau$'s are relatively rare since they require many non-trivial co-boundary maps and cohomologies. Consequently, for a large ambient space dimension $\textrm{dim}(\mathcal{A})$, the condition in \eqref{4.12} is frequently satisfied and many Yukawa couplings vanish. We stress again that vanishing due to \eqref{4.12} appears to be topological in nature, that is, these couplings vanish despite being allowed by the obvious symmetries of the four-dimensional effective theory.

Another special case of interest is for types $\tau_i$ satisfying $\tau_1 + \tau_2 + \tau_3 = \textrm{dim}(\mathcal{A})$. In this case, only one
term in \eqref{eq4.11} contributes and the integral simplifies to
\begin{equation}
\l (\nu_1,\nu_2,\nu_3) \sim \dfrac{1}{(2 \pi)^k} \int_{\mathcal{A}} \mu \wedge \hat{\nu}_{1,\tau_1} \wedge \hat{\nu}_{2,\tau_2} \wedge \hat{\nu}_{3,\tau_3} \, ,
\end{equation}
\noindent where we have dropped an overall phase factor. Note that, unlike in the general case \eqref{eq4.11}, all three forms $\hat{\nu}_{i,\tau_i}$ in the integrand are closed and represent ambient space cohomologies in $H^{\tau_i}(\mathcal{A}, \Lambda^{\tau_i-1} \mathcal{N}^* \otimes \mathcal{K})$. They can, therefore, be directly constructed from the rules given in Appendix~\ref{appendixPn}, without any need to solve Eqs.~\eqref{4.10}.


\section{Examples}
\label{chapter3examples}

In this section, we will illustrate our general statements for models on a certain co-dimension two CICY
and show that the relevant ambient space integrals can, in fact, be carried out explicitly. We begin
by introducing the specific CICY and its properties, then move on to describing line bundles and line
bundle-valued forms before we derive two more specific formulae for the Yukawa couplings for types
$(\tau_1, \tau_2, \tau_3) = (1, 1, 3)$ and $(\tau_1, \tau_2, \tau_3) = (1, 2, 2)$, respectively. These results are then applied to three examples, each defined by a certain line bundle sum on the relevant CICY.

\subsection{A co-dimension two CICY and its properties}

Our chosen CICY is a co-dimensional two manifold in the ambient space $\mathcal{A} = \mathbb{P}^{1} \times \mathbb{P}^{1} \times \mathbb{P}^{1} \times \mathbb{P}^{1} \times \mathbb{P}^{1}$, whose
homogeneous coordinates we either denote by $\mathbf{x}=(x_i^{\alpha})$, where $i=1,...,5$ and $\alpha=0,1$ or, more explicitly, by $\mathbf{x}=((x_0, x_1),(y_0, y_1),(u_0, u_1),(v_0, v_1),(w_0, w_1))$. We also introduce affine coordinates $z_i = x_i^1/x_i^0$ on the coordinate patch of $\mathcal{A}$ where all $x_i^0 \neq 0$. The CICY is defined as the common zero locus in $\mathcal{A}$ of
two homogeneous polynomials  $p = (p_1,p_2)$ with multi-degrees $\mathbf{q}_1 = (0,1,1,1,1)$ and $\mathbf{q}_2=(2,1,1,1,1)$, respectively.  This information is often summarised by the configuration matrix
\begin{eqnarray}
X=\begin{pmatrix}
\mathbb{P}^{1}& \vline & 0 & 2  \\
\mathbb{P}^{1}&\vline & 1 & 1  \\
\mathbb{P}^{1} &\vline & 1 & 1  \\
\mathbb{P}^{1} &\vline & 1 & 1 \\
\mathbb{P}^{1}& \vline & 1 & 1
\end{pmatrix}_{- 80}^{5, 45}
\label{cicy7487}
\end{eqnarray}
\noindent whose columns are given by $\mathbf{q}_1$ and $\mathbf{q}_2$. Attached as a superscript are the Hodge numbers $h^{1,1}(X)$, $h^{2,1}(X)$ and as a subscript the Euler number, $\eta(X)$.  In the standard list of Refs. \cite{Candelas:1987kf, candelascicy2}, this manifold carries the number 7487.\footnote{The reason why CICY 7487 was chosen is the large number of line bundle GUT models that can be built on this manifold. Twenty-four of those models, which are listed in Ref.~\cite{Anderson:2011ns}, were investigated during the development of this chapter. They all have a $SU(5) \times S(U(1)^5)$ gauge group and produce the Standard Model with three families upon dividing by the freely acting symmetry $\Gamma$.} The defining polynomials $p = (p_1, p_2)$ can also be viewed as a section of the line bundle sum
\begin{eqnarray}
\mathcal{N} = \mathcal{O}_{\mathcal{A}}(\mathbf{q}_1) \oplus \mathcal{O}_{\mathcal{A}}(\mathbf{q}_2) \, .
\end{eqnarray}
\noindent For later reference, we also define $\mathbf{q}=\mathbf{q}_1+\mathbf{q}_2 = (2,2,2,2,2)$ and note that
\begin{eqnarray}
\Lambda^2 \mathcal{N} = \mathcal{O}_{\mathcal{A}}(\mathbf{q}) \, .
\end{eqnarray}
\noindent In order to reduce the size of the problem, it will frequently be useful to work on a discrete quotient of
the above manifold. In fact, $X$ has a freely-acting symmetry $\Gamma = \mathbb{Z}_2 \times \mathbb{Z}_2$ whose generators act on the homogeneous coordinates as
\begin{eqnarray}
\label{Gammasym}
\gamma(g_1) = \mathbbmss{1}_5 \times \begin{pmatrix} 1 & 0 \\ 0 & -1\end{pmatrix} \, , \qquad \gamma(g_2) = \mathbbmss{1}_5 \times \begin{pmatrix} 0 & 1 \\ 1 & 0\end{pmatrix}  \, ,
\end{eqnarray}
\noindent while the action on the defining polynomials is
\begin{eqnarray}
\label{Gammarho}
\rho(g_1) = \textrm{diag}(1,-1) \, , \qquad \rho(g_2) = \textrm{diag}(1,-1) \, .
\end{eqnarray}
\noindent The quotient $\tilde{X}= X/\Gamma$ is a Calabi-Yau manifold with Euler number $\eta (\tilde{X}) = \eta(X)/\vert\Gamma\vert = -20$ and Hodge numbers $h^{1,1}(\tilde{X})=5$, $h^{2,1}(\tilde{X})=15$.

\subsection{Line bundles and line bundle-valued harmonic forms}

The CICY defined by \eqref{cicy7487} is favourable, by which we mean that the entire second cohomology of $X$
descends from the ambient space. This implies that every line bundle $L \rightarrow X$ can be obtained as a
restriction $L = \mathcal{L}\vert_X$ of an ambient space line bundle $\mathcal{L} = \mathcal{O}_{\mathcal{A}}(\mathbf{l})$, where $\mathbf{l} = (l^1,...,l^5)$. In order to compute Yukawa integrals, we need to understand the cohomology of such ambient space line bundles and
write down explicit differential forms representing these cohomologies. Since we are dealing with products of $\mathbb{P}^1$ factors, results from Chapter \ref{tetraquadricchapter} can be imported, however a generalisation to arbitrary $\mathbb{P}^n$ factors is found in Appendix \ref{appendixPn}.

As before, the cohomology dimensions for a line bundle $\mathcal{L} = \mathcal{O}_{\mathcal{A}}(\mathbf{l})$ is obtained by combining Bott’s formula for line bundle cohomology on $\mathbb{P}^1$ and the K\"unneth formula. Firstly, all cohomologies of $\mathcal{L}$ vanish if at least one of the integers $l^i$ equals $-1$.  If all $l^i \neq -1$, then there is precisely one non-vanishing cohomology $H^q(\mathcal{A},\mathcal{L})$, and $q$ equals the number of integers $l^i$ with $l^i \leq -2$. The
dimension of this one non-vanishing cohomology is given by
\begin{equation}
\label{eq5.6}
h^q(\mathcal{A},\mathcal{L}) = \prod_{i:l^i \geq 0} (l^i+1)\prod_{i:l^i \leq -2} (-l^i-1) \, .
\end{equation} 
\noindent The $\mathcal{L}$-valued $(0,q)$-forms representing $H^q(\mathcal{A},\mathcal{L})$ can be written down as
\begin{eqnarray}
\label{5.5}
\alpha_{(\mathbf{l})}=P_{(\mathbf{l})}\prod_{i: l^{i} \leq -2} \kappa_i^{l^i} d \overline{z}_i \, ,
\end{eqnarray}
\noindent where $\kappa_i = 1+\vert z_i \vert^2$ and $P_{(\mathbf{l})}$ is a polynomial of degree $l^i$ in $z_i$, if $l^i \geq 0 $, and of degree $-l^i -2$ in $\overline{z}_i$, if $l^i \leq - 2$. In fact, the above forms are harmonic (relative to the Fubini-Study metric) and are, hence, in
one-to-one correspondence with the elements of $H^q(\mathcal{A},\mathcal{L})$. In particular, note that the number of arbitrary coefficients in the polynomial $P_{(\mathbf{l})}$ equals the dimension \eqref{eq5.6} of the cohomology group.

The above differential forms have been written down in affine coordinates $z_i$. A useful equivalent
version in terms of homogeneous coordinates is given by
\begin{eqnarray}
\alpha_{(\mathbf{l})}=\tilde{P}_{(\mathbf{l})}\prod_{i: l^{i} \leq -2} \sigma_i^{l^i} d \overline{\mu}_i \, ,
\end{eqnarray}
\noindent where $\tilde{P}_{(\mathbf{l})}$ is the homogeneous counterpart of $P_{(\mathbf{l})}$ and
\begin{eqnarray}
\sigma_i = \vert x_i^0\vert^2 + \vert x_i^1\vert^2, \qquad \mu_i = \epsilon_{\alpha \beta} x_i^{\alpha}x_i^{\beta} \, .
\end{eqnarray} 

The Yukawa couplings involve wedge products of differential forms and we should, therefore, understand what happens if we form wedge products of the above forms. To be specific, let us consider a form $\alpha_{(\mathbf{l})}$ with associated polynomial $\tilde{P}_{(\mathbf{l})}$, representing the cohomology $H^p(\mathcal{A}, \mathcal{O}_{\mathcal{A}}(\mathbf{l}))$ and a form $\beta_{(\mathbf{m})}$ with associated polynomial $\tilde{Q}
_{(\mathbf{m})}$, representing the cohomology $H^q(\mathcal{A}, \mathcal{O}_{\mathcal{A}}(\mathbf{m}))$. It is clear that $\alpha_{(\mathbf{l})}\wedge \beta_{(\mathbf{m})}$ is $\overline{\partial}$–closed and represents an element of $H^{p+q}(\mathcal{A}, \mathcal{O}_{\mathcal{A}}(\mathbf{l} + \mathbf{m}))$, however, this will, in general not be the harmonic representative. We can ask how this harmonic representative, which we denote by $\gamma_{(\mathbf{l}+\mathbf{m})}$ with associated polynomial $\tilde{R}_{(\mathbf{l}+\mathbf{m})}$, can be obtained from $\alpha_{(\mathbf{l})}$ and $\beta_{(\mathbf{m})}$. Fortunately, there is a simple answer which can be expressed in terms of the associated polynomials $\tilde{P}_{(\mathbf{l})}$, $\tilde{Q}_{(\mathbf{m})}$ and $\tilde{R}_{(\mathbf{l}+\mathbf{m})}$. For a product of $\mathbb{P}^1$ spaces this has been derived in Chapter~\ref{tetraquadricchapter}. In Appendix~\ref{appendixPn}, we explain how harmonic forms on a single $\mathbb{P}^n$ are multiplied. These results can be easily applied to a product of projective spaces with arbitrary
dimensions and lead to
\begin{equation}
\label{RPQ}
\tilde{R}_{(\mathbf{l}+\mathbf{m})} = c_{\mathbf{l},\mathbf{m}} \tilde{P}_{(\mathbf{l})} \tilde{Q}_{(\mathbf{m})} \, ,
\end{equation}
\noindent where $c_{\mathbf{l},\mathbf{m}}$ is a numerical coefficient explicitly given by
\begin{equation}
\label{coefc}
c_{\mathbf{l},\mathbf{m}} = \prod_{i: l^i\leq-2} c_{l^i,m^i}\prod_{j: m^j\leq-2} c_{m^j,l^j} \, , \quad c_{l,m} = \dfrac{(-l-m-1)!}{(-l-1)!} \, .
\end{equation}
\noindent The polynomial multiplication on the RHS of Eq.~\eqref{RPQ} is understood with a replacement of coordinates by associated partial derivatives whenever positive degrees meet negative degrees. More specifically, whenever coordinates $x_i^{\alpha}$ in $\tilde{P}_{(\mathbf{l})}$ act on coordinates $\overline{x}{}_i^{\alpha}$ in $\tilde{Q}_{(\mathbf{m})}$, the former should be replaced by $\partial/\partial \overline{x}{}_i^{\alpha}$.

In the following, we would like to further evaluate the Yukawa couplings for our example manifold
and certain specific types. We will work within our familiar setting, that is, we have three line bundles
$K_i = \mathcal{O}_X (\mathbf{k}_i)$ on $X$ underlying the expression for the Yukawa couplings. These line bundles descend from their ambient space counterparts $\mathcal{K}_i = \mathcal{O}_{\mathcal{A}}(\mathbf{k}_i)$  and have to satisfy the condition
\begin{equation}
\label{conditionk1k2k3}
K_1 \otimes K_2 \otimes K_3 = \mathcal{O}_X \qquad \Longrightarrow \qquad  \mathbf{k}_1+\mathbf{k}_2+\mathbf{k}_3 = 0 \, .
\end{equation}
\noindent We would like to calculate the Yukawa couplings for three $K_i$–valued $(0, 1)$–forms $\nu_i \in H^1(X,K_i)$. From the Eqs.~\eqref{3.37.2} each of these comes with a chain of ambient space forms, namely the $(0,1)$-forms $\hat{\nu}_i$, the $(0, 2)$–forms $\hat{\omega}_i$ and the $(0, 3)$–forms $\hat{\rho}_i$ which enter the general formula \eqref{3.43} for the Yukawa couplings. In the following, we focus on certain cases where the $\nu_i$ have specific types $\tau_i$.

\subsection{Yukawa couplings of type $(1, 1, 3)$}

We now assume that two of the forms $\nu_i$, say $\nu_1$ and $\nu_2$ for definiteness, are of type 1, while $\nu_3$ is of
type 3. Note that this saturates the bound in Eq.~\eqref{theoremdim5} and constitutes one of the two simplest cases for co-dimension two to which the vanishing theorem does not apply (the other one being discussed in
the next sub-section). In this case, the Yukawa couplings are given by Eq.~\eqref{yuk113}, which only involves the
ambient space forms $\hat{\nu}_1 \in H^1(\mathcal{A},\mathcal{K}_1)$, $\hat{\nu}_2 \in H^1(\mathcal{A},\mathcal{K}_2)$ and $\hat{\rho}_3 \in H^3(\mathcal{A},\Lambda^2 \mathcal{N}^*\otimes\mathcal{K}_3)$ .

Following the rules for cohomology explained in the last sub-section, in order for $H^1(\mathcal{A},\mathcal{K}_1)$ and $H^1(\mathcal{A},\mathcal{K}_2)$ to be non-trivial, we require that $\mathbf{k}_1$ and $\mathbf{k}_2$ each have precisely one entry less than or equal to $-2$ and all other entries positive. Further, for $H^3(\mathcal{A},\Lambda^2 \mathcal{N}^*\otimes\mathcal{K}_3) = H^3(\mathcal{A}, \mathcal{O}_{\mathcal{A}}(\mathbf{k}_3 - \mathbf{q}))$ to be non-trivial, the vector $\mathbf{k}_3$ is required to have precisely three entries less than or equal to $0$ and the others greater than or equal to $2$. Due to Eq.~\eqref{conditionk1k2k3}, these non-positive entries must arise in different components of the three vectors. Without restricting generality, we can, therefore, assume that $k_1^1 \leq -2$, with all other components of $\mathbf{k}_1$ being greater than or equal to $0$, $k_2^2 \leq -2$ with all other components of $\mathbf{k}_2$ greater than
or equal to 0, $k_3^3 \leq 0$, $k_3^4 \leq 0$, $k_3^5 \leq 0$, while $k_3^1 \geq 2$, $k_3^2 \geq 2$. Using these conventions, we can specialise Eq.~\eqref{5.5} to find the following explicit expressions for the relevant ambient space forms
\begin{equation}
\begin{array}{ll}
 \hat{\nu}_1 & \!\!\! = \kappa^{k^1_1}_1 P_{(\mathbf{k_1})} d \overline{z}_1 \, , \qquad  \quad \,\,\,\, \hat{\nu}_2 = \kappa^{k^2_2}_2 Q_{(\mathbf{k_2})} d \overline{z}_2 \, , \\[1mm] 
 \hat{\rho}_3 & \!\!\! = \kappa^{k^3_3-2}_3 \kappa^{k^4_3-2}_4 \kappa^{k^5_3-2}_5 R_{(\mathbf{k_3}-\mathbf{q})} d \overline{z}_3 \wedge d \overline{z}_4 \wedge d \overline{z}_5 \, .
 \end{array}
 \label{5.9}
\end{equation}
\noindent Inserting the forms into Eq.~\ref{yuk113} leads to 
\be 
\l(\nu_1,\nu_2,\nu_3) =
\frac{1}{(2 \pi )^2}\int_{{\mathbb{C}^5}} d^5 z \ d^5\overline{z} \ \kappa^{k^1_1}_1 \kappa^{k^2_2}_2 \kappa^{k^3_3-2}_3 \kappa^{k^4_3-2}_4 \kappa^{k^5_3-2}_5 \ P_{(\mathbf{k_1})} Q_{(\mathbf{k_2})} R_{(\mathbf{k_3}-\mathbf{q})}.
 \label{5.10}
 \ee
\noindent By inserting expressions for the polynomials, this integral splits up into products of integrals over $\mathbb{P}^1$ and can be worked explicitly. Alternatively, we can proceed by noticing that the integrand $\hat{\nu}_1 \wedge \hat{\nu}_2 \wedge \hat{\rho}_3$ represents a cohomology class in $H^5(\mathcal{A},\mathcal{O}_{\mathcal{A}}(-\mathbf{q}))$, which is one-dimensional. Its harmonic representative has the form
\begin{equation}
c \mu(P,Q,R) \kappa_1^{-2} \kappa_2^{-2} \kappa_3^{-2} \kappa_4^{-2} \kappa_5^{-2} d^5 \overline{z} \, ,
\end{equation}
\noindent where
\begin{equation}
\mu (P,Q,R)= \tilde{P}\tilde{Q}\tilde{R}
\end{equation}
\noindent must be a number, since $h^5(\mathcal{A},\mathcal{O}_{\mathcal{A}}(-\mathbf{q}))=1$. This number is obtained from polynomial multiplication as discussed in the previous sub-section and $c$ is a constant obtained from \eqref{coefc},
\begin{equation}
\resizebox{0.65\hsize}{!}{$
\begin{array}{lll}
c& = &c_{k_1^1,-k_1^1-2} \ c_{k_2^2,-k_2^2-2} \ c_{k_3^3-2,-k_3^3} \ c_{k_3^4-2,-k_3^4} \ c_{k_3^5-2,-k_3^5} \\[1mm]
& = &\tfrac{1}{(-k_1^1-1)!}  \tfrac{1}{(-k_2^2-1)!}  \tfrac{1}{(-k_3^3+1)!}  \tfrac{1}{(-k_3^4+1)!}  \tfrac{1}{(-k_3^5+1)!}
\, .
\end{array}$}
\label{c113}
\end{equation}
\noindent Together with the basic identity
\begin{eqnarray}
\int_{\mathbb{C}} \dfrac{1}{\kappa^2} d z \wedge d \overline{z} = 2 \pi i \, ,
\end{eqnarray}
\noindent this leads to the final expression
\be 
\lambda(\nu_1,\nu_2,\nu_3) = 8 i \pi^3 c \ \mu(P,Q,R) \, , \qquad \mu(P,Q,R) = \tilde{P}\tilde{Q}\tilde{R} \, .
 \label{result113}
 \ee
\noindent This equation represents our final result for the Yukawa couplings in this case and it allows for an “algebraic” calculation by multiplying together the polynomials $\tilde{P}$, $\tilde{Q}$ and $\tilde{R}$. Note that, given the rules for converting coordinates into partial derivatives in these polynomials, as discussed in the last sub-section, this must always result in a number, that is, the partial derivatives remove all remaining coordinates.

\subsection{Yukawa couplings of type $(1, 2, 2)$}

The other simple case which avoids the vanishing theorem \eqref{theoremdim5}  arises if one of the forms, say $\nu_1$, is of type 1, while $\nu_2$ and $\nu_3$ are of type 2. This case can be dealt with in complete analogy with the $(1, 1, 3)$ case in the previous sub-section. The relevant formula for the Yukawa couplings in this case is Eq.~\eqref{yuk122}, which only involves the ambient space forms $\hat{\nu}_1 \in H^1(\mathcal{A},\mathcal{K}_1)$, $\hat{\omega}_2 \in H^2(\mathcal{A}, \mathcal{N}^* \otimes\mathcal{K}_2)$ and $\hat{\omega}_{3} \in H^2(\mathcal{A}, \mathcal{N}^* \otimes \mathcal{K}_{3} )$.

In order to construct these forms, it is again useful to fix our conventions. Since we require that $H^1(\mathcal{A}, \mathcal{K}_1)$ be non-trivial, we need precisely one component in $\mathbf{k}_1$ less than or equal to $-2$ (and all others non-negative) and we choose $k^1_1 \leq -2$.  The two $(0, 2)$-forms $\hat{\omega}_2$, $\hat{\omega}_3$ need to originate from different line
bundles in the rank two bundle $\mathcal{N}^*$ tensored with $\mathcal{K}_2$ and $\mathcal{K}_3$, or else the Yukawa coupling would vanish, so we assume that $\hat{\omega}_2 \in H^2(\mathcal{A},\mathcal{O}_{\mathcal{A}}(-\mathbf{q}_1)\otimes \mathcal{K}_2)$ and $\hat{\omega}_3 \in H^2(\mathcal{A},\mathcal{O}_{\mathcal{A}}(-\mathbf{q}_2)\otimes \mathcal{K}_3)$. Hence we need precisely two entries in $\mathbf{k}_2 - \mathbf{q}_1$ and in $\mathbf{k}_3 - \mathbf{q}_2$ to be less than or equal to $-2$ (with all other entries non-negative). Due to Eq.~\eqref{conditionk1k2k3}, all negative entries have to arise in different components. Hence, we can choose $k_2^2 - q_1^2 \leq -2$, $k_2^3-q_1^3 \leq -2$, $k_3^4 - q_2^4 \leq -2$ and $k_3^5 - q_2^5 \leq -2$, with all the other entries non-negative.  Applying these conventions to Eq.~\eqref{5.5} results in
\begin{equation}
\begin{array}{lll}
 \hat{\nu}_1 &=& \kappa^{k^1_1}_1 P_{(\mathbf{k_1})} d \overline{z}_1 \, , \\[1mm]  \hat{\omega}_2 &=& \kappa^{k^2_2 - q^2_1}_2 \kappa^{k^3_2 - q^3_1}_3 Q_{(\mathbf{k}_2-\mathbf{q}_1)} d \overline{z}_2 \wedge d \overline{z}_3 \, , \\ [1mm] \hat{\omega}_3 &=& \kappa^{k^4_3-q^4_2}_4 \kappa^{k^5_3-q^5_2}_5  R_{(\mathbf{k_3}-\mathbf{q}_2)} d \overline{z}_4 \wedge d \overline{z}_5 \, .
\end{array}
 \label{5.13}
\end{equation}
\noindent Inserting these forms into Eq.~\eqref{yuk122}, the integral can be carried out as in the previous subsection and results in the same formula
\be 
\lambda(\nu_1,\nu_2,\nu_3) = 8 i \pi^3 c \ \mu(P,Q,R) \, , \qquad \mu(P,Q,R) = \tilde{P}\tilde{Q}\tilde{R} \, ,
 \label{result122}
 \ee
\noindent but with the constant $c$ now given by
\begin{equation}
\resizebox{0.94\hsize}{!}{$
\begin{array}{lll}
 c& \!\!\! = &\!\!\! c_{k_1^1,-k_1^1-2} \ c_{k_2^2 - q^2_1 ,-k_2^2+q^2_1-2} \ c_{k_2^3 - q^3_1 ,-k_2^3+q^3_1-2} \ c_{k_3^4 - q^4_2 ,-k_3^4+q^4_2-2} \ c_{k_3^5 - q^5_2 ,-k_3^5+q^5_2-2} \\[1mm]
 &\!\!\! = & \!\!\! \tfrac{1}{(-k_1^1-1)!}  \tfrac{1}{(-k_2^2+q_1^2-1)!}  \tfrac{1}{(-k_2^3+q_1^3-1)!}  \tfrac{1}{(-k_3^4+q_2^4-1)!}  \tfrac{1}{(-k_3^5+q_2^5-1)!}
\, .
\end{array}$}
\label{c122}
\end{equation}

\subsection{An example with vanishing Yukawa couplings}
\noindent We consider a rank five line bundle sum on the CICY \eqref{cicy7487} specified by the following line bundles:
\begin{equation}
\!\!\!\!
\begin{array}{l}
L_1=\mathcal{O}_X(1,0,-2,0,1) \, , \quad L_2=\mathcal{O}_X(1,-2,0,1,0) \, , \quad L_3=\mathcal{O}_X(0,1,0,0,-1) \, , \\[1mm]
L_4=\mathcal{O}_X(0,0,1,-1,0) \, , \quad  L_5=\mathcal{O}_X(-2,1,1,0,0) \, .
\end{array}
\end{equation}
\noindent This model leads to a four-dimensional theory with gauge group $SU(5) \times S(U(1)^5)$. The non-vanishing
cohomologies of these line bundles and their tensor products are
\begin{equation}
\begin{array}{llllllll}
h^\bullet(X,L_1)&\! =&\!(0, 4, 0, 0) \, , & \quad& h^\bullet(X,L_2)&\!=&\!(0, 4, 0, 0) \, , \\[1mm]
h^\bullet(X,L_5)&\! =&\!(0, 4, 0, 0) \, , &\quad & h^\bullet(X,L_1 \otimes L_2)&\! =&\!(0, 4, 0, 0) \, ,  \\[1mm]
h^\bullet(X,L_1 \otimes L_3)&\!=&\!(0,4,0,0) \, , & \quad&  h^\bullet(X,L_2 \otimes L_4)&\!=&\!(0,4,0,0) \, ,    \\[1mm]
h^\bullet(X,L_3 \otimes L_4)&\!=&\!(0,1,1,0) \, , & \quad & h^\bullet(X,L_1 \otimes L_2^*)&\!=&\!(0,4,4,0) \, ,   \\[1mm]
h^\bullet(X,L_1 \otimes L_4^*)&\!=&\!(0, 16, 0, 0) \, , & \quad  & h^\bullet(X,L_2 \otimes L_3^*)&\!=&\!(0, 16, 0, 0) \, ,  \\[1mm]
h^\bullet(X,L_3 \otimes L_4^*)&\!=&\!(0, 1, 1, 0) \, , & \quad  & h^\bullet(X,L_5 \otimes L_3^*)&\!=&\!(0, 4, 0, 0) \, ,  \\[1mm]
h^\bullet(X,L_5 \otimes L_4^*)&\!=&\!(0, 4, 0, 0) \, . &\ \quad &&&
\end{array}
\end{equation}
\noindent These results imply the following upstairs spectrum
\begin{equation}
\begin{array}{l}
4 \ {\bf 10}_1,\,\, 4 \ {\bf 10}_2,\,\, 4 \ {\bf 10}_5, \\[1mm]   4 \ \overline{\bf 5}_{1,2}, \,\, 4 \ \overline{\bf 5}_{1,3}, \,\, 4 \ \overline{\bf 5}_{2,4}, \,\, \overline{\bf 5}{}^H_{3,4}, \,\,  {\bf 5}^{\overline{H}}_{3,4}, \\[1mm]   4 \ {\bf 1}_{1,2}, \,\,  4 \ {\bf 1}_{2,1}, \,\, 4 \ {\bf 1}_{1,3}, \,\, 12 \ {\bf 1}_{1,4}, \,\,  12 \ {\bf 1}_{2,3}, \,\,  4 \ {\bf 1}_{2,4}, \,\, {\bf 1}_{3,4}, \,\, {\bf 1}_{4,3}, \,\,  4 \ {\bf 1}_{5,3}, \,\,  4 \ {\bf 1}_{5,4} \, .
\end{array}
\end{equation}
\noindent Here, the bold-face numbers denote $SU(5)$ representations and the subscripts indicate under which of the five $U(1)$ symmetries a multiplet is charged. This spectrum consists of 12 families in $\overline{\mathbf{5}}\oplus \mathbf{10}$, one $\mathbf{5}$--$\overline{\mathbf{5}}$ pair of Higgs multiplets and a number of $SU(5)$ singlets. Upon dividing by the freely-acting symmetry $\Gamma = \mathbb{Z}_2 \times \mathbb{Z}_2$ in Eq.~\eqref{Gammasym}, one obtains the standard model spectrum with three families. It is important to remember, however, that only couplings which respect the $S(U(1)^5)$ symmetry are allowed in the
four-dimensional theory. One such allowed coupling is described by the following superpotential term
\begin{align}
\label{Yuk111}
W & =  \lambda_{I J K} \overline{\mathbf{5}}{}^{(I)}_{1,3} \overline{\mathbf{5}}{}^{(J)}_{2,4} \mathbf{10}^{(K)}_5 \, . 
\end{align}
\noindent In order to compute this coupling, we write down the relevant line bundles and bundle-valued forms which
are given by
\begin{equation}
\label{type111}
\!\!\!\!\!\!\! \resizebox{0.935\hsize}{!}{$
\begin{array}{ll}
\,\,\ 4 \ \overline{\mathbf{5}}_{1,3} \rightarrow K_1=L_1 \otimes L_3 = \mathcal{O}_X(1, 1, -2, 0, 0), & \hat{\nu}_1= \sigma_3^{-2} \tilde{P}_{(1, 1, -2, 0, 0)} d\bar{\mu}_3 \in H^1(\mathcal{A},\mathcal{K}_1) \, ,  \\[1mm]
\,\,\ 4 \ \overline{\mathbf{5}}_{2,4} \rightarrow K_2 = L_2 \otimes L_4 = \mathcal{O}_X(1, -2, 1, 0, 0), & \hat{\nu}_2= \sigma_2^{-2}  \tilde{Q}_{(1, -2, 1, 0, 0)} d\bar{\mu}_2 \in H^1(\mathcal{A},\mathcal{K}_2) \, , \\[1mm]
\,\,\ 4 \ \mathbf{10}_5 \rightarrow K_3 = L_5  = \mathcal{O}_X(-2,1,1,0,0) , &  \hat{\nu}_3=\sigma_1^{-2}  \tilde{R}_{(-2,1,1,0,0)} d\bar{\mu}_1 \in H^1(\mathcal{A},\mathcal{K}_3) \, , 
\end{array}$}
\end{equation}
\noindent with explicit polynomials
\begin{equation}
\begin{array}{lll}
\tilde{P} & = &  p_0 x_0 y_0+p_1 x_0 y_1 + p_2 x_1 y_0 + p_3 x_1 y_1 \, ,  \\[1mm]
\tilde{Q} & = & q_0 x_0 u_0+q_1 x_0 u_1 + q_2 x_1 u_0 + q_3 x_1 u_1 \, ,  \\[1mm]
\tilde{R} & =  &r_0 y_0 u_0+r_1 y_0 u_1 + r_2 y_1 u_0 + r_3 y_1 u_1 \, .
\end{array}
\end{equation}
\noindent Evidently, from Eq.~\eqref{type111}, all three forms $\nu_i$ are of type $\tau_i = 1$ and, hence, the Yukawa couplings $\lambda_{IJK}$
in Eq.~\eqref{Yuk111} are all zero as a consequence of the vanishing theorem~\eqref{theoremdim5}.

\subsection{An example with Yukawa couplings of type $(1, 1, 3)$}
\label{sectionexample113}

\noindent A line bundle model on the CICY \eqref{cicy7487} which realises Yukawa couplings of type $(\tau_1, \tau_2, \tau_3) = (1, 1, 3)$ is defined by the five line bundles
\begin{equation}
\!\!\!\!
\begin{array}{l}
L_1=\mathcal{O}_X(1,-2,0,0,1) \, , \quad L_2=\mathcal{O}_X(0,1,0,1,-2) \, , \quad L_3=\mathcal{O}_X(0,0,1,-2,1) \, , \\[1mm]
L_4=\mathcal{O}_X(0,0,-1,0,1) \, , \quad L_5=\mathcal{O}_X(-1,1,0,1,-1) \, .
\end{array}
\end{equation}
As before, the four-dimensional gauge group is $SU(5) \times S(U(1)^5)$ and the non-trivial cohomologies of the
above line bundles and their tensor product
\begin{equation}
\begin{array}{llllllll}
h^\bullet(X,L_1)&\!=&\!(0, 4, 0, 0) \, , &\quad & h^\bullet(X,L_2)&\!=&\! (0, 4, 0, 0) \, , \\[1mm]
h^\bullet(X,L_3)&\!=&\!(0, 4, 0, 0) \, , & \quad & h^\bullet(X,L_1 \otimes L_3)&\!=&\!(0, 4, 0, 0) \, , \\[1mm]
h^\bullet(X,L_2 \otimes L_3)&\!=&\!(0,1,1,0) \, , & \quad &  h^\bullet(X,L_2 \otimes L_4)&\!=&\!(0,1,1,0) \, , \\[1mm]
h^\bullet(X,L_2 \otimes L_5)&\!=&\!(0,8,0,0) \, ,  & \quad & h^\bullet(X,L_3 \otimes L_4)&\!=&\!(0,3,3,0) \, ,  \\[1mm]
h^\bullet(X,L_1 \otimes L_4^*)&\!=&\!(0, 4, 0, 0) \, , & \quad & h^\bullet(X,L_1 \otimes L_5^*)&\!=&\!(0, 8, 0, 0) \, , \\[1mm]
h^\bullet(X,L_2 \otimes L_3^*)&\!=&\!(0, 9, 9, 0) \, , &\quad &  h^\bullet(X,L_2 \otimes L_4^*)&\!=&\!(0, 16, 0, 0) \, , \\[1mm]
h^\bullet(X,L_3 \otimes L_4^*)&\!=&\!(0, 3, 3, 0) \, , & \quad & h^\bullet(X,L_3 \otimes L_5^*)&\!=&\!(0, 12, 0, 0) \, , \\[1mm]
h^\bullet(X,L_5 \otimes L_4^*)&\!=&\!(0, 4, 0, 0) \,  
\end{array}
\end{equation}
\noindent lead to the following spectrum:
\begin{equation}
\begin{array}{l}
 4 \ {\bf 10}_1,\,\, 4 \ {\bf 10}_2,\,\, 4 \ {\bf 10}_3, \\[1mm]  4 \ \overline{\bf 5}_{1,3},  \,\, \ \overline{\bf 5}{}^H_{2,3}, \,\,  \ {\bf 5}^{\overline{H}}_{2,3}, \,\, \ \overline{\bf 5}{}^H_{2,4}, \,\,  \ {\bf 5}^{\overline{H}}_{2,4}, \,\, 8 \ \overline{\bf 5}_{2,5},  \,\, 3 \ \overline{\bf 5}{}^H_{3,4}, \,\,  3 \ {\bf 5}^{\overline{H}}_{3,4},  \\[1mm]  4 \ {\bf 1}_{1,4}, \,\, 8 \ {\bf 1}_{1,5}, \,\, 9 \ {\bf 1}_{2,3}, \,\,  9 \ {\bf 1}_{3,2}, \,\,  16 \ {\bf 1}_{2,4}, \,\,  3 \ {\bf 1}_{3,4}, \,\,  3 \ {\bf 1}_{4,3}, \,\, 12 \ {\bf 1}_{3,5}, \,\,  4 \ {\bf 1}_{5,4} \, .
\end{array}
\end{equation}
\noindent This spectrum contains 12 families $\overline{\mathbf{5}}\oplus \mathbf{10}$, five $\mathbf{5}$--$\overline{\mathbf{5}}$ Higgs pairs and $SU(5)$-singlet multiplets and gives rise to a three-family standard model after a suitable quotient with the symmetry \eqref{Gammasym}. We are interested in the superpotential terms
\begin{align}
\label{coupling5.34}
W & =  \lambda_{I J K} \overline{\mathbf{5}}{}^{H, (I)}_{3,4} \mathbf{10}^{(J)}_1 \overline{\mathbf{5}}{}^{(K)}_{2,5} \, , 
\end{align}
\noindent which are allowed by all gauge symmetries of the model. The relevant harmonic forms are given by
\begin{equation}
\!\!\! \resizebox{1.04\hsize}{!}{$
\begin{array}{ll}
3 \ \overline{\mathbf{5}}^H_{3,4} \rightarrow K_1 = L_3 \otimes L_4 = \mathcal{O}_X(0, 0, 0, -2, 2), & \hat{\nu}_1=  \sigma_4^{-2}  \tilde{P}_{(0, 0, 0, -2, 2)} d\bar{\mu}_4 \, ,  \\[1mm]
4 \ \mathbf{10}_1 \rightarrow K_2 = L_1  = \mathcal{O}_X(1,-2,0,0,1), &  \hat{\nu}_2=\sigma_2^{-2}  \tilde{Q}_{(1,-2,0,0,1)} d\bar{\mu}_2 \, , \\[1mm]
8 \ \overline{\mathbf{5}}_{2,5} \rightarrow K_3=L_2 \otimes L_5 = \mathcal{O}_X(-1, 2, 0, 2, -3), & \hat{\rho}_3= \sigma_1^{-3} \sigma_3^{-2} \sigma_5^{-5} \tilde{R}_{(-3,0,-2,0,-5)} d\bar{\mu}_1 \wedge d\bar{\mu}_3 \wedge d\bar{\mu}_5 , \\[1mm]
\end{array}$}
\end{equation}
\noindent where $\hat{\nu}_1 \in H^1(\mathcal{A},\mathcal{K}_1)$, $\hat{\nu}_2 \in H^1(\mathcal{A},\mathcal{K}_2)$ and $\hat{\rho}_3 \in H^3(\mathcal{A},\Lambda^2 \mathcal{N}^* \otimes \mathcal{K}_3)$. The polynomials $\tilde{P}$, $\tilde{Q}$, $\tilde{R}$ can be explicitly written as
\begin{equation}
\begin{array}{lll}
\tilde{P} & = &p_0 w_0^2 + p_1 w_0 w_1 + p_2 w_1^2  \, , \\[1mm]
\tilde{Q} & = & q_0 x_0 w_0 + q_1 x_1 w_0 + q_2 x_0 w_1 + q_3 x_1 w_1 \, , \\[1mm]
\tilde{R} & = &  r_0 \overline{x}_0 \overline{w}^3_0 + r_1 \overline{x}_0 \overline{w}^2_0 \overline{w}_1 + r_2  \overline{x}_0 \overline{w}_0 \overline{w}^2_1 + r_3 \overline{x}_0 \overline{w}^3_1 \, + \\[1mm] &  & r_4 \overline{x}_1 \overline{w}^3_0 + r_5 \overline{x}_1 \overline{w}^2_0 \overline{w}_1 + r_6  \overline{x}_1 \overline{w}_0 \overline{w}^2_1 + r_7 \overline{x}_1 \overline{w}^3_1 \, .
\end{array}
\end{equation}
\noindent Note that the coefficients $p_I$ , $q_J$ and $r_K$ in these polynomials parametrise the various families. Using these polynomials we can compute the upstairs Yukawa couplings from Eq.~\eqref{result113}, which leads to
\begin{align}
\mu(\tilde{P},\tilde{Q},\tilde{R})  \, = &  \; \ 6 p_0 q_0 r_0 +  6 p_0 q_1 r_4 + 2 p_0 q_2 r_1 + 2 p_0 q_3 r_5 + 2 p_1 q_0 r_1 +  2 p_1 q_1 r_5 \, +  \notag \\ &  \; \ 2 p_1 q_2 r_2 + 2 p_1 q_3 r_6 + 2 p_2 q_0 r_2 + 2 p_2 q_1 r_6 + 6 p_2 q_2 r_3   + 6 p_2 q_3 r_7 \, .
 \end{align}
\noindent The Yukawa couplings $\lambda_{IJK}$ in Eq.~\eqref{coupling5.34} can be easily obtained from this expression by choosing a basis
for the coefficients, for example by setting each of the coefficients $p_I$, $q_J$, $r_K$ to one and the others
to zero. It is however more interesting to see what happens in the downstairs theory, obtained from the present $SU(5)$ GUT theory by a quotient with the $\Gamma = \mathbb{Z}_2 \times \mathbb{Z}_2$ symmetry \eqref{Gammasym}. The GUT multiplets branch as $\mathbf{10} \rightarrow (Q,u,e)$, $\overline{\mathbf{5}} \rightarrow (d,L) $, $\overline{\mathbf{5}}{}^H \rightarrow (T,H)$ into standard model multiplets, where $T$ is the Higgs
triplet which has to be projected out. On the quotient manifold $\tilde{X}$, we introduce a Wilson line in the
standard hypercharge direction in order to break $SU(5)$ to the standard model group. Such a Wilson line
can be described by two $\Gamma$-representations $\chi_2$, $\chi_3$ which we choose as $\chi_2 = (1, 1)$ and $\chi_3 = (0, 0)$. This induces the multiplet charges
\begin{eqnarray}
\label{charges}
  \chi_d = \chi_3^* = (0,0)\, , \qquad \chi_H = \chi_2^* = (1,1)\, , \qquad \chi_Q  = \chi_2 \otimes \chi_3 = (1,1) \, .
\end{eqnarray}
\noindent In order to determine the polynomials corresponding to the downstairs spectrum, one has to keep in mind that every differential $d \overline{\mu}_i$ has charge $(1,1)$ under $\Gamma$. Moreover, for the $(0,3)$-form $\hat{\rho}_3$, there is an additional $(1,1)$ charge flip due to the fact that the bundle $ \Lambda^2 \mathcal{N}^* \otimes \mathcal{K}_3$ transforms non-trivially under $\Gamma$ from Eq.\eqref{Gammarho}. Matching these charges up with the Wilson line charges \eqref{charges}, the representatives of the downstairs spectrum become
\begin{equation}
\begin{array}{ll}
H_{3,4} &\! : \,\, \tilde{P} \in \textrm{Span}  (w_0^2 + w_1^2 ) \, ,\\[1mm]
Q_1 &\! : \,\,  \tilde{Q} \in \textrm{Span}  (x_0 w_0 + x_1 w_1) \, , \\[1mm] 
d_{2,5} &\! : \,\,  \tilde{R} \in \textrm{Span}  (\overline{x}_0 \overline{w}_0 \overline{w}_1^2 + \overline{x}_1 \overline{w}_1 \overline{w}_0^2 \, , \, \overline{x}_0 \overline{w}_0^3 + \overline{x}_1 \overline{w}_1^3  ) \, .
\end{array}
\end{equation}
\noindent Using Eq.~\eqref{result113} the Yukawa couplings in $\lambda_K H_{3,4} Q_1 d^{(K)}_{2,5} $ become proportional to
\begin{eqnarray}
\!\!\!\! \mu(H_{3,4}, Q_1, d_{2,5})\! = \!\dfrac{1}{4} \!  \big( \partial^2_{\overline{w}_0}+ \partial^2_{\overline{w}_1}\big) \! \big( \partial_{\overline{x}_0}\partial_{\overline{w}_0}+ \partial_{\overline{x}_1}\partial_{\overline{w}_1}\big) \!\! \begin{pmatrix} \overline{x}_0 \overline{w}_0 \overline{w}_1^2 + \overline{x}_1 \overline{w}_1 \overline{w}_0^2 \\ \overline{x}_0 \overline{w}_0^3 + \overline{x}_1 \overline{w}_1^3 \end{pmatrix} \!\! = \!\! \begin{pmatrix} 1 \\ 3 \end{pmatrix}\!,
\end{eqnarray}
\noindent where we have converted the homogeneous coordinates into derivatives and introduced an additional
factor $1/4$, to account for the fact that the integral is carried out on the quotient manifold. The numerical
coefficient $c$ in Eq.~\eqref{c113} is given by
\begin{eqnarray}
c = c_{(-2,0)} c_{(-2,0)} c_{(-5,3)} c_{(-4,2)}  c_{(-7,5)} = 1\cdot 1\cdot\dfrac{1}{4!}\cdot\dfrac{1}{3!}\cdot\dfrac{1}{6!} \, .
\end{eqnarray}

\subsection{An example with Yukawa couplings of type $(1, 2, 2)$}

\noindent This example on the CICY \eqref{cicy7487} is defined by the five line bundles 
\begin{equation}
\!\!\!\!
\begin{array}{l}
L_1=\mathcal{O}_X(1,-2,0,0,1)\, , \quad L_2=\mathcal{O}_X(0,1,-2,0,1)\, , \quad L_3=\mathcal{O}_X(0,0,1,1,-2) \, , \\[1mm]
L_4=\mathcal{O}_X(0,0,1,-1,0)\, , \quad L_5=\mathcal{O}_X(-1,1,0,0,0) \, ,
\end{array}
\end{equation}
\noindent The non-vanishing cohomologies of these line bundles and their tensor products
\begin{equation}
\begin{array}{llllllll}
h^\bullet(X,L_1)&\!=&\!(0, 4, 0, 0) \, , &\quad & h^\bullet(X,L_2)&\!=&\!(0, 4, 0, 0) \, , \\[1mm]
h^\bullet(X,L_3)&\!=&\!(0, 4, 0, 0)  \, , &\quad & h^\bullet(X,L_1 \otimes L_3)&\!=&\!(0, 4, 0, 0) \, , \\[1mm]
h^\bullet(X,L_1 \otimes L_4)&\!=&\!(0,4,0,0) \, , &\quad &  h^\bullet(X,L_2 \otimes L_3)&\!=&\!(0,1,1,0) \, , \\[1mm]
h^\bullet(X,L_2 \otimes L_4)&\!=&\!(0,1,1,0)  \, , &\quad & h^\bullet(X,L_3 \otimes L_4)&\!=&\!(0,3,3,0) \, , \\[1mm]
h^\bullet(X,L_3 \otimes L_5)&\!=&\!(0, 4, 0, 0) \, , &\quad & h^\bullet(X,L_1 \otimes L_2^*)&\!=&\!(0, 12, 0, 0) \, , \\[1mm]
h^\bullet(X,L_3 \otimes L_1^*)&\!=&\!(0, 12, 0, 0) \, , &\quad & h^\bullet(X,L_1 \otimes L_4^*)&\!=&\!(0, 4, 0, 0) \, , \\[1mm]
h^\bullet(X,L_1 \otimes L_5^*)&\!=&\!(0, 12, 0, 0) \, , &\quad & h^\bullet(X,L_2 \otimes L_3^*)&\!=&\!(0, 9, 9, 0) \, , \\[1mm]
h^\bullet(X,L_2 \otimes L_4^*)&\!=&\!(0, 16, 0, 0) \, , &\quad & h^\bullet(X,L_2 \otimes L_5^*)&\!=&\!(0, 4, 0, 0) \, , \\[1mm]
h^\bullet(X,L_3 \otimes L_4^*)&\!=&\!(0, 3, 3, 0) \, , &\quad & h^\bullet(X,L_3 \otimes L_5^*)&\!=&\!(0, 4, 0, 0) \,
\end{array}
\end{equation}
\noindent lead to the upstairs spectrum
\begin{equation}
\!\!
\begin{array}{l}
 4 \ {\bf 10}_1,\,\, 4 \ {\bf 10}_2,\,\, 4 \ {\bf 10}_3, \\[1mm]   4 \ \overline{\bf 5}_{1,3}, \,\, 4 \ \overline{\bf 5}_{1,4}, \,\, \ \overline{\bf 5}{}^H_{2,3}, \,\,  \ {\bf 5}^{\overline{H}}_{2,3}, \,\, \ \overline{\bf 5}{}^H_{2,4}, \,\,  \ {\bf 5}^{\overline{H}}_{2,4},  \,\, 3 \ \overline{\bf 5}{}^H_{3,4}, \,\,  3 \ {\bf 5}^{\overline{H}}_{3,4}, \,\, 4 \ \overline{\bf 5}_{3,5}, \\[1mm] 12 \ {\bf 1}_{1,2}, \,\, 12 \ {\bf 1}_{3,1}, \,\, 4 \ {\bf 1}_{1,4}, \,\,  12 \ {\bf 1}_{1,5}, \,\,  9 \ {\bf 1}_{2,3}, \,\,  9 \ {\bf 1}_{3,2}, \,\,  16 \ {\bf 1}_{2,4}, \,\, 4 \ {\bf 1}_{2,5}, \,\,  3 \ {\bf 1}_{3,4}, \,\,  3 \ {\bf 1}_{4,3}, \,\,  4 \ {\bf 1}_{3,5} \, . \\[1mm]
\end{array}
\end{equation}
\noindent As before, this spectrum with 12 families in $\overline{\mathbf{5}} \oplus \mathbf{10}$, five $\mathbf{5}$--$\overline{\mathbf{5}}$ Higgs pairs and $SU(5)$-singlets leads to a three-family standard model after the quotient by $\Gamma = \mathbb{Z}_2 \times \mathbb{Z}_2$. We are interested in the following superpotential term:
\begin{align}
W & =  \lambda_{I J K} \mathbf{10}^{(I)}_2  \overline{\mathbf{5}}{}^{(J)}_{1,4} \overline{\mathbf{5}}{}^{(K)}_{3,5}  \, .
\end{align}
\noindent The associated harmonic forms
\begin{equation}
\!\!\!\!\!\!\! \resizebox{0.935\hsize}{!}{$
\begin{array}{ll}
\,\,\, 4 \ \mathbf{10}_2 \rightarrow K_1 = L_2  = \mathcal{O}_X(0,1,-2,0,1)\ , &  \hat{\nu}_1=\sigma_3^{-2}  P_{(0,1,-2,0,1)} d\bar{\mu}_3 \ , \\[1mm]
\,\,\, 4 \ \overline{\mathbf{5}}_{1,4} \rightarrow K_2=L_1 \otimes L_4 = \mathcal{O}_X(1,-2,1,-1,1)\ , & \hat{\omega}_2= \sigma_2^{-3} \sigma_4^{-2} Q_{(1, -3, 0, -2, 0)} d\bar{\mu}_2 \wedge d\bar{\mu}_4  \ , \\[1mm]
\,\,\, 4 \ \overline{\mathbf{5}}_{3,5} \rightarrow K_3 = L_3 \otimes L_5 = \mathcal{O}_X(-1,1,1,1,-2)\ , & \hat{\omega}_3= \sigma_1^{-3} \sigma_5^{-3}  R_{(-3, 0, 0, 0, -3)} d\bar{\mu}_1 \wedge d\bar{\mu}_5 \ ,
\end{array}$}
\end{equation}
\noindent where $\hat{\nu}_1 \in H^1(\mathcal{A},\mathcal{K}_1)$, $\hat{\omega}_2 \in H^2(\mathcal{A},\mathcal{N}^* \otimes \mathcal{K}_2)$ and $\hat{\omega}_3 \in H^2(\mathcal{A},\mathcal{N}^* \otimes \mathcal{K}_3)$ contain the polynomials
\begin{align}
\tilde{P} & = p_0 y_0 w_0 + p_1 y_1 w_0 + p_2 y_0 w_1 + p_3  y_1 w_1 \, , \notag \\
\tilde{Q} & =  q_0 x_0 \overline{y}_0 + q_1 x_1 \overline{y}_0 + q_2 x_0 \overline{y}_1  + q_3 x_1 \overline{y}_1 \, ,  \\
\tilde{R} & = r_0 \overline{x}_0 \overline{w}_0 + r_1 \overline{x}_1 \overline{w}_0  + r_2 \overline{x}_0 \overline{w}_1  + r_3 \overline{x}_1 \overline{w}_1  \, . \notag  
\end{align}
\noindent From Eq.~\eqref{result122}, this leads to upstairs Yukawa couplings
\begin{align}
\mu(\tilde{P},\tilde{Q},\tilde{R}) = & \; p_0 q_0 r_0 +  p_0 q_1 r_1 + p_1 q_2 r_0 + p_1 q_3 r_1 \, + \notag \\
& \;  p_2 q_0 r_2 + p_2 q_1 r_3 + p_3 q_2 r_2 + p_3 q_3 r_3  \, .
 \end{align}
\noindent Under the standard model group, the relevant part of the upstairs spectrum branches as $\mathbf{10}_2 \rightarrow (Q,u,e)_2$, $\overline{\mathbf{5}}_{1,4} \rightarrow (d,L)_{1,4}$, $\overline{\mathbf{5}}_{3,5} \rightarrow (T,H)_{3,5}$. We choose the same Wilson line, $\chi_2 = (1, 1)$ and $\chi_3 = (0, 0)$, as in
Section~\ref{sectionexample113}, which then leads to the same multiplet charges as in Eq.~\eqref{charges}. Once again, we have to
keep in mind that the differentials $d\overline{\mu}_i$
carry charge $(1, 1)$ under $\Gamma$. Moreover, we have to remember from
Eq.~\eqref{Gammarho} that forms which arise from $\mathcal{O}_{\mathcal{A}}(- \mathbf{q}_2) \otimes \mathcal{K}_i$ transform with an additional overall $\Gamma$-charge $(1, 1)$, while forms from $\mathcal{O}_{\mathcal{A}}(-\mathbf{q}_1)\otimes \mathcal{K}_i$ do not. With these rules, the polynomials corresponding to the downstairs
spectrum turn out to be
\begin{align}
Q_2& \,\, : \,\, \tilde{P} \, \in \, \textrm{Span} (y_0 w_0 + y_1 w_1)\, , \notag \\
d_{1,4}& \,\, : \,\, \tilde{Q} \, \in \, \textrm{Span} (x_0 \overline{y}_0 + x_1 \overline{y}_1) \, , \\
H_{3,5}& \,\, : \,\, \tilde{R}\, \in \, \textrm{Span} (\overline{x}_0 \overline{w}_0 + \overline{x}_1 \overline{w}_1) \, . \notag 
\end{align}
\noindent Then, from Eq.~\eqref{result122}, the downstairs Yukawa coupling for $Q_2 \ d_{1,4} H_{3,5}$ is proportional to
\begin{eqnarray}
\mu(Q_2, d_{1,4}, H_{3,5}) = \dfrac{1}{4}  \left( \partial_{\overline{y}_0}\partial_{\overline{w}_0}+ \partial_{\overline{y}_1} \partial_{\overline{w}_1}\right) \left(  \overline{y}_0 \partial_{\overline{x}_0} + \overline{y}_1 \partial_{\overline{x}_1} \right)  \left( \overline{x}_0 \overline{w}_0 + \overline{x}_1 \overline{w}_1 \right) = \dfrac{1}{2} \, ,
\end{eqnarray}
\noindent with the constant $c$ given by
\begin{eqnarray}
c = c_{(-2,0)} c_{(-4,2)} c_{(-3,1)} c_{(-4,2)} c_{(-5,3)} = 1 \cdot \dfrac{1}{3!} \cdot \dfrac{1}{2!} \cdot \dfrac{1}{3!} \cdot \dfrac{1}{4!} \, .
\end{eqnarray}
\section{Final remarks}
\label{chapter3conc}

In the previous chapter, methods to calculate the holomorphic Yukawa couplings have been developed for line bundle models on certain special Calabi-Yau manifolds, with a focus on the tetra-quadric Calabi-Yau manifolds defined in the ambient space $\mathbb{P}^1 \times \mathbb{P}^1 \times \mathbb{P}^1 \times \mathbb{P}^1$. This chapter generalises these methods to all CICY manifolds, and, hence, demonstrates that they are applicable to large classes of Calabi-Yau manifolds.

Our methods rely on the presence of an ambient space $\mathcal{A}$, presently a product of projective spaces,
although generalisations are likely possible, into which the Calabi-Yau manifold $X$ is embedded at co-dimension $k$. Likewise, the three line bundles $K_i \rightarrow X$ associated to a Yukawa coupling should be restrictions of ambient space line bundles $\mathcal{K}_i \rightarrow \mathcal{A}$. We have shown that, in this situation, the three $K_i$-valued $(0, 1)$–forms $\nu_i \in H^1(X, K_i)$ which enter the expression for the holomorphic Yukawa couplings can each be related to a chains $\hat{\nu}_{i,a}$ of $(0, a)$ ambient space forms, where $a = 1, ..., k + 1$. Moreover, from Eq.~\eqref{eq4.11}, the Yukawa couplings can be written in terms of these ambient space forms $\hat{\nu}_{i,a}$.

We say that a form $\nu_i$ is of type $\tau_i \in \lbrace 1, ... , k + 1\rbrace$, if it originates from the ambient space $(0, \tau_i)$-form $\hat{\nu}_{i,\tau_i} \in H^{\tau_i} (\mathcal{A},\Lambda^{\tau_i-1} \mathcal{N}^* \otimes \mathcal{K}_i)$. This means that the associated chain of ambient space forms breaks down at $a = \tau_i$, that is, if $\hat{\nu}_{i,a} = 0$ for $a > \tau_i$. One of our main results is a vanishing theorem which states that the
Yukawa coupling between three forms $\nu_i$ vanishes if their associated types satisfy $\tau_1 + \tau_2 + \tau_3 < \textrm{dim}(\mathcal{A})$. Especially for large ambient space dimension $\textrm{dim}(\mathcal{A})$, this implies the vanishing of many Yukawa couplings,
since large types $\tau_i$ tend to be rare. This vanishing is not explained by one of the obvious four-dimensional
symmetries and, therefore, appears to be topological in nature.

The nature of this vanishing statement is somewhat puzzling in that it relates a physical property --
the vanishing of Yukawa couplings -- to conditions on unphysical quantities, essentially properties of the
ambient space $\mathcal{A}$, which is auxiliary and carries no physical relevance. We do not currently know if the
vanishing statement is restricted to Calabi-Yau manifolds which can be embedded into an ambient space
in this way or if it extends to all Calabi-Yau manifolds. In the latter case, there should be an “intrinsic”
formulation of this statement which only refers to properties of the Calabi-Yau manifold.

We have illustrated our methods by computing certain holomorphic Yukawa couplings for three different line bundle standard models on a co-dimension two CICY. The most pressing issue is, of course, the
calculation of the matter field K\"ahler potential and, hence, of the physical Yukawa couplings.

\chapter{Matter Field K\"ahler Metric from Localisation}
\label{kahlerchapter}

The computation of physical Yukawa couplings from string theory is notoriously difficult, mainly
because methods to compute the matter field K\"ahler metric are lacking. In this chapter we report some progress in this direction. We outline a method to calculate
the matter field K\"ahler metric in the context of Calabi-Yau compactifications of the heterotic string with
Abelian internal gauge fluxes.

So far, the only class of heterotic Calabi-Yau models where an analytic expression for the matter field K\"ahler
metric is known is for models with standard embedding of the spin connection into the gauge connection.
In this case, the matter field K\"ahler metrics for the $(1, 1)$ and $(2, 1)$ matter fields are essentially given by
the metrics on the corresponding moduli spaces \cite{Candelas:1987se, Candelas:1990pi}. Recently, Candelas, de la Ossa and McOrist \cite{candelasmetric}
(see also Ref. \cite{mcoristeffective}) have proposed an $\alpha'$-correction of the heterotic moduli space metric, which includes bundle moduli. This information may be used to infer the K\"ahler metric of matter fields that arise from
bundle moduli. However, we will not pursue this method here, since our main interest is not in bundle moduli but in the gauge matter fields which can account for the physical particles.

There are two other avenues for calculating the matter field K\"ahler metric, suggested by results in the
literature. The first one relies on Donaldson’s numerical algorithm to determine the Ricci-flat Calabi-Yau metric \cite{donaldson1, donaldson2, donaldson3} and subsequent work applying this algorithm to various explicit examples and to the numerical calculation of the Hermitian Yang-Mills connection on vector bundles \cite{wang, headrick1, douglas1, headrick2, headrick3, douglas2, numericallara1, numericallara2}. At present, this
approach has not been pushed as far as numerically calculating physical Yukawa couplings. However, it
appears that this is possible in principle and, while constituting a very significant computational challenge,
would be very worthwhile carrying out. A disadvantage of this method is that it will only provide the
Yukawa couplings at specific points in moduli space and that extracting information about their moduli
dependence will be quite difficult.

In this chapter, we will focus on a different approach, based on localisation due to flux, which can lead to
analytic results for the matter field K\"ahler metric. This method is motivated by work in F-theory \cite{heckmanvafa, fontibanez, conlonpalti, aparicio},
where the localisation of matter fields on the intersection curves of D7-branes and Yukawa couplings on
intersections of such curves facilitates local computations of the Yukawa couplings which do not require
knowledge of the Ricci-flat Calabi-Yau metric. It is not immediately obvious whether and how this approach
might transfer to the heterotic case, since heterotic compactifications lack the intuitive local picture, related
to intersecting D-brane models, which is available in F-theory. In this chapter, we will show, using methods
from differential geometry developed in previous chapters (see also \cite{yukunification}), that localisation of wave functions can nevertheless
arise in heterotic models. The underlying mechanism is, in fact, similar to the one employed in F-theory.
Sufficiently large flux -- in the heterotic case, $E_8 \times E_8$ gauge flux -- leads to a localisation of wave functions
which allows calculating their normalisation locally, without recourse to the Ricci-flat Calabi-Yau metric.

To carry this out explicitly we will proceed in three steps. First, we derive the general formula for the
matter field K\"ahler metric for heterotic Calabi-Yau compactifications, as hinted in Section~\ref{mattersectorsection}. This formula, which provides the matter field K\"ahler metric in terms of an
integral over harmonic bundle valued forms is not, in itself, new (see, for example, Ref. \cite{boundaryinflation}). Our rederivation serves two purposes. First, we would like to fix conventions and factors as this will be required
for an accurate calculation of the physical Yukawa couplings and, secondly, we will show explicitly how
this formula for the matter field K\"ahler metric is consistent with four-dimensional $N = 1$ supergravity. We
observe that this consistency already determines the dependence of the matter field K\"ahler metric on the
T-moduli, a result which, to our knowledge, has not been pointed out in the literature so far.

The second step is to show how (Abelian) $E_8 \times E_8$ gauge flux can lead to a localisation of the matter
field wave functions around certain points of the Calabi-Yau manifold. We will first demonstrate this for toy examples based on line bundles on $\mathbb{P}^1$, as well as on products of projective spaces and then show that
the effect generalises to Calabi-Yau manifolds. As a result, we obtain local matter field wave functions on
Calabi-Yau manifolds and explicit results for their normalisation integrals. 

The final step is to express these results in terms of the global moduli of the Calabi-Yau manifold. We show
that this can indeed be accomplished by relating global to local quantities on the Calabi-Yau manifold and
by using information from four-dimensional $N = 1$ supersymmetry. In this way, we can obtain explicit
results for the matter field K\"ahler metric as a function of the Calabi-Yau moduli and this is carried out for
the Calabi-Yau hyper-surface in $\mathbb{P}^1 \times \mathbb{P}^3$. We believe this is the first time such a result for the matter field K\"ahler metric as a function of the properly defined moduli has been obtained in any geometrical string compactification, including F-theory.

The plan of the chapter is as follows. In the next section, we sketch the supergravity calculation which
leads to the general formula for the matter field K\"ahler metric and we discuss the implications from
four-dimensional $N = 1$ supersymmetry. In Section~\ref{kahlersec3}, we show how gauge flux leads to the localisation
of matter field wave functions, starting with toy examples on $\mathbb{P}^1$ and then generalising to products of
projective spaces. Section \ref{kahlersec4} contains the local calculation of the wave function normalisation on a patch of the Calabi-Yau manifold. In Section \ref{kahlersec5}, we express this result in terms of the properly defined moduli by relating global and local quantities and we obtain an explicit result for the matter field K\"ahler metric on Calabi-Yau hyper-surfaces in $\mathbb{P}^1 \times \mathbb{P}^3$. We conclude in Section \ref{kahlersec6}.

\section{The matter field K\"ahler metric in heterotic compactifications}
\label{kahlersec2}

Our first step is to derive a general formula for the matter field K\"ahler metric, in terms of the underlying
geometrical data of the Calabi-Yau manifold and the gauge bundle. The basic structure of this formula
is well-known for some time, see, for example Ref. \cite{boundaryinflation}, and our re-derivation here serves two purposes.
Firstly, we would like to fix notations and conventions so that our result is accurate, as is required for a
detailed calculation of Yukawa couplings. Secondly, we would like to explore the constraints on the matter
field K\"ahler metric which arise from four-dimensional $N = 1$ supergravity.

The starting point is the ten-dimensional $N = 1$ supergravity coupled to a ten-dimensional $E_8 \times E_8$ super Yang-Mills theory. To first order in $\alpha'$ and at the two-derivative level, the bosonic part of the action is given by Eq.~\eqref{10daction}. As in Section~\ref{n=1susy}, we consider the reduction of this action on a Calabi-Yau three-folds $X$, with Ricci-flat metric $g^{(6)}$ and a holomorphic bundle $V \rightarrow X$ with a connection $A^{(6)}$ that satisfies the Hermitian Yang-Mills equations \eqref{hermitianyangmills}. Let us introduce the K\"ahler form $J$ on $X$, related to the Ricci-flat metric $g^{(6)}$ on $X$ by $g^{(6)}_{m \overline{n}} = - i J_{m \overline{n}}$, and a basis $\lbrace J_i\rbrace$ of harmonic $(1,1)$-forms,  where $i = 1,... , h^{1,1}(X)$. The reader is reminded that $J$ and the NS two-form $B$ can be expanded as
\begin{eqnarray}
J = t^i J_i \, , \qquad B = B^{(4)} + \tau^i J_i \, ,
\end{eqnarray}
\noindent where $t^i$ are the K\"ahler moduli, $\tau^i$ are their axionic partners and $B^{(4)}$ is the four-dimensional two-form, dual to a scalar $\sigma$. In addition, we have the zero mode $\phi^{(4)}$ of the ten-dimensional dilaton $\phi$, as well as the complex structure moduli $z^a$, where $a = 1, ... , h^{2,1}(X)$. In the absence of matter fields, these bosonic fields fit into four-dimensional $N = 1$ chiral multiplets as 
\begin{eqnarray}
\label{modulichiralmultiplets}
S = e^{- 2 \phi^{(4)}} + i \sigma \, , \qquad T^i = t^i + i \tau^i \, , \qquad Z^a = z^a \, .
\end{eqnarray}
\noindent Also, it is important to notice that the Calabi-Yau volume is given by
\begin{eqnarray}
\label{cyvolume2}
\mathcal{V} = \int_X d^6 x \sqrt{ g^{(6)}} = \dfrac{1}{6} \mathcal{K}\, , \quad \mathcal{K} = d_{ijk} t^i t^j t^k \, , \quad d_{i j k} = \int_X J_i \wedge J_j \wedge J_k \, ,
\end{eqnarray}
\noindent where $d_{ijk}$ are the triple intersection numbers of $X$, thus giving the following expression for the K\"ahler moduli space metric
\begin{eqnarray}
\mathcal{G}_{ij} = - \dfrac{1}{4} \dfrac{\partial^2}{\partial t^i \partial t^j} \, \textrm{ln} \, \mathcal{K} = - \dfrac{3}{2} \left( \dfrac{\mathcal{K}_{ij}}{\mathcal{K}} - \dfrac{3}{2}\dfrac{\mathcal{K}_{i}\mathcal{K}_{j}}{\mathcal{K}^2}\right) \, ,
\end{eqnarray}
\noindent where $\mathcal{K}_i = d_{i j k} t^j t^k $ and $\mathcal{K}_{i j} = d_{i j k} t^k $. 

Next, we obtain matter fields $C^I$ by expanding the gauge field around the vacuum. The result is imported from Eq.~\eqref{expansiona} in a simplified form
\begin{eqnarray}
\label{Aexpansion}
A = A^{(6)}+C^I \nu_I \, ,
\end{eqnarray}
\noindent where $\nu_I$ are harmonic one-forms which take values in the bundle $V$. It is important to emphasise that the correct matter field metric has to be computed relative to \textit{harmonic} forms $\nu_I$ and this is, in fact, how the dependence on the Ricci-flat metric and the Hermitian Yang-Mills connection comes about. The fields $C^I$
each form the bosonic part of an $N = 1$ chiral supermultiplet. It is known that the definition of the $T^i$ superfields in Eq.~\eqref{modulichiralmultiplets} has to be adjusted in the presence of these matter fields. In the universal case with
only one T-modulus and one matter field $C$, the required correction to Eq.~\eqref{modulichiralmultiplets} has been found to be
proportional to $\vert C \vert$ (see, for example, Ref. \cite{lukas1997}). For our general case, we therefore start by modifying
the definition of the T-moduli in Eq.~\eqref{modulichiralmultiplets} by writing
\begin{eqnarray}
\label{Tagain}
T^i = t^i + i \tau^i + \alpha' \; \Gamma^i_{I J} C^I \overline{C}^J \, ,
\end{eqnarray}
\noindent where $\Gamma^i_{IJ}$ is a set of (potentially moduli-dependent) coefficients to be determined.\footnote{The dilaton superfield $S$ receives a similar correction in the presence of matter fields \cite{lukas1997}, but this arises at one-loop level and will not be of relevance here.} To our knowledge, no general expression for $\Gamma^i_{I J}$ has been obtained in the literature so far.

The kinetic terms of the above superfields derive from a K\"ahler potential of the general form
\begin{eqnarray}
\label{Kaehlercomplete}
K = - \textrm{ln} \; (S+\overline{S}) + K_{\textrm{cs}} - \textrm{ln} \; (d_{ijk}(T^i+\overline{T}{}^i)(T^j+\overline{T}{}^j)(T^k+\overline{T}{}^k)) + \alpha' G_{I J} C^I \overline{C}{}^J ,
\end{eqnarray}
\noindent where $K_{\textrm{cs}}$ is the K\"ahler potential for the complex structure moduli $Z^a$ whose explicit form is well-known but is not relevant to our present discussion and $G_{IJ}$ is the (moduli-dependent) matter field K\"ahler metric we would like to determine. The general task is now to compute the kinetic terms which result from this
K\"ahler potential, insert the definitions of $S$ in Eq.~\eqref{modulichiralmultiplets} and of $T^i$ in Eq. \eqref{Tagain} and compare the result with what has been obtained from the reduction of the ten-dimensional action \eqref{10daction}. This comparison should
lead to explicit expressions for $G_{IJ}$ and $\Gamma^i_{IJ}$.

A quick look at the K\"ahler potential \eqref{Kaehlercomplete} shows that achieving this match is by no means a trivial matter. The matter field K\"ahler metric $G_{IJ}$ depends on the $T$-moduli and, hence, the kinetic terms from \eqref{Kaehlercomplete} can
be expected to include cross-terms of the form $\partial^{\mu}t^i \partial_{\mu} C^I$. However, such cross-terms can clearly not arise
from the dimensional reduction of the 10-dimensional action \eqref{10daction} and, hence, there must be non-trivial
cancellations which involve the derivatives of $G_{IJ}$ and $\Gamma^i_{IJ}$. We find that this issue can be resolved and
indeed a complete match between the reduced ten-dimensional action \eqref{10daction} and the four-dimensional K\"ahler
potential \eqref{Kaehlercomplete} can be achieved provided the following three requirements are satisfied.

\begin{itemize}

\item The coefficients $\Gamma^i_{IJ}$ which appear in the definition \eqref{Tagain} of the $T^i$ superfields are given by
\begin{eqnarray}
\label{Gammarestated}
\Gamma^i_{IJ} = - \dfrac{1}{2} \mathcal{G}^{i j} \dfrac{\partial G_{I J}}{\partial \overline{T}{}^j} \, ,
\end{eqnarray}
\noindent where $\mathcal{G}^{ij}$ is the inverse of the K\"ahler moduli space metric $\mathcal{G}_{ij}$.

 \item The matter field K\"ahler metric is given by
\begin{eqnarray}
\label{GIJrestated}
G_{I J} = \dfrac{1}{2\mathcal{V}} \int_X \nu_I \wedge \bar{\star}_V (\nu_J) \, ,
\end{eqnarray}
\noindent where $\bar{\star}_V$ refers to a Hodge dual combined with a complex conjugation and an action of the hermitian bundle metric on $V$.

\item Since the Hodge dual on a Calabi-Yau manifold acting on a $(1, 0)$-form $\rho$ can be carried out as $\star \rho = - \tfrac{i}{2} J \wedge J \wedge \rho$, the result \eqref{GIJrestated} for the matter field K\"ahler metric can be re-written as
\begin{eqnarray}
\label{tdependence}
G_{IJ} = - \dfrac{3 i t^i t^j}{2 \mathcal{K}} \Lambda_{ijIJ} \, , \qquad \Lambda_{ijIJ} = \int_X J_i \wedge J_j \wedge \nu_I \wedge (H \overline{\nu}_J) \, ,
\end{eqnarray}
\noindent where $H$ is the hermitian bundle metric on $V$ . The final requirement for a match between the
dimensionally reduced ten-dimensional and the four-dimensional theory \eqref{Kaehlercomplete} can then be stated by
saying that the above integrals $\Lambda_{ijIJ}$ do not explicitly depend on the K\"ahler moduli $t^i$.
\end{itemize}

\noindent The above result means that the K\"ahler moduli dependence of the matter field metric is completely
determined as indicated in the first equation \eqref{tdependence}, while the remaining integrals $\Lambda_{ijIJ}$ are $t^i$-independent but can still be functions of the complex structure moduli. To our knowledge this is a new result which is of considerable relevance for the structure of the matter field K\"ahler metric and the physical Yukawa couplings. Note that the $t^i$ dependence of $G_{IJ}$ in Eq. \eqref{tdependence} is homogeneous of degree $-1$, as expected on general grounds.

It is worth noting that the K\"ahler potential \eqref{Kaehlercomplete} with the matter field K\"ahler metric as given in Eq. \eqref{tdependence} can, alternatively, also be written in the form
\begin{equation}
\label{altK}
\begin{array}{l}
K = - \textrm{ln} \; (S+\overline{S}) + K_{\textrm{cs}} - \textrm{ln} \; (d_{ijk}(T^i+\overline{T}{}^i - \gamma^i)(T^j+\overline{T}{}^j- \gamma^j)(T^k+\overline{T}{}^k- \gamma^k)) \, \\[1mm]
\gamma^i = 2 \alpha' \, \Gamma^{i}_{I J} C^I \overline{C}{}^J \, ,
\end{array}
\end{equation}
\noindent  provided that terms of higher than quadratic order in the matter field $C^I$ are neglected. This can be
seen by expanding the logarithm in Eq.~\eqref{altK} to leading order in $\gamma^i$  and by using $\tfrac{3 \mathcal{K}_i}{\mathcal{K}} \Gamma^i_{I J} = G_{I J}$. (The
latter identity follows from $\mathcal{G}^{ij} \tfrac{3 \mathcal{K}_j}{4 \mathcal{K}} = t^i$, the fact that $G_{IJ}$ is homogeneous of degree $-1$ in $t^i$ and the result \eqref{Gammarestated} for $\Gamma^i_{IJ}$). This form of the K\"ahler potential, together with the definition \eqref{Tagain} of the fields $T^i$, means that, in terms of the underlying geometrical K\"ahler moduli $t^i$, the dependence on the matter fields
$C^I$ cancels. Indeed, inserting the definition \eqref{Tagain} of the $T^i$ moduli into Eq. \eqref{altK} turns the last logarithm into $- \textrm{ln} \,(8\mathcal{K})$. That this part of the K\"ahler potential can be written as the negative logarithm of the Calabi-Yau volume is in fact expected and provides a check of our calculation.

\section{Localisation of matter field wave functions on projective spaces}
\label{kahlersec3}

As a warm-up, we first discuss wave function normalisation on $\mathbb{P}^n$ and products of projective spaces, beginning with the simplest case of $\mathbb{P}^1$. (For a related discussion, in the context of F-theory, see Ref. \cite{paltiwavefunctions}.) In doing so we have two basic motivations in mind. First, considering projective space and $\mathbb{P}^1$ in particular provides us with a toy model for the actual Calabi-Yau case which we will tackle later. From this point of view, the following discussion will provide some intuition as to when wave function localisation occurs
and when it leads to a good approximation for the normalisation integrals. On the other hand, projective
spaces and their products provide the ambient spaces for the Calabi-Yau manifolds of interest and, hence,
this chapter will be setting up some of the requisite notation and results we will be using later.

\subsection{Wave functions on $\mathbb{P}^1$}

Homogeneous coordinates on $\mathbb{P}^1$ are denoted by $x_0$, $x_1$, the affine coordinates on the patch $\lbrace x_0 \neq 0\rbrace$ by $z = x_1/x_0$ and we also define $\kappa = 1 + \vert z\vert^2$. For simplicity, we will write all quantities in terms of the affine coordinate $z$ and we will ensure they are globally well-defined by demanding the correct transformation property under the transition $z \rightarrow 1/z$. In terms of $z$, the standard Fubini-Study K\"ahler potential and K\"ahler form can be written as
\begin{eqnarray}
\label{F-SKpotential}
K = \dfrac{i}{2 \pi}\textrm{ln} \, \kappa \, , \qquad J = \partial \overline{\partial} K = \dfrac{i}{2 \pi \kappa^2} d z \wedge d \overline{z} \, .
\end{eqnarray}
\noindent Here, $J$ has the standard normalisation, that is, $\int_{\mathbb{P}^1} J =1$. The associated Fubini-Study metric is K\"ahler-Einstein and, hence, the closest analogue of a Ricci-flat Calabi-Yau metric we can hope for on $\mathbb{P}^1$.

We are interested in line bundles $L = \mathcal{O}_{\mathbb{P}^1}(k)$ on $\mathbb{P}^1$ with first Chern class $c_1(L) = k J$ on which we introduce a hermitian structure with the bundle metric and the associated (Chern) connection and field strength given by
\begin{eqnarray}
H=\kappa^{-k} \, , \qquad A = \partial \, \textrm{ln} \bar{H} = - \dfrac{k \overline{z}}{\kappa} dz \, , \qquad  F= d A = \overline{\partial} \partial \, \textrm{ln} \bar{H} = - 2 \pi i k J \, .
\end{eqnarray}
\noindent The analogue of the harmonic forms $\nu_I$ in Eq.~\eqref{Aexpansion} associated to matter fields are harmonic $L$-valued $\alpha$, that is, forms satisfying the equations
\begin{eqnarray}
\label{harmoniceqs}
\overline{\partial} \alpha = 0 \, , \qquad \partial(\bar{H}\star \alpha) = 0 \, ,
\end{eqnarray}
\noindent where the Hodge star is taken with respect to the Fubini-Study metric. We would like to compute their
normalisation integrals
\begin{eqnarray}
\label{normalisationintegral}
\langle \alpha, \beta \rangle = \int_{\mathbb{P}^1} \alpha \wedge \star (H \overline{\beta}) \, ,
\end{eqnarray}
\noindent the analogue of the matter field K\"ahler metric \eqref{GIJrestated}. These harmonic forms are in one-to-one correspondence with the bundle cohomologies $H^p(\mathbb{P}^1, L)$ and, depending on the value of $k$, we should distinguish three cases.

\begin{itemize}
\item $k \geq 0$: In this case, the only non-vanishing cohomology of $L$ is $h^0(\mathbb{P}^1, L) = k + 1$, so that the relevant harmonic forms $\alpha$ are $L$-valued zero forms. The relevant solutions to Eqs.~\eqref{harmoniceqs} are explicitly given by the degree $k$ polynomials in $z$.

\item $k = -1$: In this case, all cohomologies of $L$ vanish so there are no harmonic forms.

\item $k \leq -2$: In this case, the only non-vanishing cohomology of $L$ is $h^1(\mathbb{P}^1,L) = -k - 1$ and the
corresponding $L$-valued $(0, 1)$-forms which solve Eqs.~\eqref{harmoniceqs} can be written as $\alpha = \kappa^k h(\overline{z})d \overline{z}$, where $h$ is a polynomial of degree $-k -2$ in $\overline{z}$. In the following, it is useful to work with the monomial basis
\begin{eqnarray}
\label{monomialbasis}
\alpha_q = \kappa^k \overline{z}{}^q d \overline{z} \, , \qquad q = 0, ... , -k - 2
\end{eqnarray}
\noindent for these forms.
\end{itemize}
\noindent Given that the forms $\nu_I$ which appear in the actual reduction \eqref{Aexpansion} are $(0, 1)$-forms the most relevant case is the last one for $k \leq -2$. In this case, inserting the forms \eqref{monomialbasis} into the the normalisation integral \eqref{normalisationintegral} leads to
\begin{eqnarray}
\langle \alpha_q, \alpha_p \rangle = - i \int_{\mathbb{C}} z^q \overline{z}{}^p \kappa^k d z \wedge d \overline{z} = \dfrac{2 \pi q!}{(-k-1)...(-k-1-q)} \delta_{q p} \, .
\end{eqnarray}
\noindent In physical terminology, the integer $k$ quantifies the flux and the integer $q$ labels the families of matter
fields. It is clear that the above integrals receive their main contribution from a patch near the affine origin
$z \simeq 0$, provided that the flux $\vert k \vert$ is sufficiently large and the family number $q$ is sufficiently small. In this case, it seems that the above integrals can be approximately evaluated locally near $z \simeq 0$, by using the flat metric instead of the Fubini-Study metric as well as the corresponding flat counterparts of the bundle
metric and the harmonic forms. Formally, these flat space quantities can be obtained from the exact ones
by setting $\kappa$ to one in the expression \eqref{F-SKpotential} for the K\"ahler form and by the replacement $\kappa^k \rightarrow e^{k \vert z \vert^2}$ in the
other quantities. That is, we use the replacements
\begin{equation}
\begin{array}{l}
J  = \dfrac{i}{2 \pi \kappa^2} d z \wedge d \overline{z} \rightarrow  \dfrac{i}{2 \pi} d z \wedge d \overline{z} \, , \qquad H = \kappa^{-k} \rightarrow e^{-k\vert z\vert^2} \, , \\[3mm] \alpha_q = \kappa^k \overline{z}{}^q d \overline{z} \rightarrow e^{k \vert z\vert^2} \overline{z}{}^q d \overline{z} \, ,
\end{array}
\end{equation}
\noindent to work out the local version of the normalisation integrals, which leads to
\begin{eqnarray}
\langle \alpha_q, \alpha_p \rangle_{\textrm{loc}} = -i \int_{\mathbb{C}} z^q \overline{z}{}^p e^{k \vert z \vert^2} d z \wedge d \overline{z} = \dfrac{2 \pi q!}{(-k-1)^{q+1}} \delta_{q p} \, .
\end{eqnarray} 
\noindent For the ratio of local to exact normalisation this implies
\begin{eqnarray}
\dfrac{\langle \alpha_q, \alpha_q \rangle_{\textrm{loc}}}{\langle \alpha_q, \alpha_q \rangle} = \dfrac{(-k-1)...(-k-1-q)}{(-k-1)^{q+1}} = 1- \mathcal{O} \left( \dfrac{q^2}{-k-1}\right) \, .
\end{eqnarray}
\noindent Hence, as long as the flux $\vert k \vert$ is sufficiently large and the family number satisfies $q^2 \ll \vert k \vert$, the local versions of these integrals do indeed provide a good approximation. The above ratio expressing the accuracy of the local correction can be roughly evaluated by computing the coefficient of the $\mathcal{O} \big( \tfrac{q^2}{-k-1}\big)$ term. For example, if $\vert k \vert =11$ and $q=1$, then the ratio is $0.9$ and the error is $10 \%$. If $\vert k \vert=101$ and $q=2$, then the ratio is close to $0.96$ and the error is approximately $4 \%$ .

It is worth noting that a transformation $z \rightarrow 1/z$ to the other standard coordinate patch of $\mathbb{P}^1$ transforms the monomial basis forms $\alpha_q$ into forms of the same type but with the family number changing as $q \rightarrow (-k - 1) - q$. This means that families with a large family number $q$ close to $-k - 1$ in the patch $\lbrace x_0 \neq 0\rbrace$ acquire a small family number when transformed to the patch $\lbrace x_1 \neq 0\rbrace$ and, hence, localise at the affine origin of this patch, that is near $z = \infty$. From this point of view it is not surprising that families with large $q$ in the patch $\lbrace x_0 \neq 0\rbrace$ cannot be dealt with by a local calculation near $z \simeq 0$. Instead, for such modes, we can carry out a local calculation analogous to the above one but near the affine origin of the patch $\lbrace x_1 \neq 0 \rbrace$.

\vspace{3mm}

In summary, the harmonic bundle valued $(0, 1)$-forms for $L = \mathcal{O}_{\mathbb{P}^1}(k)$, where $k \leq -2$, are given by $\alpha_q$ as in
Eq.~\eqref{monomialbasis}. For sufficiently large flux $\vert k \vert$, the modes with small family number $q$ localise near the affine origin of the patch $\lbrace x_0 \neq 0 \rbrace$, that is at $z \simeq 0$ and their normalisation can be obtained from a local calculation near this point. The modes with large family number $q$ localise near the affine origin of the other patch $\lbrace x_1 \neq 0\rbrace$, that is, near $z = \infty$ and their normalisation can be obtained by a similar local calculation around this point.

\subsection{Wave functions on products of projective spaces}
\label{introducingP1P3}
The previous discussion for line bundles on $\mathbb{P}^1$ can be straightforwardly generalised to line bundles on
arbitrary products of projective spaces. For the sake of keeping notation simple, we will now illustrate this
for the case of $\mathcal{A} = \mathbb{P}^1 \times \mathbb{P}^3$ which is, in fact, the ambient space of the Calabi-Yau manifold on which we focus later. The situation for general products of projective spaces is easily inferred from this discussion.

Homogeneous coordinates on $\mathcal{A} = \mathbb{P}^1 \times \mathbb{P}^3$ are denoted by $x_0$, $x_1$ for the $\mathbb{P}^1$ factor and by $y_0$, $y_1$, $y_2$, $y_3$ for $\mathbb{P}^3$. The associated affine coordinates on the patch $\lbrace x_0 \neq 0, y_0 \neq 0\rbrace$ are $z_1 = x_1/x_0$ and $z_{\alpha+1} = y_{\alpha}/y_0$ for $\alpha = 1, 2, 3$, and we define $\kappa_1 = 1 + \vert z_1\vert^2$ and $\kappa_2 = 1 + \sum_{\alpha=2}^4 \vert z_{\alpha} \vert^2$. The Fubini-Study K\"ahler forms for the two projective factors are\footnote{As in Chapters~\ref{tetraquadricchapter} and \ref{chaptern>1codimension}, we will denote quantities defined on the “ambient space" $\mathcal{A}$ by a hat, in order to distinguish them from their Calabi-Yau counterparts to be introduced later.}
\begin{equation}
\begin{array}{l}
\hat{J}_1 = \dfrac{i}{2 \pi} \partial \overline{\partial} \, \textrm{ln} \kappa_1 = \dfrac{i}{2 \pi \kappa_1^2} d z_1 \wedge d \overline{z}_1 \, , \\[2mm]
\hat{J}_2 = \dfrac{i}{2 \pi} \partial \overline{\partial} \, \textrm{ln} \kappa_2  = \dfrac{i}{2 \pi \kappa_2^2} \sum_{\alpha, \beta = 2}^4 (\kappa_2 \delta_{\alpha \beta} - \overline{z}_{\alpha} z_{\beta}) d z_{\alpha} \wedge d \overline{z}_{\beta} \, ,
\end{array}
\end{equation}
\noindent and, more generally, we can introduce the K\"ahler forms
\begin{equation}
\hat{J} = t_1\hat{J}_1 + t_2\hat{J}_2 \, ,
\end{equation}
\noindent with K\"ahler parameters $t_1 > 0$, $t_2 > 0$ on $\mathcal{A}$. Line bundles $\hat{L} = \mathcal{O}_{\mathcal{A}}(k_1, k_2)$ with first Chern class $c_1(\hat{L}) = k_1\hat{J}_1+k_2\hat{J}_2$ can be equipped with the hermitian bundle metric
\begin{eqnarray}
\label{hermitianmetriceq3.12}
\hat{H}= \kappa_1^{-k_1} \kappa_2^{-k_2} \quad \Rightarrow \quad \hat{F} = \overline{\partial} \partial \, \textrm{ln} \bar{\hat{H}} = - 2 \pi i (k_1\hat{J}_1+k_2\hat{J}_2) \, .
\end{eqnarray}
\noindent Specifically, we are interested in those line bundles $\hat{L}$ with a non-vanishing first cohomology which are precisely those with $k_1 \leq -2$ and $k_2 \geq 0$. In these cases (see also Eq.~\eqref{generalizedbott}),
\begin{eqnarray}
\label{dimeq3.13}
h^1(\mathcal{A},\mathcal{O}_{\mathcal{A}}(k_1,k_2)) = (- k_1 -1) \dfrac{(k_2+3)(k_2+2)(k_2+1)}{6} \, ,
\end{eqnarray}
\noindent and a basis for the associated harmonic $\hat{L}$-valued $(0, 1)$-forms is provided by
\begin{eqnarray}
\label{basiseq3.14}
\hat{\nu}_{\mathbf{q}} = \kappa_1^{k_1} \overline{z}{}^{\hat{q}_1}_1 z^{\hat{q}_2}_2 z^{\hat{q}_3}_3 z^{\hat{q}_4}_4 d \overline{z}_1 \, ,
\end{eqnarray}
\noindent where $\hat{\mathbf{q}} = (\hat{q}_1, \hat{q}_2, \hat{q}_3, \hat{q}_4)$ is a positive integer vector which labels the families and whose entries are constrained by $\hat{q}_1 = 0, ..., -k_1-2$ and $\hat{q}_2+\hat{q}_3+\hat{q}_4 \leq k_2$. Given these quantities, the integrand of the
normalisation integral is proportional to
\begin{eqnarray}
\hat{\nu}_{\hat{\mathbf{q}}} \wedge \star (\hat{H} \bar{\hat{\nu}}_{\hat{\mathbf{q}}} ) \sim \kappa_1^{k_1} \kappa_2^{-k_2} \prod_{\alpha=1}^4 \vert z_{\alpha}\vert^{2 \hat{q}_{\alpha}} \, .
\end{eqnarray}
\noindent Hence, provided the fluxes $\vert k_1 \vert$ and $k_2$ are sufficiently large and the family numbers $\hat{q}_{\alpha}$ sufficiently small, we expect localisation on a patch $\hat{U}$ around the affine origin $z_{\alpha} \simeq 0$. In this case, we can again work with the flat limit where the above quantities turn into
\begin{equation}
\label{localapprox}
\begin{array}{lll}
\hat{J}_1 \rightarrow \dfrac{i}{2\pi} dz_1 \wedge d \overline{z}_1 \, , & \qquad & \hat{J}_2 \rightarrow \dfrac{i}{2\pi} \sum_{\alpha=2}^4 dz_{\alpha} \wedge d \overline{z}_{\alpha} \, ,\\[3mm]
\hat{H} \rightarrow e^{-k_1 \vert z_1 \vert^2 -k_2 \sum_{\alpha=2}^4 \vert z_{\alpha} \vert^2} \, , & \qquad & \hat{\nu}_{\mathbf{q}} \rightarrow e^{k_1 \vert z_1 \vert^2} \overline{z}{}^{\hat{q}_1}_1 z^{\hat{q}_2}_2 z^{\hat{q}_3}_3 z^{\hat{q}_4}_4 d \overline{z}_1 \, .
\end{array}
\end{equation}
\noindent A few general conclusions can be drawn from this. First, localisation near a point in $\mathcal{A}$ does require all fluxes $\vert k_i\vert$ to be large. If one of the fluxes is not large then localisation will happen near a higher-dimensional variety in $\mathcal{A}$. For example, if $\vert k_1\vert$ is not large then the wave function will localise near $\mathbb{P}^1$ times a point in $\mathbb{P}^3$. We note that such a partial localisation may actually be sufficient when we come to discuss Calabi-Yau manifolds embedded in $\mathcal{A}$. For example, localisation near a curve in $\mathcal{A}$  will typically lead to localisation near a point on a Calabi-Yau hyper-surface embedded in $\mathcal{A}$. Secondly, provided all $\vert k_i\vert$ are indeed large, localisation on $\hat{U}$ near the affine origin $z_{\alpha} \simeq 0$, for $\alpha = 1, 2, 3, 4$, requires all $\hat{q}_{\alpha}$ to be sufficiently small. If a certain $\hat{q}_{\alpha}$ is large, localisation may still arise near another point in $\mathcal{A}$. For example,
if $\hat{q}_{1}$ is large while the other $\hat{q}_{\alpha}$ are small, then localisation occurs near $z_1 = \infty$, $z_2 = z_3 = z_4 = 0$.

\section{A local Calabi-Yau calculation}
\label{kahlersec4}

So far, we have approached the problem of computing wave function normalisations on Calabi-Yau manifolds from the viewpoint of the prospective ambient embedding spaces. In this section, we will take the complementary point of view and carry out a local calculation on a Calabi-Yau manifold. In the next section, we will show how to connect this local Calabi-Yau calculation with the ambient space point of
view in order to obtain results as functions of globally defined moduli.

We start with a Calabi-Yau three-fold $X$ and a line bundle $L \rightarrow X$ with a non-vanishing first cohomology
and associated $L$-valued harmonic $(0, 1)$-forms. Our goal is to determine the normalisation of these harmonic forms by a local calculation, assuming, at this stage, that localisation indeed occurs. To do this, we
focus on a patch $U \subset X$ with local complex coordinates $Z_a$, where $a = 1, 2, 3$, chosen such that the K\"ahler
form $J$, associated to the Ricci-flat Calabi-Yau metric, is locally on $U$ well approximated by\footnote{We will denote local quantities, defined on the patch $U$, by script symbols.}
\begin{eqnarray}
\label{eqlocalJ}
\mathcal{J} = \dfrac{i}{2\pi} \sum_{a=1}^3 \beta_a d Z_a \wedge d \bar{Z}_a \, ,
\end{eqnarray}
\noindent where the $\beta_a$ are positive constants. (It is, of course, possible to set $\beta_a$ equal to one by further coordinate re-definitions but, for later purposes, we find it useful to keep these explicitly.) On $U$, we can approximate
the hermitian bundle metric $H$ and the associated field strength $F$ of $L$ by
\begin{eqnarray}
\label{eqlocalH}
\mathcal{H} = e^{-\sum_{a=1}^3 K_a \vert Z_a\vert^2} \quad \Rightarrow \quad \mathcal{F} = \overline{\partial} \partial \, \textrm{ln} \, \bar{\mathcal{H}} = \sum_{a=1}^3 K_a d Z_a \wedge d \bar{Z}_a \, ,
\end{eqnarray}
\noindent where $K_a$ are constants which will ultimately become functions of the Calabi-Yau moduli. The Hermitian
Yang-Mills equation, $J \wedge J \wedge F = 0$, should be satisfied locally which leads to
\begin{eqnarray}
\mathcal{J} \wedge \mathcal{J} \wedge \mathcal{F} \quad  \Leftrightarrow \quad \beta_1 \beta_2 K_3 + \beta_1 \beta_3 K_2 +\beta_2 \beta_3 K_1 = 0 \, .
\end{eqnarray}
\noindent The resulting equation for the $K_a$ will translate into a constraint on the Calabi-Yau moduli in a way that
will become more explicit later. For now we should note that it implies not all $K_a$ can have the same sign
(given that the $\beta_a$ need to be positive). Consider harmonic $(0, 1)$-forms $v \in H^1(X,L)$. On $U$, they are
approximated by $(0, 1)$-forms $\nu$, which must satisfy the local version of the harmonic equations
\begin{eqnarray}
\overline{\partial} \nu = 0 \, , \qquad \mathcal{J} \wedge \mathcal{J} \wedge \partial (\mathcal{H} \nu) = 0 \, .
\end{eqnarray}
\noindent In analogy with the projective case, specifically Eq.~\eqref{localapprox}, we assume that $K_1 < 0$ and $K_2, K_3 > 0$. Whether these sign choices are actually realised cannot be checked locally but requires making contact with
the global picture -- we will come back to this later. If they are, potentially localising solutions to these equations are of the form $\nu = e^{K_1\vert Z_1 \vert^2} P(\bar{Z}_1, Z_2, Z_3)d \bar{Z}_1$, where $P$ is an arbitrary function of the variables indicated. Localisation of these solution still depends on the precise form of the function $P$, which cannot
be determined from a local calculation. We will return to this issue in the next section when we discuss
the relation to the global picture. For now, we take a practical approach and work with a monomial basis
of solutions given by
\begin{eqnarray}
\label{monomialbasis2}
\nu_{\mathbf{q}} = e^{K_1 \vert Z_1 \vert^2} \bar{Z}{}^{q_1}_1 Z_2^{q_2} Z_3^{q_3} d \bar{Z}_1 \, ,
\end{eqnarray}
\noindent where $\mathbf{q} = (q_1, q_2, q_3)$ is a vector with non-negative integers. The normalisation of these monomial solutions can be explicitly computed and is given by
\begin{align}
M_{\mathbf{q},\mathbf{p}}& \; := \; \langle \nu_{\mathbf{q}}, \nu_{\mathbf{p}}\rangle_{\textrm{loc}} = \int_U \nu_{\mathbf{q}} \wedge \star (\mathcal{H} \overline{\nu}_{\mathbf{p}}) = \dfrac{i}{2} \delta_{\mathbf{q},\mathbf{p}} \int_U \mathcal{J} \wedge \mathcal{J} \wedge \nu_{\mathbf{q}} \wedge (\mathcal{H} \overline{\nu}_{\mathbf{q}}) \notag \\ &\; \simeq \; \dfrac{i}{4 \pi^2} \beta_2 \beta_3 \delta_{\mathbf{q},\mathbf{p}} \prod_{a=1}^3 \int_{\mathbb{C}} d Z_a \wedge d \bar{Z}_a \vert Z_a \vert^{2 q_a} e^{-\vert K_a\vert \vert Z_a\vert^2} \, .
\end{align}
\noindent After performing the integration, we find for the locally-computed normalisation
\begin{eqnarray}
\label{Mlocalresult}
M_{\mathbf{q},\mathbf{p}} = \langle \nu_{\mathbf{q}}, \nu_{\mathbf{p}}\rangle_{\textrm{loc}} = 2 \pi \beta_2 \beta_3 \delta_{\mathbf{q},\mathbf{p}} \prod_{a=1}^3 q_a! \, \vert K_a \vert^{ - q_a - 1} \, .
\end{eqnarray}
\noindent The appearance of the exponential in each of the integrals in the second line indicates that there is indeed
a chance for localisation to occur. However, the validity and practical usefulness of this result depends on a
number of factors which are impossible to determine in the local picture. First of all, we should indeed have
$K_1 < 0$ and $K_2, K_3 > 0$ for localisation to happen, but these conditions can only be verified by relating
to the global picture. Secondly, families are defined as cohomology classes in $H^1(X, L)$ and at this stage
it is not clear precisely how these relate to the monomial basis forms \eqref{monomialbasis2}. The above calculation shows that the smaller the integers in $\mathbf{q} = (q_1, q_2, q_3)$ the better the localisation and this ties in with the result on projective spaces in the previous section. Finding the relation between the elements of $H^1(X, L)$ and
the local basis forms $\nu_{\mathbf{q}}$ is, therefore, crucial in deciding the validity and accuracy of the approximation
for the physical families. Finally, we would like to express the local result \eqref{Mlocalresult} in terms of the properly
defined global Calabi-Yau moduli. We will now address these issues by relating the above local calculation to the full Calabi-Yau manifold.

\section{Relating local and global quantities}
\label{kahlersec5}
We will start by relating the local quantities which have entered the previous calculation to global quantities
on the Calabi-Yau manifold, starting with the K\"ahler form and the connection on the bundle and then
proceeding to bundle-valued forms. This will allows us to express the result \eqref{Mlocalresult} for the wave function
normalisation in terms of properly defined moduli.

\subsection{K\"ahler form and connection}
We begin, somewhat generally, with a Calabi-Yau three-fold $X$, a basis $\lbrace J_i \rbrace$ of its second cohomology, where $i = 1,... , h^{1,1}(X)$, and K\"ahler forms
\begin{eqnarray}
\label{kaehlerconeagain}
J=\sum_i t^i J_i \, ,
\end{eqnarray}
\noindent with the K\"ahler moduli $\mathbf{t} = (t^i)$ restricted to the K\"ahler cone. Further, we assume that all the forms $J_i$, and, hence, $J$ are chosen to be harmonic relative to the Ricci-flat metric on $X$ specified by the K\"ahler class $[J]$. Note that, despite what Eq. \eqref{kaehlerconeagain} might seem to suggest, the harmonic forms $J_i$ are typically $t^i$-dependent – all we know is that their cohomology classes $[J_i]$ do not change with the K\"ahler class, so they are allowed to vary by exact forms.

On a small patch $U \subset X$, we would like to introduce the forms $\mathcal{J}_i$, where $i = 1, . . . , h^{1,1}(X)$, and
\begin{eqnarray}
\label{kaehlerconeagain2}
\mathcal{J}=\sum_i t^i \mathcal{J}_i \, ,
\end{eqnarray}
\noindent which are local $(1, 1)$-forms with constant coefficients which approximate their global counterparts $J_i$ and $J$ on $U$. How are these global and local forms related? We first note that the top forms $J \wedge J \wedge J$ and $J_i \wedge J \wedge J$ are harmonic and must therefore be proportional
\begin{eqnarray}
\label{relation5.3}
J_i \wedge J \wedge J = c_i (\mathbf{t}) J \wedge J \wedge J \, ,
\end{eqnarray}
\noindent where $c_i (\mathbf{t})$ are functions of the K\"ahler moduli but independent of the coordinates of $X$. By inserting Eq.~\eqref{kaehlerconeagain} and integrating over $X$ we can easily compute these constants as
\begin{eqnarray}
\label{Ki/Kratio}
c_i (\mathbf{t}) = \dfrac{\mathcal{K}_i}{\mathcal{K}} \, ,
\end{eqnarray}
\noindent where the quantities $\mathcal{K}$ and $\mathcal{K}_i$ were defined in and around Eq. \eqref{cyvolume2}. On the other hand, the relation \eqref{relation5.3} holds point-wise and, hence, has a local counterpart
\begin{eqnarray}
\label{relation5.5}
\mathcal{J}_i \wedge \mathcal{J} \wedge \mathcal{J} = c_i (\mathbf{t}) \mathcal{J} \wedge \mathcal{J} \wedge \mathcal{J} \, ,
\end{eqnarray}
\noindent which must involve the same constants $c_i (\mathbf{t})$. Inserting flat forms into Eq.~\eqref{relation5.5} then allows us to determine
the $c_i (\mathbf{t})$ in terms of the parameters in these forms and equating these expressions to the global result \eqref{Ki/Kratio} leads to constraints on the local forms $\mathcal{J}_i$.

This global-local correspondence has an immediate implication for bundles on $X$ and their local counterparts on $U$. Consider a line bundle $L \rightarrow X$ with first Chern class $c_1(L) = \sum_i k^iJ_i$ and field strength $F = -2\pi i \sum_i k^i J_i$. Then, for the local version $\mathcal{F} = -2\pi i \sum_i k^i \mathcal{J}_i$ of the field strength we find, using Eqs. \eqref{Ki/Kratio} and \eqref{relation5.5}, that
\begin{eqnarray}
\label{relation5.6}
\mathcal{F} \wedge \mathcal{J} \wedge \mathcal{J} = - 2 \pi i\, \dfrac{k^i \mathcal{K}_i}{\mathcal{K}} \mathcal{J} \wedge \mathcal{J} \wedge \mathcal{J}
\end{eqnarray}
\noindent and, hence, that the local version of the Hermitian Yang-Mills equation is satisfied as long as the slope $\mu(L)= k^i \mathcal{K}_i$ of $L$ vanishes.

To work out the above global-local correspondence more explicitly, we consider a case with two K\"ahler
moduli, so $h^{1,1}(X) = 2$. In this case, we can choose complex coordinates $z_a$, where $a = 1, 2, 3$, on the patch
$U \subset X$ such that
\begin{equation}
\label{eq5.7J}
\begin{array}{l}
\mathcal{J}_1 = \dfrac{i}{2\pi} \sum_{a=1}^3 \lambda_a d z_a \wedge d \overline{z}_a \, , \qquad \mathcal{J}_2 = \dfrac{i}{2\pi} \sum_{a=1}^3  d z_a \wedge d \overline{z}_a \, , 
\\[4mm]
\mathcal{J} = \dfrac{i}{2\pi} \sum_{a=1}^3 (\lambda_a t_1 + t_2) d z_a \wedge d \overline{z}_a \, , 
\end{array}
\end{equation}
\noindent where the $\lambda_a$ are constants. (More specifically, starting with two arbitrary $(1, 1)$-forms $\mathcal{J}_1$ and $\mathcal{J}_2$ with constant coefficients, by standard linear algebra, we can always diagonalise $\mathcal{J}_2$ into “unit matrix form" and then further diagonalise $\mathcal{J}_1$ without affecting $\mathcal{J}_2$.) Inserting the above forms into Eq. \eqref{relation5.5} gives
\begin{eqnarray}
\label{coef5.8}
c_1(\mathbf{t}) = \dfrac{\sum_a \lambda_a \prod_{b \neq a} (\lambda_b t_1+t_2)}{3 \prod_c (\lambda_c t_1 + t_2)} \, , \qquad c_2(\mathbf{t}) =\dfrac{\sum_a \prod_{b \neq a} (\lambda_b t_1+t_2)}{3 \prod_c (\lambda_c t_1 + t_2)} \, ,
\end{eqnarray}
\noindent and equating these results to the global ones in Eq.~\eqref{Ki/Kratio} imposes constraints on the unknown local coefficients $\lambda_a$. However, it is not obvious that the $\lambda_a$ are K\"ahler moduli independent, particularly since the forms $J_i$ do, in general, depend on K\"ahler moduli. In the following, we will assume that this is indeed the case, although we do not, at present, have a clear-cut proof. There are two pieces of evidence which
support this assumption. First, it is not obvious that equating \eqref{coef5.8} with \eqref{Ki/Kratio} allows for a solution with constant $\lambda_a$ (valid for all $\mathbf{t}$) but we find that, in all cases which we have checked, that it does. Secondly, it is hard to see how a local calculation of the integrals in Eq. \eqref{tdependence} can lead to K\"ahler moduli independent results for $\Lambda_{ijIJ}$, as four-dimensional supersymmetry demands, if the $\lambda_a$ are $t^i$-dependent. In the following, we will proceed on the assumption that the $\lambda_a$ are indeed $t^i$-independent.

\subsection{An example}

To complete the above calculation we should consider a specific Calabi-Yau manifold. As before, we focus
on the ambient space $\mathcal{A} = \mathbb{P}^1 \times \mathbb{P}^3$, discussed in Section \ref{introducingP1P3}, and use the same notation for coordinates, K\"ahler forms and K\"ahler potentials as introduced there. The Calabi-Yau hyper-surfaces $X \subset \mathcal{A}$ we would
like to consider are then defined as the zero loci of bi-degree $(2, 4)$ polynomials $p$, that is sections of the
bundle $\hat{N} = \mathcal{O}_{\mathcal{A}}(2, 4)$. This manifold has Hodge numbers $h^{1,1}(X) = 2$, $h^{2,1}(X) = 86$ and Euler number $\eta(X) = -168$. Its second cohomology is spanned by the restrictions $\hat{J}_i\vert_X$, where $i = 1, 2$, of the two ambient space K\"ahler forms and, relative to this basis, the second Chern class of the tangent bundle is $c_2(T X) = (24, 44)$. The K\"ahler class on X can be parametrised by the restricted ambient space K\"ahler
forms
\begin{eqnarray}
\hat{J}\vert_X = t_1 \hat{J}_1\vert_X +  t_2 \hat{J}_2\vert_X \, ,
\end{eqnarray}
\noindent where $t_1, t_2 > 0$ are the two K\"ahler parameters. Of course neither of these forms is harmonic relative to the Ricci-flat metric on $X$ associated to the class $[\hat{J}\vert_X]$ (as they are obtained by restricting the ambient space Fubini-Study K\"ahler forms) but there exist forms $J_i$ and $J$ in the same cohomology classes which are. In other words, $J$ and $J_i$ are the harmonic forms introduced in Eq.~\eqref{kaehlerconeagain} and we demand that their cohomology classes satisfy $[J]=[\hat{J}\vert_X]$, $[J_i]=[\hat{J}_i\vert_X]$.

The non-vanishing triple intersection numbers of this manifold are given by
\begin{eqnarray}
\label{exampleintnumbers}
d_{122}=4\, , \quad d_{222} = 2 \quad \Rightarrow \quad \mathcal{K} = d_{ijk} t^i t^j t^k = 2 t_2^2 (6 t_1 +t_2) \, .
\end{eqnarray}
\noindent Inserting these results into Eq.~\eqref{Ki/Kratio} we find
\begin{eqnarray}
c_1 (\mathbf{t}) = \dfrac{2}{6 t_1 + t_2} \, , \qquad c_2 (\mathbf{t}) = \dfrac{4 t_1+t_2}{t_2 (6 t_1+t_2)} \, ,
\end{eqnarray}
\noindent and equating these expressions to the local results \eqref{coef5.8} leads to the solution
\begin{eqnarray}
\lambda_1 = 6\, , \qquad \lambda_2=\lambda_3 =0 \, ,
\end{eqnarray}
\noindent which is unique, up to permutations of the coordinates $z_a$. This means, from Eqs.~\eqref{eq5.7J}, the local forms $\mathcal{J}_i$ and $\mathcal{J}$ can (after another coordinate re-scaling $z_1 \rightarrow z_1/\sqrt{6}$) be written as
\begin{align}
\mathcal{J}_1 & = \dfrac{i}{2\pi} d z_1 \wedge d \overline{z}_1 \, , \\
\mathcal{J}_2 & = \dfrac{i}{2\pi} \left( \dfrac{1}{6} d z_1 \wedge d \overline{z}_1 + d z_2 \wedge d \overline{z}_2 + d z_3 \wedge d \overline{z}_3  \right) \, , \\
\label{Jeq5.15}
\mathcal{J} & = \dfrac{i}{2\pi} \left(t_1 d z_1 \wedge d \overline{z}_1 + t_2 \left( \dfrac{1}{6} d z_1 \wedge d \overline{z}_1 + d z_2 \wedge d \overline{z}_2 + d z_3 \wedge d \overline{z}_3  \right)\right) \, .
\end{align}
\noindent We note that $\mathcal{J}$ is of the form \eqref{eqlocalJ} used in our local calculation and we can match expressions by setting $z_a = Z_a$ and
\begin{eqnarray}
\label{beta123}
\beta_1 = t_1 + \dfrac{1}{6} t_2\, , \qquad \beta_2 = \beta_3 = t_2 \, .
\end{eqnarray}
\noindent Another interesting observation is that these forms satisfy
\begin{eqnarray}
\mathcal{J}_i \wedge \mathcal{J}_j \wedge \mathcal{J}_k = - \dfrac{1}{16 \pi^3} d_{ijk} \bigwedge_{a=1}^3 d z_a \wedge d \overline{z}_a \, ,
\end{eqnarray}
\noindent where $d_{ijk}$ are the intersection numbers \eqref{exampleintnumbers} of the manifold in question, that is, our local forms “intersect" on the global intersection numbers. They also relate in an interesting way to the ambient space K\"ahler forms $\hat{J}_i$. So far, we have considered an arbitrary patch $U$ on $X$, but from now on let us focus on a specific choice, starting with the ambient space patch $\hat{U} \subset \mathcal{A}$ near the affine origin $z_{\alpha} \simeq 0$. This patch is of obvious
interest since we know from the ambient space discussion in Section \ref{introducingP1P3} that some wave functions localise on it. If it is sufficiently small, the defining equation of the Calabi-Yau manifold on $\hat{U}$ can be approximated by
\begin{eqnarray}
\label{polyapprox}
p = p_0 + \sum_{\alpha=1}^4 p_{\alpha} z_{\alpha} + \mathcal{O}(z^2) \, ,
\end{eqnarray}
\noindent where $p_0$ and $p_{\alpha}$ are some of the parameters in $p$. It is possible, by linear transformations of the homogeneous coordinates on $\mathbb{P}^1$ and $\mathbb{P}^3$, to eliminate the $p_0$ term and, in the following, we assume that this has been done. Then, the Calabi-Yau manifold $X = \lbrace p = 0\rbrace$ intersects the patch $\hat{U}$ at the affine origin and near it $X$ is approximately given by the hyper-plane equation $\sum_{\alpha=1}^4 p_{\alpha}z_{\alpha} = 0$. By a further linear re-definition of coordinates on the $\mathbb{P}^3$ factor of the ambient space, this equation can be brought into the simpler form
\begin{eqnarray}
\label{simpleeq5.19}
z_4 = a z_1 \, ,
\end{eqnarray}
\noindent where $a$ is a constant. If we restrict the flat versions of the ambient space K\"ahler forms, as given in
Eq.~\eqref{localapprox}, to $U$ using Eq.~\eqref{simpleeq5.19}, we find that
\begin{eqnarray}
\hat{J}_i\vert_U = \mathcal{J}_i \, ,
\end{eqnarray}
\noindent provided we set $a=1/\sqrt{6}$. This means on the patch $U$ we understand the relation between ambient space K\"ahler forms $\hat{J}_i$, local K\"ahler forms $\mathcal{J}_i$ and their global counterparts $J_i$ on $X$.

We can now extend this correspondence to (line) bundles and their connections. As in Section~\ref{introducingP1P3}, we
consider line bundles $\hat{L} = \mathcal{O}_{\mathcal{A}}(k_1, k_2)$ and we restrict these to line bundles $L = \mathcal{O}_X (k_1, k_2) := \hat{L}\vert_X$ on the
Calabi-Yau manifold $X$. (Of course, the line bundle $L$ should be thought of as merely part of the full
vector bundle of the compactification in question.) The hermitian bundle metric $\hat{H}$ for $\hat{L}$ was given in
Eq.~\eqref{hermitianmetriceq3.12} and its local approximation on $\hat{U}$ in Eq.~\eqref{localapprox}. If we restrict this local bundle metric on $\hat{U}$ to $U$, using the defining equation \eqref{simpleeq5.19} with $a = 1/\sqrt{6}$, we find
\begin{equation}
\begin{array}{c}
\mathcal{H} = \hat{H}\vert_U = \textrm{exp} \, (-(k_1 + k_2/6)\vert z_1\vert^2 - k_2\vert z_2\vert^2 - k_2 \vert z_3 \vert^2) \\[1mm]
\Downarrow \\[1mm]
\mathcal{F} = \overline{\partial} \partial \, \textrm{ln} \,\bar{\mathcal{H}} = - 2 \pi i (k_1 \mathcal{J}_1 + k_2 \mathcal{J}_2) \, .
\end{array}
\end{equation}
\noindent We note that this expression of $\mathcal{H}$ is of the general form \eqref{eqlocalH} used in the local calculation, provided we set $z_a = Z_a$ and identify
\begin{eqnarray}
\label{identificationeq5.22}
K_1 = k_1 + \dfrac{1}{6} k_2 \, , \qquad K_2 = K_3 = k_2 \, .
\end{eqnarray}
\noindent 
From the discussion around Eq.~\eqref{relation5.6} we also conclude that the Hermitian Yang-Mills equation is locally
satisfied for $\mathcal{F}$, provided that the slope $\mu(L) = d_{ijk} k^i t^j t^k = 2t_2(2k_1 t_2 + k_2(4t_1 + t_2))$ vanishes. As usual, this is the case on a certain sub-locus of K\"ahler moduli space, provided that $k_1$ and $k_2$ have opposite signs.

\subsection{Wave functions and the matter field K\"ahler metric}

As the last step, we should work out the global-local correspondence for wave functions. As in Section \ref{introducingP1P3}, we consider line bundles $\hat{L} = \mathcal{O}_{\mathcal{A}}(k_1, k_2)$ with $k_1 \leq -2$ and $k_2 > 0$ with a non-zero first cohomology $H^1(\mathcal{A},\hat{L})$, whose dimension is given in Eq.~\eqref{dimeq3.13} and with harmonic basis forms $\hat{\nu}_{\hat{\mathbf{q}}}$ introduced in Eq.~\eqref{basiseq3.14}. These line bundles restrict to line bundle $L = \mathcal{O}_X (k_1, k_2) := \hat{L}\vert_X$ on the Calabi-Yau manifold $X$ with a
non-vanishing first cohomology (see, for example, Ref.~\cite{yukunification})
\begin{eqnarray}
\label{quotienteq5.23}
H^1(X,L)\cong \dfrac{H^1(\mathcal{A}, \hat{L})}{p(H^1(\mathcal{A},\hat{N}^* \otimes \hat{L}))} \; .
\end{eqnarray}
\noindent Explicit representatives for this cohomology can be obtained by restrictions $\hat{\nu}_{\hat{\mathbf{q}}}\vert_X$, although these forms are
not necessarily harmonic with respect to any particular metric. (Also, they have to be suitably identified
due to the quotient in Eq.~\eqref{quotienteq5.23}. As long as $k_2 < 4$, the cohomology in the denominator of Eq.~\eqref{quotienteq5.23} vanishes, so that the quotient is trivial and the restrictions $\hat{\nu}_{\hat{\mathbf{q}}}\vert_X$ form a basis of $H^1(X, L)$ as stands.) Finally, we have the monomial basis $\nu_{\mathbf{q}}$ of locally harmonic forms defined in Eq.~\eqref{monomialbasis2}. In summary, we are dealing with three sets of basis forms and their linear combinations, namely
\begin{equation}
\begin{array}{ll}
\hat{\nu}_{\hat{\mathbf{q}}} = e^{k_1 \vert z_1 \vert^2} \overline{z}{}^{\hat{q}_1}_1 z^{\hat{q}_2}_2 z^{\hat{q}_3}_3 z^{\hat{q}_4}_4 d \overline{z}_1 \, , &  \qquad \hat{\nu}(\hat{\mathbf{a}})= \sum_{\hat{\mathbf{q}}} \hat{a}_{\hat{\mathbf{q}}} \hat{\nu}_{\hat{\mathbf{q}}} \, ,
\\[1mm]
\tilde{\nu}_{\tilde{\mathbf{q}}} = e^{k_1 \vert z_1 \vert^2} \overline{z}{}^{\tilde{q}_1}_1 z^{\tilde{q}_2}_2 z^{\tilde{q}_3}_3 z^{\tilde{q}_4}_1 d \overline{z}_1 \, , & \qquad \tilde{\nu}(\tilde{\mathbf{a}})= \sum_{\tilde{\mathbf{q}}} \tilde{a}_{\tilde{\mathbf{q}}} \tilde{\nu}_{\tilde{\mathbf{q}}} \, , \\[1mm]
\nu_{\mathbf{q}} = e^{K_1 \vert z_1 \vert^2} \overline{z}{}^{q_1}_1 z_2^{q_2} z_3^{q_3} d \overline{z}_1 \, , & \qquad \nu(\mathbf{a})= \sum_{\mathbf{q}} a_{\mathbf{q}} \nu_{\mathbf{q}} \, .
\end{array}
\end{equation}
\noindent To be clear, hatted wave functions $\hat{\nu}_{\hat{\mathbf{q}}}$ are defined on the ambient space $\mathcal{A}$, wave functions $\tilde{\nu}_{\tilde{\mathbf{q}}}$ refer to their
restrictions to the Calabi-Yau patch $U$ and the $\nu_{\mathbf{q}}$ are the harmonic wave functions on the patch $U$.

Recall that we need $K_1 < 0$ as a necessary condition for the harmonic solutions $\nu_{\mathbf{q}}$ to have a finite norm and, by virtue of the identification \eqref{identificationeq5.22}, this translates into
\begin{eqnarray}
\label{conditioneq5.25}
K_1 < 0 \quad \Leftrightarrow \quad -k_1>\dfrac{k_2}{6} \, .
\end{eqnarray}
\noindent Hence, for this particular example, the condition $K_1 < 0$ is not moduli-dependent and can be satisfied by
a suitable choice of line bundle.

We would like to determine the relation between the above three types of forms, or, equivalently, the
relation between the coefficients $\hat{\mathbf{a}}$, $\tilde{\mathbf{a}}$ and $\mathbf{a}$, given that $\tilde{\nu}(\tilde{\mathbf{a}}) = \hat{\nu}(\hat{\mathbf{a}})\vert_U$ are related by restriction and that $\tilde{\nu}(\tilde{\mathbf{a}})$ and $\nu(\mathbf{a})$ are in the same cohomology class, so must differ by a $\overline{\partial}$-exact $L$-valued $(0, 1)$-form.

The first of these correspondences, between $\hat{\mathbf{a}}$ and $\tilde{\mathbf{a}}$, is easy to establish. Given the relation is by restriction, there is a matrix $\mathcal{S}$ such that $\tilde{\mathbf{a}}= \mathcal{S} \hat{\mathbf{a}}$, and using the approximate defining equation \eqref{simpleeq5.19}, we find that
\begin{eqnarray}
\mathcal{S}_{\tilde{\mathbf{q}},\hat{\mathbf{p}}} = \delta_{\tilde{\mathbf{q}},\hat{\mathbf{p}}} 6^{\hat{q}_4/2} \, .
\end{eqnarray}
\noindent To establish the correspondence between $\mathbf{a}$ and $\tilde{\mathbf{a}}$, we first define the matrix $\mathcal{T}$ by
\begin{eqnarray}
\label{Tdef5.27}
\langle \nu_{\mathbf{q}} , \tilde{\nu}_{\tilde{\mathbf{p}}}\rangle = (M \mathcal{T})_{\mathbf{q},\tilde{\mathbf{p}}} \, ,
\end{eqnarray}
\noindent where $M$ is the local normalisation matrix computed in Eq.~\eqref{Mlocalresult}. Since $\nu(\mathbf{a})$ and $\tilde{\nu}(\tilde{\mathbf{a}})$ differ by an exact form, we know that $\langle \nu(\mathbf{a}), \nu(\mathbf{b}) \rangle = \mathbf{a}^{\dagger} M \mathbf{b}$ and $ \langle \nu(\mathbf{a}), \tilde{\nu}(\tilde{\mathbf{b}}) \rangle = \mathbf{a}^{\dagger} M \mathcal{T} \tilde{\mathbf{b}}$ must be equal to each other and, since this holds for all $\mathbf{a}$, it follows that
\begin{eqnarray}
\mathbf{b} = \mathcal{T} \tilde{\mathbf{b}} \, .
\end{eqnarray}
\noindent The explicit form of the matrix $\mathcal{T}$, from its definition \eqref{Tdef5.27}, is
\begin{eqnarray}
\mathcal{T}_{\mathbf{q},\tilde{\mathbf{p}}} = \delta_{q_1 , \tilde{p}_1 - \tilde{p}_4} \delta_{q_2 , \tilde{p}_2} \delta_{q_3 , \tilde{p}_3} \dfrac{\tilde{p}_1!\vert k_1\vert^{-\tilde{p}_1 -1}}{q_1 ! \vert K_1\vert^{-q_1 -1}} \; .
\end{eqnarray}
\noindent As discussed earlier, the families correspond to cohomology classes in $H^1(X, L)$ and, in view of Eq.~\eqref{quotienteq5.23} and subject to possible identifications, it makes sense to label families by the hatted basis $\hat{\nu}_{\hat{\mathbf{q}}}$ on the ambient space. For simplicity of notation, we write the hatted indices as $\mathbf{I} = \hat{\mathbf{q}}$ from now on. We also recall from Section~\ref{introducingP1P3} that these indices are non-negative and further constrained by $I_1 = 0,..., - k_1 - 2$ and $I_2 +I_3 +I_4 \leq k_2$. With this notation, the matter field K\"ahler metric is given by the general expression
\begin{eqnarray}
\label{generalexpressionkahler}
G_{\mathbf{I},\mathbf{J}} := \dfrac{1}{2 \mathcal{V}} (\mathcal{S}^{\dagger}\mathcal{T}^{\dagger} M \mathcal{T} \mathcal{S})_{\mathbf{I},\mathbf{J}} \; .
\end{eqnarray}
\noindent Inserting the above results for $\mathcal{S}$ and $\mathcal{T}$ as well as the local normalisation matrix \eqref{Mlocalresult}, we find explicitly
\begin{eqnarray}
\label{resultdependence}
G_{\mathbf{I},\mathbf{J}} = \dfrac{\mathcal{N}_{\mathbf{I},\mathbf{J}} }{6t_1+t_2} \, ,
\end{eqnarray}
\noindent where the constants $\mathcal{N}_{\mathbf{I},\mathbf{J}}$ are given by
\begin{eqnarray}
\label{longeq5.32}
\!\!\!\!\! \mathcal{N}_{\mathbf{I},\mathbf{J}} = \dfrac{\pi J_1! I_1! I_2! I_3!\vert k_1 +k_2/6\vert^{I_1-I_4+1}6^{I_4/2+J_4/2+1}}{2(I_1-I_4)!\vert k_1\vert^{I_1+J_1+2}k_2^{I_2+I_3+2}} \theta (I_1-I_4)\delta_{I_1-I_4, J_1-J_4} \delta_{I_2,J_2} \delta_{I_3,J_3} .
\end{eqnarray}
\noindent For the lowest mode, $\mathbf{I} = \mathbf{0}$, this number specialises to
\begin{eqnarray}
\mathcal{N}_{\mathbf{0},\mathbf{0}} = 3 \pi \dfrac{\vert k_1+k_2/6\vert}{\vert k_1\vert^2 k_2^2} \, .
\end{eqnarray}
\noindent A few remarks about this result are in order. First, we note that the K\"ahler moduli dependence in
Eq.~\eqref{resultdependence} is in line with the result \eqref{tdependence} from dimensional reduction, as homogeneity of degree $-1$ is expected. What is surprising however is that Eq.~\eqref{generalexpressionkahler}, involving $\mathcal{V}$, a cubic function of K\"ahler moduli, reduces to Eq.~\eqref{resultdependence}, an inverse linear function of K\"ahler moduli. This cancellation may just be a property of our particular example, stemming from the fact that the parameters $\beta_i$ in Eq.~\eqref{beta123} are proportional to the factors in $\mathcal{K}$ in Eq.~\eqref{exampleintnumbers}, which comes out of the global-local matching. Typically, one would expect quadratic over cubic functions of the K\"ahler moduli.

In general, the matter field K\"ahler metric is also a function of complex structure moduli. For our example, this dependence has dropped
out completely, that is, the quantities $N_{\mathbf{I},\mathbf{J}}$ are constants. This feature results from our linearised local
approximation $\eqref{simpleeq5.19}$ of the Calabi-Yau manifold, where all remaining complex structure parameters can be absorbed into coordinate re-definitions. We do expect complex structure dependence to appear at the
next order, that is, if we approximate the defining equation locally by a quadric in affine coordinates. Also,
our result \eqref{resultdependence} has an implicit complex structure dependence, in that its validity depends on the choice of complex structure. Whether neglecting the quadratic and higher terms in $z$ in Eq.~\eqref{polyapprox} does indeed
provide a good approximation depends, among other things, on the choice of coefficient in the defining
equation $p$, that is, on the choice of complex structure. Another feature of our result \eqref{resultdependence} is that it is diagonal in family space and, formally, this happens because the matrices $M$, $\mathcal{S}$ and $\mathcal{T}$ are all diagonal. We have seen in Section~\ref{kahlersec4} that this is a general feature of the matrix $M$. However, $\mathcal{S}$ and $\mathcal{T}$ do not need to
be diagonal in general. In our example, this happens due to the simple form \eqref{simpleeq5.19} of the local Calabi-Yau
defining equation and the resulting diagonal form of the local K\"ahler form $\mathcal{J}$ in Eq.~\eqref{Jeq5.15}. Finally, we
remind the reader that the result \eqref{resultdependence} can only be trusted if the line bundle $L = \mathcal{O}_X(k_1, k_2)$ satisfies the
condition \eqref{conditioneq5.25}, if the flux parameters $\vert k_i\vert$ are sufficiently large and if the family numbers $\mathbf{I}$ are sufficiently small, in line with our discussion in Section~\ref{kahlersec3}.

\section{Final remarks}
\label{kahlersec6}
In this chapter, we have reported progress on computing the matter field K\"ahler metric in heterotic Calabi-Yau compactifications. Three main results have been obtained. First, by dimensional reduction we have
derived a general formula \eqref{tdependence} for the matter field K\"ahler metric and we have argued that constraints from four-dimensional supersymmetry already fully determine the K\"ahler moduli dependence of this metric.

Secondly, provided large flux leads to localisation of the matter field wave function, we have shown how the
matter field K\"ahler metric can be obtained from a local computation on the Calabi-Yau manifold, leading
to the general result \eqref{Mlocalresult}. This result, while quite general, is unfortunately of limited use, mainly since it is not expressed in terms of the global moduli of the Calabi-Yau manifold. This makes it difficult to
identify the conditions for its validity and it falls short of the ultimate goal of obtaining the matter field
K\"ahler metric as a function of the properly defined moduli superfields.

We have attempted to address these problems by working out a global-local relationship and by expressing
the local result in terms of global quantities. This has been explicitly carried out for the example of Calabi-Yau hyper-surfaces $X$ in the ambient space $\mathbb{P}^1 \times \mathbb{P}^3$, but the method can be applied to other Calabi-Yau hyper-surfaces (and, possibly complete intersections) as well. Our main result in this context is the K\"ahler
metric for matter fields from line bundles $L = \mathcal{O}_X (k_1, k_2)$ on $X$ given in Eqs.~\eqref{resultdependence}, \eqref{longeq5.32}, and is expressed as a function of the proper four-dimensional moduli fields. We have also stated the conditions for this result to be trustworthy, namely the constraint \eqref{conditioneq5.25} on the line bundle $L$ as well as large fluxes $\vert k_i \vert$ and small family numbers. 

The global-local relationship established in this way points to two problems of localised calculations both of
which are intuitively plausible. First, the large flux values demanded by localisation typically also lead to
large numbers of families. For this reason, there is a tension between localisation and the phenomenological
requirement of three families. Secondly, large flux typically leads to a “large" second Chern class $c_2(V)$
of the vector bundle, which might violate the anomaly constraint $c_2(V ) \leq c_2(TX)$. Hence, there is also
a tension between localisation and consistency of the models. It remains to be seen and is a matter of ongoing research whether consistent three-family models with localisation of all relevant matter fields can be constructed. It is likely that some of our methods can be applied to F-theory and be used to express local F-theory results in terms of global moduli of the underlying four-fold. It would be interesting to carry this out explicitly and check if the tension between localisation on the one hand and the phenomenological requirement of three families and cancelation of anomalies on the other hand persists in the F-theory context.

\chapter{Conclusion}

\noindent The purpose of this thesis was to expand the area of string phenomenology by proposing methods to calculate holomorphic Yukawa couplings for a specific class of $E_8 \times E_8$ heterotic models, namely for line bundle models on Complete Intersection Calabi-Yau manifolds. In addition, we identified a method to evaluate the matter field K\"ahler metric for models where sufficiently large gauge fluxes permit the localisation of matter fields around certain points. Both the holomorphic Yukawa couplings and the matter field K\"ahler metric are required to compute the physical Yukawa couplings of a given heterotic model, so that it can be eventually compared to measurable physics.

The line bundle models that we considered give rise to the correct MSSM spectrum, with some additional gauge-neutral bundle moduli. They were borrowed from a rich database of quasi-realistic models, that was generated several years before this thesis, through an automated scan \cite{Anderson:2011ns,Anderson:2012yf,Anderson:2013xka}. At various energy levels, we wanted the Standard Model, the MSSM and a SUSY GUT to be naturally embedded in the string model, and similarly, General Relativity and $N=1$ supergravity to be low-energy limits of the gravity sector.  For this reason, we compactified the $E_8 \times E_8$  string theory over a Calabi-Yau manifold $X$ with holomorphic poly-stable vector bundle $V$, and we ensured that the resulting 4d action matched the standard $N=1$ supergravity action in Ref.~\cite{wessandbagger} (plus some kinetic terms for the moduli).

From this point forward, we evaluated the holomorphic Yukawa coupling, using the simplifications that our class of models provided. We started with the well-known integral $ \int_X \Omega \wedge \nu_1 \wedge \nu_2 \wedge \nu_3$ and expressed line bundle-valued $(0,1)$-forms $\nu_i$ in terms of projective ambient space $(0,a)$-forms $\hat{\nu}_{i,a}$, where $a=1,...,k+1$ and $k$ is the co-dimension of $X$. By defining the \textit{type} of a form $\nu_i$ as the number $\tau_i$ for which $\hat{\nu}_{i,\tau_i} \neq 0$ and $\hat{\nu}_{i,a} = 0$, for all $ \ a > \tau_i$, we were able to formulate a vanishing theorem, Eq.~\eqref{4.12}, according to which the holomorphic Yukawa couplings are zero if $\tau_1 + \tau_2 + \tau_3$ is smaller than the ambient space dimension.
In the non-vanishing case, the Yukawa couplings are calculated as ambient space integrals over products of forms $\hat{\nu}_{i,a}$ and we have showed that the result can also be obtained algebraically, in a way that relates our method to Refs.~\cite{Candelas:1987se, Anderson:2009ge}. Explicit results were obtained for line-bundle models on the tetra-quadric (Chapter~\ref{tetraquadricchapter}) and on a co-dimension two CICY (Chapter~\ref{chaptern>1codimension}), although the method is general enough to be applied to any CICY in the database.

Altogether, our computational techniques have revealed some interesting phenomenological features. For example, topological constraints such as theorem \eqref{4.12} give a condition for the vanishing of Yukawa couplings that is not based on symmetry (only topological reasoning is involved). This has been expected since the early days of heterotic model building (see, for example, Ref.~\cite{GSW}), and can provide an explanation for the relatively light masses of the electron and first-generation quarks. In some examples discussed in Chapter~\ref{tetraquadricchapter}, it was found that the holomorphic Yukawa couplings depend explicitly on the complex structure moduli and their rank is reduced for certain regions of the moduli space. Such a dependence can be used to fine-tune the model according to observation. In addition, global $U(1)$ symmetries arising from the line bundle sum construction can motivate why certain proton decay operators are forbidden in the MSSM or various Grand Unified models. The same symmetry criteria can also be applied perturbatively to models with non-Abelian bundle structure group that are obtained through smooth deformations from the Abelian locus.

In Chapter~\ref{kahlerchapter}, our computation of the matter field K\"ahler metric was supported by the observation that for large gauge flux, the integral $\int_X \nu_I \wedge \bar{\star}_V \nu_J$ is localised on a patch $U$, so that precise knowledge of the Calabi-Yau metric is not needed. The computation was performed locally and then re-expressed in terms of the global moduli of the Calabi-Yau manifold, via a global-local relationship. It has to be noted however that the requirement of a large gauge flux may often be in conflict with the anomaly cancellation condition $c_2(V) \leq c_2(TX)$ or the phenomenological requirement for three families. It constitutes an object of future research to establish whether a consistent and realistic model can be built using the localisation method.

Finally, despite the progress reported in this thesis, a lot of problems in string phenomenology still remain unresolved. Ideally, we would have liked to construct a method to evaluate the matter field K\"ahler metric for general line bundle models, and not only for models with large flux. It is uncertain however whether this goal is achievable, given that the knowledge of the specific Calabi-Yau metric is lacking. 
An alternative research avenue would be to compactify on other classes of Calabi-Yau manifolds, such as hypersurfaces in toric varieties, for which similar methods for calculating Yukawa couplings could be constructed. In addition, one could investigate how to apply these methods to vector bundles with non-Abelian structure group, and in particular to monad bundles, which are built from sums of line bundles \cite{Anderson:2008uw, Anderson:2009mh}. In the end, a complete string model has to also contain mechanisms for moduli stabilisation and soft supersymmetry breaking. It is only when these problems are solved that the model can be proposed as ``realistic".

\appendix
\chapter{Bundles on K\"ahler manifolds}
\label{appendixvectorbundles}
\label{app:Kbundle}
In this appendix, we review some standard mathematics for K\"ahler manifolds and holomorphic vector bundles thereon, which we rely on in the main part of the text. The exposition mainly follows Refs.~\cite{H}, and more details can also be found in Refs.~\cite{Candelas:1987is,GH}.

Let $M$ be a K\"ahler manifold of dimension $n$ and $E\rightarrow M$ be a rank $r$ holomorphic vector bundle over $M$ with fibres $E_x$, where $x \in M$. The space of $E$-valued $(p,q)$ forms on $M$ is denoted by ${\cal A}^{p, q} (E)$. The usual operator $\bar{\partial}:{\cal A}^{p, q}\rightarrow {\cal A}^{p, q+1}$ for differential forms can be generalised to $E$-valued forms
\begin{equation}
 {\bar \pt}_E : {\cal A}^{p, q} (E) \to {\cal A}^{p, q+1} (E)
\end{equation} 
mapping bundle-valued $(p,q)$-forms to bundle-valued $(p,q+1)$-forms. Explicitly, this operator is defined as follows. For a local holomorphic trivialisation $s = (s_1, s_2, \dots,  s_r)$ of $E$ we can write a vector bundle-valued $(p, q)$-form $\a \in {\cal A}^{p, q} (E)$ as $\a= \sum_{i=1}^r \a^i \otimes s_i$, where  $\a^i \in {\cal A}^{p, q}$ are regular $(p, q)$-forms. Then ${\bar \pt}_E$ acts as
\begin{equation}
 {\bar \pt}_E \a=  \sum_{i=1}^r {\bar \pt} \a^i \otimes s_i\; . \label{defbp}
\end{equation} 
Since the transition functions are holomorphic, this definition is independent of the chosen trivialisation, as it should be. It is straightforward to show from this definition that ${\bar \pt}_E^2 =0$ and that the Leibniz rule
\begin{equation}
{\bar \pt}_E (f \a) = {\bar \pt} (f) \wedge \a + f {\bar \pt}_E (\a)
\end{equation}
holds (here, $f$ is a differentiable function on $M$).\\[4mm]
A Hermitian structure on $E$ (which can also be de defined more generally over complex vector bundles) is defined
by providing  a Hermitian scalar product $h_x$ on each fibre $E_x$. Let $\s$ and $\r$ be two sections of $E$ 
which, for the aforementioned trivialisation of $E$, are expanded  as $\s= \sum_{i=1}^r \s^i s_i$ and  $\r= \sum_{i=1}^r \r^i s_i$.
Then, the Hermitian structure, acting on $\s$ and $\r$, can be written out as
\be 
h (\s, \r)= H_{ij} \s^i {\bar \r}^j  =  \s^{{\rm T}} H {\bar \rho}\,, \quad H_{i j}= h (s_i, s_j)\,. 
\label{C1}
\ee
In other words, locally, we can think of the Hermitian structure as being described by Hermitian $r \times r$ matrices $H$.
For a different local trivialisation $s' = (s'_1, s'_2, \dots,  s'_r)$ related to the original one by $s'_i = \phi^j_{\ i} s_j$, it follows that $H$ transforms as 
\be 
H'= \phi^{{\rm T}} H {\bar \phi}\,. 
\label{C2}
\ee
The Hermitian structure $h$ can also be viewed as an  isomorphism between the vector bundle $E$ and its dual $E^*$, so $h: E \stackrel{\simeq}{\rightarrow} E^*$. 
This isomorphism can be written more explicitly by introducing a ``dual" trivialisation $s_*=(s_*^1,\ldots ,s_*^r)$ of the dual bundle $E^*$, defined by the relations $s_*^i(s_j)=\delta^i_j$.
If we further denote the inverse map of $h$ by $h^*:E^* \stackrel{\simeq}{\rightarrow} E$ then we have
\begin{equation}
 h(s_i)=H_{ji}s^j_*\,,\qquad h^*(s^i_*)=\bar{H}^{ji}s_j\,,\qquad H^{ij}H_{jk}=\delta^i_k\; .
\end{equation}
A Hermitian structure allows one to define a generalisation of the Hodge dual operation ${\bar \star}_E : {\cal A}^{p, q} (E) \to  {\cal A}^{n-p, n-q} (E^*)$ to vector bundle-valued forms by setting
\be 
{\bar \star}_E (\a \otimes s) = \star ({\bar \a}) \otimes h(s)\,,
\label{C11}
\ee
where $\star$ is the regular Hodge star operation on forms. It follows that ${\bar \star}_E \circ {\bar \star}_E = (-1)^{p+q}$, in analogy with corresponding rule for the regular Hodge star. Using this generalised Hodge dual, one can define the scalar product
\be 
(\a, \b)= \int_M \a \wedge {\bar \star}_E (\b)\,
\label{C12}
\ee
on  ${\cal A}^{p, q} (E)$. The adjoint operator ${\bar \pt}_E^{\dagger}: {\cal A}^{p, q} (E) \to  {\cal A}^{p, q-1} (E)$ of $\bar{\partial}_E$ relative to this scalar product satisfies
\be 
({\bar \pt}_E \a, \b)= (\a, {\bar \pt}_E^{\dagger} \b)\,, 
\label{C13}
\ee
and takes the form
\be 
{\bar \pt}_E^{\dagger}= - {\bar \star}_E \circ {\bar \pt}_{E^*} \circ {\bar \star}_E\,, 
\label{C14}
\ee
as can be seen explicitly from Eqs.~\eqref{C11}, \eqref{C12} and \eqref{C13}. Furthermore, one can define the generalised Laplacian
\be 
\D_E = {\bar \pt}_E^{\dagger} {\bar \pt}_{E} + {\bar \pt}_{E} {\bar \pt}_E^{\dagger} \,, 
\label{C15}
\ee
which is self-adjoint under the above scalar product. Bundle-valued forms  $\a \in {\cal A}^{p, q} (E)$ satisfying $\D_E \a =0$ are called harmonic with respect to the Hermitian structure $h$. For a compact manifold, the harmonic forms $\a$ are precisely the closed and co-closed forms, so the forms satisfying
\be
{\bar \pt}_E\a=0, \ \quad {\bar \pt}_E^{\dagger} \a=0\,.
\label{C16}
\ee
These forms are in one-to-one correspondence with the cohomology groups $H^{p, q}(M, E)\cong H^q (M, E \otimes \L^p \O_M)$. Finally, there is a generalisation of the Hodge decomposition which states that every form $\a \in {\cal A}^{p, q} (E)$ can be written as a unique sum $\a =\eta +  {\bar \pt}_{E} \b +{\bar \pt}_E^{\dagger}\g$, where $\eta$ is harmonic. 
\vskip 4mm\noindent
A connection, $\nabla$, on $E$ is a map $\nabla: {\cal A}^0 (E) \to  {\cal A}^1 (E)$ which satisfies the Leibniz rule 
\be 
\nabla (f\s)= d (f) \otimes \s + f \nabla (\s)
\label{C5}
\ee
for local sections $\s$ and local functions $f$. Writing $ \s =\sum_{i=1}^r \s^i s_i$ in terms of a local trivialisation $s=(s_1,\ldots ,s_r)$ we have 
\be 
\nabla (\s)= (d \s^i + A^{i}_{\ j} \s^j) \otimes s_i\,, \quad \nabla (s_j)= A^{i}_{\ j} s_i\,, 
\label{C6}
\ee
where $A$ is the gauge field. In short, locally, the connection can be written as $\nabla=d+A$, with the gauge field transforming as
\be 
A'= \phi^{-1} A \phi + \phi^{-1} d\phi\, 
\label{C7}
\ee
under a change of trivialisation, $s'_i =\phi^j_{\ i} s_j$. The curvature $F_{\nabla} \in {\cal A}^2 ({\rm End} (E))$ is defined by $F_{\nabla} =\nabla \circ \nabla$. For a given trivialisation, its local form is
\be 
F_{\nabla} = dA + A \wedge A\,. 
\label{C8}
\ee
A connection is called compatible with the holomorphic structure if $\nabla^{0,1}={\bar \pt}$ and it is called Hermitian if it satisfies $d (h (\s, \r))= h (\nabla (\s), \r) +
 h (\s, \nabla (\r))$ for any two sections $\s$ and $\r$. For a holomorphic vector bundle, there exists a unique Hermitian connection compatible with the 
 holomorphic structure which is called the Chern connection. In a local frame, the gauge field associated to the Chern connection is given by
 \be 
 A ={\bar H}^{-1} \pt {\bar H}\,.
 \label{C9}
\ee
For a holomorphic change of the trivialisation, $s'_i =\phi^j_{\ i} s_j$,
it is straightforward to verify that Eq.~\eqref{C9} is consistent with the transformation laws~\eqref{C2} and \eqref{C7}. 
It can be shown, using Eq.~\eqref{C8}, that the curvature of the Chern connection is a $(1, 1)$-form and, locally, is explicitly given by 
\be 
F_{\nabla} = {\bar \pt} ({\bar H}^{-1} \pt {\bar H})\,. 
\label{C10}
\ee
\vskip 4mm\noindent
In the main part of the thesis, we are calculating certain bundle-valued harmonic forms and it is, therefore, important to re-write the defining Eqs.~\eqref{C16} for such forms in a simple and explicit way. As before, we introduce local trivialisations $s=(s_1,\ldots ,s_r)$ and $s_*=(s_*^1,\ldots ,s_*^r)$  on $E$ and $E^*$, satisfying $s^i_* (s_j)= \d^i_{\ j}$. We start with two $(p,q)$-forms $\a =\a^i s_i$ and $\b =\b_i s^i_*$ taking values in $E$ and $E^*$, respectively. Then from the definition~\eqref{defbp} of ${\bar \pt}_E$, we have 
\be 
{\bar \pt}_E (\a)= ({\bar \pt} \a^i) \otimes s_i\,, \quad 
{\bar \pt}_{E^*} (\b)= ({\bar \pt} \b_i) \otimes s^i_*\,.
\label{C17}
\ee
For the generalised Hodge star operation~\eqref{C11}, we get 
\be
{\bar \star}_{E}  (\a) = (* {\bar \a}^i)  \otimes h(s_i)=
H_{ji} (* {\bar \a}^i) \otimes s_*^j \,, \quad 
{\bar \star}_{E^*}  (\b) = (* {\bar \b}_i)  \otimes h^*(s^i_*)=
\bar{H}^{ji} (* {\bar \b}_i) \otimes s_j \,. 
\label{C18}
\ee
Combining these equations, we obtain
\be
{\bar \pt}_E^{\dagger} \a = -\star (\d^k_{\ i} \pt + {\bar H}^{kj} \pt {\bar H}_{ji}) \star \a^i \otimes s_k
= -\star  (\d^k_{\ i} \pt + A^k_{\ i}) \star \a^i \otimes s_k\,, 
\label{C19}
\ee
where $A$ is the  Chern connection~\eqref{C9}. Hence, ${\bar \pt}_E^{\dagger}$ corresponds to the 
dual of the $\nabla^{1, 0}$ part of the Chern connection. From the above 
argument, we conclude that a harmonic bundle-valued form $\a$, written as ${\boldsymbol \a}= (\a^1, \ldots ,\a^r)^{{\rm T}}$ relative to a local frame, is characterised by
\be 
{\bar \pt} \boldsymbol { \a} =0\,, \quad  (\pt + A) \star {\boldsymbol \a}=0\,, 
\label{C20}
\ee
where $A$ is the gauge field associated to the Chern connection on the bundle. Using the explicit expression~\eqref{C9} for the Chern connection, these equations can be cast into the somewhat more convenient form
\be 
{\bar \pt} \boldsymbol { \a} =0\,, \quad  \pt ({\bar H} \star {\boldsymbol \a})=0\, ,
\label{C21}
\ee
with the Hermitian structure $H$ on the bundle. 
\chapter{The coboundary map}
\label{coboundarymapappendix}

It is well-known that for every short exact sequence of sheaves there is an associated long exact sequence in sheaf cohomology. A crucial ingredient in this correspondence is the co-boundary map whose construction can be found in standard textbooks, see for example \cite{GH}, page 40. Since the co-boundary map plays an important role for our discussion in the main part of the thesis, we now briefly review its construction. 

\noindent We start with the short exact sequence
\be 
0 \longrightarrow A \stackrel{g}{\longrightarrow} B  \stackrel{f}{\longrightarrow} C  \longrightarrow 0
\label{B1}
\ee
of sheaves $A$, $B$, $C$ and sheave morphisms $f$, $g$, satisfying $f \circ g = 0$. The associated long exact sequence in cohomology has the form
\bea
 \cdots&\longrightarrow &  H^i (A) \stackrel{g}{\longrightarrow} H^i (B) \stackrel{f}{\longrightarrow} H^i (C)\nonumber \\
&\stackrel{\delta}{\longrightarrow} &H^{i+1} (A) \stackrel{g}{\longrightarrow} H^{i+1} (B) \stackrel{f}{\longrightarrow} H^{i+1} (C) 
\longrightarrow \dots\;,
\label{B2}
\eea
where $f$ and $g$ are the induced maps in cohomology and $\delta$ is the co-boundary map which needs to be constructed. To be in line with the main part of the thesis, we will use the language appropriate for vector bundles, rather than more general sheaves, from now on.

To derive $\delta$, we start with a differential $(0,i)$-form $\nu \in H^i
(C)$ taking values in $C$. Since the map $f : B \to C$ in~\eqref{B1} is surjective it follows that $\nu$ can always be written as $\nu = f(\hat{\nu})$ for some form $\hat{\nu} \in \Omega^i(B)$. However, if $H^{i+1} (A) \neq 0$, the induced map $f: H^i (B) \to H^i (C)$ is not surjective, which implies that 
the form  $\hat{\nu}$ is not necessarily closed. 
Now we consider 
${\bar \pt} \hat{\nu} \in \Omega^{i+1} (B)$. We get 
\be 
f ({\bar \pt} \hat{\nu}) = {\bar \pt} (f ( \hat{\nu})) =  {\bar \pt} \nu =0\,, 
\label{B3}
\ee
where we have used the fact that the map $f$ is holomorphic. This implies that ${\bar \pt} \hat{\nu}$ is in the kernel of $f$
and, by the exactness of the sequence~\eqref{B2}, it is in the image of $g$. That is, there exists an element $\hat{\o}\in 
\Omega^{i+1} (A)$ such that $g \hat{\o} = {\bar \pt} \hat{\nu}$. Moreover, since $g {\bar \pt} \hat{\o} = {\bar \pt}g \hat{\o} = {\bar \pt}^2 \hat{\nu} = 0$ and $g$ is injective, we have ${\bar \pt} \hat{\o} = 0$. Hence, $\hat{\o}$ represents an element of $H^{i+1}(A)$ and we can define the co-boundary map by
\be 
\delta(\nu) = \hat{\omega} \,. 
\label{B4}
\ee
\noindent In summary, the main features of the short exact sequence \eqref{B1} and its long exact counterpart \eqref{B2} that we will require are as follows. For a $(0,i)$-form $\nu \in H^i(C)$ and its image $\hat{\o} = \delta(\nu)\in H^{i+1}(A)$ under the co-boundary map, there exists a $(0,i)$–form $\hat{\nu} \in \Omega^i(B)$ such that
\be 
\nu = f (\hat{\nu})\,, \quad {\bar \pt} \hat{\nu} = g \hat{\o}\,. 
\label{B5}
\ee
%


\chapter{Harmonic line bundle-valued forms on $\mathbb{P}^n$}
\label{appendixPn}

One of the main ingredients of our calculation of Yukawa couplings is the explicit construction of bundle-valued forms, representing line bundle cohomologies on the ambient space. Since the ambient spaces under consideration are products of projective spaces, it is sufficient to discuss a single projective space $\mathbb{P}^n$.

We begin by setting up and reviewing standard facts about projective space including the Fubini-Study metric. One way to obtain a one-to-one correspondence between cohomology and forms is to focus on harmonic forms and we will do this relative to the Fubini-Study metric. Line bundles, their Chern connections and cohomology are the subject of the next two parts of the appendix. Most of this material can be found in standard textbooks, such as Refs.~\cite{H,GH,hartshorne}. Finally, we explain how harmonic line-bundle valued forms are related under multiplication.

\section{Basics of projective space}

As explained in Section \ref{complexmanifoldssection}, the complex projective space $\mathbb{P}^n$ is the set of complex lines through the origin in $\mathbb{C}^{n+1}$. We denote coordinates on $\mathbb{C}^{n+1}$ by $x_{\alpha}$, where $\alpha = 0, 1, ... , n$. The element of $\mathbb{P}^n$ given by the line through the origin and a point $(x_0, x_1, ... , x_n)$ (with at least one $x_{\alpha} \neq 0$) is denoted by $(x_0\!:\!x_1\!:\!...\!:\!x_n) \in \mathbb{P}^n$. The standard open patches on $\mathbb{P}^n$ are $U_{\alpha} = \lbrace(x_0\!:\!x_1\!:\!...\!:\!x_n)\vert x_{\alpha} \neq 0 \rbrace$, where $\alpha = 0, ... , n+1$, with associated charts $(U_{\alpha}, \phi_{\alpha})$ and maps $\phi_{\alpha} : U_{\alpha} \rightarrow \mathbb{C}^n$ defined by $\phi_{\alpha}(x_0\!:\!x_1\!:\!...\!:\!x_n) = (\xi^{\alpha}_0, \xi^{\alpha}_1, ..., \widehat{\xi^{\alpha}_{\alpha}}, ...,\xi^{\alpha}_n)$.  Here, $\xi^{\alpha}_{\mu} = x_{\mu}/x_{\alpha}$ are the coordinates on $\mathbb{C}^n$ and it is understood that $\xi^{\alpha}_{\alpha}=1$ is discarded. For an overlap $U_{\alpha} \cap U_{\beta} \neq \emptyset$, the transition functions $\phi_{\beta \alpha} = \phi_{\beta} \circ \phi^{-1}_{\alpha}:\mathbb{C}^n \rightarrow \mathbb{C}^n$ takes the form $\xi^{\alpha}_{\mu} \mapsto \xi^{\beta}_{\mu} = \tfrac{x_{\alpha}}{x_{\beta}}\xi^{\alpha}_{\mu}$.

On each patch $U_{\alpha}$, the Fubini-Study K\"ahler potential can be written as
\begin{eqnarray}
K_{\alpha} = \dfrac{i}{2\pi} \ \textrm{ln}(\kappa_{\alpha})\, , \qquad \kappa_{\alpha} = \sum_{\mu=0}^n \vert\xi^{\alpha}_{\mu}\vert^2 \, .
\end{eqnarray}
\noindent The associated Fubini-Study K\"ahler form is given by
\begin{eqnarray}
J = \partial \overline{\partial} K_{\alpha} \, 
\end{eqnarray}
\noindent as usual and it is easy to check that this definition is independent of $\alpha$ on the overlaps and, hence, gives a globally defined form on $\mathbb{P}^n$. The above K\"ahler form is normalised such that
\begin{eqnarray}
\int_{\mathbb{P}^n} J^n = 1 \, .
\end{eqnarray}
\noindent It will frequently be convenient to work on the patch $U_0 = \mathbb{C}^n$ whose coordinates we also denoted by $z_{\mu} = x_{\mu}/x_0$, where $\mu = 1,...,n$, and we write $\kappa = \kappa_0 = 1 + \sum_{\mu=1}^n \vert z_{\mu} \vert^2$. 

\section{Line bundles on projective space}

\noindent The $k^{\textrm{th}}$ power of the hyperplane bundle on $\mathbb{P}^n$ is denoted by $L = \mathcal{O}_{\mathbb{P}^n} (k)$. For each patch $U_{\alpha}$, a hermitian bundle metric on $L$ is given by
\begin{eqnarray}
\label{hermitianbundlemetric}
H_{\alpha}=\kappa_{\alpha}^{-k} \, .
\end{eqnarray}
\noindent On the patch $U_0$, we also write $H = H_0 = \kappa^{-k}$. The associated Chern connection $\nabla^{0,1}=\bar{\partial}$ and $\nabla^{1,0}=\partial+A_{\alpha}$ is specified by the gauge field
\begin{eqnarray}
A_{\alpha}= \partial \textrm{ln}  \bar{H}_{\alpha} = - k \partial  \textrm{ln} \kappa_{\alpha} = 2 \pi i k \partial K_{\alpha} \, ,
\end{eqnarray}
\noindent whose curvature $F_{\alpha} = d A_{\alpha} = - \partial \overline{\partial} \textrm{ln} \bar{H}_{\alpha}$ is explicitly given by
\begin{eqnarray}
F_{\alpha} =  k \partial \bar{\partial} \textrm{ln} \kappa_{\alpha} = - 2 \pi i k \partial \bar{\partial} K_{\alpha} = - 2 \pi i k J \, .
\end{eqnarray}
\noindent For the first Chern class of $L=\mathcal{O}_{\mathbb{P}^n}(k)$, this implies
\begin{eqnarray}
c_1(\mathcal{O}_{\mathbb{P}^n}(k))=\dfrac{i}{2 \pi} F = k J \, ,
\end{eqnarray}
\noindent as expected.

\section{Line bundle cohomology}

The dimension of line bundle cohomology for a line bundle $\mathcal{K}=\mathcal{O}_{\mathbb{P}^n}(k)$ is described by Bott’s formula
\begin{eqnarray}
\label{generalizedbott}
h^q(\mathbb{P}^n,\mathcal{O}_{\mathbb{P}^n}(k))=\begin{cases} \dfrac{(n+k)!}{n!k!}\, , & \textrm{ for } q=0, \;\; k \geq 0 \, . \\[2.2mm] \dfrac{(-k-1)!}{n! (-k-n-1)!}\, ,  & \textrm{ for } q=n, \;\; k\leq -(n+1) \, . \\[3mm] 0\, , & \textrm{ otherwise} \ . \end{cases}
\end{eqnarray}
\noindent This means that line bundles $\mathcal{O}_{\mathbb{P}^n}(k)$ in the “gap” $-n + 1 < k < 0$ only have trivial cohomologies, while all other line bundles have precisely one non-trivial cohomology. For $k \geq 0$, this non-trivial cohomology is the zeroth cohomology with dimension given in the first row of Eq.~\eqref{generalizedbott}. For $k \leq (-n - 1)$, on the other
hand, only the highest, $n^{\textrm{th}}$ cohomology is non-trivial with dimension given in the second row of Eq.~\eqref{generalizedbott}.

We would like to represent these cohomologies by line bundle valued $(0, q)$–forms which are harmonic relative to the Fubini-Study metric. Such forms $\nu_{\alpha}$ should, on each patch $U_{\alpha}$ satisfy the equations (see
Appendix~\ref{appendixvectorbundles} for details)
\begin{eqnarray}
\label{2equations}
\bar{\partial} \nu_{\alpha} =0 \, , \qquad \partial(\bar{H}_{\alpha} \ast \nu_{\alpha})=0 \, , 
\end{eqnarray}
\noindent where $H_{\alpha}$ is the hermitian bundle metric \eqref{hermitianbundlemetric}. To solve these equations, we should distinguish the different cases displayed in the Bott formula \eqref{generalizedbott}.

\begin{enumerate}
\item \underline{$\mathcal{K} =\mathcal{O}_{\mathbb{P}^n}(k)$ with $k \geq 0$:}

\noindent In this case, $H^0(\mathbb{P}^n, \mathcal{O}_{\mathbb{P}^n}(k))$ is the only non-zero cohomology, so we are looking for sections, that is harmonic $(0, 0)$–forms. On the patch $U_0$ they are given by 
\begin{eqnarray}
\nu_{(k)} = P_{(k)}(z_1, ..., z_n) \, , 
\end{eqnarray}
\noindent where $P_{(k)}$ are polynomials of degree $k$ in $z_{\mu}$. It is straightforward to check that these have the
correct transition functions upon transformation to another patch. Note that the dimension of the space of degree $k$ polynomials in $n$ variables is indeed given by the first line in the Bott formula \eqref{generalizedbott}, as required.

\item \underline{$\mathcal{K} =\mathcal{O}_{\mathbb{P}^n}(k)$ with $-(n + 1) < k < 0$:}

\noindent In this case, all cohomologies vanish and there are no harmonic forms to construct.

\item \underline{$\mathcal{K} =\mathcal{O}_{\mathbb{P}^n}(k)$ with $k \leq -(n + 1)$:}

\noindent In this case, $H^n(\mathbb{P}^n, \mathcal{O}_{\mathbb{P}^n}(k))$ is the only non-vanishing cohomology, so we are looking for harmonic $(0, n)$–forms. It is straightforward to verify that, on the patch $U_0$, these can be written as
\begin{eqnarray}
\nu_{(k)} = \kappa^k P_{(k)}(\bar{z}_1,...,\bar{z}_n) d \bar{z}_1 \wedge ... \wedge d \bar{z}_n \, ,
\end{eqnarray}
\noindent where $P_{(k)}$ are polynomials of degree $-k-n-1$ in the $n$ variables $\overline{z}_{\mu}$. Note that the dimension of
this polynomial space equals the value in the second row of the Bott formula \eqref{generalizedbott}, as it should.
\end{enumerate}
\noindent For uniformity of notation, in the following $P_{(k)}$ for $k \geq 0$ denotes a polynomial of degree $k$ in $z_{\mu}$, while $P_{(k)}$ for $k \leq -n-1$ denotes a polynomial of degree $-k - n - 1$ in $\overline{z}_{\mu}$.

\section{Multiplication of harmonic forms}

Calculating Yukawa couplings requires performing wedge products of harmonic bundle-valued forms on $\mathbb{P}^n$ (or on products of projective spaces) and we would like to understand in detail how this works. As we have seen, on $\mathbb{P}^n$, we have harmonic bundle-valued $(0, 0)$-forms $\nu_{(k)} = P_{(k)}$, which represent the cohomology $H^{0}(\mathbb{P}^n, \mathcal{O}_{\mathbb{P}^n}(k))$ for $k \geq 0$ and harmonic bundle-valued $(0, n)$ forms $\nu_{(k)} = \kappa^k P_{(z)} d \overline{z}_1 \wedge ... \wedge d \overline{z}_n$, which represent the cohomology $H^n(\mathbb{P}^n, \mathcal{O}_{\mathbb{P}^n}(k))$ for $k \leq -n-1$. Performing a wedge product between
any two of those forms clearly produces a $\overline{\partial}$–closed form which is a representative of the appropriate
cohomology. If this wedge product is between two harmonic $(0, 0)$–forms the result is clearly again a
harmonic $(0, 0)$–form. However, the situation is more complicated for a product of a harmonic $(0, 0)$–
form and a harmonic $(0, n)$–form. The result is a $\overline{\partial}$–closed $(0, n)$–form which, however, is generally not harmonic. An obvious problem is to find the harmonic $(0, n)$–form in the same cohomology class as this
product.

To discuss this in detail, we start with a harmonic $(0, 0)$–form $p_{(\delta)}$ representing a class in $H^0(\mathbb{P}^n, \mathcal{O}_{\mathbb{P}^n}(\delta))$ and a harmonic $(0, n)$–form
\begin{eqnarray}
\label{nuk-delta}
\nu_{(k-\delta)}=\kappa^{k-\delta} P_{(k-\delta)} d \overline{z}_1 \wedge ... \wedge d \overline{z}_n \, 
\end{eqnarray}
\noindent representing a class in $H^n(\mathbb{P}^n, \mathcal{O}_{\mathbb{P}^n}(k-\delta))$, where $k \leq -n-1$. The product $p_{(\delta)}\nu_{(k-\delta)}$ is $\overline{\partial}$-closed, but not generally harmonic, and defines a class in $H^n(\mathbb{P}^n, \mathcal{O}_{\mathbb{P}^n}(k))$, whose harmonic representative we denote by
\begin{eqnarray}
\label{nuk}
\nu_{(k)}=\kappa^{k} Q_{(k)} d \overline{z}_1 \wedge ... \wedge d \overline{z}_n \, .
\end{eqnarray}
\noindent This harmonic representative differs from the original product by an exact piece, so we have an equation
of the form
\begin{eqnarray}
\label{mainequation}
p_{(\delta)} \nu_{(k-\delta)} + \overline{\partial} s = \nu_{(k)} \, ,
\end{eqnarray}
\noindent where $s$ is a section of $\mathcal{O}_{\mathbb{P}^n}(k)$. It turns out, and will be shown below, that the correct ansatz for $s$ is
\begin{equation}
s = \kappa^{k-\delta+1}( S^{(1)} d \overline{z}_2 \wedge ... \wedge d \overline{z}_n - S^{(2)} d \overline{z}_1\wedge d\overline{z}_3 \wedge ... \wedge d \overline{z}_n+...+ (-1)^{n-1} S^{(n)} d \overline{z}_1 \wedge  ... \wedge d \overline{z}_{n-1} ) ,
\end{equation}
\noindent where the $S^{(i)}$ are multivariate polynomials of degree $\delta-1$ in $z_i$ and of degree $-k +\delta -n$ in $\overline{z}_i$. Eq.~\eqref{mainequation} can be solved by inserting the various differential forms including the most general polynomials of the appropriate degrees and then matching polynomials coefficients. In this way, given $p_{(\delta)}$ and $\nu_{(k-\delta)}$, both $s$ and $\nu_{(k)}$ can be determined as we will see below. While this is straightforward in principle, the details are complicated. However, the main result can be stated in a simple way and we would like to do this upfront. It turns out that the polynomial $Q_{(k)}$ which determines $\nu_{(k)}$ is given by
\begin{eqnarray}
\label{resultappC}
\tilde{Q}_{(k)} = c \tilde{p}_{(\delta)} \tilde{P}_{(k-\delta)} \, , \quad \textrm{where} \quad c = \dfrac{(-k-1)!}{(-k+\delta - 1)!} \, .
\end{eqnarray}
\noindent We recall that the tilde denotes the homogenous counterparts of the various polynomials, so all polynomials
in the above equation depend on the homogeneous coordinates $x_{\mu}$, where $\mu = 0, 1, ... , n$. The polynomial
“multiplication” on the RHS of this equation should be carried out by converting the coordinates $x_{\mu}$ in $\tilde{p}_{(\delta)}$ into the partial derivatives $\partial/\partial\overline{x}_{\mu}$ which then, in turn, act on $\tilde{P}_{(k-\delta)}$ which depends on $\overline{x}_{\mu}$. Note that this leads to the correct degree required for the polynomial $\tilde{Q}_{(k)}$. This remarkably simple solution to
Eq.~\eqref{mainequation} is the key to converting the calculation of Yukawa couplings into an “algebraic” calculation. From this result, the wedge products of harmonic forms which appears in the Yukawa integral can simple
be converted into polynomial multiplication, with the appropriate conversion of coordinates into partial
derivatives, as discussed. Although $s$ is determined by Eq.~\eqref{mainequation}, we are unfortunately not aware of a
formula for s as simple as Eq.~\eqref{resultappC}.

In order to prove Eq.~\eqref{resultappC}, we first note the derivative
\begin{align}
\overline{\partial} s  = & \, \kappa^{k-\delta+1} \left( \partial_{\overline{z}_1} S^{(1)}+\partial_{\overline{z}_2} S^{(2)}+...+ \partial_{\overline{z}_n} S^{(n)}\right) d \overline{z}_1 \wedge  ... \wedge d \overline{z}_{n} \ + \notag \\& \, (k-\delta+1)\kappa^{k-\delta}\left(z_1 S^{(1)}+...+z_n S^{(n)}\right)d \overline{z}_1 \wedge  ... \wedge d \overline{z}_{n} \, .
\end{align}
\noindent Inserting this together with Eqs.~\eqref{nuk-delta} and \eqref{nuk} into Eq.~\eqref{mainequation} leads to
\begin{eqnarray}
\label{maineq}
p P+ \kappa \sum^n_{i=1} \partial_{\overline{z}_i} S^{(i)} - (-k+\delta-1) \sum^n_{i=1} z_i S^{(i)} = \kappa^{\delta} Q \, .
\end{eqnarray}
\noindent Next, we should write out each of the polynomials explicitly. For each $S^{(i)}$ we have
\begin{eqnarray}
\label{expansion1}
S^{(i)}= \sum_{\lbrace 0 \leq i_1+...+i_n\leq \delta -1\rbrace } \sum_{ \lbrace 0 \leq j_1+...+ j_n \leq -k+\delta-n \rbrace } c^{(i)}_{i_1...i_n; j_1 ... j_n}z_1^{i_1} ... z_n^{i_n} \overline{z}{}^{j_1}_1...\overline{z}{}^{j_n}_n \, ,
\end{eqnarray}
\noindent with coefficients $c^{(i)}_{i_1...i_n; j_1 ... j_n}$ such that $(i_1,...,i_n; j_1, ...,j_n)$ represents any index combination satisfying $0 \leq i_1+...+i_n\leq \delta -1$ and $0 \leq j_1+...+ j_n \leq -k+\delta-n$. Similarly, we can expand the other polynomials
\begin{align}
\label{expansion2}
p_{(\delta)} \; & \, = \sum_{0\leq i_1+...+i_n \leq \delta} a_{i_1 ... i_n} z^{i_1}_1...z^{i_n}_n \, ,\\
\label{expansion3}
P_{(k-\delta)} \; & \, =\sum_{0\leq j_1+...+j_n\leq -k+\delta-n-1} b_{j_1...j_n} \overline{z}{}^{j_1}_1...\overline{z}{}^{j_n}_n \, , \\
\label{expansion4}
Q_{(k)} \; &= \, \sum_{0\leq j_1+...+j_n\leq -k-n-1} q_{j_1...j_n} \overline{z}{}^{j_1}_1...\overline{z}{}^{j_n}_n \, .
\end{align}
\noindent A useful polynomial expansion of $\kappa = 1+z_1 \overline{z}_1+...+z_n \overline{z}_n$ is given by
\begin{eqnarray}
\label{expansion5}
\kappa^{\delta}=\sum_{0\leq i_1+...+i_n\leq \delta } \dfrac{\delta !}{i_1 ! i_2! ... i_n! (\delta-i_1-...-i_n)!} z_1^{i_1}...z_n^{i_n} \overline{z}{}^{i_1}_1...\overline{z}{}^{i_n}_n \, .
\end{eqnarray}
\noindent Now substituting the polynomials from Eq.~\eqref{expansion1}-\eqref{expansion5} into Eq.~\eqref{maineq}, one can derive the following identity, by extracting the coefficient of the $z_1^{i_1}...z_n^{i_n} \overline{z}{}^{i_1+j_1}_1...\overline{z}{}^{i_n+j_n}_n$ term
\begin{align}
\label{bigbadvoodoo}
\dfrac{\delta!}{i_1!...i_n!(\delta-i_1-...-i_n)!}q_{j_1...j_n}= & \;\, a_{i_1...i_n}b_{l_1...l_n}+\sum_{s=1}^{n} (l_s+1)c^{(s)}_{i_1...i_n;l_1...l_s+1...l_n}+ \notag \\ & \;\, \sum_{s=1}^n (k-\delta+1+l_s)c^{(s)}_{i_1...i_s-1,...,i_n;l_1...l_n}+ \notag \\ & \;\,  \sum_{s=1}^n \sum_{\substack{r=1 \\ r \neq s}}^n (l_s+1) c^{(s)}_{i_1...i_r-1...i_n;l_1...l_r-1...l_s+1...l_n} ,
\end{align}
\noindent where we have denoted $l_s=i_s+j_s$, for all $s=1,...,n$. Note, however, that Eq.~\eqref{bigbadvoodoo} is true only if all $i_s$ are strictly positive and strictly smaller than $\delta-\sum^n_{r \neq s} i_r$. For $i_s = 0$ the $c^{(s)}_{i_1,...,i_s-1,...,i_n;l_1...l_n}$ term is not present, because the polynomial expansion of $S^{(s)}$ contains only positive exponents. For $i_s = \delta-\sum^n_{r \neq s} i_r$, the term $c^{(s)}_{i_1...i_n;l_1...,l_s+1,...,l_n}$ is missing, because it does not respect the summation rule of Eq.~\eqref{expansion1}. However, we can conventionally define all these unwanted $c^{(s)}$ coefficients to be zero, so that Eq.~\eqref{bigbadvoodoo} is valid for any $i_s \geq 0$.

In order to solve the above set of equations for $q_{j_1...j_n}$, it is useful to define the quantities
\begin{equation}
\label{defofbeta}
\beta_{i_1 ... i_n} = \dfrac{(-k+\delta-n-1-(i_1+...+i_n)-(j_1+...+j_n))!}{(-k-n-1-(j_1+...+j_n))!}\dfrac{(i_1+j_1)!}{j_1!}...\dfrac{(i_n+j_n)!}{j_n!} \, ,
\end{equation}
\noindent which satisfy the following combinatorial identity
\begin{eqnarray}
\label{tobeproven}
\sum_{0\leq i_1+...+i_n\leq \delta}\beta_{i_1 ... i_n}  \dfrac{\delta!}{i_1!...i_n! (\delta-i_1-...-i_n)!}=\dfrac{(-k+\delta -1 )!}{(-k-1)!} \, .
\end{eqnarray}
\noindent A proof of this identity can be found at the end of this appendix. Next, we multiply both sides of Eq.~\eqref{bigbadvoodoo} by $\beta_{i_1 ... i_n} $ and then sum over all indices $\lbrace i_1, ..., i_n \rbrace$ with $0 \leq i_1+...+i_n \leq \delta$. This trick removes all coefficients $c^{(s)}$ from our equation, as a result of the identity
\begin{multline}
\label{relief}
\sum_{0 \leq i_1+...+i_n \leq \delta} \beta_{i_1 ... i_n} \biggl(\sum_{s=1}^{n} (l_s+1)c^{(s)}_{i_1...i_n;l_1...l_s+1...l_n}+\sum_{s=1}^n (k-\delta+1+l_s)c^{(s)}_{i_1...i_s-1...i_n;l_1...l_n}\\+ \sum_{s=1}^n \sum_{\substack{r=1 \\ r \neq s}}^n (l_s+1) c^{(s)}_{i_1...i_r-1...i_n;l_1...l_r-1...l_s+1...l_n}\biggr)=0 \, .
\end{multline}
\noindent To see this, consider the weight $w$ of an arbitrary coefficient $c^{(n)}_{i_1...i_n; l_1...l_{s}+1...l_n}$ in the above sum, defined as
\begin{eqnarray}
w = (k-\delta +2+l_s) \beta_{i_1...i_s+1...i_n} + (l_s+1) \beta_{i_1...i_n} + (l_s+1)\sum_{\substack{r=1 \\ r \neq s}}^{n} \beta_{i_1...i_r+1...i_n} \, .
\end{eqnarray}
\noindent Starting from the definition of $\beta$ in Eq.~\eqref{defofbeta}, we notice that
\begin{eqnarray}
(l_s+1) \beta_{i_1...i_r+1...i_n} = (l_r+1)\beta_{i_1...i_s+1...i_n}, \,\,\,\,\,\ \forall r \neq s \, .
\end{eqnarray}
\noindent Therefore, the weight of $c^{(n)}_{i_1...i_n; l_1...l_{s}+1...l_n}$ becomes
\begin{eqnarray}
w=(k-\delta+\sum_{r=1}^n l_r + n+1) \beta_{i_1...i_s+1...i_n} + (l_s+1) \beta_{i_1...i_n} ,
\end{eqnarray}
\noindent which vanishes.  Coming back to Eq.~\eqref{bigbadvoodoo}, we multiply with $\beta_{i_1 ... i_n}$ and sum over all $\lbrace i_1,...,i_n\rbrace$ with $0 \leq i_1+...+i_n \leq \delta$. This removes $c^{(i)}$ and leads to an equation for the coefficients of $Q$, namely
\begin{eqnarray}
\label{proofderiv}
 q_{j_1...j_n} = \dfrac{(-k-1)!}{(-k+\delta -1 )!} \sum_{0 \leq i_1+...+i_n\leq \delta} \beta_{i_1 ... i_n} a_{i_1...i_n}b_{l_1,...,l_n} \, .
\end{eqnarray}
\noindent We should now compare this result for $Q$, obtained by solving Eq.~\eqref{mainequation}, with the proposed solution \eqref{resultappC}. To this end, we convert all relevant polynomials into their homogeneous counterparts and
also convert the coordinates in $\tilde{p}_{(\delta)}$ into derivatives. This leads to
\begin{align}
\tilde{p}_{(\delta)} \; &= \, \sum_{i_0+...+i_n = \delta} a_{i_1...i_n}\left( \dfrac{\partial}{\partial \overline{x}_0}\right)^{i_0}\left( \dfrac{\partial}{\partial \overline{x}_1}\right)^{i_1}... \ \left(\dfrac{\partial}{\partial \overline{x}_n}\right)^{i_n} \, ,\\
\tilde{P}_{(k-\delta)} \;&=\, \sum_{j_0+ ...+j_n = -k+\delta-n-1} b_{j_1...j_n} \overline{x}_0^{j_0} \overline{x}_1^{j_1}... \ \overline{x}_n^{j_n} \, , \\
\tilde{Q}_{(k)} \; &=\, \sum_{ j_0 + ...+j_n= -k-n-1} q_{j_1...j_n} \overline{x}_0^{j_0} \overline{x}_1^{j_1}...\ \overline{x}_n^{j_n} \, .
\end{align}
\noindent Inserting this into the RHS of Eq.~\eqref{resultappC} gives
\begin{equation}
\resizebox{1.01\hsize}{!}{$
\tilde{p}_{(\delta)} \tilde{P}_{(k-\delta)} =  \!\! \mathlarger{\mathlarger{\sum}}_{ \lbrace i_0+...+i_n = \delta \rbrace } \, \mathlarger{\mathlarger{\sum}}_{\lbrace j_0 + ...+j_n= -k-n-1 \rbrace} \underbrace{\dfrac{(i_0+j_0)!}{j_0!}...\dfrac{(i_n+j_n)!}{j_n!}}_{\beta_{i_1 ... i_n}} a_{i_1...i_n} b_{(i_1+j_1)...(i_n+j_n)} \overline{x}_0^{j_0} \overline{x}_1^{j_1}... \overline{x}_n^{j_n} ,$}
\end{equation}
\noindent and inserting the result~\eqref{proofderiv} for the coefficients of $Q$ proves Eq.~\eqref{resultappC}.

\vspace{4mm}

\noindent \underline{Proof of Eq.~\eqref{tobeproven}}: We start from the $n = 1$ equation
\begin{eqnarray}
\label{n1case}
\sum_{i=0}^{\delta} \dfrac{(-k+\delta -2-i-j)!(i+j)!}{(-k-j-2)!j!} \dfrac{\delta !}{i! (\delta -i)!} = \dfrac{(-k+\delta -1)!}{(-k-1)!} \, ,
\end{eqnarray}
\noindent which can be proven by explicit calculation. It is then useful to write the sum for $n>1$ as
\begin{equation}
\!\!\! \sum_{i_1=0}^{\delta} \sum_{i_2=0}^{\delta-i_1}...\!\!\sum_{i_n=0}^{\delta-i_1-...-i_{n-1}} \dfrac{(-k+\delta-n-1-\sum_{s=1}^n l_s)!}{(-k-n-1-\sum_{s=1}^n j_s)!} \dfrac{l_1 !}{j_1 !}... \dfrac{l_n !}{j_n !} \dfrac{\delta!}{i_1!...i_n!(\delta-i_1-...-i_n)!} \, ,
\end{equation}
\noindent and to perform the summation step by step, starting from $i_n$ and ending with $i_1$, while using Eq.~\eqref{n1case} every time. For $i_n$, we use Eq.~\eqref{n1case} with $\delta_n=\delta - \sum_{s=1}^{n-1} i_s$ instead of $\delta$ and $k_n=k+n-1+\sum_{s=1}^{n-1} j_s$ instead of $k$, which leads to

\begin{multline}
\sum^{\delta-i_1-...-i_{n-1}}_{i_n=0} \dfrac{(-k+\delta-n-1-\sum_{s=1}^n l_s)!}{(-k-n-1-\sum_{s=1}^n j_s)!} \dfrac{l_n!}{j_n!} \dfrac{\delta !}{i_n! (\delta-i_1-...-i_n)!}=\\=\dfrac{(-k+\delta-n-\sum_{s=1}^{n-1} l_s)!}{(-k-n-\sum_{s=1}^{n-1} j_s)!} \dfrac{\delta!}{(\delta-\sum_{s=1}^{n-1}i_s)!}=\dfrac{(-k_{n-1}+\delta_{n-1}-2-l_{n-1})!}{(-k_{n-1}-2-j_{n-1})!}\dfrac{\delta!}{(\delta_{n-1}-i_{n-1})!} \, .
\end{multline}

\noindent After performing all the sums, we obtain the required result, \eqref{tobeproven}.

\chapter{The boundary integral}
\label{appendixboundary}


When deriving the Yukawa coupling in the main text, in particular converting Eq.~\eqref{Yukamb} into Eq.~\eqref{Yukamb1} in Chapter~\ref{tetraquadricchapter} and Eq.~\eqref{3.8} into Eq.~\eqref{3.9} in Chapter~\ref{chaptern>1codimension}, we have neglected the boundary term which arises from the partial integration. In this appendix, we show that this boundary term does indeed vanish for the cases discussed. 

Before we get to Yukawa couplings, it might be useful to note that this boundary term can indeed be important for certain integrals of interest. Consider the tetra-quadric in the ambient space ${\cal A}=\mathbb{P}^1\times\mathbb{P}^1\times\mathbb{P}^1\times\mathbb{P}^1$, with the four ambient space K\"ahler forms $\hat{J}_i$, where $i=1,2,3,4$, normalised as $\int_{\mathbb{P}^1}\hat{J}_i=1$ and their restrictions $J_i=\hat{J}_i|_X$ to the tetra-quadric. An object of interest are the triple-intersection numbers of the tetra-quadric, for example
\begin{equation}
 d_{123}=\int_XJ_1\wedge J_2\wedge J_3\; . \label{d123}
\end{equation} 
It is well-known~\cite{Hubsch:1992nu} how to compute these intersection numbers by introducing the two-form $\mu=2\sum_{i=1}^4\hat{J}_i$ and re-writing the above expression as an ambient space integral. This leads to
\begin{equation}
 d_{123}=\int_{{\cal A}}\hat{J}_1\wedge\hat{J}_2\wedge\hat{J}_3\wedge\mu=2\; . \label{d123res}
\end{equation}
This method is applicable since the ambient space version $\hat{J}_1\wedge\hat{J}_2\wedge\hat{J}_3$ of the integrand is a closed form. However, alternatively, we may proceed to evaluate the integral~\eqref{d123} by inserting a $\delta$-function, as we did for Eq.~\eqref{Yukamb} and, subsequently, using the current identity~\eqref{10.5}. This leads to
\begin{equation}
 d_{123}=\frac{1}{2\pi i}\int_{\cal A}\hat{J_1}\wedge\hat{J}_2\wedge\hat{J}_3\wedge\bar{\partial}\left(\frac{1}{p}\right)\wedge dp
 =\frac{1}{2\pi i}\int_{\cal A}\hat{J_1}\wedge\hat{J}_2\wedge\hat{J}_3\wedge\left(\bar{\partial}_{\bar{z}_4}\left(\frac{1}{p}\right)d\bar{z}_4\right)\wedge dp \, .
\end{equation} 
Since the K\"ahler forms $\hat{J}_i$ are $\bar{\partial}$-closed, integration by parts and neglecting the boundary term leads to $d_{123}=0$, in contradiction with~\eqref{d123res}. Hence, in this case, the result comes entirely from the boundary term
\begin{equation}
 d_{123}=\frac{1}{2\pi i}\int_{\mathbb{P}_1\times\mathbb{P}_1\times\mathbb{P}^1\times \gamma_4}\hat{J_1}\wedge\hat{J}_2\wedge\hat{J}_3\wedge \frac{dp}{p}\; .
\end{equation} 
where $\gamma_4$ is a contour with $|z_4|\rightarrow\infty$. In this limit, $p\sim z_4^2$ and $p^{-1}dp\sim 2z_4^{-1}dz_4$, which leads to the correct answer $d_{123}=2$.

For Yukawa integrals, the integrand is typically not a closed form, so the $\delta$-function current should be
used to re-write them as ambient space integrals. As the above example indicates, we should be careful about the boundary term. 

\section{The co-dimension one case}
We start with the ambient space

\be 
{\cal A}= {\mathbb P}^{n_1} \times {\mathbb P}^{n_2} \times \dots \times {\mathbb P}^{n_m}\,, \quad 
\sum_{i=1}^m n_i = 4
\label{A0}
\ee
and a Calabi-Yau hypersurface $X \subset \mathcal{A}$ defined as the zero locus of a polynomial $p$ of multi-degree $(n_1 +
1, ... , n_m + 1)$. The relevant integral for the Yukawa couplings, before the integration by parts, reads\footnote{In this Appendix we ignore various numeric prefactors since they do not matter for our discussion.}
\be 
\l (\nu_1, \nu_2, \nu_3) = \int_{X} \Omega \wedge \nu_1 \wedge \nu_2 \wedge \nu_3 \sim \int_{{\mathbb C}^4} d^4 z \wedge \hat{\nu}_1 \wedge \hat{\nu}_2 \wedge \hat{\nu}_3 
\wedge {\bar \pt} \Big(\frac{1}{p}\Big)\, ,
\label{A1}
\ee
\noindent where $z_1, ... , z_4$ are affine coordinates on a patch $\mathbb{C}^4$ of $\mathcal{A}$. Let us introduce the $(0, 3)$-form
\be 
\hat{\a} =  \hat{\nu}_1 \wedge \hat{\nu}_2 \wedge \hat{\nu}_3  \in \Omega^3 ({\cal A}, {\cal O}_{{\cal A}})\, , 
\label{A2}
\ee
\noindent which takes values in the trivial bundle. Further, we define the form $\hat{\beta}$ by
\be 
\bar \pt \hat{\a} = p\hat{\b}\,. 
\label{A3}
\ee
\noindent Note that $\hat{\beta}\in H^4(\mathcal{A}, \mathcal{O}_{\mathcal{A}}(-n_1 - 1, ... , -n_m - 1)) \cong \mathbb{C}$ and, hence, that $\hat{\beta}$ is uniquely fixed up to an overall constant and an exact form, both of which are irrelevant for the present purposes. A harmonic
representative for $\hat{\beta}$ can be written down following the rules in Appendix~\ref{appendixPn} (see also Section~\ref{maps1} for the case $\mathcal{A} = \mathbb{P}^1 \times \mathbb{P}^1\times \mathbb{P}^1 \times \mathbb{P}^1$) and this leads to
\be 
\hat{\b} \sim \frac{d^4 {\bar z}}{ \kappa_1^{n_1+1} \dots \kappa_m^{n_m+1}} \, . 
\label{A5}
\ee
In order to understand the boundary integral, we need to study the limit when the modulus of one of the
coordinates, say $z_1$, goes to infinity. Let us assume that $z_1$ is an affine coordinate of the first projective
factor $\mathbb{P}^{n_1}$. Then, for large $\vert z_1 \vert$, we have
\be 
\hat{\b} \sim \frac{d^4 \bar z}{z_1^{n_1+1} {\bar z}_1^{n_1+1}} \,, \quad p\hat{\b} \sim \frac{d^4 \bar z}{ {\bar z}_1^{n_1+1}}\,. 
\label{A6}
\ee
Let us solve Eq.~\eqref{A3} for $\hat{\a}$ in this limit. The general solution for $\hat{\a}$ is given by 
$\hat{\a}=\hat{\a}_0+ \hat{\a}_1$, where $\hat{\a}_0$ is the general solution to the homogeneous equation $\bar{\partial}\hat{\alpha} = 0$ and $\hat{\a}_1$ is a 
partial solution to the inhomogeneous equation~\eqref{A3}. For a four-dimensional ambient space of the form~\eqref{A0}, we have $H^3 ({\cal A}, {\cal O}_{{\cal A}})=0$,  and,
hence, $\hat{\alpha}_0$ is exact and, therefore, irrelevant for the integral. From Eq.~\eqref{A6} we conclude that
\be 
\hat{\a}=\hat{\a}_1 = \frac{1}{{\bar z}_1^{n_1}} \hat{\a}'\,, 
\label{A7}
\ee
\noindent where $\hat{\a}'$ is  a $(0, 3)$-form independent of $z_1, {\bar z}_1$ and $d {\bar z}_1$.  Note that $\hat{\a} \to 0$ for large $|z_1|$. From Eq.~\eqref{A1} we
find that the boundary term in the limit $\vert z_1\vert \rightarrow \infty$ behaves as
\be 
\int_{{\mathbb C}^3 \times \gamma_1} d^4 z \wedge \frac{\hat{\a}}{p}\Big|_{|z_1| \to \infty}\,, 
\label{A8}
\ee
where $\gamma_1$ is the circle at infinity in the complex plane parameterised by $z_1$. This contour integral is zero since, generically, $p \sim z_1^{n_1+1}$ and $\hat{\a} \to 0$ for large $|z_1|$. 


\section{The co-dimension two case}

We will now repeat this discussion for a co-dimension two CICY with ambient space
\be 
{\cal A}= {\mathbb P}^{n_1} \times {\mathbb P}^{n_2} \times \dots \times {\mathbb P}^{n_m}\,, \quad 
\sum_{i=1}^m n_i = 5 \, .
\label{A9}
\ee
\noindent The CICY $X \subset \mathcal{A}$ is defined as the common zero locus of a pair of polynomials $p = (p_1, p_2)$ with multidegrees $\mathbf{q}_1 = (q_1^1, ..., q_1^m)$ and $\mathbf{q}_2 = (q_2^1, ..., q_2^m)$, satisfying the Calabi-Yau condition $q_1^i+q_2^i = n_i+1$, for all $i=1,...,m$. Introducing affine coordinates $(z_1, ... , z_5)$ on a patch in $\mathcal{A}$, the formula for the Yukawa
coupling can be written as
\be 
\l \sim \int_{{\mathbb C}^5} d^5 z \wedge \hat{\a} \wedge \bar \pt \Big( \frac{1}{p_1}\Big) \wedge \bar \pt \Big( \frac{1}{p_2}\Big)\,, 
\label{A10}
\ee
where $\hat{\a}$ is given by~\eqref{A2}. Using the results from Section~\ref{derivation}, we obtain 
\bea
&& 
\bar \pt \hat{\a} = p \hat{\b} = p_1 \hat{\b}^{1} + p_2 \hat{\b}^{2}\,, \nonumber \\
&&
\bar \pt \hat{\b}^{1}= - p_2 \hat{\eta}\,, \quad \bar \pt \hat{\b}^{2}=  p_1 \hat{\eta}\, , \label{A13}  \\
&&
\bar \pt \hat{\eta} =0\,.
\nonumber
\eea
From Eqs.~\eqref{A2}, \eqref{A13} it follows that 
\be
\hat{\b}^{a} \in \Omega^4 ({\cal A}, {\cal O}_{{\cal A}}  (-{\bf q}_{a}))\,, \quad
\hat{\eta} \in H^5 ({{\cal A}}, {\cal O}_{{\cal A}}  (-{\bf q}_{1}-{\bf q}_{2} )) = H^5 ({{\cal A}}, \Lambda^2 {\cal N}^*)\cong \mathbb{C}\,. 
\label{A14}
\ee
\noindent This means that the form $\hat{\eta}$ is unique up to a multiplicative coefficient and an exact form, both irrelevant in the present context. As in the previous subsection, we can use the results from Appendix~\ref{appendixPn} to write
down the harmonic representative
\be 
\hat{\eta} \sim
\frac{d^5 \bar z}{\kappa_1^{n_1+1} \dots \kappa_m^{n_m+1}}\,. 
\label{A15}
\ee
\noindent To compute the boundary integrals we need to study the behaviour in the limit when the modulus of one
of the affine coordinates, say $z_1$, goes to infinity. Let us assume that $z_1$ is an affine coordinate of the first
projective factor $\mathbb{P}^{n_1}$. In the large $|z_1|$ limit we obtain
\begin{eqnarray}
\hat{\eta} \sim \frac{d^5 \bar z}{z_1^{n_1+1} {\bar z}{}^{n_1+1}_1}\,, \qquad
p_1 \hat{\eta}  \sim \frac{d^5 \bar z}{z^{q_2^{1}}_1 {\bar z}^{n_1+1}_1}\,, \qquad 
p_2 \hat{\eta}  \sim \frac{d^5 \bar z}{z^{q_1^{1}}_1 {\bar z}{}^{n_1+1}_1}\,. 
\label{A16}
\end{eqnarray}
Using Eq.~\eqref{A13}, we can now obtain the behaviour of $\hat{\b}^{a}$ and $\hat{\a}$ in the limit of large $|z_1|$. Their general solution is given 
by 
\be 
\hat{\b}^{a}= \hat{\b}^{a}_0+ \hat{\b}^{a}_1 \,, \qquad \hat{\a}= \hat{\a}_0+ \hat{\a}_1\,, 
\label{A17}
\ee
where $\hat{\b}^{a}_0$, $\hat{\a}_0$ are the general solutions to the corresponding homogeneous equations and $\hat{\b}^{a}_1$, $\hat{\a}_1$
are partial solutions to the inhomogeneous equations. For a 5-dimensional ambient space of the form~\eqref{A9}, we have
$H^{3}({\cal A}, {\cal O}_{{\cal A}})=0$ and $H^{4}({\cal A}, {\cal O}_{{\cal A}} (-{\bf q}_{a}))=0$, so that $\hat{\a}_0$ and $\hat{\b}^{a}_0$ are both exact and can be discarded. Solving for $\hat{\b}^{a}_1$ and $\hat{\a}_1$ yields
\bea
&& 
\hat{\b}^{1}= \hat{\b}^{1}_1 \sim \frac{d {\bar  z}_2 \wedge \dots \wedge  d {\bar  z}_5}{z_1^{q_1^{1}}  {\bar z}_1^{n_1}}\,, \quad 
\hat{\b}^{2}= \hat{\b}^{2}_1 \sim \frac{d {\bar  z}_2 \wedge \dots \wedge  d {\bar  z}_5}{z_1^{q_2^{1}}  {\bar z}_1^{n_1}}\,, 
\nonumber \\
&&
\hat{\a}=\hat{\a}_1 =\frac{1}{{\bar z}^{n_1}_1} \hat{\a}'\,, 
\label{A19}
\eea
where $\hat{\a}'$ is a $(0, 3)$-form independent of $z_1, {\bar z}_1$ and $d {\bar z}_1$.

Now we have all the ingredients to integrate by parts in~\eqref{A10}. Doing this once leads to
\be 
\l \sim \int_{{\mathbb C}^5} d^5 z \wedge \hat{\b}^{1} \wedge {\bar \pt} \Big(\frac{1}{p_2}\Big) + 
{\rm boundary} \   {\rm terms}\,. 
\label{A20}
\ee 
We focus on the boundary terms in this expression for $|z_1|\to \infty$ and first note that
\be 
\frac{\pt}{\pt {\bar z}_1}  \Big(\frac{1}{p_1}\Big) d {\bar z}_1 \wedge {\bar \pt} \Big(\frac{1}{p_2}\Big) = 
\frac{\pt}{\pt {\bar z}_1}  \Big(\frac{1}{p_1}\Big) d {\bar z}_1 \wedge {\bar \pt}_{\hat{1}} \Big(\frac{1}{p_2}\Big)\,, 
\label{A21}
\ee
where ${\bar \pt}_{\hat{1}} $ is the Dolbeault operator with the derivative over ${\bar z}_1$ omitted. 
Then the boundary term for $|z_1|\to \infty$ turns into
\be 
\int_{{\mathbb C}^4 \times \gamma_1} d^5 z \wedge \frac{\hat{\a}}{p_1} \wedge {\bar \pt}_{\hat{1}} \Big(\frac{1}{p_2}\Big)\Big|_{|z_1|\to \infty}\,. 
\label{A22}
\ee
In the limit of large $|z_1|$, we generically have $p_1 \sim z_1^{q_1^{1}} p_1^{\prime}$, 
$p_2 \sim z_1^{q_2^{1}} p_2^{\prime}$, where $p_1^{\prime}$, $p_2^{\prime}$ are holomorphic polynomials independent of $z_1$. Inserting this into Eq.~\eqref{A22} gives
\be 
\int_{{\mathbb C}^4 \times \gamma_1} d^5 z \wedge \frac{\hat{\a}}{z^{n_1+1}_1} \frac{1}{p_1^{\prime}}
\wedge {\bar \pt}_{\hat{1}} \Big(\frac{1}{p_2^{\prime}}\Big)\Big|_{|z_1|\to \infty}\,. 
\label{A23}
\ee
This integral is indeed zero, because  $n_1 >0$ and $\hat{\a} \to 0$ at infinity. 

Finally, we need to perform the second integration by parts in the first term in Eq.~\eqref{A20}. As before,
we focus on the boundary term for $|z_1|\to \infty$, which is given by
\be 
\int_{{\mathbb C}^4 \times \gamma_1} d^5 z \wedge \frac{\hat{\b}^{1}}{p_2} \Big|_{|z_1|\to \infty}
\sim 
\int_{{\mathbb C}^4 \times \gamma_1} d^5 z  \wedge 
\frac{d {\bar  z}_2 \wedge \dots \wedge  d {\bar  z}_5}{z_1^{n_1+1}  {\bar z}_1^{n_1} p_2^{\prime}} \Big|_{|z_1|\to \infty} =0\,. 
\label{A24}
\ee
%

\addcontentsline{toc}{chapter}{Bibliography}


\begin{thebibliography}{99}
\ifx\doiref\asklfhas\newcommand{\doiref}[2]{\href{http://dx.doi.org/#1}{#2}}\fi
\raggedright 
\ifx\arxivref\asklfhas\newcommand{\arxivref}[2]{\href{http://arxiv.org/abs/#1}{arXiv:#1}}\fi
\raggedright

\bibliographystyle{plain}  


\bibitem{greenschwarzarticle} M. B. Green and J. H. Schwarz,  \textit{Anomaly cancellations in supersymmetric D=10 gauge theory and superstring theory}, \textsf{\doiref{10.1016/0370-2693(84)91565-X}{Phys. Lett. {\bf B149} (1984) 117--122}}.
\bibitem{stringquartet1} D. J. Gross, J. A. Harvey, E. Martinec and R. Rohm, \textit{Heterotic string theory: (I). The free heterotic string}, \textsf{\doiref{10.1016/0550-3213(85)90394-3}{Nucl. Phys. {\bf B256} (1985) 253--284}}.
\bibitem{stringquartet2} D. J. Gross, J. A. Harvey, E. Martinec and R. Rohm, \textit{Heterotic string theory: (II). The interacting heterotic string}, \textsf{\doiref{10.1016/0550-3213(86)90146-X}{Nucl. Phys. {\bf B267} (1986) 75--124}}.
\bibitem{stringquartet0} D. J. Gross, J. A. Harvey, E. Martinec and R. Rohm, \textit{Heterotic String}, \textsf{\doiref{10.1103/PhysRevLett.54.502}{Phys. Rev. Lett. {\bf 54} (1985) 502--505}}.

\bibitem{Candelas:1985en}
P.~Candelas, G.~T. Horowitz, A.~Strominger, and E.~Witten, \textit{Vacuum Configurations for Superstrings}, \textsf{\doiref{10.1016/0550-3213(85)90602-9}{Nucl.Phys. {\bf B258} (1985) 46--74}}

\bibitem{Strominger:1985it}
A.~Strominger and E.~Witten, \textit{New Manifolds for Superstring Compactification}, \textsf{\doiref{10.1007/BF01216094}{Commun.\ Math.\ Phys.\  {\bf 101} (1985) 341}}.

\bibitem{Witten:1985xc}
E.~Witten,  \textit{Symmetry Breaking Patterns in Superstring Models}, \textsf{\doiref{10.1016/0550-3213(85)90603-0}{Nucl.\ Phys.\ B {\bf 258} (1985) 75}}.
 
\bibitem{greene1986} B. R. Greene, K. H. Kirklin, P. J. Miron and G. G. Ross, \textit{A Superstring Inspired Standard
Model}, \textsf{\doiref{10.1016/0370-2693(86)90137-1}{Phys. Lett. {\bf B180}, 69 (1986)}}. 

\bibitem{GSW} M. B. Green, J. H. Schwarz and E. Witten, \textit{Supersting Theory. Vol. 2: Loop Amplitudes, Anomalies and Phenomenology}, \textsf{\doiref{DOI:10.1002/zamm.19880680631}{Cambridge University Press 1987}}.
\bibitem{Candelas:1987kf}
  P.~Candelas, A.~M.~Dale, C.~A.~L\"utken and R.~Schimmrigk,
  \textit{Complete Intersection Calabi-Yau Manifolds}, \textsf{\doiref{10.1016/0550-3213(88)90352-5}{Nucl.\ Phys.\ B {\bf 298} (1988) 493}}.


\bibitem{candelascicy2} P. Candelas, C. A. L\"utken and R. Schimmrigk, \textit{Complete Intersection Calabi-Yau Manifolds. 2.
Three Generation Manifolds}, \textsf{\doiref{10.1016/0550-3213(88)90173-3}{Nucl. Phys. B {\bf 306} (1988) 113}}.
\bibitem{7890} A.-M. He and P. Candelas, \textit{On the Number of Complete Intersection Calabi-Yau Manifolds}, \textsf{\doiref{10.1007/BF02097661}{Commun. Math. Phys. {\bf 135} (1990) 193--199}}.
\bibitem{Candelas:1990pi}
P.~Candelas and X.~de la Ossa,  \textit{Moduli Space of {Calabi-Yau} Manifolds}, \textsf{\doiref{10.1016/0550-3213(91)90122-E}{Nucl.\ Phys.\ B {\bf 355} (1991) 455}}.

\bibitem{Braun:2005ux}
V.~Braun, Y.-H. He, B.~A. Ovrut, and T.~Pantev, \textit{A Heterotic standard model},
\textsf{\doiref{10.1016/j.physletb.2005.05.007}{Phys.Lett. {\bf B618} (2005) 252--258}, \arxivref{hep-th/0501070}}.

\bibitem{Braun:2005bw}
V.~Braun, Y.-H. He, B.~A. Ovrut, and T.~Pantev, \textit{A Standard model from the E(8) $\times$ E(8) heterotic superstring}, \textsf{\doiref{10.1088/1126-6708/2005/06/039}{ JHEP {\bf 0506} (2005) 039}, \arxivref{hep-th/0502155}}.

\bibitem{Braun:2005nv}
V.~Braun, Y.-H. He, B.~A. Ovrut, and T.~Pantev, \textit{The Exact MSSM spectrum from string theory},
\textsf{\doiref{10.1088/1126-6708/2006/05/043}{JHEP {\bf 0605} (2006) 043}, \arxivref{hep-th/0512177}}.

\bibitem{Bouchard:2005ag}
V.~Bouchard and R.~Donagi, \textit{An SU(5) heterotic standard model},
\textsf{\doiref{10.1016/j.physletb.2005.12.042}{Phys.Lett. {\bf B633} (2006) 783--791}, \arxivref{hep-th/0512149}}.

\bibitem{Blumenhagen:2006ux} 
  R.~Blumenhagen, S.~Moster and T.~Weigand,
  \textit{Heterotic GUT and standard model vacua from simply connected Calabi-Yau manifolds},  \textsf{\doiref{10.1016/j.nuclphysb.2006.06.005}{Nucl.\ Phys.\ B {\bf 751}, 186 (2006)}, \arxivref{hep-th/0603015}}.
   
\bibitem{Blumenhagen:2006wj} 
  R.~Blumenhagen, S.~Moster, R.~Reinbacher and T.~Weigand,
  \textit{Massless Spectra of Three Generation U(N) Heterotic String Vacua},  \textsf{\doiref{10.1088/1126-6708/2007/05/041}{JHEP {\bf 0705}, 041 (2007)}, \arxivref{hep-th/0612039}}.
 
\bibitem{Anderson:2007nc}
L.~B. Anderson, Y.-H. He, and A.~Lukas, \textit{Heterotic Compactification, An  Algorithmic Approach}, \textsf{\doiref{10.1088/1126-6708/2007/07/049}{JHEP {\bf 0707} (2007) 049}, \arxivref{hep-th/0702210}}.

\bibitem{Anderson:2008uw}
L.~B. Anderson, Y.-H. He, and A.~Lukas, \textit{Monad Bundles in Heterotic String Compactifications}, \textsf{\doiref{10.1088/1126-6708/2008/07/104}{JHEP {\bf 0807} (2008) 104}, \arxivref{0805.2875}}.

\bibitem{Anderson:2009mh}
L.~B. Anderson, J.~Gray, Y.-H. He, and A.~Lukas, \textit{Exploring Positive Monad Bundles And a New Heterotic Standard Model},
\textsf{\doiref{10.1007/JHEP02(2010)054}{JHEP {\bf 1002} (2010) 054}, \arxivref{0911.1569}}.
  
\bibitem{Braun:2009qy} 
  V.~Braun, P.~Candelas and R.~Davies,
  \textit{A Three-Generation Calabi-Yau Manifold with Small Hodge Numbers},
  \textsf{\doiref{10.1002/prop.200900106}{Fortsch.\ Phys.\  {\bf 58}, 467 (2010)}, \arxivref{0910.5464}}.
    
\bibitem{Braun:2011ni}
V.~Braun, P.~Candelas, R.~Davies, and R.~Donagi, \textit{The MSSM Spectrum from (0,2)-Deformations of the Heterotic Standard Embedding},
\textsf{\doiref{10.1007/JHEP05(2012)127}{JHEP {\bf 1205} (2012) 127}, \arxivref{1112.1097}}.

 \bibitem{Anderson:2011ns}
  L.~B.~Anderson, J.~Gray, A.~Lukas and E.~Palti,
  \textit{Two Hundred Heterotic Standard Models on Smooth Calabi-Yau Threefolds}, \textsf{\doiref{10.1103/PhysRevD.84.106005}{Phys.\ Rev.\ D {\bf 84} (2011) 106005}, \arxivref{1106.4804}}.
 
 \bibitem{Anderson:2012yf}
  L.~B.~Anderson, J.~Gray, A.~Lukas and E.~Palti,
  \textit{Heterotic Line Bundle Standard Models},
  \textsf{\doiref{10.1007/JHEP06(2012)113}{JHEP {\bf 1206} (2012) 113}, \arxivref{1202.1757}}.

\bibitem{Anderson:2013xka}
L.~B.~Anderson, A.~Constantin, J.~Gray, A.~Lukas and E.~Palti,
\textit{A Comprehensive Scan for Heterotic SU(5) GUT models},
\textsf{\doiref{10.1007/JHEP01(2014)047}{JHEP {\bf 1401} (2014) 047}, \arxivref{1307.4787}}.

\bibitem{Strominger:1985ks}
A.~Strominger,  \textit{Yukawa Couplings in Superstring Compactification},
\textsf{\doiref{10.1103/PhysRevLett.55.2547}{Phys.\ Rev.\ Lett.\  {\bf 55} (1985) 2547}}.
   
\bibitem{Candelas:1987se}
P.~Candelas,  \textit{Yukawa Couplings Between (2,1) Forms},
\textsf{\doiref{10.1016/0550-3213(88)90351-3}{Nucl.\ Phys.\ B {\bf 298} (1988) 458}}.

\bibitem{greene1} B. R. Greene, K. H. Kirklin, P. J. Miron and G. G. Ross, \textit{A Three Generation Superstring Model.
1. Compactification and Discrete Symmetries}, \textsf{\doiref{10.1016/0550-3213(86)90057-X}{Nucl. Phys. B {\bf 278} (1986) 667}}.

\bibitem{greene2} B. R. Greene, K. H. Kirklin, P. J. Miron and G. G. Ross, \textit{A Three Generation Superstring Model.
2. Symmetry Breaking and the Low-Energy Theory}, \textsf{\doiref{10.1016/0550-3213(87)90662-6}{Nucl. Phys. B {\bf 292} (1987) 606}}.

\bibitem{greene3} B. R. Greene, K. H. Kirklin, P. J. Miron and G. G. Ross, \textit{27**3 Yukawa Couplings for a Three
Generation Superstring Model}, \textsf{\doiref{10.1016/0370-2693(87)91151-8}{Phys. Lett. B {\bf 192} (1987) 111}}.

\bibitem{braunheovrut} V. Braun, Y. H. He and B. A. Ovrut, \textit{Yukawa couplings in heterotic standard models},
\textsf{\doiref{10.1088/1126-6708/2006/04/019}{JHEP {\bf 0604} (2006) 019}, \arxivref{hep-th/0601204}}

\bibitem{bouchardcvetic} V. Bouchard, M. Cvetic and R. Donagi, \textit{Tri-linear couplings in an heterotic minimal
supersymmetric standard model}, \textsf{\doiref{10.1016/j.nuclphysb.2006.03.032}{Nucl. Phys. B {\bf 745} (2006) 62}, \arxivref{hep-th/0602096}}.

\bibitem{Anderson:2009ge}
  L.~B.~Anderson, J.~Gray, D.~Grayson, Y.~H.~He and A.~Lukas,
  \textit{Yukawa Couplings in Heterotic Compactification},
\textsf{\doiref{10.1007/s00220-010-1033-8}{Commun.\ Math.\ Phys.\  {\bf 297} (2010) 95}, \arxivref{0904.2186}}.

\bibitem{textures} L. B. Anderson, J. Gray, and B. Ovrut, \textit{Yukawa Textures From Heterotic Stability Walls}, \textsf{\doiref{10.1007/JHEP05(2010)086}{JHEP
{\bf 1005} (2010) 086}, \arxivref{1001.2317}}.

 \bibitem{Green:1986ck}
  P.~Green and T.~H\"ubsch,
  \textit{Calabi-Yau Manifolds As Complete Intersections In Products Of Complex Projective Spaces}, \textsf{\doiref{10.1007/BF01205673}{Commun.\ Math.\ Phys.\ {\bf 109} (1987) 99}}.
  
\bibitem{Hubsch:1992nu}
T.~H\"ubsch,  \textit{Calabi-Yau manifolds: A Bestiary for physicists},
\textsf{\doiref{10.1142/1410}{World Scientific, Singapore  (1992)}}.  

\bibitem{Buchbinder:2013dna}
  E.~I.~Buchbinder, A.~Constantin and A.~Lukas,
  \textit{The Moduli Space of Heterotic Line Bundle Models: a Case Study for the Tetra-Quadric}, \textsf{\doiref{10.1007/JHEP03(2014)025}{JHEP {\bf 1403} (2014) 025}, \arxivref{1311.1941}}.


\bibitem{Buchbinder:2014qda}
  E.~I.~Buchbinder, A.~Constantin and A.~Lukas,
  \textit{A heterotic standard model with $B - L$ symmetry and a stable proton}, \textsf{\doiref{10.1007/JHEP06(2014)100}{JHEP {\bf 1406} (2014) 100}, \arxivref{1404.2767}}.

\bibitem{Buchbinder:2014sya}
E.~I.~Buchbinder, A.~Constantin and A.~Lukas,  \textit{Non-generic Couplings in Supersymmetric Standard Models},
\textsf{\doiref{10.1016/j.physletb.2015.07.012}{Phys.\ Lett.\ B {\bf 748} (2015) 251}, \arxivref{1409.2412}}.

\bibitem{Buchbinder:2014qca}
E.~I.~Buchbinder, A.~Constantin and A.~Lukas, \textit{Heterotic QCD axion},
\textsf{\doiref{10.1103/PhysRevD.91.046010}{Phys.\ Rev.\ D {\bf 91} (2015) 4,  046010}, \arxivref{1412.8696}}. 

\bibitem{H}
D. Huybrechts, \textit{Complex Geometry: An Introduction}, \textsf{\doiref{10.1007/b137952}{Springer, Berlin (2004)}}.



\bibitem{Candelas:1987is}
P.~Candelas,  \textit{Lectures On Complex Manifolds}, Published in Trieste 1987, Proceedings, Superstrings 87, 1-88.

\bibitem{GH}
P. Griffiths and J. Harris, \textit{Principles of algebraic geometry}, \textsf{\doiref{10.1002/9781118032527}{Wiley Classics Library (2011)}}.

\bibitem{hartshorne} R. Hartshorne, \textit{Algebraic Geometry}, \textsf{\doiref{10.1007/978-1-4757-3849-0}{Graduate Texts in Mathematics, Springer, New York (1977)}}.


\bibitem{aglomoduli} L. B. Anderson, J. Gray, A. Lukas and B. Ovrut, \textit{Stabilizing the Complex Structure in Heterotic Calabi-Yau Vacua}, \textsf{\doiref{10.1007/JHEP02(2011)088}{JHEP {\bf 1102} (2011) 088}, \arxivref{1010.0255}}.
\bibitem{aglomoduli2} L. B. Anderson, J. Gray, A. Lukas and B. Ovrut, \textit{Stabilizing All Geometric Moduli in Heterotic Calabi-Yau Vacua}, \textsf{\doiref{10.1103/PhysRevD.83.106011}{Phys. Rev. {\bf D83} (2011) 106011}, \arxivref{1102.0011}}.


\bibitem{srednicki} M. Srednicki, \textit{Quantum Field Theory}, \textsf{\doiref{10.1017/CBO9780511813917}{Cambridge University Press, New York (2007)}}.
\bibitem{johnellis} J. Ellis, \textit{Limits of the Standard Model}, CERN lectures (2002), \textsf{\arxivref{hep-ph/0211168}}.
\bibitem{haaglopuszanski} R. Haag, J. $\slashed{\textrm{L}}$opusza\' nski and M. Sohnius, \textit{All Possible Generators of Supersymmetries of the $S$-Matrix}, \textsf{\doiref{10.1016/0550-3213(75)90279-5}{Nucl. Phys. B {\bf 88} (1975) 257}}.
\bibitem{colemanmandula} S. Coleman and J. Mandula, \textit{All Possible Symmetries of the $S$ Matrix}, \textsf{\doiref{10.1103/PhysRev.159.1251}{Phys. Rev. {\bf 159} (1967) 1251}}. 
\bibitem{cohen} A. G. Cohen, D. B. Kaplan and A. E. Nelson, \textit{The More Minimal Supersymmetric Standard Model}, \textsf{\doiref{10.1016/S0370-2693(96)01183-5}{Phys. Lett. {\bf B388} (1996) 588-598}, \arxivref{hep-ph/9607394}}.
\bibitem{grisaru1979} M. T. Grisaru, W. Siegel, M. Rocek, \textit{Improved Methods for Supergraphs}, \textsf{\doiref{10.1016/0550-3213(79)90344-4}{Nucl. Phys. {\bf B159} (1979) 429}}.
\bibitem{seiberg1993} N. Seiberg, \textit{Naturalness Versus Supersymmetric Non-renormalization Theorems}, 	\textsf{\doiref{10.1016/0370-2693(93)91541-T}{Phys. Lett. {\bf B318} (1993) 469--475}, \arxivref{hep-ph/9309335}}.
\bibitem{wessandbagger} J. Wess and J. Bagger, \textit{Supersymmetry and Supergravity}, \textsf{\doiref{10.2307/j.ctvzxx9rz}{Princeton Univ. Press (1992)}}.
\bibitem{djhchung} D. J. H. Chung et al., \textit{The Soft Supersymmetry-Breaking Lagrangian: Theory and Applications}, \textsf{\doiref{10.1016/j.physrep.2004.08.032}{Phys. Rept. {\bf 407} (2005) 1--203}, \arxivref{hep-ph/0312378}}.
\bibitem{georgiglashow} H. Georgi and S. Glashow, \textit{Unity of All Elementary-Particle Forces}, \textsf{\doiref{10.1103/PhysRevLett.32.438}{Phys.
Rev. Lett. {\bf 32} (1974) 438}}.
\bibitem{ksbabu} K. S. Babu et al., \textit{Baryon Number Violation}, group report for the Community Planning Study (Snowmass 2013), \textsf{\arxivref{1311.5285}}.
\bibitem{jhisano} J. Hisano, \textit{Proton Decay in the Supersymmetric Grand Unified Models}, \textsf{\arxivref{hep-ph/0004266}}.
\bibitem{wdeboer} W. de Boer, \textit{Grand Unified Theories and Supersymmetry in Particle Physics and Cosmology}, \textsf{\doiref{10.1016/0146-6410(94)90045-0}{Prog. Part. Nucl. Phys. {\bf 33} (1994) 201--302}, \arxivref{hep-ph/9402266}}.
\bibitem{dboer} D. Boer and R. Peeters, \textit{Fine-tuning and the doublet-triplet splitting problem in the minimal SU(5) GUT}, \textsf{\arxivref{1912.09369}}.
\bibitem{patisalam} J. C. Pati and A. Salam, \textit{Unified Lepton-Hadron Symmetry and a Gauge Theory of the Basic Interactions}, \textsf{\doiref{10.1103/PhysRevD.8.1240}{Phys. Rev. {\bf D8} (1973) 1240}}; 
\bibitem{sundermeyer} K. Sundermeyer, \textit{Symmetries in Fundamental Physics}, \textsf{\doiref{10.1007/978-3-319-06581-6}{Springer International Publishing, Switzerland (2014)}}.
\bibitem{pathron} P. Athron, S.F. King, D.J. Miller, S. Moretti and R. Nevzorov, \textit{Predictions of the Constrained Exceptional Supersymmetric Standard Model}, \textsf{\doiref{10.1016/j.physletb.2009.10.051}{Phys. Lett {\bf B681} (2009) 448--456}, \arxivref{0901.1192}}.
\bibitem{mohapatra} R. N. Mohapatra, \textit{Supersymmetric Grand Unification}, lectures at TASI97, \textsf{\arxivref{hep-ph/9801235}}, (1999 update) \textsf{\arxivref{hep-ph/9911272}}.
\bibitem{raby} S. Raby, \textit{Supersymmetric Grand Unified Theories}, \textsf{\doiref{10.1007/978-3-319-55255-2}{Springer International Publishing (2017)}}.
\bibitem{pwest} P. C. West, \textit{Supergravity, Brane Dynamics and String Duality}, \textsf{\arxivref{hep-th/9811101}}. 
\bibitem{kiritsislecture} E. Kiritsis, \textit{Introduction to Superstring Theory}, Leuven notes in mathematical and theoretical physics, B9 (1997), \textsf{\arxivref{hep-th/9709062}}. 
\bibitem{bbschwarz} K. Becker, M. Becker and J.H. Schwarz, \textit{String theory and M-theory}, \textsf{\doiref{10.1017/CBO9780511816086}{Cambridge University Press, New York (2007)}}.
\bibitem{fabianruehle} F. Ruehle, \textit{Exploring the Web of Heterotic String Theories using Anomalies}, PhD Thesis, University of Bonn (2013).
\bibitem{kkltref} S. Kachru, R. Kallosh, A. Linde and S. P. Trivedi, \textit{de Sitter Vacua in String Theory}, \textsf{\doiref{	10.1103/PhysRevD.68.046005}{Phys. Rev. D {\bf 68} (2003) 046005}, \arxivref{hep-th/0301240}}.
\bibitem{calabiconj1} E. Calabi, \textit{The space of K\"ahler metrics}, Proc. Int. Congr.
Math. Amsterdam {\bf 2} (1954) 206--207.
\bibitem{calabiconj2} E. Calabi, \textit{On K\"ahler manifolds with vanishing canonical class}, Algebraic geometry and topology, Symposium in honor of S. Lefschetz, Princeton Univ. Press,
Princeton (1957), 78--89.
\bibitem{yautheorem} S.-T. Yau, \textit{Calabi's conjecture and some new results in algebraic geometry}, \textsf{\doiref{10.1073/pnas.74.5.1798}{Proc. Natl. Acad. Sci. USA, {\bf 74} ({\bf 5}) (1977) 1798--1799}}.
\bibitem{duycitation1}  S. Donaldson, \textit{Anti-self-dual Yang-Mills connections over complex algebraic surfaces and stable vector bundles}, \textsf{\doiref{10.1112/plms/s3-50.1.1}{Proc. London Math. Soc. {\bf 50} (1985) 1--26}}.
\bibitem{duycitation2} K. Uhlenbeck and S.-T. Yau, \textit{On the existence of Hermitian-Yang-Mills connections in stable vector bundles}, \textsf{\doiref{10.1002/cpa.3160390714}{Comm. Pure Appl. Math. {\bf 39} (1986) 257--293}}.
\bibitem{timoweigandunitary} T. Weigand, \textit{Compactifications of the heterotic string with unitary bundles}, \textsf{\doiref{10.1002/prop.200610327}{Fortsch. Phys. {\bf 54} (2006) 963--1077}}.
\bibitem{dominicjoyce}  D. Joyce, \textit{Compact Manifolds with Special Holonomy}, Oxford University Press (2000).
\bibitem{marianagrana} M. Gra\~na, H. Triendl, \textit{String Theory
Compactifications}, \textsf{\doiref{10.1007/978-3-319-54316-1}{SpringerBriefs in Physics (2017)}}.
\bibitem{dsfreed} D. S. Freed, \textit{Geometry of Dirac Operators}, University of Chicago, lecture notes (1987).
\bibitem{benmachiche} I. Benmachiche, J. Louis and D. Martinez-Pedrera, \textit{The effective action of the heterotic string compactified on
manifolds with SU(3) structure}, \textsf{\doiref{10.1088/0264-9381/25/13/135006}{Class.Quant.Grav. {\bf 25} (2008) 135006}, \arxivref{0802.0410}}. 
\bibitem{lukas1997} A. Lukas, B.~A. Ovrut and D. Waldram, \textit{On the Four-Dimensional Effective Action of Strongly Coupled Heterotic String Theory}, \textsf{\doiref{10.1016/S0550-3213(98)00463-5}{Nucl. Phys. B {\bf 532} (1998) 43--82}, \arxivref{hep-th/9710208}}.
\bibitem{Polchinski} J. Polchinski, \textit{String theory. Vol.2. Superstring theory and beyond}, \textsf{\doiref{10.1017/CBO9780511618123}{Cambridge University Press (2001)}}.



\bibitem{delaossahardy} X. de la Ossa, E. Hardy and E. E. Svanes, \textit{The Heterotic Superpotential and Moduli}, \textsf{\doiref{10.1007/JHEP01(2016)049}{JHEP {\bf 1601} (2016) 049}, \arxivref{1509.08724}}.
\bibitem{candelasmetric} P. Candelas, X. de la Ossa and J. McOrist, \textit{A Metric for Heterotic Moduli},
\textsf{\doiref{10.1007/s00220-017-2978-7}{Commun. Math. Phys. {\bf 356} (2017) 2, 567--612}, \arxivref{1605.05256}}.
\bibitem{mcoristeffective} J. McOrist, \textit{On the Effective Field Theory of Heterotic Vacua}, \textsf{\doiref{10.1007/s11005-017-1025-0}{Lett. Math. Phys. {\bf 108} (2018) 4, 1031-1081}, \arxivref{1606.05221}}.
\bibitem{braunquotients} V. Braun, \textit{On Free Quotients of Complete Intersection Calabi-Yau Manifolds}, \textsf{\doiref{10.1007/JHEP04(2011)005}{JHEP {\bf 1104} (2011) 005}, \arxivref{1003.3235}}.



\bibitem{distler1} J. Distler and B. R. Greene, \textit{Aspects of (2,0) String Compactifications}, \textsf{\doiref{10.1016/0550-3213(88)90619-0}{Nucl. Phys. {\bf B304} (1988) 1--62}}.
\bibitem{distler2} J. Distler and S. Kachru, \textit{(0,2) Landau-Ginzburg theory}, \textsf{\doiref{10.1016/0550-3213(94)90619-X}{Nucl. Phys. {\bf B413} (1994) 213--243}, \arxivref{hep-th/9309110}}.
\bibitem{kachru} S. Kachru, \textit{Some three generation (0,2) Calabi-Yau models}, \textsf{\doiref{10.1016/0370-2693(95)00259-N}{Phys.Lett. {\bf B349} (1995) 76--82}, \arxivref{hep-th/9501131}}.
\bibitem{heleesun} Y.-H. He, S.-J. Lee, A. Lukas, and C. Sun, \textit{Heterotic Model Building: 16 Special Manifolds},
\textsf{\doiref{10.1007/JHEP06(2014)077}{JHEP {\bf 1406} (2014) 077}, \arxivref{1309.0223}}.
\bibitem{constantinmishra} A. Constantin, A. Lukas, and C. Mishra, \textit{The Family Problem: Hints from Heterotic Line Bundle Models}, \textsf{\doiref{10.1007/JHEP03(2016)173}{JHEP {\bf 1603} (2016) 173}, \arxivref{1509.02729}}.
\bibitem{braunbrodie} A. P. Braun, C. R. Brodie, and A. Lukas, \textit{Heterotic Line Bundle Models on Elliptically Fibered Calabi-Yau Three-folds}, \textsf{\doiref{10.1007/JHEP04(2018)087}{JHEP {\bf 1804} (2018) 087}, \arxivref{1706.07688}}.
\bibitem{yukunification} E. I. Buchbinder, A. Constantin, J. Gray, and A. Lukas, \textit{Yukawa Unification in Heterotic String Theory}, \textsf{\doiref{10.1103/PhysRevD.94.046005}{Phys. Rev. {\bf D94} (2016) 046005}, \arxivref{1606.04032}}.
\bibitem{donaldson1} S. K. Donaldson, \textit{Scalar curvature and projective embeddings. I}, \textsf{\doiref{10.4310/jdg/1090349449}{J. Differential Geom. {\bf 59} (2001) 479--522}}.
\bibitem{donaldson2} S. K. Donaldson, \textit{Scalar curvature and projective embeddings. II}, \textsf{\doiref{10.1093/qmath/hah044}{Q. J. Math. {\bf 56} (2005) 345--356}, \arxivref{math/0407534}}.
\bibitem{donaldson3} S. K. Donaldson, \textit{Some numerical results in complex differential geometry}, \textsf{\doiref{10.4310/PAMQ.2009.v5.n2.a2}{Pure and Applied Mathematics Quarterly {\bf 5} (2009) 571--618}, \arxivref{math/0512625}}.
\bibitem{wang} X. Wang, \textit{Canonical metrics on stable vector bundles}, \textsf{\doiref{10.4310/CAG.2005.v13.n2.a1}{Commun. Anal. Geom. {\bf 13} (2005) 253--285}}.
\bibitem{headrick1} M. Headrick and T. Wiseman, \textit{Numerical Ricci-flat metrics on K3}, \textsf{\doiref{	10.1088/0264-9381/22/23/002}{Class. Quant. Grav. {\bf 22} (2005) 4931--4960}, \arxivref{hep-th/0506129}}.
\bibitem{douglas1} M. R. Douglas, R. L. Karp, S. Lukic, and R. Reinbacher, \textit{Numerical solution to the hermitian
Yang-Mills equation on the Fermat quintic}, \textsf{\doiref{10.1088/1126-6708/2007/12/083}{JHEP {\bf 0712} (2007) 083}, \arxivref{hep-th/0606261}}.
\bibitem{headrick2} C. Doran, M. Headrick, C. P. Herzog, J. Kantor, and T. Wiseman, \textit{Numerical K\"ahler-Einstein
metric on the third del Pezzo}, \textsf{\doiref{10.1007/s00220-008-0558-6}{Commun. Math. Phys. {\bf 282} (2008) 357--393}, \arxivref{hep-th/0703057}}.
\bibitem{headrick3} M. Headrick and A. Nassar, \textit{Energy functionals for Calabi-Yau metrics}, \textsf{\doiref{10.4310/ATMP.2013.v17.n5.a1}{Adv. Theor. Math. Phys.
{\bf 17} (2013) no. 5, 867--902}, \arxivref{0908.2635}}.
\bibitem{douglas2} M. R. Douglas and S. Klevtsov, \textit{Black holes and balanced metrics}, \textsf{\arxivref{0811.0367}}.
\bibitem{numericallara1} L. B. Anderson, V. Braun, R. L. Karp, and B. A. Ovrut, \textit{Numerical Hermitian Yang-Mills
Connections and Vector Bundle Stability in Heterotic Theories}, \textsf{\doiref{10.1007/JHEP06(2010)107}{JHEP {\bf 1006} (2010) 107}, \arxivref{1004.4399}}.
\bibitem{numericallara2} L. B. Anderson, V. Braun, and B. A. Ovrut, \textit{Numerical Hermitian Yang-Mills Connections and
K\"ahler Cone Substructure}, \textsf{\doiref{10.1007/JHEP01(2012)014}{JHEP {\bf 1201} (2012) 014}, \arxivref{1103.3041}}.
\bibitem{heckmanvafa} J. J. Heckman and C. Vafa, \textit{Flavor Hierarchy From F-theory}, \textsf{\doiref{10.1016/j.nuclphysb.2010.05.009}{Nucl. Phys. {\bf B837} (2010) 137--151}, \arxivref{0811.2417}}.
\bibitem{fontibanez} A. Font and L. E. Ib\'a\~nez, \textit{Matter wave functions and Yukawa couplings in F-theory Grand
Unification}, \textsf{\doiref{10.1088/1126-6708/2009/09/036}{JHEP {\bf 0909} (2009) 036}, \arxivref{0907.4895}}.
\bibitem{conlonpalti} J. P. Conlon and E. Palti, \textit{Aspects of Flavour and Supersymmetry in F-theory GUTs}, \textsf{\doiref{10.1007/JHEP01(2010)029}{JHEP {\bf 1001} (2010) 029}, \arxivref{0910.2413}}.
\bibitem{aparicio} L. Aparicio, A. Font, L. E. Ib\'a\~nez, and F. Marchesano, \textit{Flux and Instanton Effects in Local
F-theory Models and Hierarchical Fermion Masses}, \textsf{\doiref{10.1007/JHEP08(2011)152}{JHEP {\bf 1108} (2011) 152}, \arxivref{1104.2609}}.
\bibitem{paltiwavefunctions} E. Palti, \textit{Wavefunctions and the Point of E8 in F-theory}, \textsf{\doiref{10.1007/JHEP07(2012)065}{JHEP {\bf 1207} (2012) 065}, \arxivref{1203.4490}}.
\bibitem{boundaryinflation} A. Lukas, B. A. Ovrut, and D. Waldram, \textit{Boundary inflation}, \textsf{\doiref{10.1103/PhysRevD.61.023506}{Phys. Rev. {\bf D61} (1999) 023506}, \arxivref{hep-th/9902071}}.


\bibitem{nastase} H. N$\breve{\textrm{a}}$stase, \textit{Introduction to Supergravity} (2011), \textsf{\arxivref{1112.3502}}.

\bibitem{ibanezuranga} L. E. Ib\'a\~nez and A. M. Uranga, \textit{String Theory and Particle Physics: An introduction to string phenomenology}, \textsf{\doiref{10.1017/CBO9781139018951}{Cambridge University Press (2012)}}.

\bibitem{quevedo} F. Quevedo, \textit{Superstring phenomenology: An Overview}, \textsf{\doiref{10.1016/S0920-5632(97)00650-6}{Nucl. Phys. Proc. Suppl. {\bf 62} (1998) 134--143}, \arxivref{hep-ph/9707434}}.

\bibitem{andreyscrucca} C. Andrey and C. A. Scrucca, \textit{Sequestering by global symmetries in Calabi-Yau string models}, \textsf{\doiref{10.1016/j.nuclphysb.2011.05.020}{Nucl.  Phys. {\bf B851} (2011) 245--288}, \arxivref{1104.4061}}.

\bibitem{half-flat} S. Gurrieri, A. Lukas and A. Micu, \textit{Heterotic on Half-flat}, \textsf{\doiref{10.1103/PhysRevD.70.126009}{Phys. \ Rev.\ {\bf D70} (2004) 126009}, \arxivref{hep-th/0408121}}.

\bibitem{half-flat2} S. Gurrieri, A. Lukas and A. Micu, \textit{Heterotic String Compactifications on Half-flat Manifolds II}, \textsf{\doiref{10.1088/1126-6708/2007/12/081}{JHEP {\bf 0712} (2007) 081}, \arxivref{0709.1932}}.

\bibitem{blumenhagenbook} R. Blumenhagen, D. L\"ust, S. Theisen, \textit{Basic Concepts of String Theory}, \textsf{\doiref{10.1007/978-3-642-29497-6}{Springer-Verlag, Berlin (2013)}}.




\bibitem{paper1} S.~Blesneag, E.~I.~Buchbinder, P.~Candelas and A.~Lukas, \textit{Holomorphic Yukawa Couplings in Heterotic String Theory},
  \textsf{\doiref{10.1007/JHEP01(2016)152}{JHEP {\bf 1601} (2016) 152}, \arxivref{1512.05322}}.

\bibitem{paper2} S. Blesneag, E. I. Buchbinder and A. Lukas, \textit{Holomorphic Yukawa Couplings for Complete
Intersection Calabi-Yau Manifolds}, \textsf{\doiref{	10.1007/JHEP01(2017)119}{JHEP {\bf 1701} (2017) 119}, \arxivref{1607.03461}}.

\bibitem{paper3} S. Blesneag, E. I. Buchbinder, A. Constantin, A. Lukas, E. Palti, \textit{Matter Field K\"ahler Metric in Heterotic String Theory from Localisation}, \textsf{\doiref{10.1007/JHEP04(2018)139}{JHEP {\bf 1804} (2018) 139}, \arxivref{1801.09645}}.


\end{thebibliography}
\end{document}